\documentclass{jfm}

\usepackage{afterpage}
\usepackage{longtable}
\usepackage[english]{babel}
\usepackage{dcolumn}
\usepackage{bm}
\usepackage{pifont}
\usepackage{amsfonts}
\usepackage{amsmath}
\usepackage{upmath}
\usepackage{amssymb}
\usepackage{wasysym}
\usepackage{amsbsy}
\usepackage{graphicx}
\usepackage{natbib}
\usepackage{psfrag}
\usepackage{lineno}
\usepackage{color}
\usepackage{multirow}
\usepackage{cases}
\usepackage{stackengine}
\usepackage{float}
\usepackage{blkarray}
\usepackage{cancel}
\usepackage{mathrsfs}
\usepackage{empheq}
\usepackage{tikz}
\usetikzlibrary{arrows}
\usetikzlibrary{decorations.pathmorphing}
\usepackage{pgfplots}
\usepackage{psfrag}
\usepackage{subfig}

\newcommand{\V}{\mathcal{V}}
\newcommand{\Surf}{\partial \V}

\newcommand{\vzero}{\boldsymbol{0}}

\newcommand{\ve}{\boldsymbol{e}}
\newcommand{\vX}{\boldsymbol{X}}
\newcommand{\vx}{\boldsymbol{x}}
\newcommand{\vu}{\boldsymbol{u}}

\providecommand\bnabla{\boldsymbol{\nabla}}
\providecommand\bcdot{\boldsymbol{\cdot}}
\newcommand{\boldm}[1]{\boldsymbol{#1}}
\newcommand{\der}[2]{\frac{\partial #1}{\partial #2}}
\newcommand{\Vsat}{V_{\text{sat}}}
\newcommand{\Ssat}{\Sigma_{\text{sat}}}
\newcommand{\DtNL}{\Delta t_{\text{NL}}}
\newcommand{\Spher}{\mathcal{S}}

\newcommand{\Reyn}{\text{\textit{Re}}} 		
\newcommand{\Web}{\mbox{\textit{We}}} 		
\newcommand{\Ca}{\mbox{\textit{Ca}}} 		
\newcommand{\Oh}{\mbox{\textit{Oh}}} 		
\newcommand{\Lap}{\mbox{\textit{La}}} 		
\newcommand{\Bou}{\mbox{$\mathscr{B}$}} 	
\newcommand{\Ela}{\mbox{$\beta$}} 		    

\title[Satellite formation regimes of surfactant-laden liquid threads]%
{Natural break-up and satellite formation regimes of surfactant-laden liquid threads}

\author%
[A. Mart\'inez-Calvo, J. Rivero-Rodr\'iguez, B. Scheid and A. Sevilla]%
{A. Mart\'inez-Calvo$^1$\thanks{Email address for correspondence: amcalvo@ing.uc3m.es},\ns
J. Rivero-Rodr\'iguez$^2$,\ns
B. Scheid$^2$,\ns and \\ A. Sevilla$^1$}

\affiliation{%
$^1$Grupo de Mec\'anica de Fluidos,
Departamento de Ingenier\'ia T\'ermica y de Fluidos,
Universidad Carlos III de Madrid,
Av.~Universidad 30,
28911 Legan\'es (Madrid),
Spain\\[\affilskip]
$^2$TIPs, Universit\'e Libre de Bruxelles, C.P. 165/67, Avenue F. D. Roosevelt 50, 1050 Bruxelles, Belgium}

\begin{document}
\maketitle

\begin{abstract}
We report a numerical analysis of the unforced break-up of free cylindrical threads of viscous Newtonian liquid whose interface is coated with insoluble surfactants, focusing on the formation of satellite droplets. The initial conditions are harmonic disturbances of the cylindrical shape with a small amplitude $\epsilon$, and whose wavelength is the most unstable one deduced from linear stability theory. We demonstrate that, in the limit $\epsilon \to 0$, the problem depends on two dimensionless parameters, namely the Laplace number, $\Lap=\rho\sigma_0\bar{R}/\mu^2$, and the elasticity parameter, $\Ela=E/\sigma_0$, where $\rho$, $\mu$ and $\sigma_0$ are the liquid density, viscosity and initial surface tension, respectively, $E$ is the Gibbs elasticity and $\bar{R}$ is the unperturbed thread radius. A parametric study is presented to quantify the influence of $\Lap$ and $\Ela$ on two key quantities: the satellite droplet volume and the mass of surfactant trapped at the satellite's surface just prior to pinch-off, $\Vsat$ and $\Ssat$, respectively. We identify a weak-elasticity regime, $\Ela \lesssim 0.05$, in which the satellite volume and the associated mass of surfactant obey the scaling law $\Vsat = \Ssat = 0.0042 La^{1.64}$ for $\Lap \lesssim 2$. For $La \gtrsim 10$, $V_{sat}$ and $\Sigma_{sat}$ reach a plateau of about $3 \%$ and $2.9 \%$ respectively, $V_{sat}$ being in close agreement with previous experiments of low-viscosity threads with clean interfaces. For $La<7.5$, we reveal the existence of a discontinuous transition in $\Vsat$ and $\Ssat$ at a critical elasticity, $\Ela_c (\Lap)$, with $\Ela_c \to 0.98$ for $\Lap \lesssim 0.2$, such that $\Vsat$ and $\Ssat$ abruptly increase at $\Ela=\Ela_c$ for increasing $\Ela$. The jumps experienced by both quantities reach a plateau when $\Lap \lesssim 0.2$, while they decrease monotonically as $\Lap$ increases up to $\Lap = 7.5$, where both become zero.



\end{abstract}



\section{Introduction}\label{sec:intro}

The break-up of free-surface flows has been investigated for a long time. The first quantitative studies of the instability responsible for the spontaneous break-up of cylindrical liquid threads date back to the 19th century; the correct physical description of the instability mechanism was due to~\citet{Plateau}, who deduced that a small perturbation with a wavelength larger than the circumference of the unperturbed column is unstable, finally breaking up into main drops and smaller satellite droplets in between. A few years later,~\citet{Rayleigh1, Rayleigh4} calculated the most unstable wavelength using a linear temporal stability analysis. The subject experienced a renaissance 50 years ago that has lasted to the present due to its central role in industrial and medical applications such as chemical reactors, ink-jet and three-dimensional printing, additive manufacturing, drug and protein encapsulation, and cytometry, to cite a few~\citep[the reader is referred to the reviews of][]{Bogy79, Eggers97, Christopher_Anna07, EGG08, Derby10, Anna16}.

The theoretical approach to the study of the dynamics of jet break-up was first based on the linear stability analysis of infinite liquid threads. As already mentioned, the local temporal approach was pioneered by~\citet{Rayleigh4}. About 80 years later, the local spatial and spatiotemporal problems, in which the liquid jet moves with uniform velocity $U$ with respect to the injector, were solved~\citep{Keller1973,LeibyGoldsteinAC,LeibyGoldstein}. In particular, it was demonstrated by~\citet{Keller1973} that the spatial and temporal stability analyses are equivalent if $U$ is sufficiently larger than the speed of small-amplitude capillary instability waves, $U_{\sigma}$. In the spatial setting, the latter condition means that the relative growth of the wave amplitude along one wavelength is small. Thus, in a frame of reference moving with the jet, the amplitude growth is spatially uniform to a first approximation, which explains the equivalence of the temporal and spatial approaches if $U\gg U_{\sigma}$. Since the wavelength of the unstable capillary waves is much larger than the unperturbed cylinder radius, $\bar{R}$, the scaling of $U_{\sigma}$ depends on the value of the associated Reynolds number, $\Reyn_{\sigma}=\rho U_{\sigma} \bar{R}/\mu$, where $\rho$ and $\mu$ are the liquid density and viscosity, respectively. In the limit of Euler flow, $\Reyn_{\sigma}\gg 1$, the value of $U_{\sigma}$ is given by the balance $\sigma_0/\bar{R}\sim \rho U_{\sigma}^2$, where $\sigma_0$ is the surface tension, yielding $U_{\sigma}\sim\sqrt{\sigma_0/(\rho \bar{R})}$, usually referred to as the \emph{capillary velocity}, and $U/U_{\sigma}\sim \sqrt{\Web}$, where $\Web=\rho U^2 \bar{R}/\sigma_0$ is the Weber number. Note that, in this case, $\Reyn_{\sigma}=\sqrt{\Lap}\gg 1$, where $\Lap=\rho\bar{R}\sigma_0/\mu^2$ is the Laplace number, which may also be written as $\Lap=\Oh^{-2}$ in terms of the usual Ohnesorge number, $\Oh=\mu/\sqrt{\rho\bar{R}\sigma_0}$. In the opposite limit of Stokes flow, $\Reyn_{\sigma}\ll 1$, the appropriate balance is $\sigma_0/\bar{R}\sim \mu U_{\sigma}/\bar{R}$, whence $U_{\sigma}\sim \sigma_0/\mu$, usually referred to as the visco-capillary velocity. In this limit, $U/U_{\sigma}\sim \Ca$, where $\Ca=\mu U/\sigma_0$ is the capillary number, and $\Reyn_{\sigma}=\Lap\ll 1$. Therefore, the condition that must be satisfied for the temporal and spatial approaches to be equivalent is that $\sqrt{\Web}\gg 1$ when $\sqrt{\Lap}\gg 1$, or that $\Ca\gg 1$ when $\Lap\ll 1$. It is also important to point out that the formation of a slender jet from a nozzle requires that $\Web>\Web_c\sim O(1)$ when $\sqrt{\Lap}\gg 1$, or that $\Ca>\Ca_c\sim O(1)$ when $\Lap\ll 1$, where $\Web_c$ and $\Ca_c$ are the critical Weber and capillary numbers for the transition from convective to absolute instability~\citep{LeibyGoldsteinAC,LeibyGoldstein}.

Many experimental studies have been carried out, from the first investigations of~\cite{Savart},~\cite{Magnus},~\cite{Plateau},~\cite{Rayleigh5}, and~\cite{DonnellyGlaberson}, to the highly accurate measurements of~\cite{GonzalezGarcia2009}, whose aim was to describe the mechanism of instability and to measure the growth rate of the associated waves in the linear regime. These experiments have shown an excellent agreement with the dispersion relation obtained by Rayleigh~\citep{Rayleigh1,Rayleigh4} and by~\citet{Chandrasekhar}. It is important to emphasise that, although linear stability theory cannot describe the final stages of the dynamics prior to pinch-off, it can be used to predict the break-up time $\bar{t}_b$ with small relative errors, provided that the initial amplitude of the disturbance, $\varepsilon$, satisfies $\epsilon= \varepsilon/\bar{R}\ll 1$. In the spatial setting, this fact can be used to estimate the break-up length as $U\bar{t}_b$, in close agreement with experiments~\citep{Kalaaji2003,GonzalezGarcia2009}.

However, to describe the satellite formation process, which is the main objective of the present study, a nonlinear approach is needed. In particular,~\cite{Goedde} first investigated such nonlinear effects in detail, comparing their experiments with the weakly nonlinear theory of~\cite{Yuen1968}. The satellite drop formation process was first quantified by~\cite{Rutland1970} and~\cite{Lafrance1975}, while~\cite{Chaudhary1980Part3} studied how satellite drop formation is affected by forcing the liquid jet with different harmonics, revealing the conditions needed to inhibit their formation. These efforts to control drop formation were mainly motivated by the practical need of increasing the performance of the ink-jet printing devices under development at that time. The increase in computational power finally allowed a fully nonlinear approach by means of direct numerical simulations of the axisymmetric Navier-Stokes equations. In particular~\cite{Mansour1990} and~\cite{Ashgriz1995} computed the satellite droplet radii just prior to pinch-off, finding an excellent agreement with the experiments of~\cite{Rutland1970} and~\cite{Lafrance1975}. Due to the high numerical cost of accurately solving the Navier-Stokes equations with a free boundary up to times close to the break-up singularity, several works have been devoted to develop one-dimensional approximations by expanding the flow variables as powers of the radial coordinate~\citep[see e.g.][]{lee1974drop, EggersDupont, GyC}. These models have been shown to work reasonably well in different configurations~\citep[see e.g.][]{Ambravaneswaran2004, Mariano, Martinez2018}. Of particular importance is the fact that the leading-order model allowed the unravelling of the universal self-similar structure of the local flow close to the singularity~\citep{Eggers1993,Papageorgiou1995}.

The presence of surfactant molecules at an interface induces an effective surface rheology by means of Marangoni stresses and surface viscosities~\cite[for reviews, see][]{FullerVermant2012,LangevinARFM}, leading to substantial changes in the dynamics with respect to the case of clean interfaces. Indeed, different flow configurations of technological interest are affected by surfactants, for instance liquid bridges~\citep{Liao2006}, dip coating~\citep{Scheid10,Delacotte2012,Champougny2015} or drop break-up~\citep{Roche2009, Ponce17, Kamat2018}, to cite a few. Regarding liquid threads, the effect of surfactants has been explored by means of theory~\citep{Timmermans02, MartinezSevilla2018} and numerical simulations~\citep{Campana2006, Dravid2006, Mcgough2006, Kamat2018}. In particular, the two latter works focused on the micro-thread cascade that appears close to break-up due to the presence of surfactants. These works also analyse the different scalings close to pinch-off and the evolution of the minimum radius of the thread and its axial position during the unfolding of the micro-cascade. Moreover,~\cite{Kamat2018} revealed that the mechanism responsible for the dynamical surface tension effects induced by surfactants in filament break-up is the action of Marangoni stresses rather than the lowering of surface tension. In the case of a surfactant-free liquid thread,~\cite{Ashgriz1995} already reported the axial movement of the location of minimum radius for low-viscosity filaments, $\Lap \gg 1$. More recently,~\cite{Castrejon2015} generalised this result, showing that this translation occurs for any finite value of $\Lap$, leading to the asymptotic inertial--viscous regime~\citep{Eggers1993}. Furthermore, by means of experiments and high-precision numerical simulations of the full axisymmetric Navier-Stokes equations,~\cite{Castrejon2015} demonstrated that, depending on the value of $\Lap$, the thinning of the filament experiences different transitions that delay the occurrence of the universal inertial--viscous regime. In contrast with the latter studies, which focused on a detailed description of the transitions between the different scaling laws prior to pinch-off, the present contribution aims at providing a global parametric description of the satellite formation process when the interface is coated with an insoluble surfactant monolayer.

Given the success of the leading-order one-dimensional approximation in capturing the nonlinear dynamics of clean interfaces under certain configurations and values of $La$~\citep{EggersDupont,Ambra2002,Notz2004,Subramani06,Yildirim2005,Mariano}, similar models have been derived that account for the presence of surfactants, and numerically solved for different flow configurations~\citep{Ambra1999,Craster2002,Craster2009,Xu2007}. However, as pointed out by~\citet{Timmermans02} and~\citet{MartinezSevilla2018}, a higher-order approximation is needed when the surface viscoelastic stresses are large enough. The failure of the leading-order one-dimensional models to describe the flow for large enough elastic and surface viscous stresses is due to the fact that the velocity profile is uniform in the leading-order equations, and cannot accomodate the shear induced by tangential interfacial stresses. Therefore, following the same strategy as~\cite{Mansour1990} and~\cite{Ashgriz1995} for a clean interface, and~\cite{Dravid2006} and~\cite{Mcgough2006} for a surfactant-laden interface, in the present contribution our approach is to numerically integrate the Navier-Stokes equations in a temporal setting, thereby avoiding the approximations involved in one-dimensional models. Unlike~\cite{Dravid2006}, we use a nonlinear equation of state to relate the surface tension to the surfactant concentration, derived from first principles, that leads to substantial differences calling out for a careful experimental analysis. Moreover, we perform an exhaustive parametric study, accurately quantifying the volume of the satellite droplet prior to pinch-off and the amount of surfactant trapped at its surface, as a function of the two dimensionless governing parameters, namely the elasticity parameter and the Laplace number.

The remainder of the paper is organised as follows. In~\S\ref{sec:formulation} we describe the mathematical model and the numerical method employed for the simulations. In~\S\ref{sec:results} we first validate the simulations by comparing the initial growth rate of small harmonic disturbances with the results provided by a temporal stability analysis. We then unravel the structure of the parameter plane spanned by the Laplace and elasticity numbers in terms of the satellite formation process, followed by a detailed analysis of the volume of the satellite droplets and their shape at break-up, the mass of surfactant trapped at their surface and the nonlinear correction to the linear break-up time. The detailed time evolution is studied in several representative cases to provide physical explanations of the results obtained. Conclusions are drawn in~\S\ref{sec:conclusions}. Finally, a stringent validation of our numerical technique is presented in the Appendix.


\section{Mathematical model and numerical method}\label{sec:formulation}

\begin{figure}
  \begin{center}
    \input{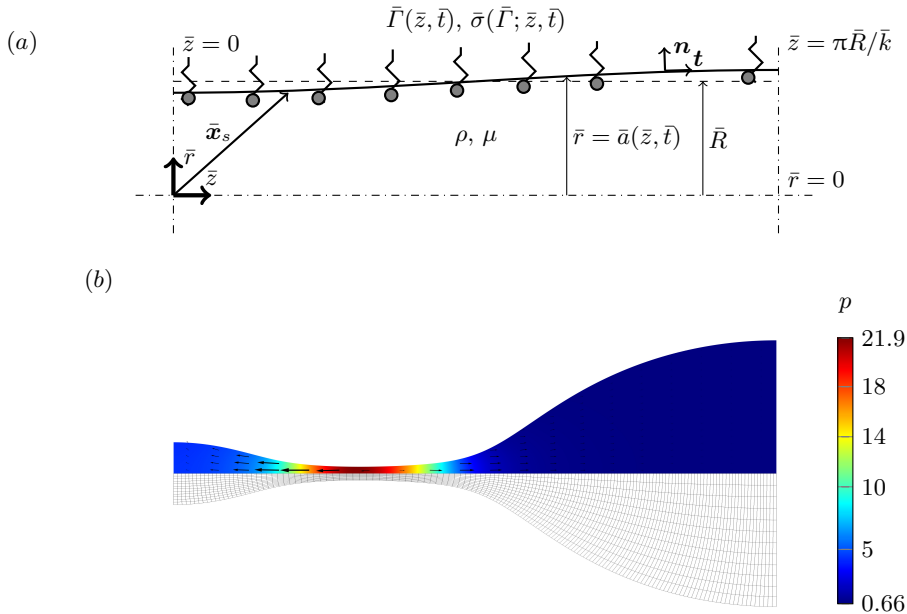}
    \caption{\label{fig:figure1} (Colour online) ($a$) Dimensional sketch of the flow configuration. ($b$) Example of a liquid thread approaching pinch-off for $\Lap = 0.01$, $\Ela = 1$, $\epsilon = 10^{-3}$ and $k = k_m = 0.516$ at time $t = 123$. The contour map represents the dimensionless pressure field $p$, and the arrows show the dimensionless velocity field $\boldm{u}$, both at the top, while the deformed mesh is shown at the bottom.}
  \end{center}
\end{figure}

We consider the axisymmetric motion of an infinitely long liquid thread of density $\rho$, viscosity $\mu$, surface tension $\bar{\sigma}$ and unperturbed radius $\bar{R}$, which occupies a volume $\V(\bar{t})$ and is embedded in a passive ambient at constant pressure $p_a$ in the absence of gravity. The interface $\Surf(\bar{t})$, placed at a radial position $\bar{r} = \bar{a}(\bar{z},\bar{t})$, is coated with a superficial concentration $\bar{\Gamma}$ of insoluble surfactant molecules (see figure~\ref{fig:figure1}$a$). Note that $\bar{r}$, $\bar{z}$ and $\bar{t}$ stand for the radial and axial coordinates and time, respectively. Henceforth an overbar bar will denote dimensional variables if not specified otherwise. The effect of surfactant adsorbed at the interface is to reduce the effective surface tension by an amount that depends on $\bar{\Gamma}$, and thus any disturbance of the interface shape generates an imbalance in $\bar{\Gamma}$ that produces a surface stress due to gradients of $\bar{\sigma}(\bar{\Gamma})$. For simplicity, in the present work we assume that surface viscous stresses can be neglected, thus disregarding the role of the surface shear and dilatational viscosities, $\mu_s$ and $\kappa_s$, respectively. The latter approximation is accurate provided that the corresponding Boussinesq numbers are small, namely $\Bou_{\mu}=\mu_s/(\mu\bar{R})\ll 1$ and $\Bou_{\kappa}=\kappa_s/(\mu\bar{R})\ll 1$~\citep{MartinezSevilla2018}. The problem is non-dimensionalised with the visco-capillary time, $\mu \bar{R}/\sigma_0$, as characteristic time and with $\bar{R}$ as characteristic length, $\sigma_0$ being the surface tension associated with the initial concentration of insoluble surfactant at the interface $\bar{\Gamma}(\bar{z},0) = \Gamma_0$, which are used to scale the surface tension and the surface concentration, respectively.

The flow is governed by the dimensionless Navier-Stokes equations
\begin{equation}\label{eq:continuity}
\bnabla \bcdot \boldm{u} = 0 \qquad \mbox{at } \V,
\end{equation}
\begin{equation}\label{eq:momentum}
\Lap \left( \frac{\partial \boldm{u}}{\partial t} + \boldm{u} \bcdot \bnabla \boldm{u} \right) = \bnabla \bcdot \mathsfbi{T}
 \qquad \mbox{at } \V,
\end{equation}
where $\boldm{u}(\vx,t) = u \, \ve_r + w \, \ve_z$ is the velocity field, and $u$, $w$, and $\ve_r$, $\ve_z$ are the radial and axial velocity components and the corresponding unit vectors, respectively. In equation~\eqref{eq:momentum}, $\mathsfbi{T} = - p \mathsfbi{I} + [\bnabla \boldm{u} + (\bnabla \boldm{u})^{\rm{T}} ]$ is the stress tensor for an incompressible Newtonian liquid, $\mathsfbi{I}$ is the standard identity tensor and $p(\vx,t)$ is the pressure field. The numerical simulations reported herein were performed using an arbitrary Lagrangian--Eulerian (ALE) method, in which the domain $\vx(\vX,t) \in \V(t)$ is parametrised by the initial position $\vX=\vx(\vX,0) \in \V(0)$, defining a time-dependent displacement field, $\boldsymbol{x}-\boldsymbol{X}$ which is enforced to satisfy the Laplace equation with proper boundary conditions specified below. The local time derivatives are evaluated in the spatial reference frame as
\begin{equation}\label{eq:bulkmaterial}
\der{\vu}{t} = \der{\hat{\vu}}{t} - \der{\vx}{t} \cdot \bnabla \vu  \qquad \mbox{at } \V,
\end{equation}
where $\hat{\vu}(\vX,t)=\vu(\vx,t)$ is the velocity in the material reference frame.

Since the interface $\Surf$ is coated with surfactant, a surface transport equation is needed for $\Gamma(\vx,t)$:
\begin{equation}\label{eq:surfactant}
\der{\Gamma}{t}
+ \bnabla_{\! s} \bcdot (\Gamma \, \boldm{u}_s) = 0  \qquad \mbox{at } \Surf,
\end{equation}
where $\boldm{u}_s = \boldm{u} (\boldm{x}_s)$ is the liquid velocity at the interface and $\boldm{x}_s$ represents any position at the surface $\vx_s(\vX_s,t) \in \partial \V(t)$, which is parametrised by its initial position $\vX_s=\vx_s(\vX_s,0) \in \partial \V(0)$. Here $\bnabla_{\! s} = \mathsfbi{I}_s \bcdot \bnabla$ is the surface gradient operator, where $\mathsfbi{I}_s = \mathsfbi{I} - \boldm{n} \boldm{n}$ is the surface projection tensor and $\boldm{n}$ is the outer unit normal vector at the surface. The local time derivatives at the interface are evaluated in the spatial reference frame as
\begin{equation}\label{eq:surfmaterial}
\der{\Gamma}{t} = \der{\hat{\Gamma}}{t} - \der{\vx_s}{t} \cdot \bnabla_s \Gamma 
 \qquad \mbox{at } \Surf,
\end{equation}
where $\hat{\Gamma}(\vX_s,t)=\Gamma(\vx_s,t)$ is the concentration of surfactant in the material frame of reference, which is needed in order to be implemented with the ALE method that is used in the present work. The reader is referred to the works of~\cite{Stone90},~\cite{Wong1996} and~\cite{Pereira2008} for further details of the time derivative of a surface quantity.

Note that the surface diffusion of surfactant has been neglected in the transport equation~\eqref{eq:surfactant}. Indeed, in the present work we only consider the limit where the surface P\'eclet number $Pe_s = U_{sc} \bar{R}/D_s \to \infty$, where $D_s$ is the surface diffusion coefficient and $U_{sc}$ is the characteristic liquid velocity at the free surface. The correct scaling for $U_{sc}$ depends on the value of $\Lap$. In the limit of dominant inertia, $\Lap\gg 1$, the appropriate velocity scale is the capillary velocity, $[\sigma_0/(\rho \bar{R}^3)]^{1/2}$, so that $Pe_s = [\sigma_0 \bar{R}/(\rho D_s^2)]^{1/2}$. For instance, if we consider a water thread of radius within the range $1$--$100$ $\mu$m, the corresponding Laplace numbers lie in the range $10^2\lesssim \Lap\lesssim 10^4$. Typical values of $D_s$ for SDS, SB12 and other monomers in aqueous solution are within the range $10^{-9}\lesssim D_s \lesssim 10^{-8}$ m$^2$ s$^{-1}$ when $\Gamma$ is below the critical micelle concentration (CMC)~\citep{Siderius2002}, providing values of the surface P\'eclet number in the range $10^4\lesssim Pe_s \lesssim 10^5$. Therefore, in configurations where $\Lap \gg 1$, it is expected that surface diffusion has a very small effect. In the opposite limit of dominant viscous forces, $\Lap\lesssim 1$, the appropriate velocity scale is the visco-capillary velocity, $\sigma_0/\mu$, leading to $Pe_s = \sigma_0 \bar{R}/(\mu D_s)$. Considering, for instance, a polydimethylsiloxane silicon oil of dynamic viscosity in the range $0.1$--$10$ Pa s, density $\rho \approx 970$ kg m$^{-3}$ and surface tension $\sigma_0 \approx 21.1$ mN m$^{-1}$, the Laplace number takes values in the range $10^{-4}\lesssim \Lap \lesssim 1$. Although we are not aware of experimental studies reporting typical values of $D_s$ in highly viscous solutions, if we assume that they are of the same order of magnitude as those of aqueous solutions, the P\'eclet number lies in the range $1\lesssim Pe_s \lesssim 10^6$. It is thereby deduced that, when $\Lap \lesssim 1$, there may be cases where surface diffusion cannot be neglected in the analysis. Therefore, although the influence of surface diffusion on the satellite droplet formation process is not addressed in the present work, it clearly deserves further study, particularly in the case of highly viscous threads.

The presence of surfactant at the interface modifies $\sigma$ by decreasing its value as $\Gamma$ increases, and thus the stress balance at the interface takes the following form in the limit $\Bou_{\mu} \ll 1$, $\Bou_{\kappa} \ll 1$~\citep{MartinezSevilla2018}:
\begin{equation}\label{eq:stressbalance}
\mathsfbi{T}  \bcdot \boldm{n} = \bnabla_{\! s} \sigma - \boldm{n} (\bnabla_{\! s} \bcdot \boldm{n}) \sigma \qquad \mbox{at } \Surf \,,
\end{equation}
where the viscous stress exerted by the ambient fluid on the interface has been neglected and the ambient pressure $p_a$ has been set to zero without loss of generality. Additionally, the kinematic condition must also hold at the interface
\begin{equation}\label{eq:kinematic}
\boldm{u}_s  \bcdot \boldm{n}   = \frac{\partial \boldm{x}_s}{\partial t} \bcdot \boldm{n}
 \qquad \mbox{at } \Surf.
\end{equation}

Finally, an equation of state that relates the surface tension, $\sigma$, to the surface concentration of surfactant, $\Gamma$, is also needed. Surface-active molecules at the interface induce a surface pressure $\bar{\Pi}$ which depends on the surfactant concentration, $\bar{\Pi} = \bar{\Pi} (\bar{\Gamma})$. The surface pressure is defined as the difference in the surface tension due to the presence of surfactant, $\bar{\Pi}(\bar{\Gamma}) = \sigma_{\text{clean}} - \bar{\sigma}(\bar{\Gamma})$, and thus $\bnabla_s \bar{\Pi} = - \bnabla_s \bar{\sigma}$. In addition, the Gibbs elasticity $E$ relates the changes of interface area, $\bar{A}$, to the surface pressure through the surface compressibility $1/E = -(1/\bar{A}) (\partial \bar{A}/\partial \bar{\Pi})_{\bar{T}}$, where $\bar{T}$ is the temperature at the interface, which is assumed to remain constant. Hence,
\begin{equation}\label{eq:eqstate1}
E = - \bar{A} \frac{\partial \bar{\Pi}}{ \partial \bar{A}} = \bar{A} \frac{\partial \bar{\sigma}}{ \partial \bar{A}}= - \bar{\Gamma} \frac{\partial \bar{\sigma}}{\partial \bar{\Gamma}},
\end{equation}
where in the last equation it has been taken into account that, in the insoluble case considered in the present work, the number of surfactant molecules is conserved at the interface. Equation~\eqref{eq:eqstate1} can be used to relate $\bar{\sigma}$ and $\bar{\Gamma}$:
\begin{equation}\label{eq:eqstate2}
\bnabla_s \bar{\sigma} = \frac{\partial \bar{\sigma}}{\partial \bar{\Gamma}} \bnabla_s \bar{\Gamma} = - \frac{E}{\bar{\Gamma}} \bnabla_s \bar{\Gamma}.
\end{equation}
Making $\bar{\sigma}$ and $\bar{\Gamma}$ dimensionless with $\sigma_0$ and $\Gamma_0$, respectively, equation~\eqref{eq:eqstate2} finally yields the dimensionless equation of state
\begin{equation}\label{eq:sigma_gamma}
\sigma = 1 - \Ela \ln \Gamma,
\end{equation}
where $\Ela = E/\sigma_0$ is the so-called elasticity parameter, also referred to as the Marangoni number~\citep{Champougny2015}. Note that, in the limit of small surface concentration variations around the initial value, $\bar{\Gamma}=\Gamma_0+\delta\bar{\Gamma}$, with $\delta\bar{\Gamma}\ll \Gamma_0$, one has $\Gamma=1+\delta\Gamma$ with $\delta\Gamma\ll 1$, and the equation of state~\eqref{eq:sigma_gamma} can be linearised to yield $\sigma = 1-\Ela \delta\Gamma$, which is equivalent to the equation of state employed by~\citet{Dravid2006}. However, it is important to emphasise that the relative variations of $\Gamma$ during the thread break-up process are not small, as demonstrated in \S\ref{sec:results}. Therefore, the use of a linearised equation of state introduces considerable errors and is not justified. At this point, the limitations of the nonlinear equation of state~\eqref{eq:sigma_gamma} should be clearly stated. Indeed, the main shortcoming of equation~\eqref{eq:sigma_gamma} is that $\sigma\to\infty$ as $\Gamma \to 0$, which eventually occurs as the surfactant is depleted from the pinch-off region due to the local advection out of the collapsing neck. Hence, a different equation of state is required to properly model the dynamics of the smallest scales close to break-up, which properly captures the saturation of $\sigma$ to the clean interface value as $\Gamma \to 0$~\citep[see][among others]{Mcgough2006,Kamat2018}. Nevertheless, for the purposes of the present contribution, the equation of state~\eqref{eq:sigma_gamma} is perfectly valid. In effect, at the smallest scales that need to be resolved to provide robust measures of $\Vsat$ and $\Ssat$, the relative variations of $\Gamma$ and $\sigma$ are small enough as to guarantee the validity of equation~\eqref{eq:sigma_gamma} in all the results reported herein.

The surface stress balance~\eqref{eq:stressbalance} can be rewritten in terms of $\Gamma$ as 
\begin{equation}
\mathsfbi{T}  \bcdot \boldm{n} =  - \frac{\Ela}{\Gamma}\, \bnabla_s \Gamma - \boldm{n} \left(\bnabla_s \bcdot \boldm{n}\right)\left(1 - \Ela \ln \Gamma\right) \qquad \mbox{at } \Surf,
\end{equation}
which, together with equations~\eqref{eq:continuity}--\eqref{eq:surfmaterial} and~\eqref{eq:kinematic}, form a closed system to determine $\boldm{u}$, $p$, $\Gamma$ and $\vx_s$.

Concerning the computational domain and the corresponding boundary conditions, in the temporal approach adopted herein we only consider half a perturbation wavelength subjected to the following symmetry conditions:
\begin{equation}\label{eq:axial_symmetry}
w = 0, \quad \frac{\partial u}{\partial z} = 0, \quad \text{and} \quad \frac{\partial \Gamma}{\partial z} = 0 \qquad \mbox{at } z = 0, \upi/k.
\end{equation}
where $k$ is the dimensionless axial wavenumber, together with the axisymmetry condition
\begin{equation}\label{eq:radial_symmetry}
\frac{\partial w}{\partial r} = 0, \quad \text{and} \quad u = 0 \qquad \mbox{at } r = 0.
\end{equation}

Finally, regarding the initial conditions imposed at $t=0$, we perturb the position of the liquid cylinder with a harmonic disturbance of amplitude $\epsilon$:
\begin{equation}\label{eq:initial_a}
\vx_s= z \ve_z + [R - \epsilon \cos (k z)] \ve_r,
\end{equation}
where $R = (1-\epsilon^2/2)^{1/2}$ is a dimensionless radius defined in terms of $\epsilon$, such that the liquid volume remains constant as $\epsilon$ varies~\citep{Ashgriz1995}. We also assume that the liquid thread is initially at rest and that the surfactant concentration is uniform
\begin{equation}\label{eq:initial_Gamma}
\vu(\vx,0)= \vzero, \quad \Gamma(\vx_s,0)= 1.
\end{equation}
Note that the assumption of a uniform initial concentration of surfactant is a good approximation, since our main results have been obtained in the limit $\epsilon \ll 1$ in which the deviations from a uniform concentration can be neglected. As explained in~\S\ref{sec:intro}, our results can also be applied to describe the spatial instability and subsequent downstream break-up of liquid jets moving with uniform velocity $U$ with respect to the injector reference frame, provided that $U\gg U_{\sigma}$, where $U_{\sigma}$ is the speed of small-amplitude capillary waves. If the latter condition is satisfied, the spatial evolution of the jet is obtained by the downstream advection of the temporal results presented herein with a uniform velocity $U$. In particular, the jet break-up length is given by $U \bar{t}_{b}$ to a first approximation.

The problem depends on four dimensionless parameters, namely the Laplace number, $\Lap$, the elasticity parameter, $\Ela$, the axial wavenumber, $k$, and the amplitude of the initial perturbation, $\epsilon$. However, in the present work we are concerned with the unforced break-up of cylindrical threads due to small-radius disturbances. Therefore, all the results were obtained by setting $k=k_m$, where $k_m(\Lap,\Ela)$ is the most unstable wavenumber (see~\S\ref{subsec:lsa}). Moreover, it will be shown that, in the small-disturbance limit, $\epsilon\ll 1$, the only result that depends on $\epsilon$ is the break-up time of the thread, $t_b(\Lap,\Ela,\epsilon)$. However, our results have revealed that the functional dependence of $t_b$ can be split into a contribution predicted by linear theory in explicit form, $t_{b,L}(\Lap,\Ela,\epsilon)$, plus a nonlinear correction, $\Delta t_{\text{NL}}(\Lap,\Ela)$, which does not depend on $\epsilon$. Consequently, only two independent dimensionless parameters appear in our formulation, namely $\Lap$ and $\Ela$.

To perform the numerical simulations, the liquid domain $0 \leq r \leq a(z,t), 0 \leq z \leq \upi/k$ is partitioned into a rectangular or triangular finite-element mesh which is dynamically deformed using the ALE method. In particular, the displacement field, $\boldsymbol{x}-\boldsymbol{X}$, is enforced to satisfy the Laplace equation, and the normal mesh velocity, $\boldsymbol{n}\boldsymbol{n} \cdot \boldm{u}$, solves the kinematic condition~\eqref{eq:kinematic}. To that end, equations~\eqref{eq:continuity}--\eqref{eq:surfactant}, together with the boundary and initial conditions~\eqref{eq:axial_symmetry}--\eqref{eq:initial_Gamma}, are written in weak form following the methodology described by~\citet{RiveroScheid2018,RiveroScheid2018b}, and the spatial discretisation is carried out using the finite-element method (FEM) provided by COMSOL, where Lagrange linear (P1) elements are used for $p$ and quadratic (P2) elements are used for $\boldsymbol{x}$, $\boldm{u}$ and $\Gamma$. The time discretisation was performed using the first-order backward Euler method with adaptive time stepping. Figure~\ref{fig:figure1}($b$) shows a representative deformed mesh for a simulation with $\Lap = 0.01$, $\Ela = 1$, $\epsilon = 10^{-3}$ and $k = k_{m} = 0.516$ at time $t = 123$, together with the pressure field as a contour plot and the velocity field represented by arrows. All the results reported were carefully checked as being mesh-independent, with an integration tolerance of the order of $10^{-6}$--$10^{-7}$. In addition, it was checked that the relative variations of liquid volume and surfactant mass where smaller than $10^{-5}$ during each simulation. The numerical code has been validated with the linear theory in~\S\ref{subsec:lsa}. In the nonlinear regime, the validation was performed by comparing our results with those of~\cite{Ashgriz1995} for a clean interface and with those of~\cite{Mcgough2006} and~\cite{Kamat2018} for a surfactant-laden thread (not shown). In particular, the Appendix is devoted to show the performance of our numerical framework close to pinch-off, comparing our results with the different theoretical scalings of the minimum radius as a function of time to break-up.

\section{Results and discussion}\label{sec:results}

Since we are interested in the spontaneous break-up of the surfactant-laden thread, all the results were computed from an initial condition where the liquid cylinder is perturbed with the wavenumber of maximum amplification, $k_m(\Lap,\Ela)$. Hence, the results of a linear stability analysis are first summarised in~\S\ref{subsec:lsa} to obtain $k_m$ and $\omega_m$, the latter being the maximum temporal growth rate. Note that $k_m$ is needed to define the initial geometry and the initial condition~\eqref{eq:initial_a}, while $\omega_m$ is used to compute the nonlinear correction to the linear break-up time, which is defined in~\S\ref{subsec:parametric}. In addition, the linear theory has also been used to validate the numerical code by comparing the associated maximum temporal growth rate, $\omega_m$, with the results extracted from the numerical simulations during the initial transient of exponential amplitude growth. Sections~\S\ref{subsec:parametric} and~\ref{subsec:nonlinear_surfactant} are devoted to the analysis of the nonlinear break-up and the satellite formation dynamics, separating the weak-elasticity limit, and the surfactant-laden case. To that end, we have performed direct numerical simulations of equations~\eqref{eq:continuity}--\eqref{eq:initial_Gamma} until times very close to pinch-off. In particular, we report a parametric study for different values of $\Lap$ and $\Ela$, computing the volume of the satellite droplet, the mass of surfactant trapped at its interface, the satellite shape at pinch-off, and the break-up time.

At this point, it has to be pointed out that a similar phenomenology was previously reported by~\cite{Dravid2006} for $\Lap = 0.01$ and $100$, although using the linearised equation of state $\sigma(\Gamma)= 1 - \Ela(\Gamma -1)$. In addition, those authors did not consider the natural break-up of the thread, since the disturbance wavenumber $k$ was restricted to fixed values different from the most amplified one, $k_m$.

\subsection{Linear stability analysis}\label{subsec:lsa}

To obtain the dispersion relation $D(\omega,k)=0$ relating the temporal growth rate $\omega$ and the axial wavenumber $k$, all the flow variables are slightly perturbed around a uniform stationary state and decomposed as temporal normal modes:
\begin{equation}\label{eq:normalmodes}
(u,w,p,a,\sigma,\Gamma) = (0,0,1,1,1,1) + \epsilon (\hat{u},\hat{v},\hat{p},\hat{a},\hat{\sigma},\hat{\Gamma}) \exp(ikz +\omega t).
\end{equation}
Introducing~\eqref{eq:normalmodes} into the system~\eqref{eq:continuity}--\eqref{eq:surfactant} and keeping terms proportional to $\epsilon$, the following dispersion relation is obtained:
\begin{align} 
& \Lap \, \omega^2 F(k) - k^2(1-k^2)+ \Ela k^2[1+F(k)(F(\tilde{k})-2)]  \nonumber & \\
& +\frac{k^4}{\Lap}\left[4 - \frac{\Ela}{\omega} \left(2 - \frac{1-k^2}{\omega} \right) \right][F(k)-F(\tilde{k})] +2 \omega k^2 (2F(k) -1) = 0, \label{eq:DR}
\end{align}
where $\tilde{k} = \sqrt{k^2 + \Lap \, \omega}$ and $F(x) = x I_0(x)/I_1(x)$. Here, $I_n(x)$ denotes the $n$th-order modified Bessel function of the first kind. Note that dispersion relation~\eqref{eq:DR} is exactly the same as the one deduced by~\citet{Timmermans02}, and is also a particular case of the one provided by~\citet{MartinezSevilla2018} in the limit of negligible surface viscosities. The Rayleigh--Chandrasekhar dispersion relation is recovered when $\Ela \to 0$~\citep{Rayleigh4,Chandrasekhar}.

\begin{figure}
\centering
\hspace{-0.95cm}
\includegraphics[height=0.45\textwidth]{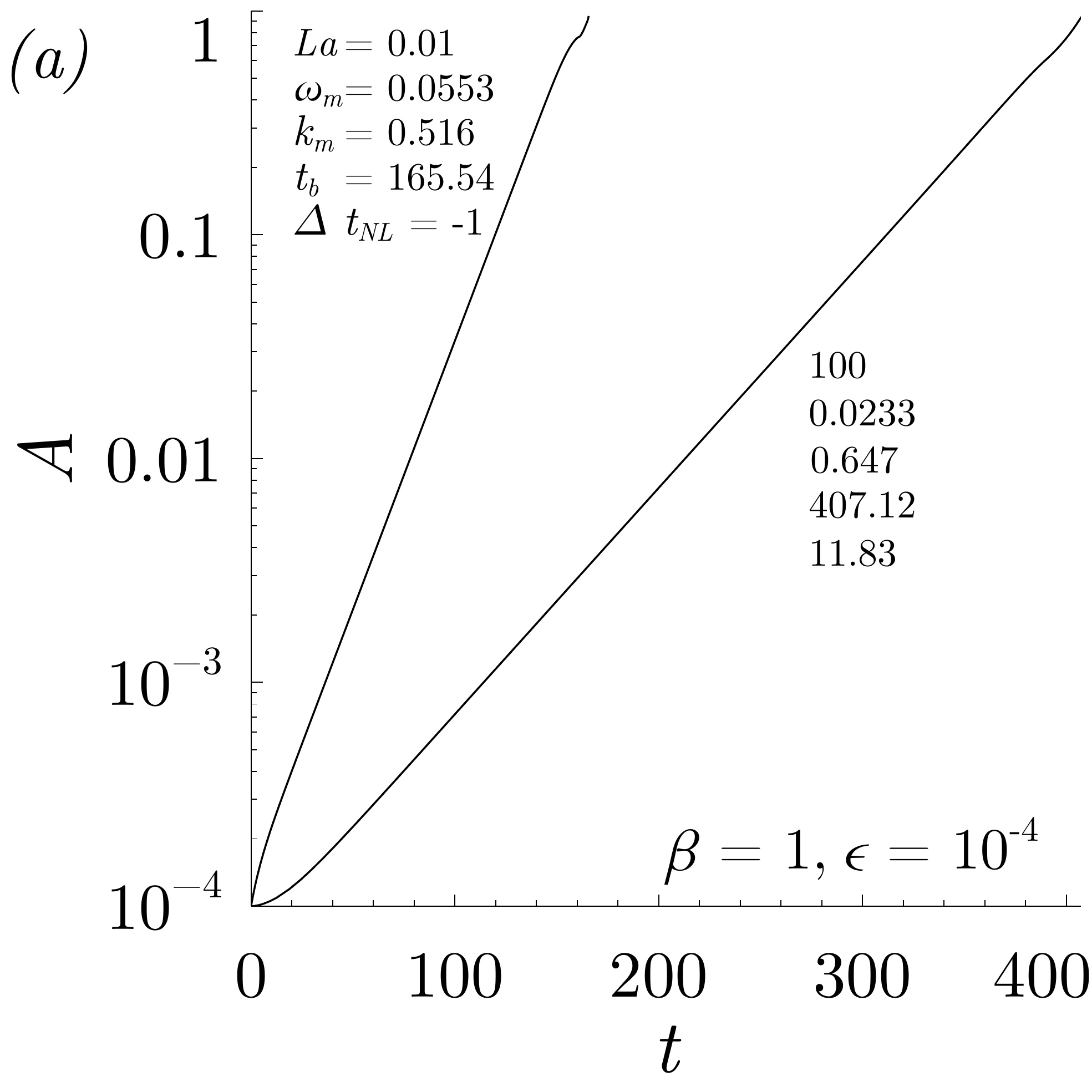} 
\hspace{0.5cm} 
\includegraphics[height=0.45\textwidth]{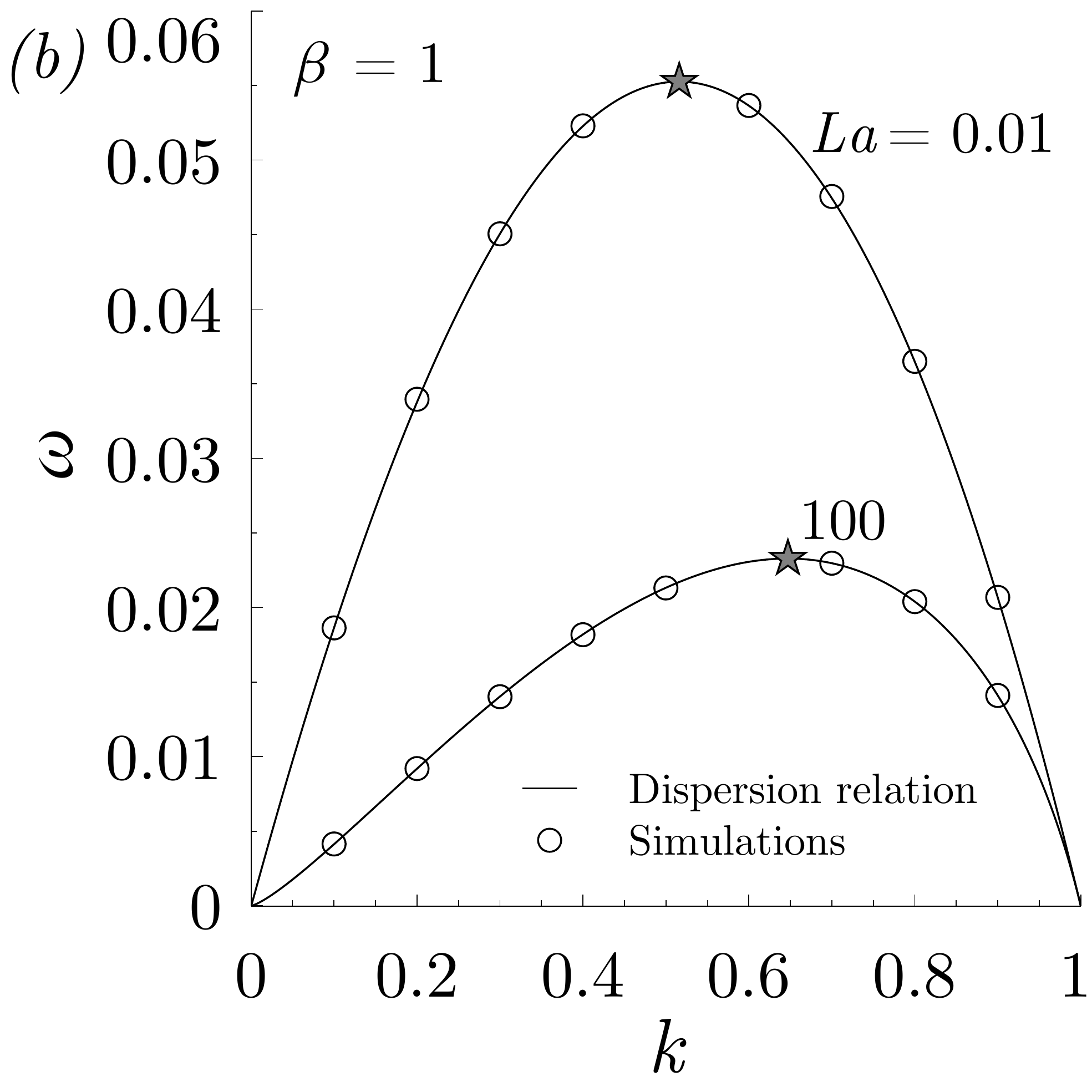}\\ 
\includegraphics[height=0.44\textwidth]{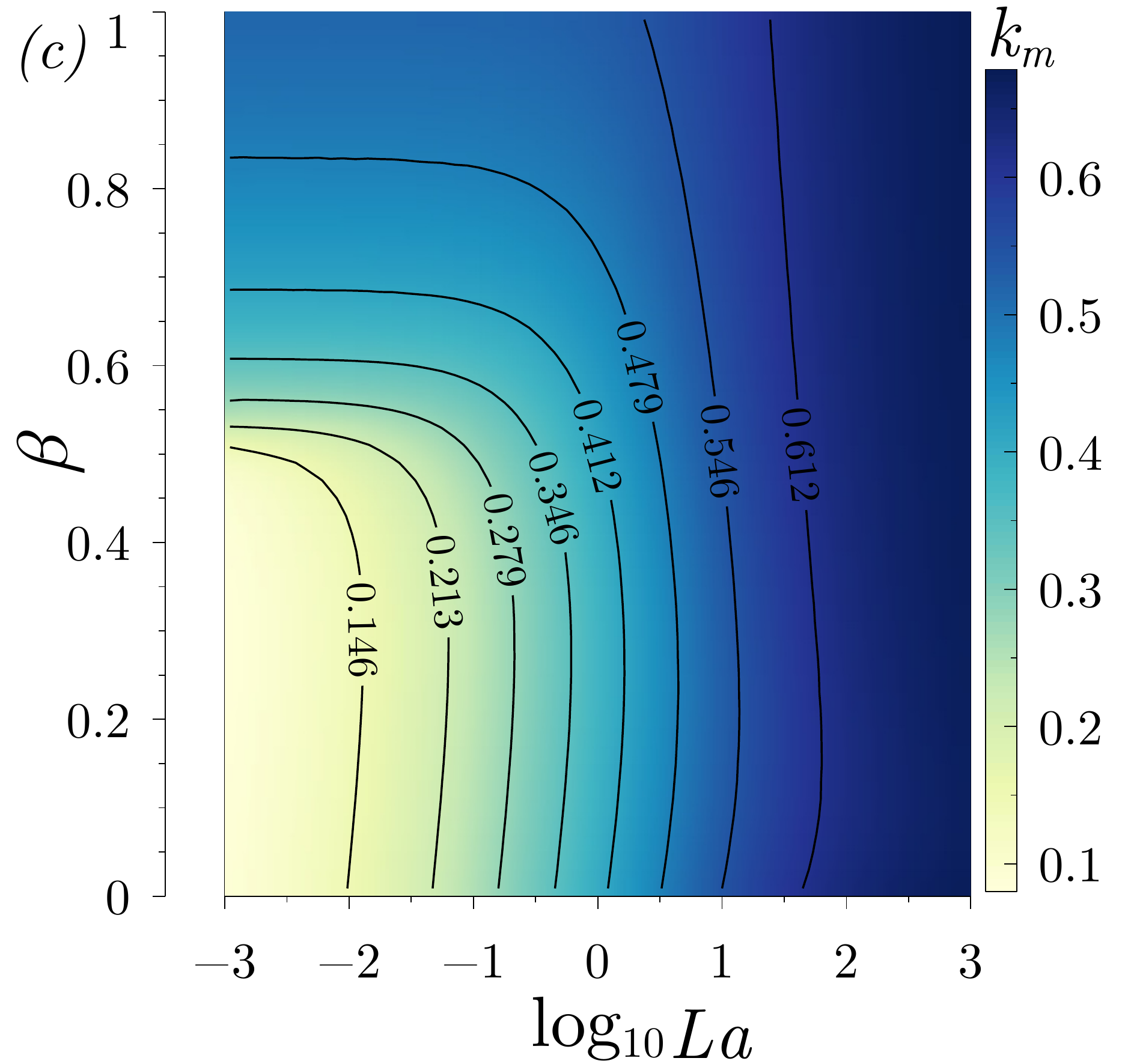}
\includegraphics[height=0.44\textwidth]{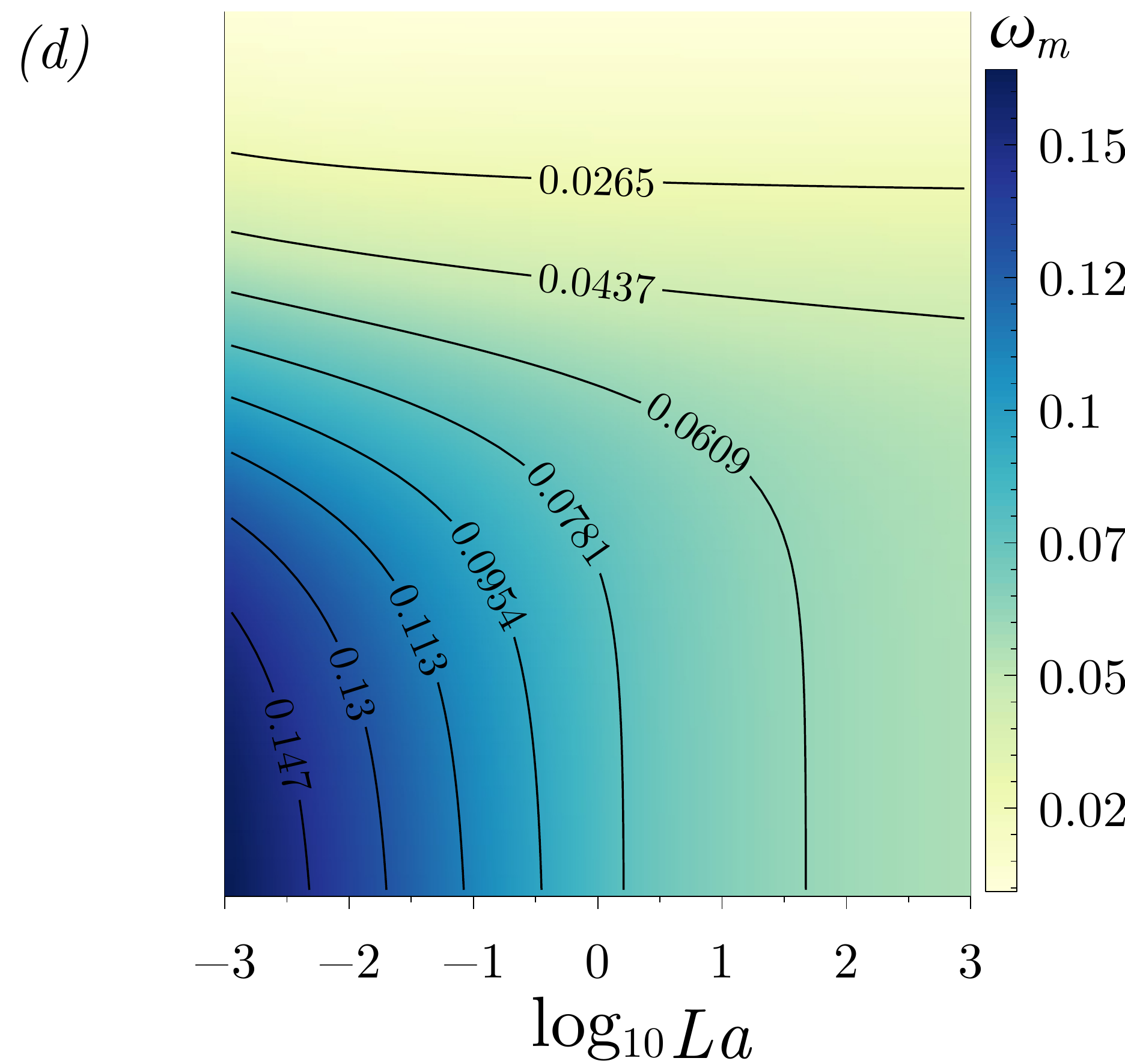}
\caption{\label{fig:figure2} (Colour online) ($a$) Semi-logarithmic plot of the radius amplitude $A(t)$ as a function of time, extracted from two numerical simulations for $\epsilon = 10^{-4}$, $\Ela = 1$ and two values of the Laplace number, namely $\Lap = (0.01, 100)$. The corresponding optimal wavenumbers, $k_m(\Lap, \Ela)$, highlighted in ($b$) with stars, are used to build the initial conditions, and their values are indicated near each curve together with the associated linear temporal growth rates, $\omega_m(\Lap, \Ela)$ and $\Lap$. ($b$) Temporal growth rate $\omega$ as a function of the axial wavenumber $k$, computed with the dispersion relation~\eqref{eq:DR} (solid lines) and with the numerical simulations (circles), for $\Ela = 1$ and two different values of $\Lap = (0.01,100)$, indicated near each curve. The maximum growth rates $\omega_m$ computed in ($a$) are marked with stars. (c) Isocontours of the most amplified wavenumber $k_m(\Lap,\Ela)$ and its corresponding growth rate $\omega_m(\Lap,\Ela)$ in ($d$).}
\end{figure}

As shown experimentally by~\cite{Goedde}, and numerically by~\cite{Mansour1990} and~\cite{Ashgriz1995}, a convenient way to compute the temporal growth rate of small disturbances is through the radius amplitude, extracted from the present simulations as $A(t) = (\max_z [a(z,t)] - \min_z [a(z,t)])/2$. Figure~\ref{fig:figure2}($a$) shows the temporal evolution of $A(t)$ in semi-logarithmic scale, extracted from two numerical simulations for an initial perturbation amplitude $\epsilon = 10^{-4}$, an elasticity parameter $\Ela = 1$, and two values of the Laplace number, $\Lap = 0.01$ and $\Lap = 100$, close to the Stokes and Euler regimes, respectively. In each case, the most amplified wavenumber, $k_m(\Lap, \Ela)$, is used to build the initial condition. As expected due to the smallness of $\epsilon$, figure~\ref{fig:figure2}($a$) shows that during most of the time the amplitude grows exponentially, i.e. $A \propto \exp(\omega_m t)$, and thus the maximum temporal growth rate, $\omega_m(\Lap,\Ela)$, can be easily computed as the slope of the linear region in the semi-logarithmic plot, $\omega_m = \mathrm{d} \ln (A) /\mathrm{d} t$. It can also be deduced from figure~\ref{fig:figure2}($a$) that there is an initial transient during which the growth of $A(t)$ is not exponential, which can be explained by the fact that the initial conditions in the numerical simulations are imposed on the shape of the interface, but disregard the associated disturbances in the velocity, pressure and surfactant concentration fields. As shown in figure~\ref{fig:figure2}($b$), this procedure was used to obtain $\omega$ for different values of $k$ (symbols), and the results are compared with the amplification curves $\omega(k)$ computed from the dispersion relation~\eqref{eq:DR} (solid lines), affording an excellent agreement that validates the numerical code in the linear regime. Finally, figures~\ref{fig:figure2}($c$, $d$) show the isocontours of $k_m$ and $\omega_m$, respectively, as a function of $\Lap$ and $\Ela$ extracted from equation~\eqref{eq:DR}, whose values will be used hereafter.


\begin{figure}
\centering
\includegraphics[height=0.7\textwidth]{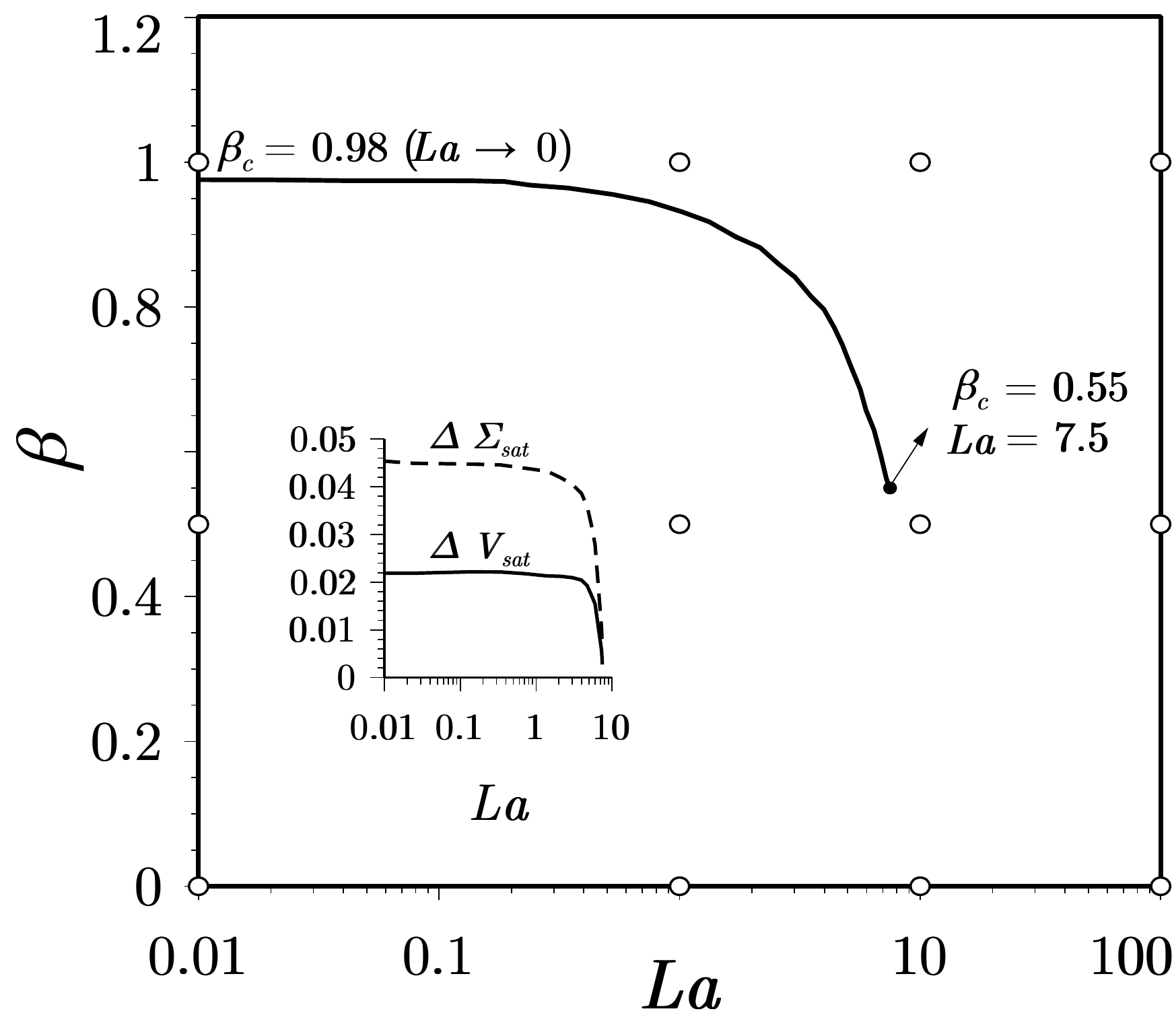}
\caption{\label{fig:figure3} The structure of the $(\Lap,\Ela)$ parameter plane. An abrupt transition takes place along the solid line, $\Ela = \Ela_c(\Lap)$, across which both the satellite volume and the entrapped mass of surfactant experience a discontinuous jump, such that both magnitudes are larger above the solid line. The inset shows the jumps in the satellite volume, $\Delta \Vsat(\Lap)=\Vsat(\Ela-\Ela_c\to 0^+ )-\Vsat(\Ela-\Ela_c\to 0^-)$, and in the associated entrapped mass of surfactant, $\Delta \Ssat$. Both jumps, together with $\Ela_c$, increase monotonically as $\Lap$ decreases, and reach respective Stokes asymptotes as $\Lap\to 0$, namely $\Delta \Vsat \to 0.022$, $\Delta \Ssat \to 0.045$, and $\beta_c \to 0.98$. The filled circle indicates the origin of the discontinuous transition, $(\Lap,\Ela_c)=(7.5,0.55)$, at which both jumps become zero. For $\Lap > 7.5$, the satellite volume is a continuous function of $\Ela$. The open circles correspond to the values of $\Lap$ and $\Ela$ of the shapes just before pinch-off shown in figure~\ref{fig:figure4}.}
\end{figure}

\begin{figure}
\hspace{-8mm}
\begin{tikzpicture}[scale=0.7, every node/.style={scale=0.7}]
	\draw[->,line width=0.25mm] (-7,-1.75) -- coordinate (x axis mid) (8,-1.75);
    \draw[->,line width=0.25mm] (-10,-1.5) -- coordinate (y axis mid) (-10,13);
    	\draw[line width=0.25mm] (-7,-1.75) -- (-7,-1.85)
    	node[anchor=north] {\Large 0};
    	     				
			\draw[line width=0.25mm] (-0.68,-1.75) -- (-0.68,-1.85)
			node[anchor=north] {\Large 0.5};
			
		    \draw[line width=0.25mm] (5.64,-1.75) -- (5.64,-1.85)
			node[anchor=north] {\Large 1};
    
    
     		\draw[line width=0.25mm] (-10,-1.5) -- (-10.1,-1.5) 
     			node[anchor=east] {};     
    
     		\draw[line width=0.25mm] (-10,0) -- (-10.1,0) 
     			node[anchor=east] {\Large 0.01}; 
     			
            \draw[line width=0.25mm] (-10,3.65) -- (-10.1,3.65) 
     			node[anchor=east] {\Large 1}; 
     			
            \draw[line width=0.25mm] (-10,7.3) -- (-10.1,7.3) 
     			node[anchor=east] {\Large 10}; 
     			
     	     \draw[line width=0.25mm] (-10,10.94999) -- (-10.1,10.94999) 
     			node[anchor=east] {\Large 100};
	\node[below=0.8cm] at (x axis mid) {\huge $\Ela$};
	\node[rotate=90, above=0.8cm] at (y axis mid) {\huge $\Lap$};
  
 \node at (-7,-0.1) {
        \includegraphics[width=58mm]{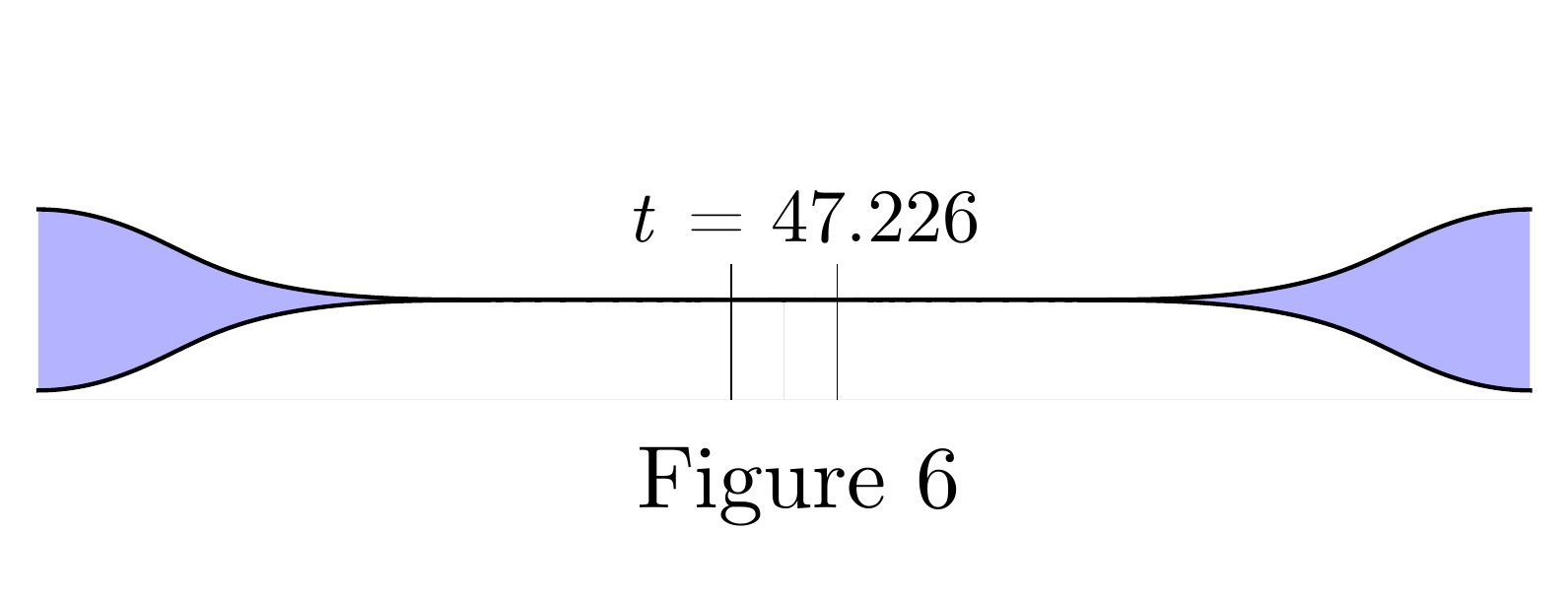}
    };
    
      \node at (-7,3.5) {
        \includegraphics[width=58mm]{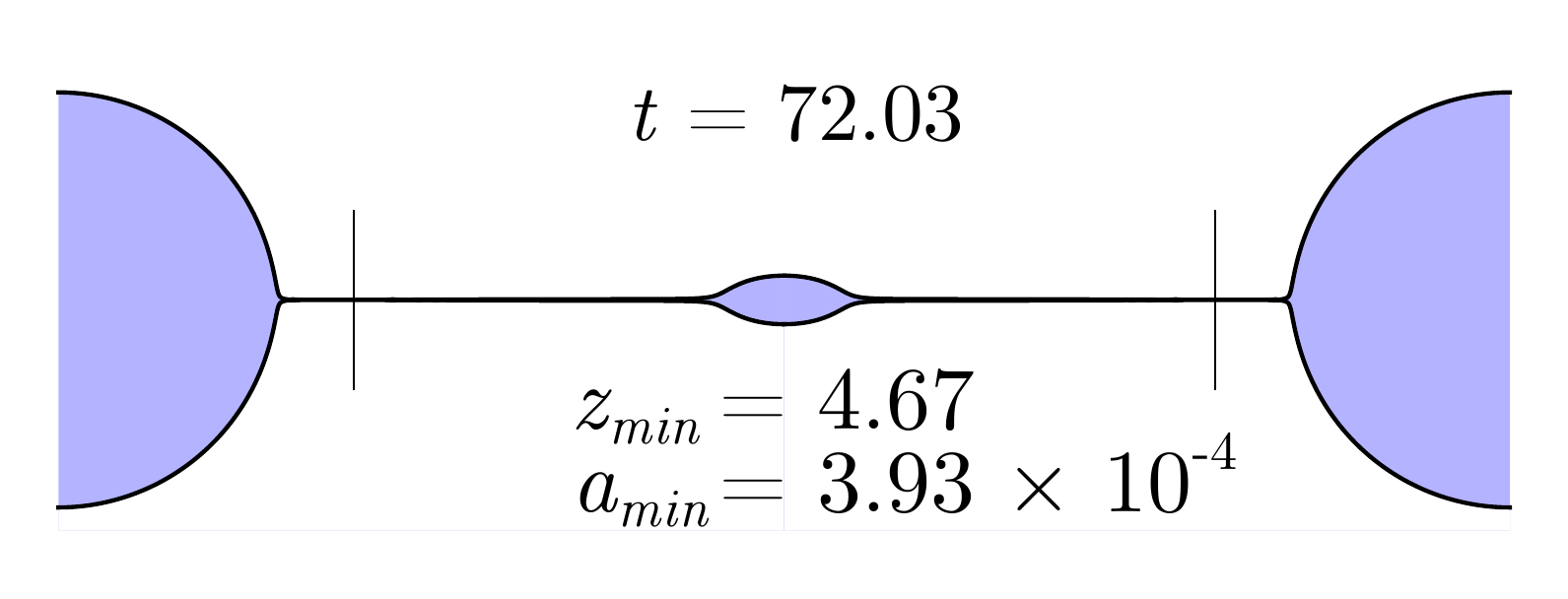}
    };
    
          \node at (-7,7.1) {
        \includegraphics[width=58mm]{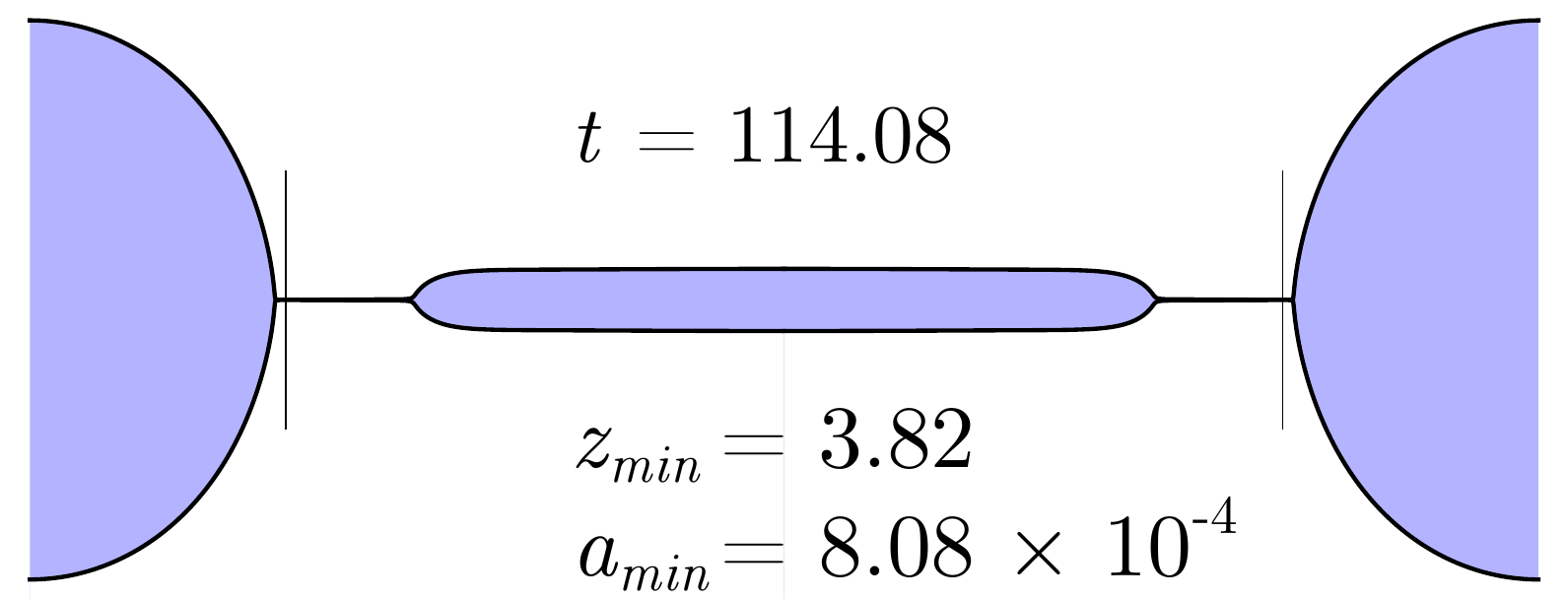}
    };
    
   \node at (-7,10.75) {
        \includegraphics[width=58mm]{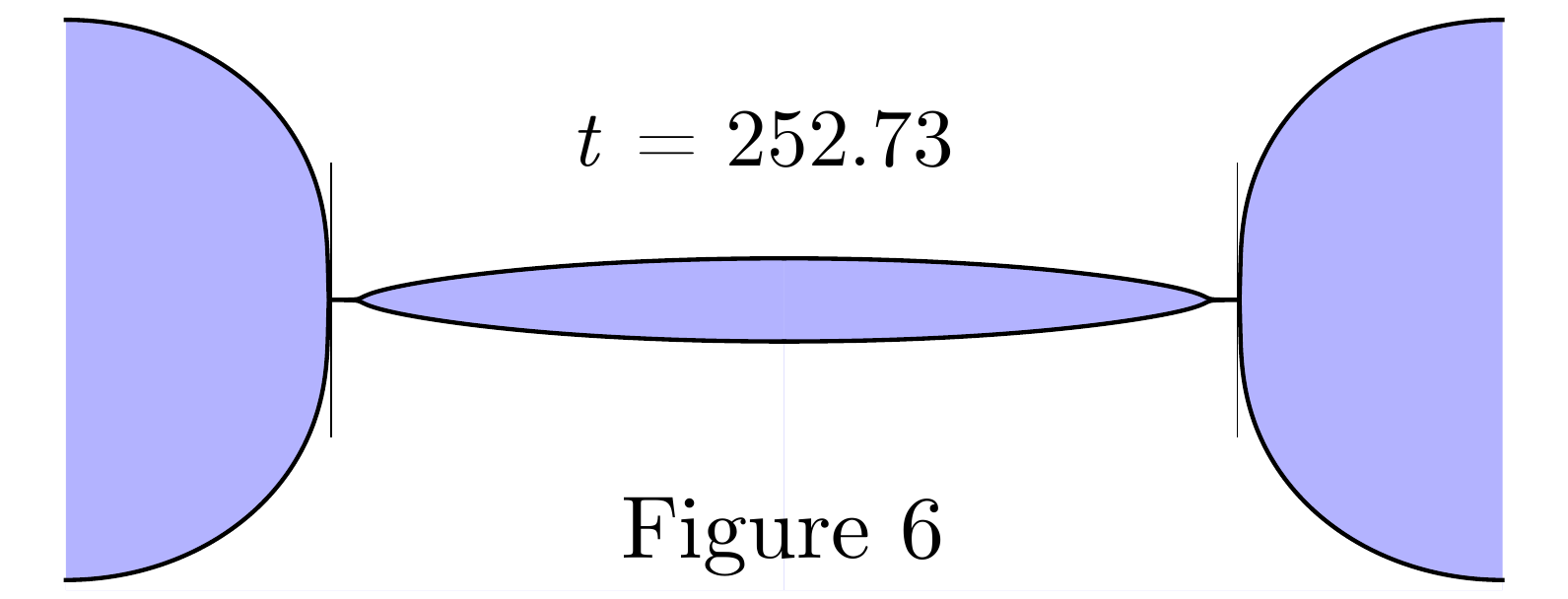}
    };
    
          \node at (-1,-0.1) {
        \includegraphics[width=58mm]{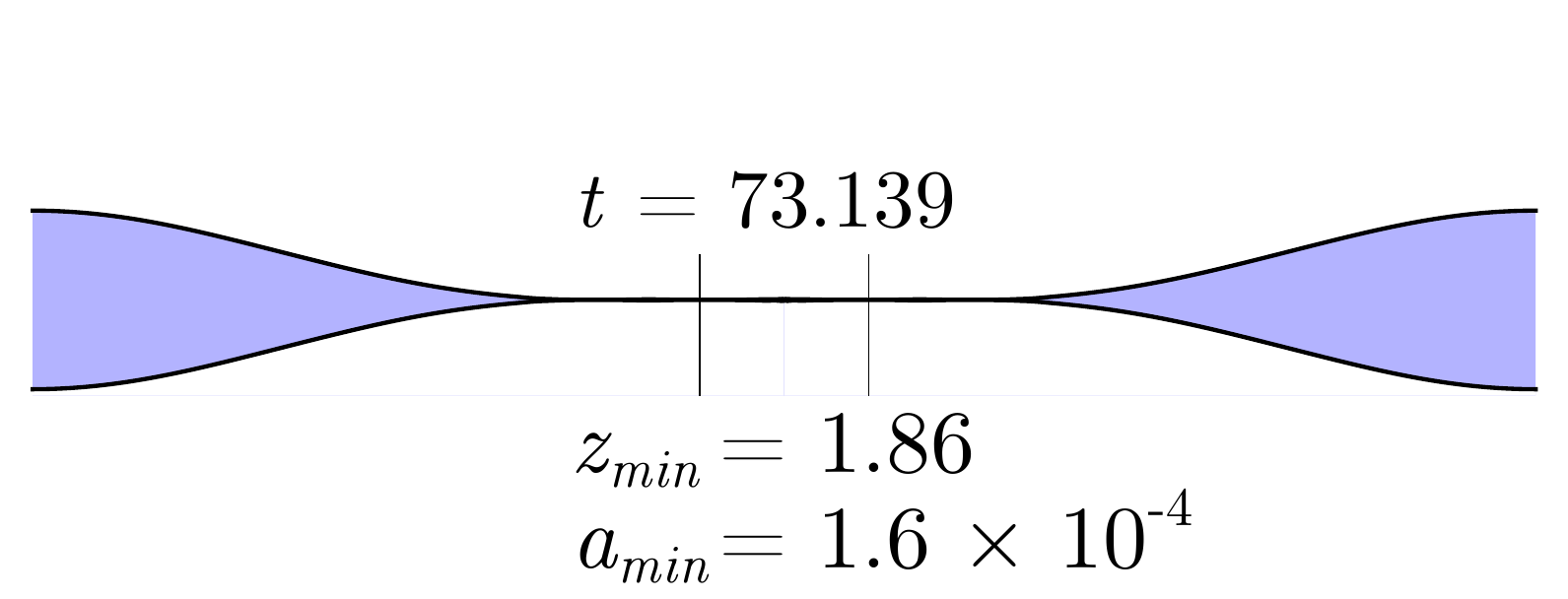}
    };

              \node at (-1,3.5) {
        \includegraphics[width=58mm]{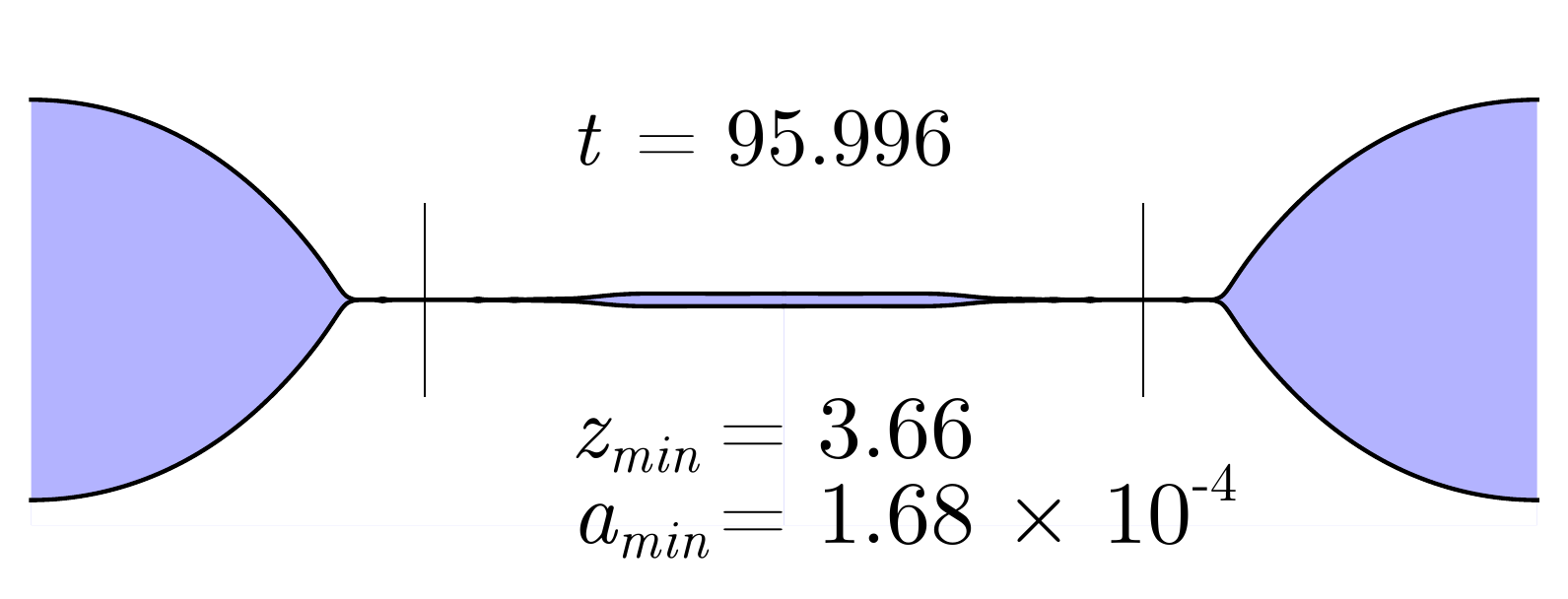}
    };
    
                  \node at (-1,7.1) {
        \includegraphics[width=58mm]{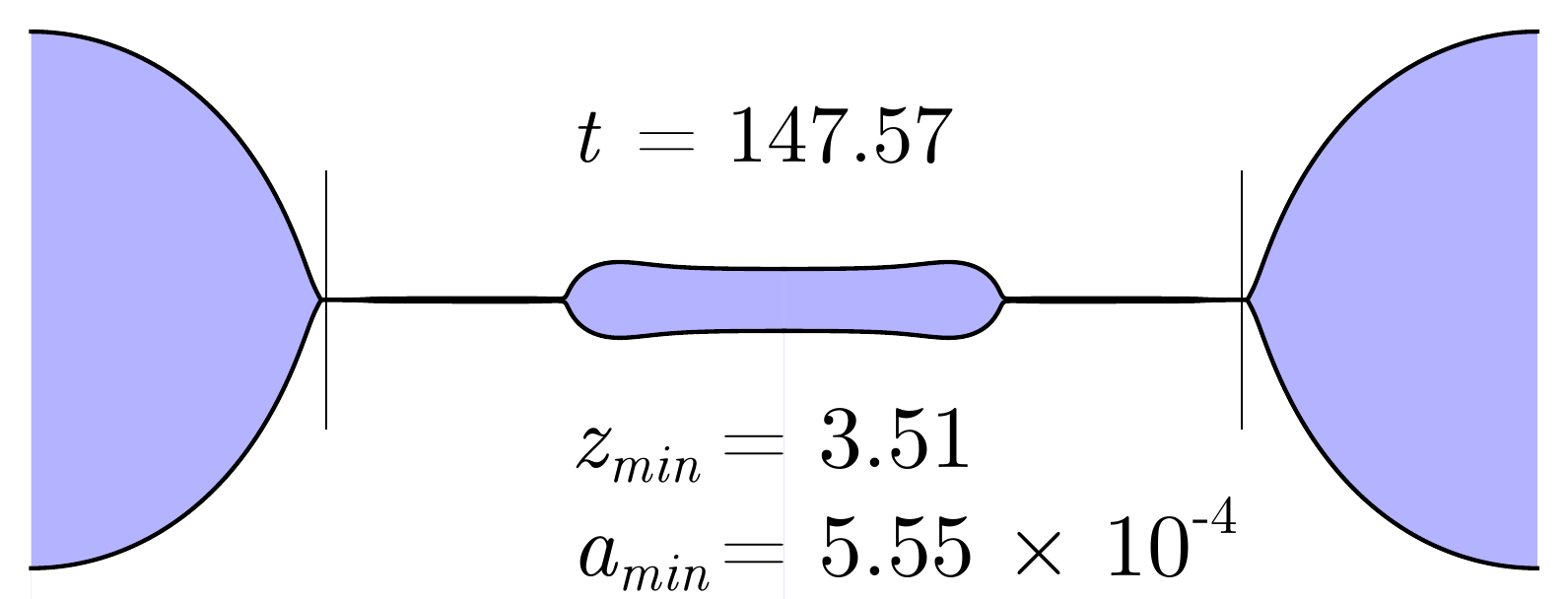}
    };

                \node at (-1,10.75) {
        \includegraphics[width=58mm]{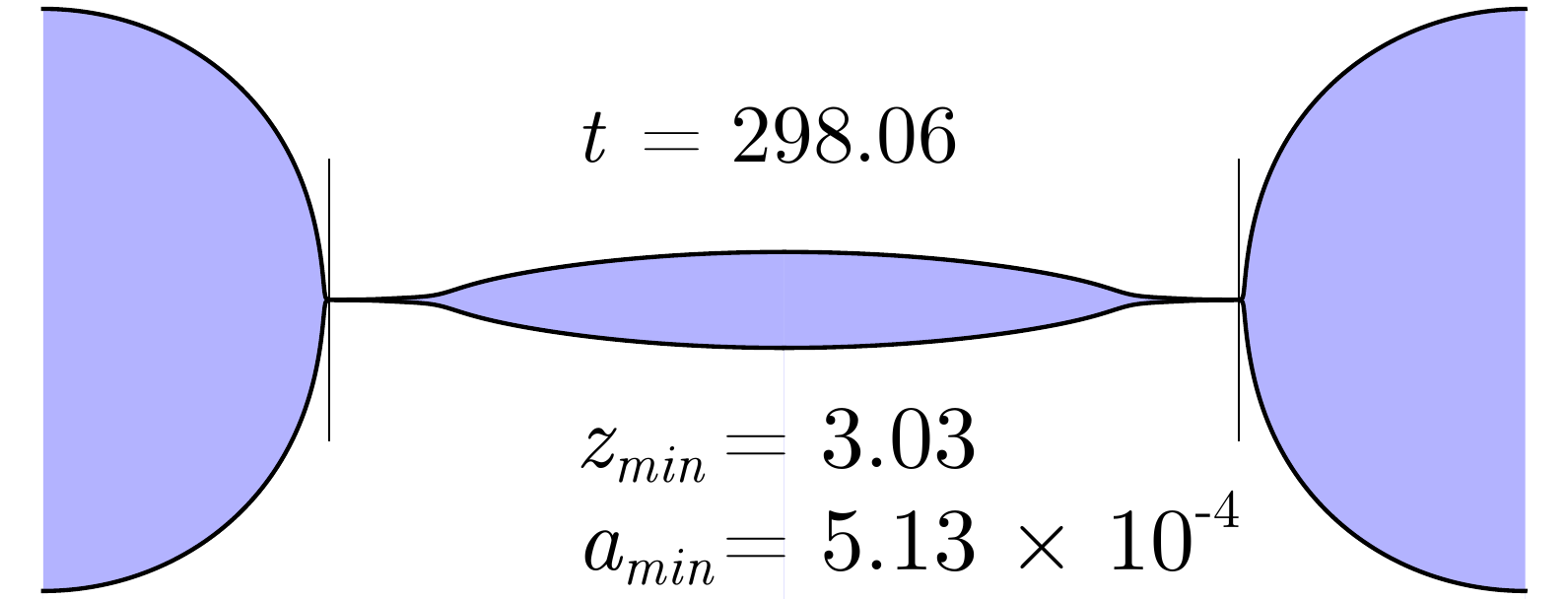}
    };
    
      \node at (5.33,-0.1) {
        \includegraphics[width=58mm]{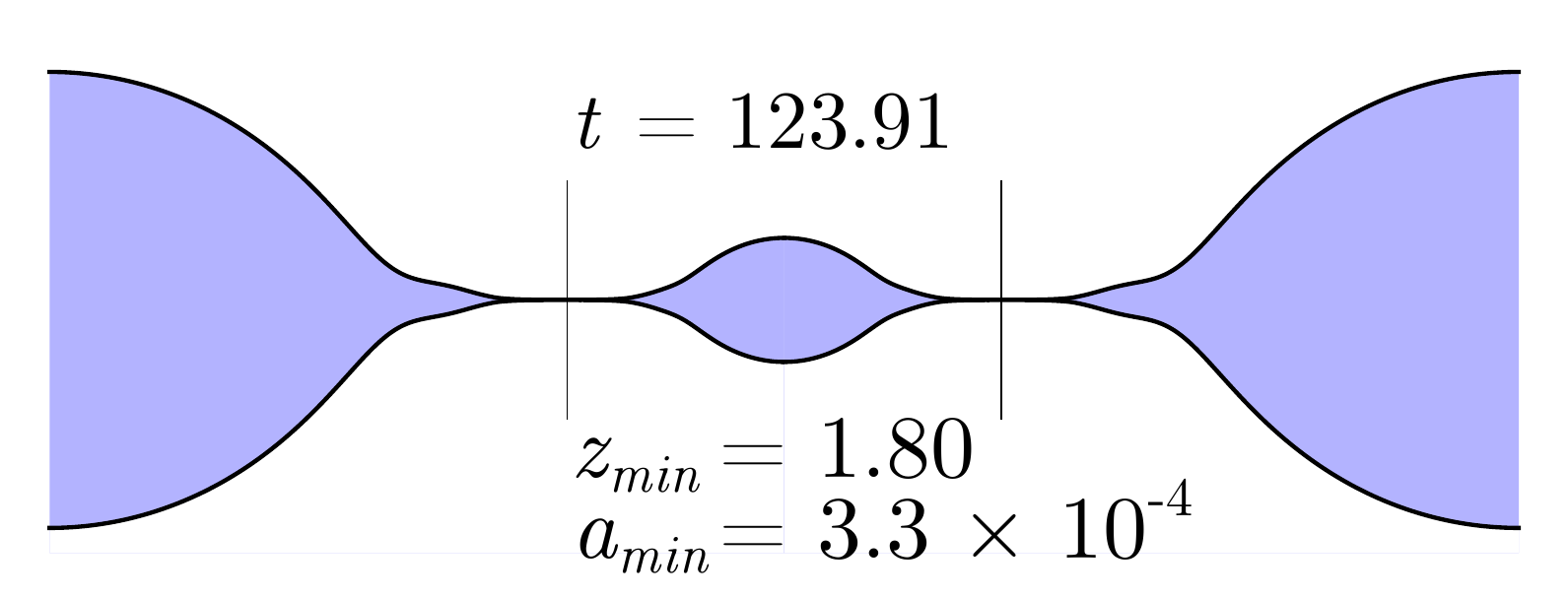}
    };
    
    \node at (5.33,3.5) {
        \includegraphics[width=58mm]{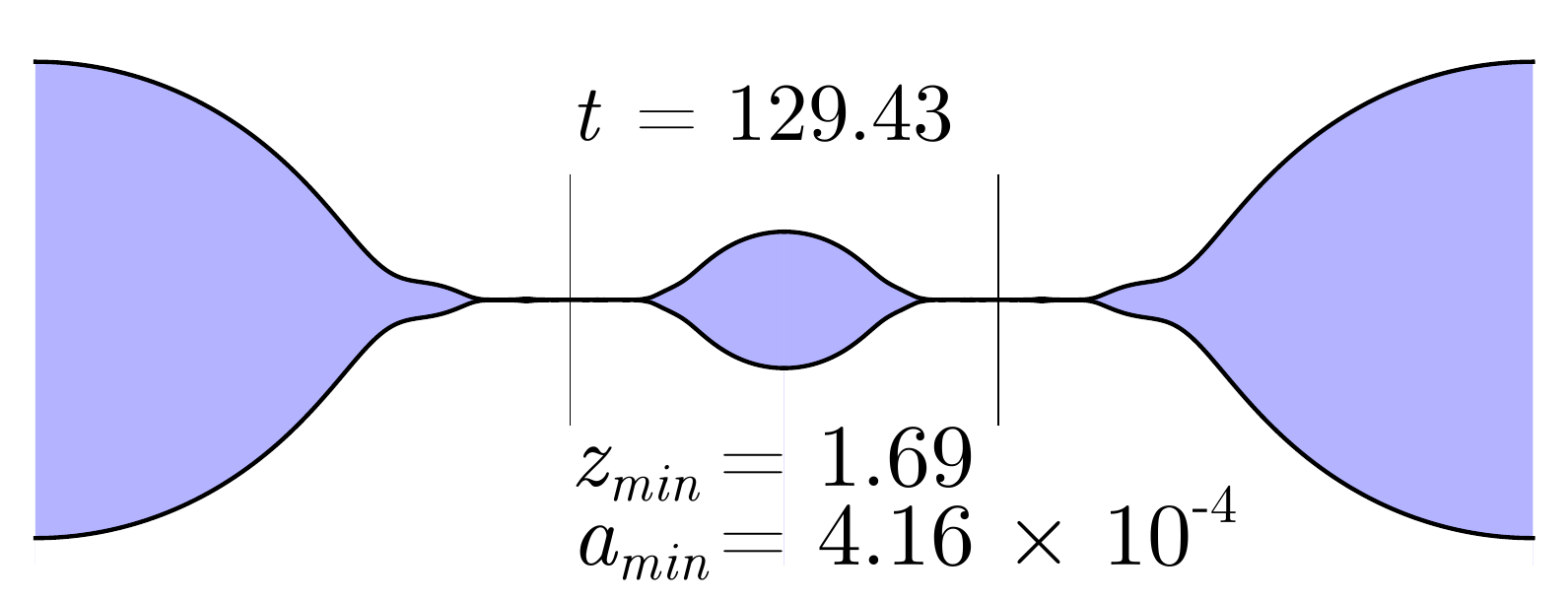}
    }; 
    
        \node at (5.33,7.1) {
        \includegraphics[width=58mm]{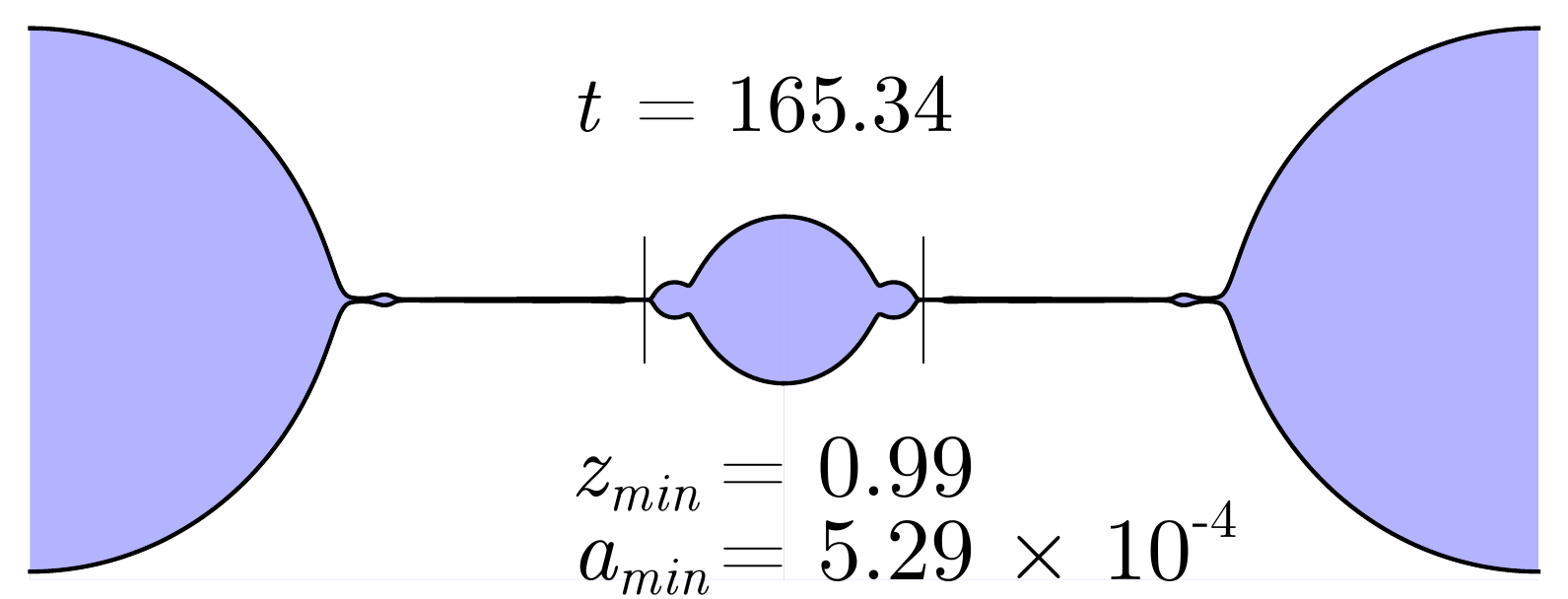}
    };     
    
        \node at (5.33,10.75) {
        \includegraphics[width=58mm]{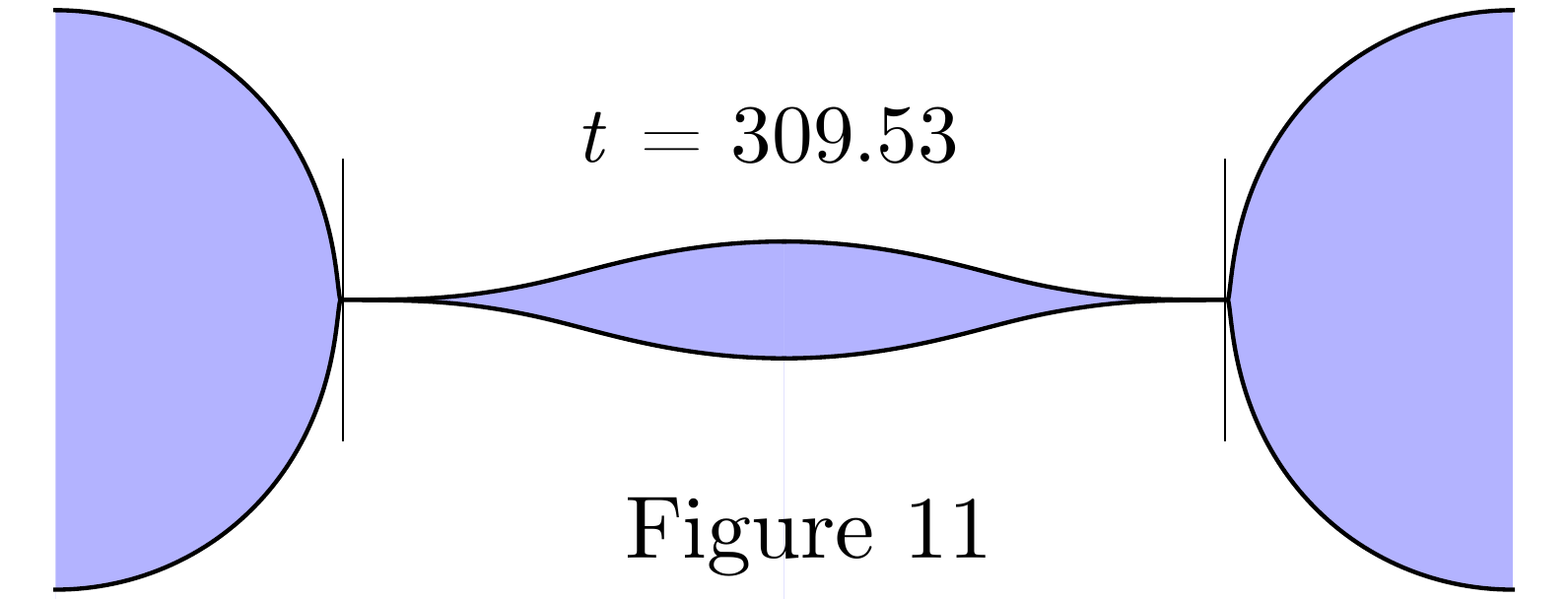}
    };
    
\end{tikzpicture}
\caption{\label{fig:figure4} (Colour online) The satellite shapes just prior to pinch-off in the $(\Lap,\Ela)$ parameter plane (see open circles in figure~\ref{fig:figure3}). The vertical lines indicate the axial positions, $z_{\min}$, of the minimum thread radii, $a_{\min}$.}
\end{figure}

\subsection{Satellite formation regimes and transitions in the $(\Lap,\Ela)$ parameter plane}\label{subsec:parametric}

\begin{figure}
\includegraphics[width=0.55\textwidth]{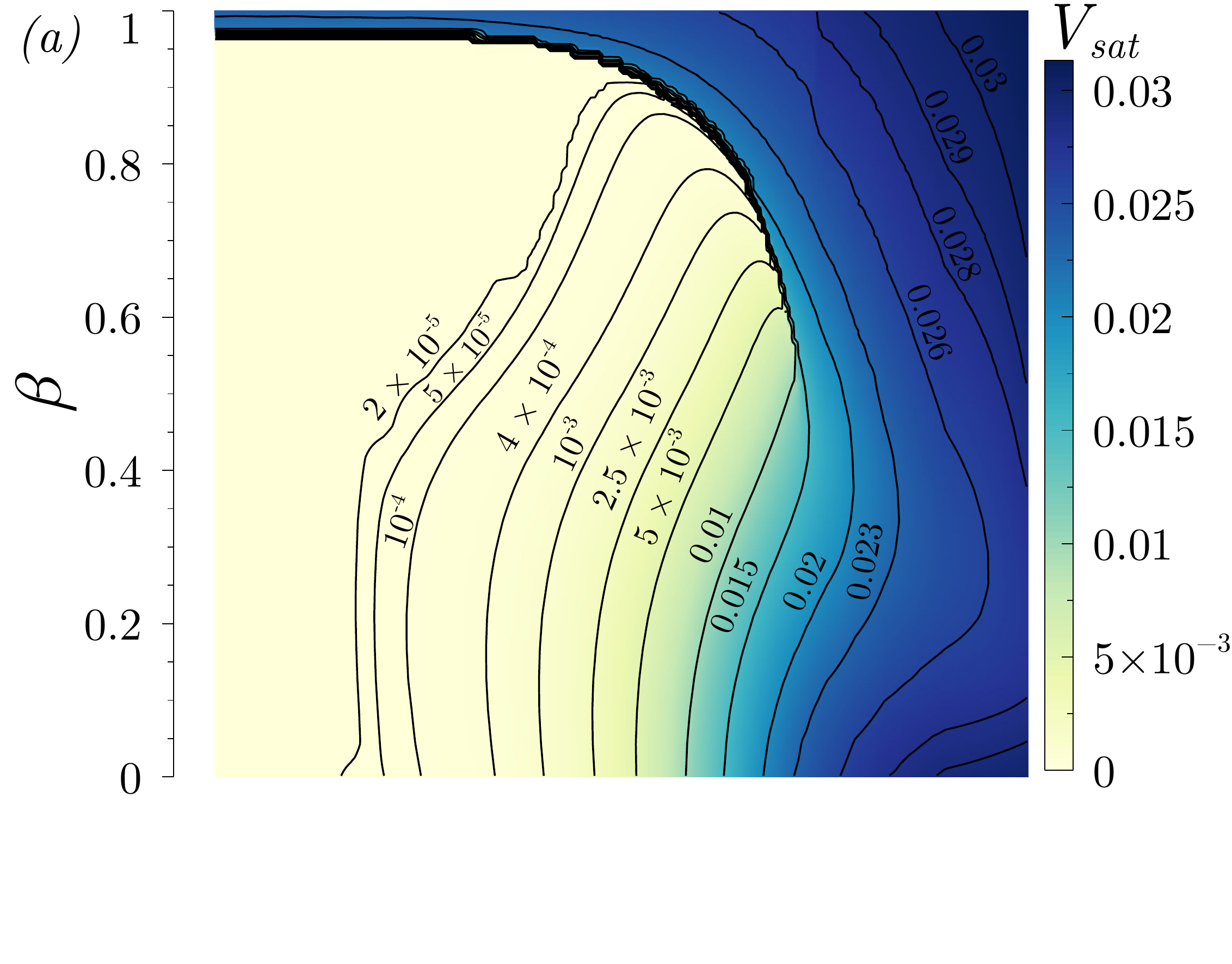} \hspace{-0.9cm}
\includegraphics[width=0.55\textwidth]{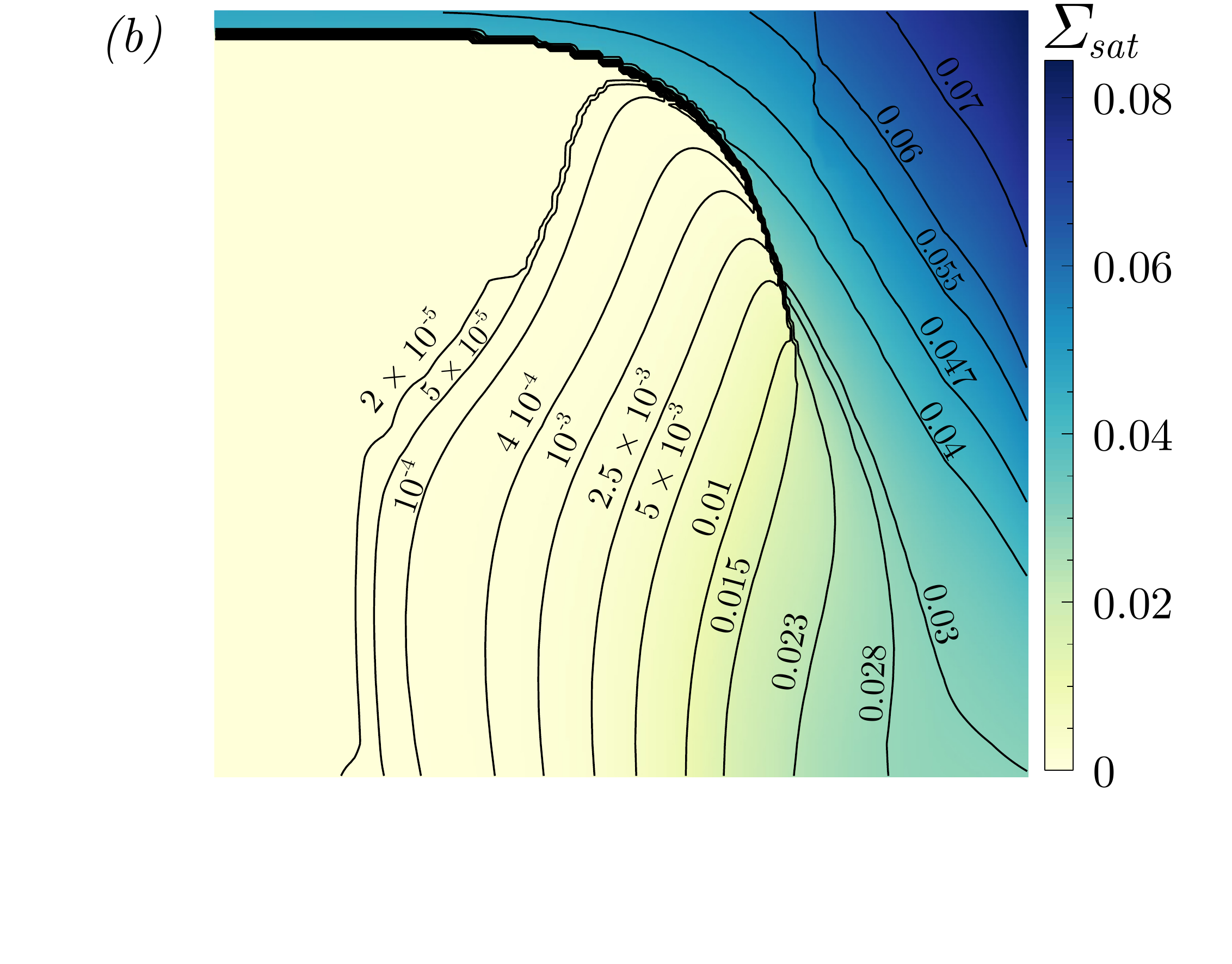}\\
\vspace{-1cm}\\
\includegraphics[width=0.55\textwidth]{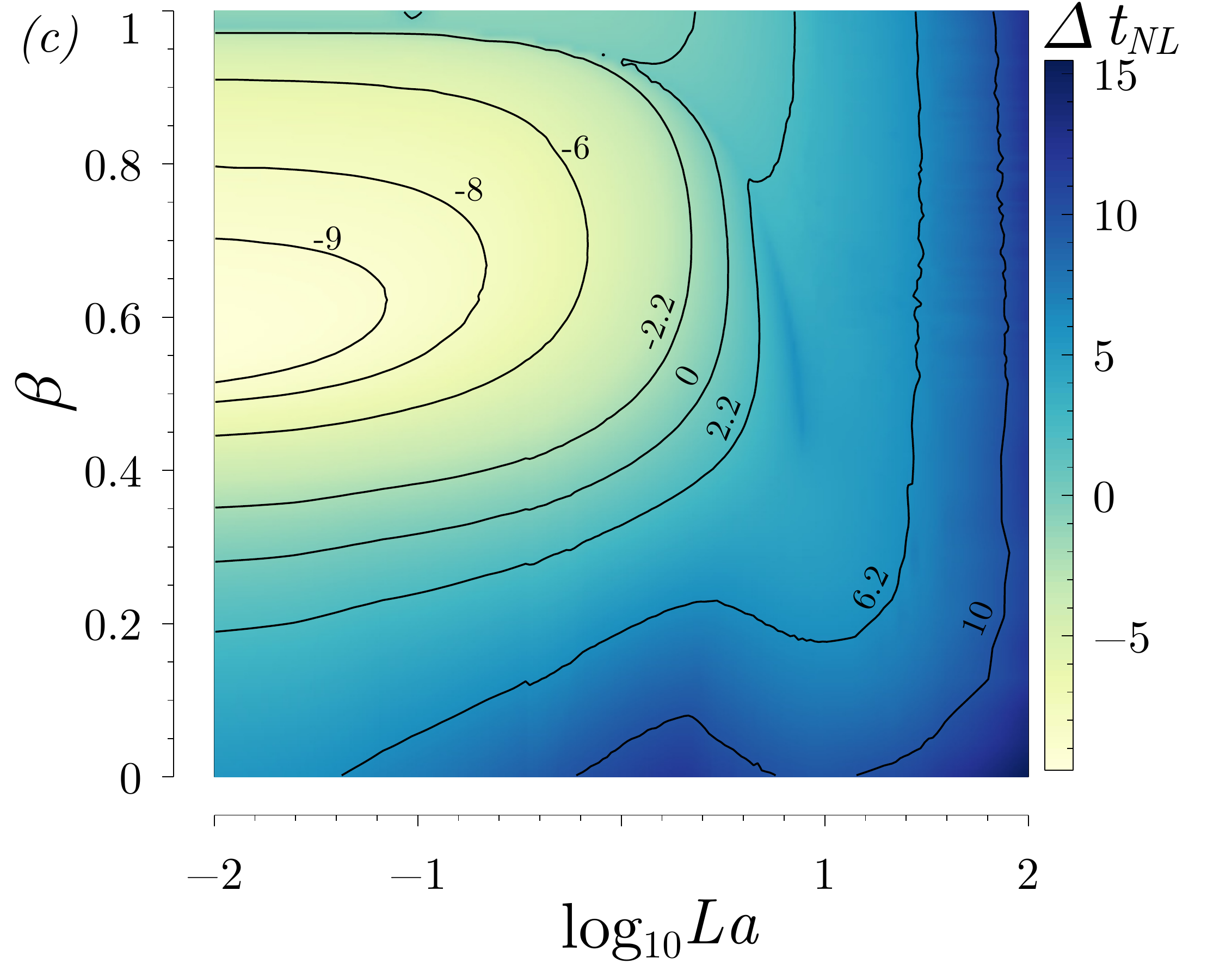} \hspace{-0.9cm}
\includegraphics[width=0.55\textwidth]{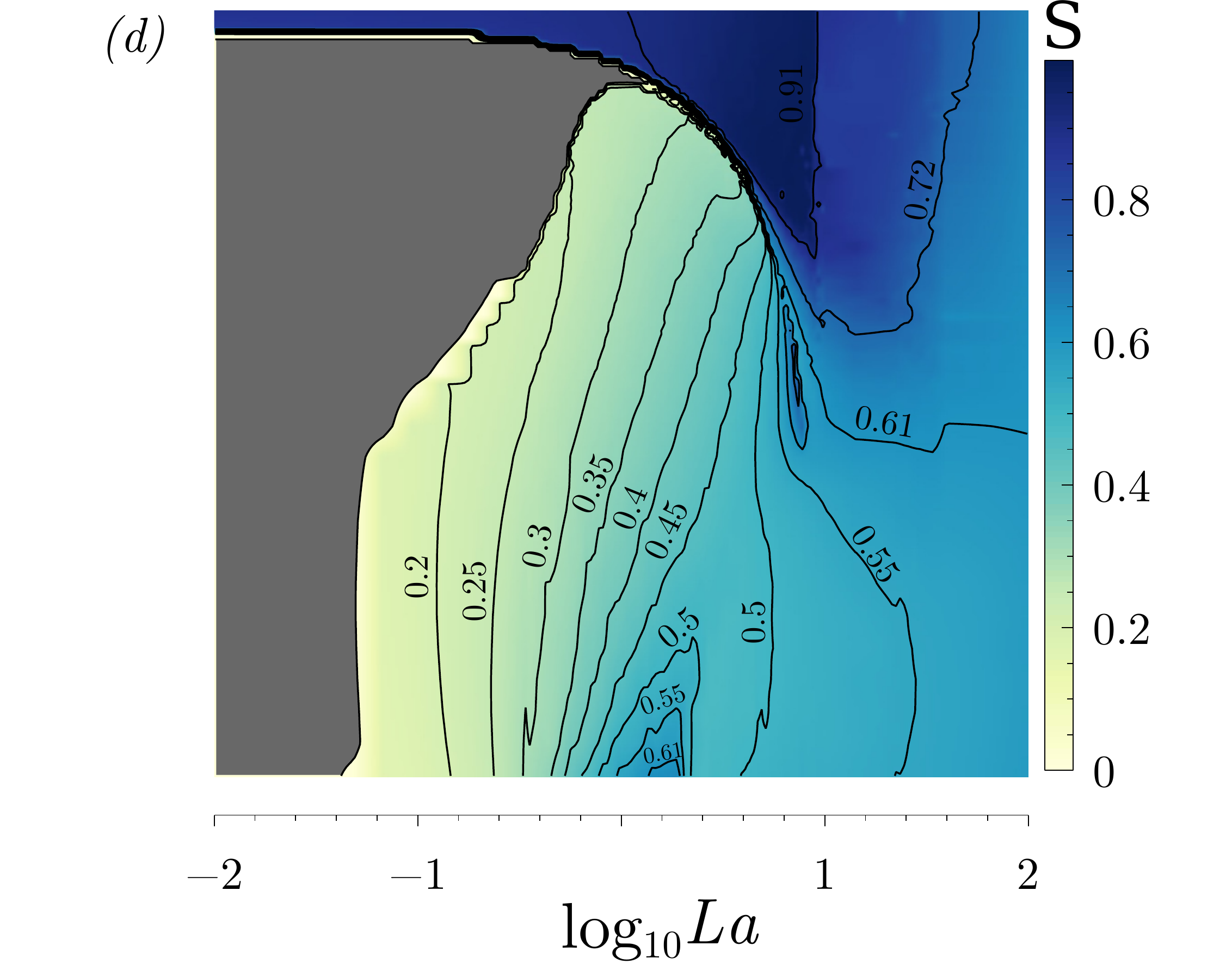}
\caption{\label{fig:figure5} (Colour online) Isocontours in the $(\Lap,\Ela)$ parameter plane of ($a$) the normalised satellite volume $\Vsat$, ($b$) the normalised mass of surfactant trapped at its interface $\Ssat$, ($c$) the nonlinear correction to the break-up time, $\DtNL$, and ($d$) the sphericity of the satellite droplet, $\mathcal{S}$.}
\end{figure}

Let us first present the structure of the $(\Lap,\Ela)$ parameter plane in terms of the satellite formation process. To that end, we conducted an exhaustive parametric study in which the Laplace and elasticity parameters were varied in small steps within wide ranges, namely $0.01\leq\Lap\leq 100$ and $0\leq\Ela\leq 1$. Thus, for each pair of values of $\Lap$ and $\Ela$, we simulated the instability-driven time evolution of the thread from an initial condition with $\epsilon\ll 1$ until a time $t_b$ very close to break-up. In total, around $10^4$ time-dependent simulations were carried out to characterise the $(\Lap,\Ela)$ parameter plane shown in figures~\ref{fig:figure3} and~\ref{fig:figure5}.

At this point, it is important to emphasise that the fate of the main and satellite drops after pinch-off is outside the scope of the present work, and therefore we do not explore the possible successive break-up events that may take place and lead to the formation of sub-satellites. Keeping this in mind, we have extracted the satellite volume at the last numerical step, $t=t_b$. Normalising its value with the total volume provides the definition
\begin{equation}\label{eq:Vsat}
\Vsat =  \frac{\int_0^{z_{\min}}  a^2 \rm{d} \it{z}}{\int_0^{\upi /k_m}  a^2 \rm{d} \it{z}},
\end{equation}
where $z_{\min}$ is the axial position where the liquid column reaches its minimum radius, $a_{\min}$, at $t=t_b$. A more common measure of the satellite size is its equivalent radius, $R_{\text{sat}}$, which is the radius of a spherical drop of the same volume as the satellite~\citep{RutlandJameson,Mansour1990,Ashgriz1995,Mashayek1995}. All the results reported herein in terms of $\Vsat$ can be easily converted to $R_{\text{sat}}$ through the equation $R_{\text{sat}} = [3 \upi V_{\text{sat}}/(2 k_m)]^{1/3}$. Following the same procedure, we have also computed the mass of surfactant trapped at the satellite surface which, normalised with the total mass of surfactant, provides the definition
\begin{equation}\label{eq:Ssat}
\Ssat = \frac{\int_0^{z_{\min}}   a \Gamma \sqrt{1+\left(\frac{\partial a}{\partial z}\right)^2} \rm{d} \it{z}}{\int_0^{\upi/k_m}   a \Gamma \sqrt{1+\left(\frac{\partial a}{\partial z}\right)^2} \rm{d} \it{z}}.
\end{equation}
We point out that, since $V_{\text{sat}}$ and $\Sigma_{\text{sat}}$ are always obtained when $a_{\min}$ is within the range $a_{\min} \sim 10^{-4}-8 \times 10^{-3}$, the sensitivity of these magnitudes to the exact value of $a_{\min}$ is negligible, such that both represent very robust measures. Similarly, the corresponding break-up time $t_b$ is barely sensitive to the value of $a_{\min}$.

In contrast with $V_{\text{sat}}$ and $\Sigma_{\text{sat}}$, which do not depend on the initial amplitude in the limit $\epsilon\ll 1$, the break-up time is a function of the form $t_b(\Lap,\Ela,\epsilon)$ such that $t_b\to \infty$ as $\epsilon\to 0$. Indeed, the break-up time can be easily estimated from linear theory through the equation $a_{\min}(t)\sim 1-\epsilon\exp{(\omega_m t)}$, where $\omega_m$ is the growth rate associated with the most amplified wavenumber $k_m$ shown in figures~\ref{fig:figure2} ($c$, $d$), leading to the estimation $t_b \sim \ln{(\epsilon^{-1})}/\omega_m$. Based on the latter result, we define the nonlinear correction to the linear break-up time as
\begin{equation}\label{eq:DtNL}
\DtNL = t_b - \frac{\ln (\epsilon^{-1})}{\omega_m},
\end{equation}
where $t_b$ is obtained by extrapolating $a_{\min}$ to zero using the last few computed time steps. Unlike $t_b$, $\Delta t_{\text{NL}}$ only depends on $\Lap$ and $\Ela$, but not on $\epsilon$, provided only that $\epsilon \ll 1$. The latter fact is demonstrated in~\S\ref{subsec:nonlinear_surfactant}. Finally, we have also computed the sphericity of the satellite droplet at pinch-off as
\begin{equation}\label{eq:Spher}
\Spher = \frac{2 \left( \frac{3}{4} \int_0^{z_{\min}} a^2 \rm{d} \it{z} \right)^{2/3}}{\int_0^{z_{\min}} a \sqrt{1+\left(\frac{\partial a}{\partial z}\right)^2}\rm{d} \it{z}},
\end{equation}
which is the ratio between the surface of a sphere of the same volume as the satellite and its actual surface. The quantification of the satellite formation process will be based on the four functions $\Vsat$, $\Ssat$, $\DtNL$ and $\Spher$, extracted from the numerical simulations. These four functions only depend on $\Lap$ and $\Ela$ when $\epsilon$ is sufficiently small, as is demonstrated in~\S\ref{subsec:nonlinear_surfactant}. Thus, the main results reported herein have been computed in the limit $\epsilon\to 0$.

The structure of the ($\Lap$,$\Ela$) parameter plane is summarised in figures~\ref{fig:figure3} and~\ref{fig:figure4} in terms of the satellite formation process. In particular, figure~\ref{fig:figure3} depicts the most salient features of the parameter plane, and figure~\ref{fig:figure4} displays several satellite shapes at the last computed numerical step just prior to pinch-off, whose associated values of $\Lap$ and $\Ela$ are indicated with circles in figure~\ref{fig:figure3}. The most important feature of the parameter plane is the solid line shown in figure~\ref{fig:figure3}, which represents a discontinuous transition that takes place for a critical elasticity, $\Ela = \Ela_c(\Lap)$ for $\Lap<7.5$. In particular, both the satellite volume and the associated entrapped mass of surfactant experience sudden jumps from certain values $\Vsat(\Ela-\Ela_c\to 0^-)$ and $\Ssat(\Ela-\Ela_c\to 0^-)$ to larger values $\Vsat(\Ela-\Ela_c\to 0^+)$ and $\Ssat(\Ela-\Ela_c\to 0^+)$. Indeed, the inset of figure~\ref{fig:figure3} shows the jumps experienced by the satellite volume, $\Delta \Vsat(\Lap)=\Vsat(\Ela-\Ela_c\to 0^+ )-\Vsat(\Ela-\Ela_c\to 0^-)$, and by the associated entrapped mass of surfactant, $\Delta \Ssat$. Both jumps and $\Ela_c$ increase monotonically as $\Lap$ decreases, and reach respective Stokes asymptotes as $\Lap\to 0$, namely $\Delta \Vsat \to 0.022$, $\Delta \Ssat \to 0.045$, and $\beta_c \to 0.98$. The filled circle in figure~\ref{fig:figure3} indicates the origin of the discontinuous transition, $(\Lap,\Ela_c)=(7.5,0.55)$, at which both jumps become zero. For values of $\Lap>7.5$, $\Vsat$ and $\Ssat$ are continuous functions of $\Lap$ and $\Ela$.

As shown in figure~\ref{fig:figure4}, for values of $\Ela=0<\Ela_c$ and $\Ela=0.5<\Ela_c$ the sequence of interface shapes at pinch-off depends continuously on $\Lap$, with the trend that larger satellites are formed as $\Lap$ increases, reaching the regular limit of inviscid flow as $\Lap\to \infty$. For $\Ela<\Ela_c$ and small values of $\Lap$, figure~\ref{fig:figure4} reveals that the main drops are separated by very thin threads of tiny volume whose break-up behaviour has been characterised in previous studies~\citep[see e.g.][]{Kowalewski1996}. For $\Ela<\Ela_c$ and intermediate values of $\Lap$, the main drops are separated by a satellite centred at $z=0$ that is connected to the main drops by very thin threads (see e.g. the case for $\Lap=1$ and $\Ela=0$ in figure~\ref{fig:figure4}). Finally, for $\Ela<\Ela_c$ and large values of $\Lap$, the satellite drop is directly connected to the main drops. In contrast, when $\Ela=1>\Ela_c$, figure~\ref{fig:figure4} shows a different picture, where large satellites are formed for all values of $\Lap$. These results have also been analysed quantitatively, and are discussed in detail below.

From figures~\ref{fig:figure3} and~\ref{fig:figure4} it is deduced that, although the physical mechanisms are different, both the liquid inertia and the interfacial elastic stress favour the formation of satellites. In particular, surface elasticity tends to form spherical-shaped satellites at pinch-off, whereas the increase of the liquid inertia generates oval-shaped satellites. In the set of shapes close to pinch-off shown in figure~\ref{fig:figure4}, a discontinuous transition is observed for $\Lap = 0.01$ and 1, as $\Ela$ increases. However, for $\Lap = 10>7.5$ a continuous transition of the thread shape is observed as $\Ela$ increases. Finally, for $\Lap=100$, the upper row evidences that the influence of the elastic stress on the shape of the thread is much weaker when the value of $\Lap$ is large enough. The physics underlying these transitions can be explained in terms of the coupling between the liquid inertia, the viscous stress, the surface tension, and the interfacial elastic stress. The competition between these forces is discussed in~\S\ref{subsec:nonlinear_surfactant}, based on the trends exhibited by the functions $\Vsat$, $\Ssat$, $\DtNL$ and $\Spher$, and also by analysing the temporal evolution of the interface shapes starting from small disturbances, depending on the values of $\Lap$ and $\Ela$.

\subsection{Nonlinear dynamics of a surfactant-laden interface: satellite drop formation}\label{subsec:nonlinear_surfactant}

To unveil the effect of liquid inertia, viscous stresses and surface elasticity on the satellite droplet formation regimes, here we present and discuss the quantitative results of the detailed numerical analysis that has been carried out in the present work.

Figure~\ref{fig:figure5} shows the isocontours of $\Vsat$, $\Ssat$, $\DtNL$ and $\Spher$ in the $(\Lap,\Ela)$ parameter plane. We first observe that, at the discontinuous transition that occurs for $\Lap < 7.5$, the value of $\Vsat$ increases from 10$^{-3}$--1.5 \% to 2--2.3 \%, whereas $\Ssat$ increases from 10$^{-3}$--1.5~\% to 3.5--4.7 \%. The exact value of both jumps as functions of $\Lap$ can be seen in the inset of figure~\ref{fig:figure3}. In contrast, for $\Lap > 7.5$ or $\Ela>\Ela_c(\Lap)$, the values of $\Vsat$, $\Ssat$, $\DtNL$ and $\Spher$ vary continuously.

As a first general observation, it is deduced from figure~\ref{fig:figure5} that the linear theory may either underestimate or overestimate the break-up time, in a way that does not necessarily coincide with the transitions in the satellite formation process. Indeed, $t_b$ is underestimated for $La\gg 1$ independently of the value of $\Ela$. However, for $La\ll 1$, $t_b$ is overestimated for $0.28\lesssim\Ela\lesssim 1$, while it is underestimated outside this range. Regarding the sphericity $\Spher$, figure~\ref{fig:figure5} confirms the trend deduced from figure~\ref{fig:figure4}: the most spherical satellite shapes, with $\Spher \gtrsim 0.9$, take place for $\Ela\gtrsim \Ela_c$ and $\Lap\lesssim 10$. In contrast, the shapes become most elongated, with $\Spher\lesssim 0.2$ , when $\Ela<\Ela_c$ and $\Lap\lesssim 0.1$ (grey area in figure~\ref{fig:figure5}$d$).


\begin{figure}
\hspace{-1cm}
\begin{tabular}{cccc}
  \includegraphics[width=0.25\textwidth]{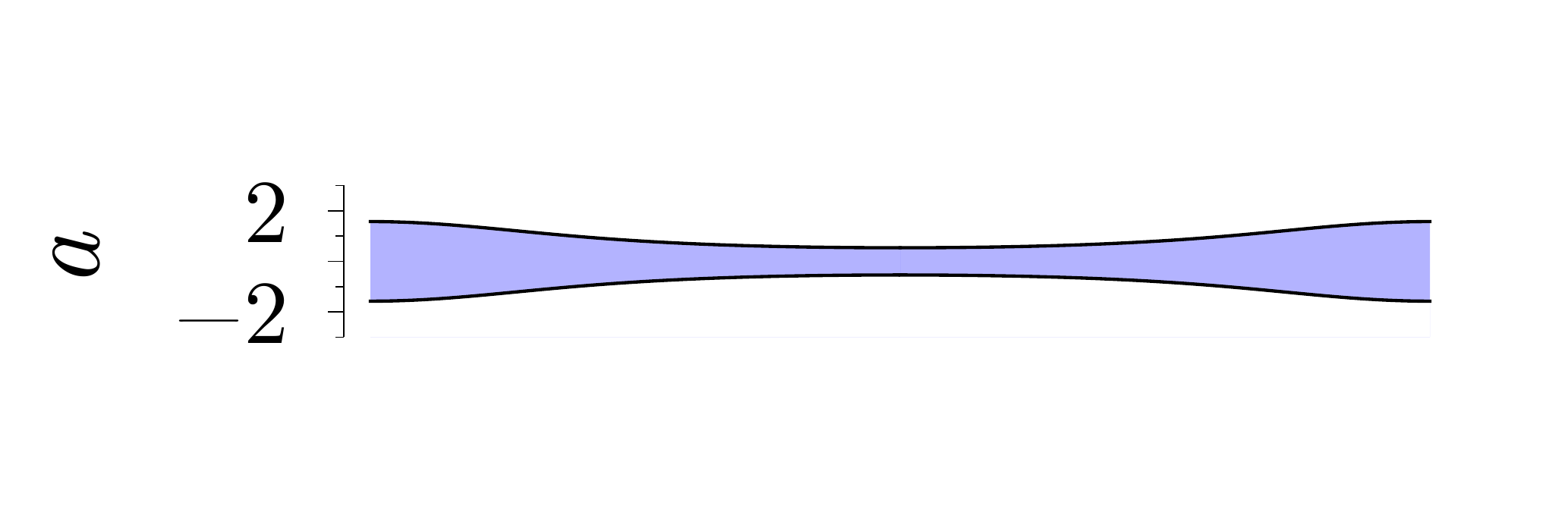} &   \includegraphics[width=0.25\textwidth]{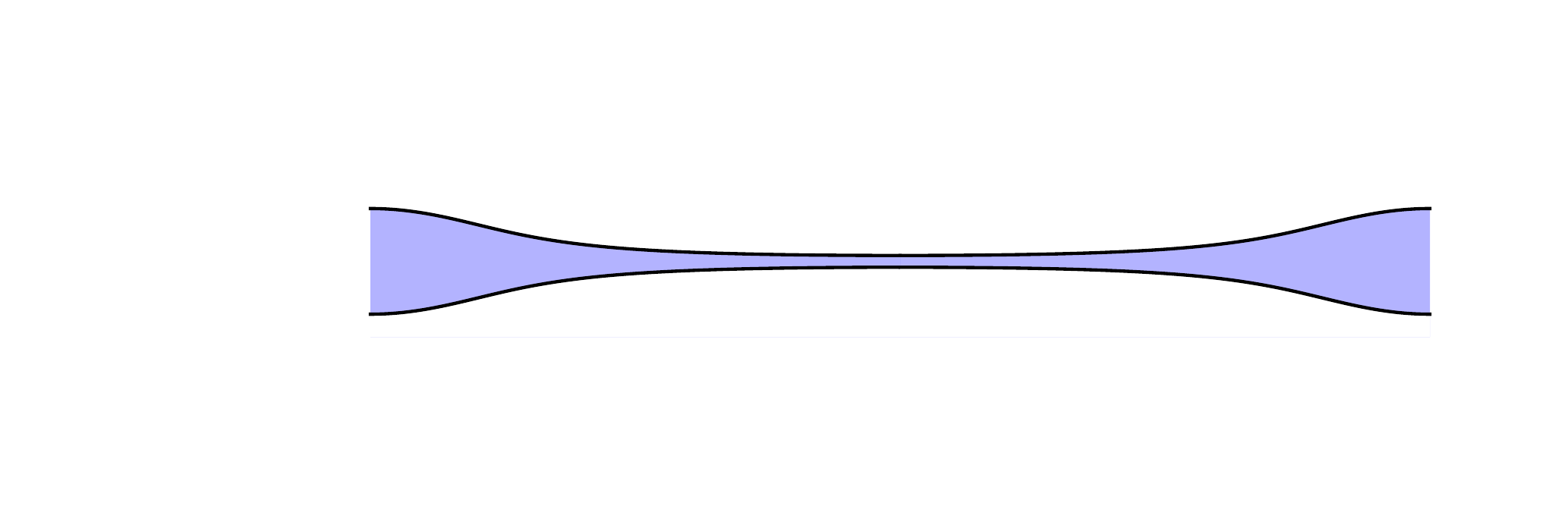} &  \includegraphics[width=0.25\textwidth]{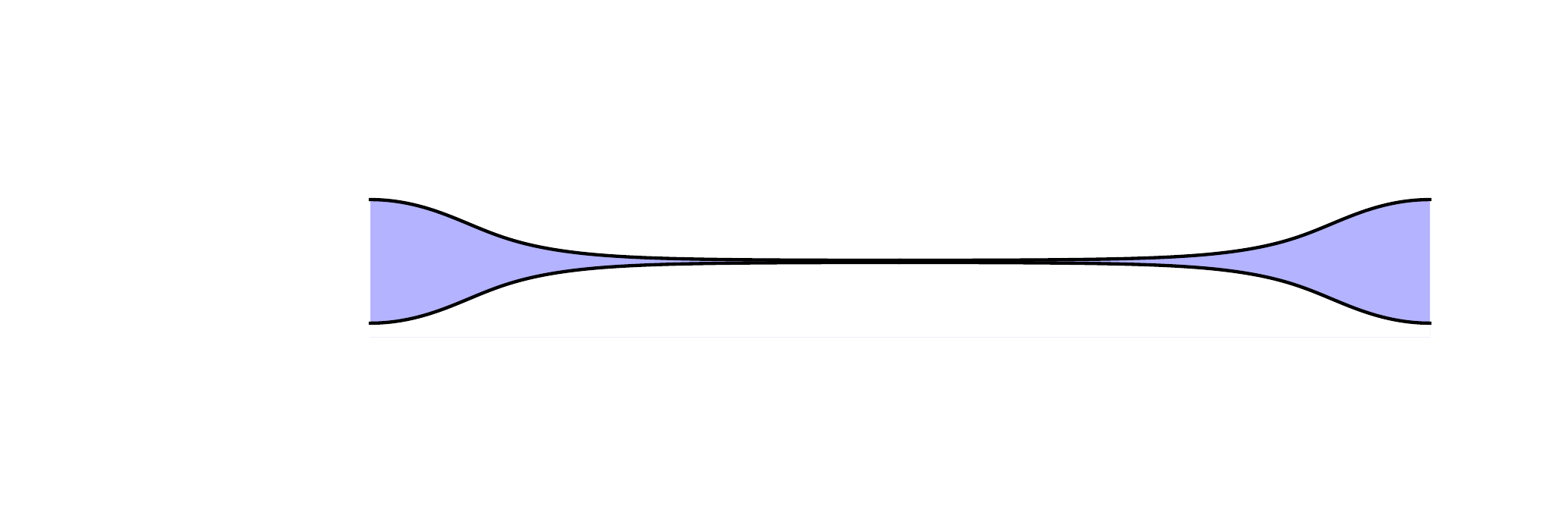}  & \includegraphics[width=0.25\textwidth]{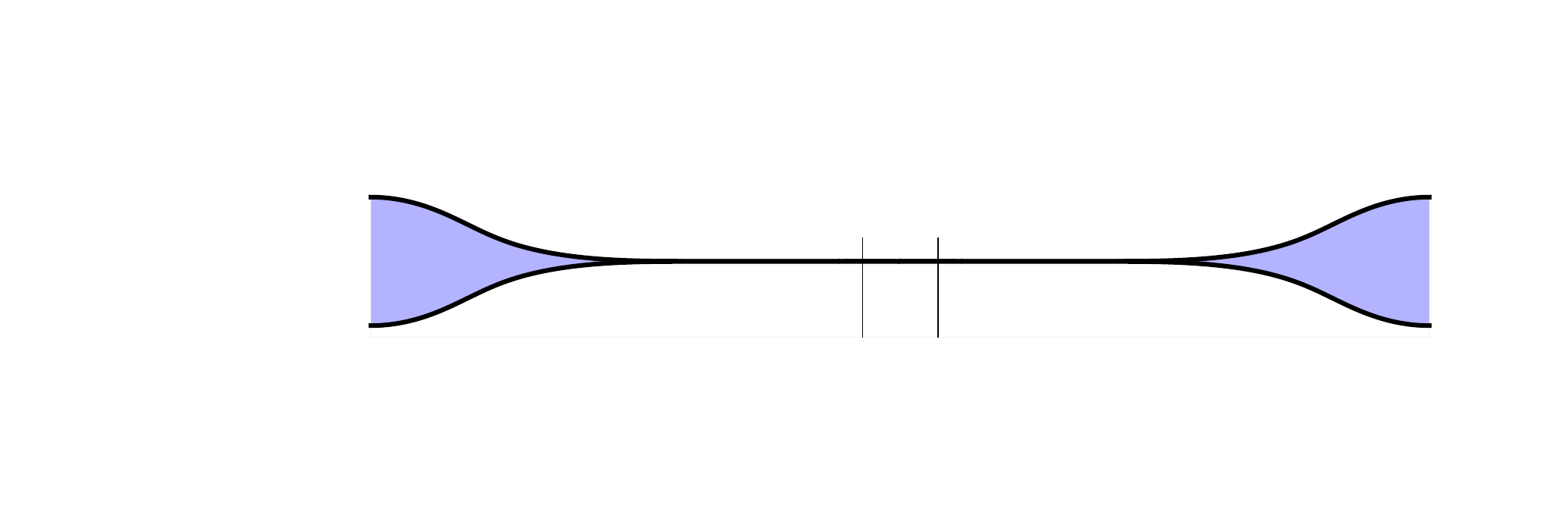}  \\
  \vspace{-0.8cm} & & & \\
      \includegraphics[width=0.25\textwidth]{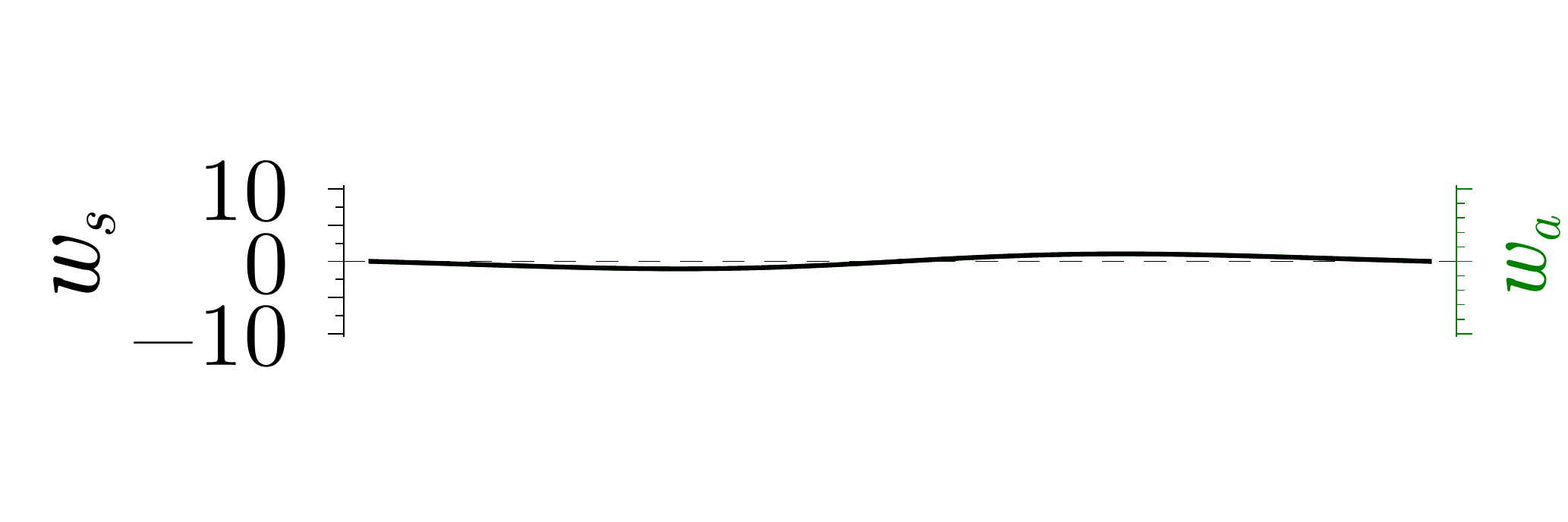} &   \includegraphics[width=0.25\textwidth]{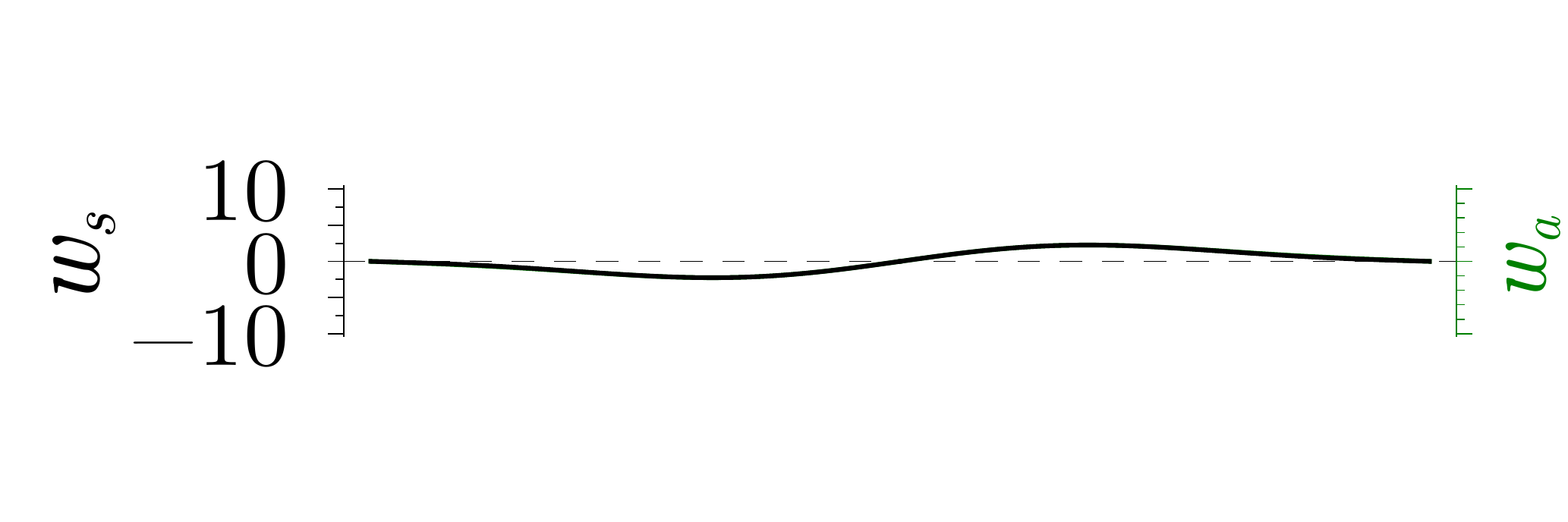} &   \includegraphics[width=0.25\textwidth]{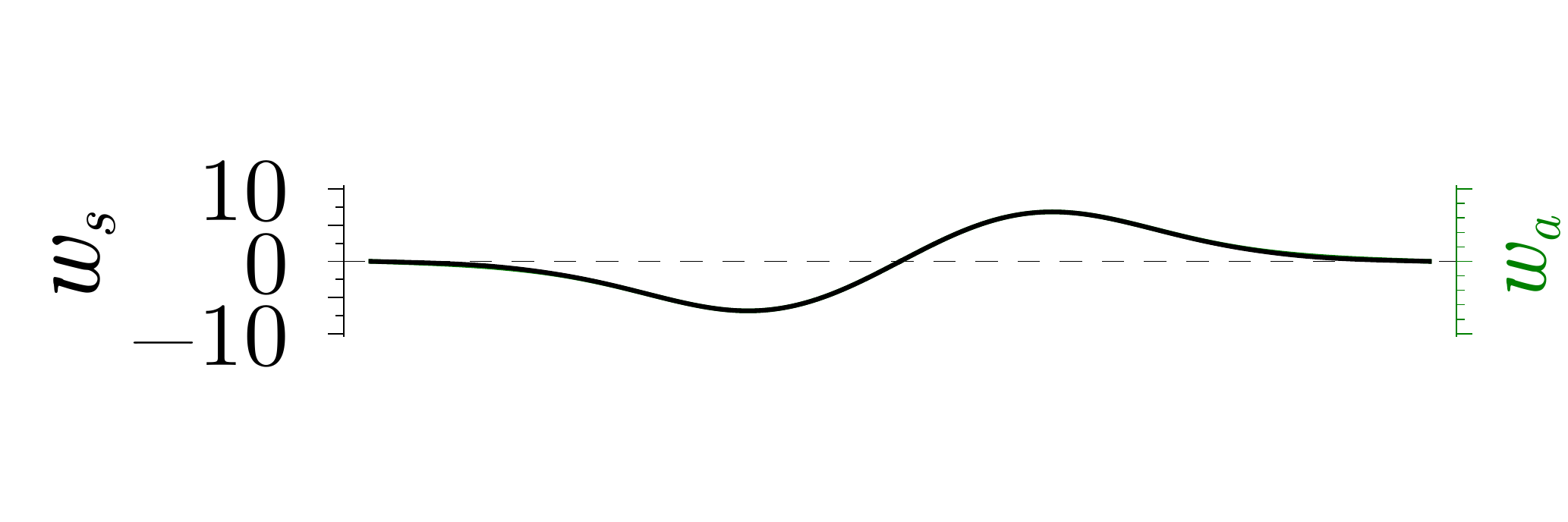} &   \includegraphics[width=0.25\textwidth]{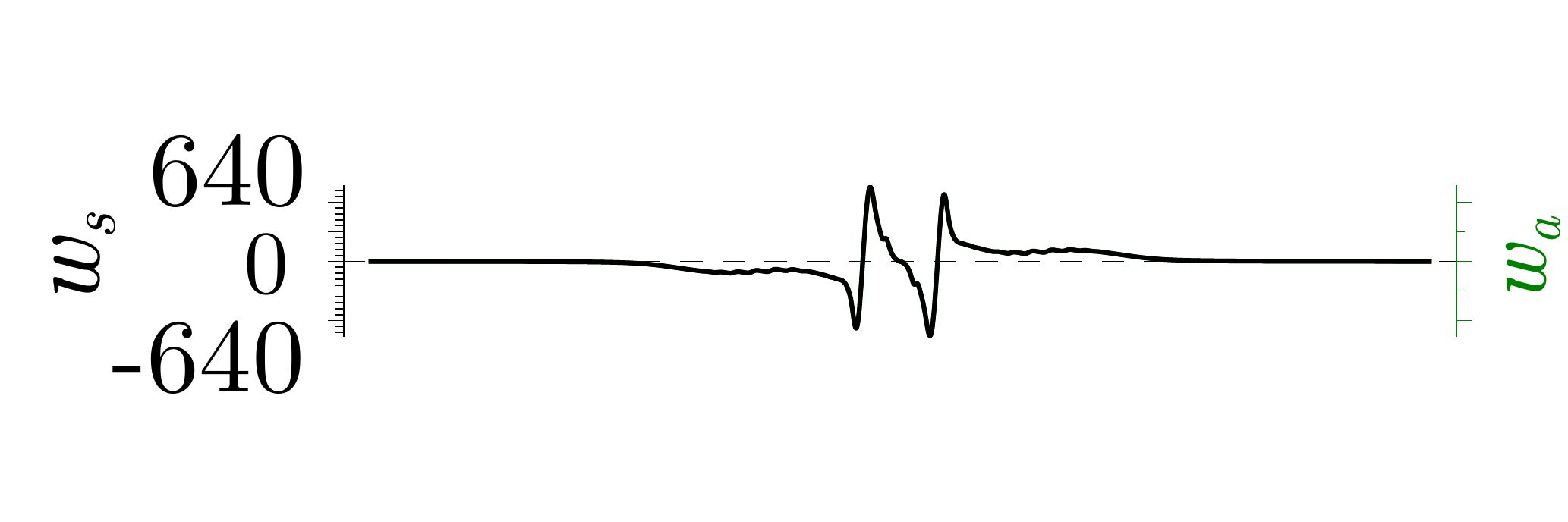} \\
        \vspace{-0.8cm} & & & \\
           \includegraphics[width=0.25\textwidth]{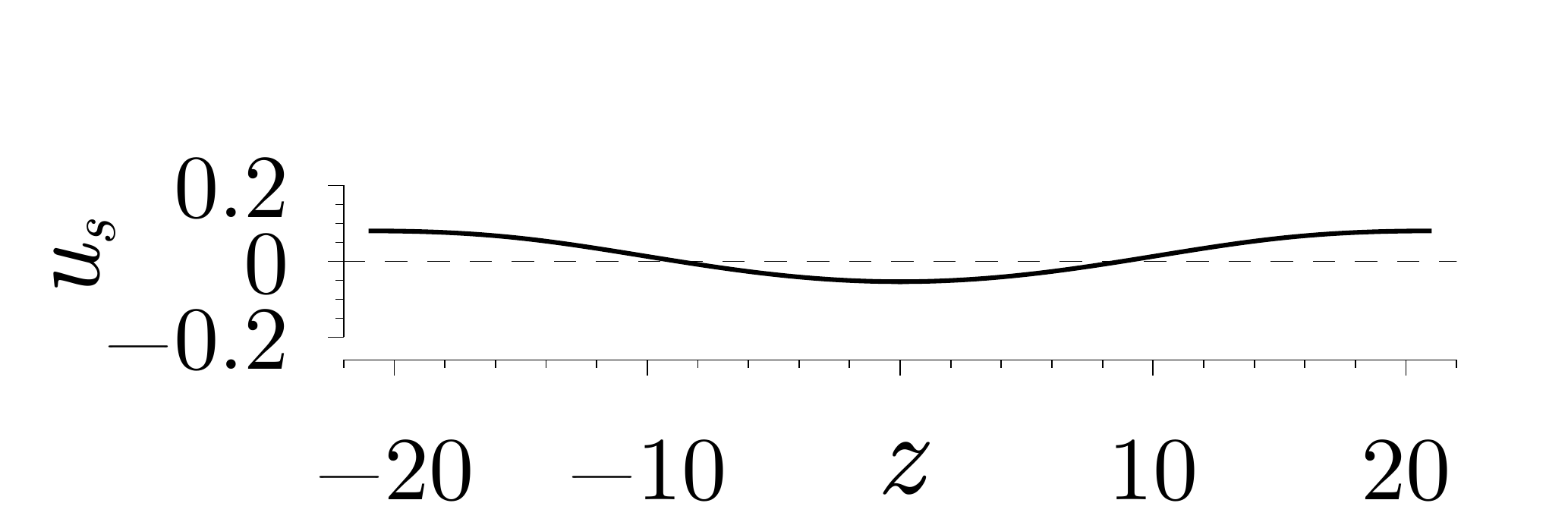} &   \includegraphics[width=0.25\textwidth]{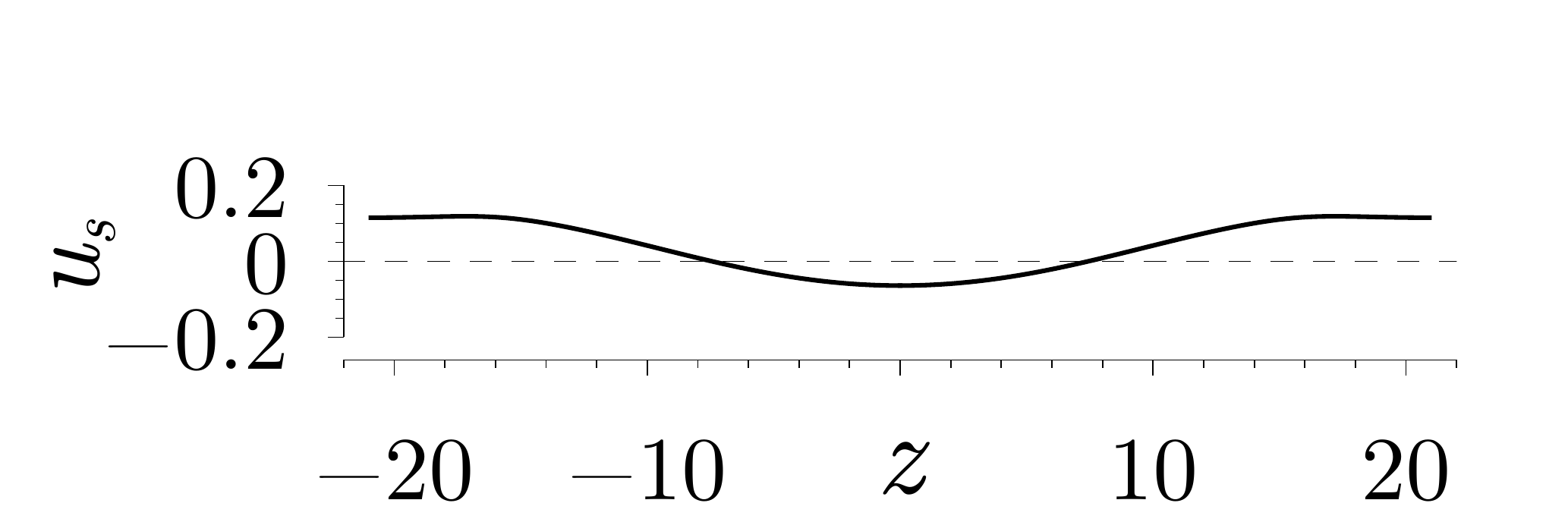} &   \includegraphics[width=0.25\textwidth]{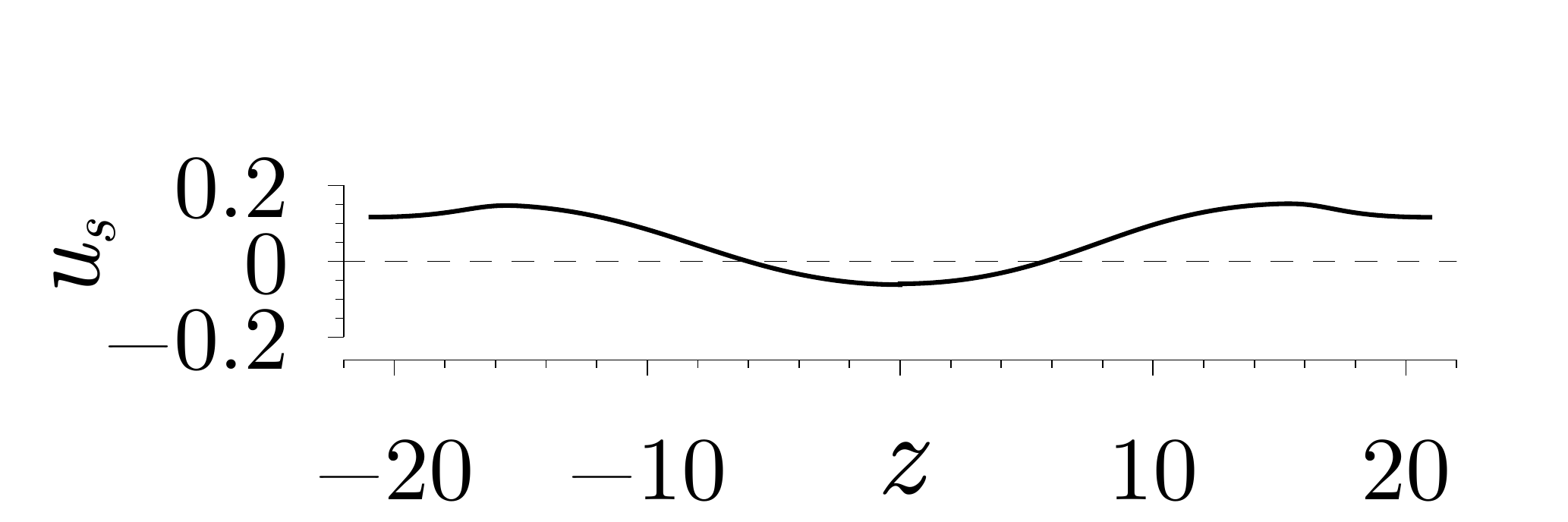} &   \includegraphics[width=0.25\textwidth]{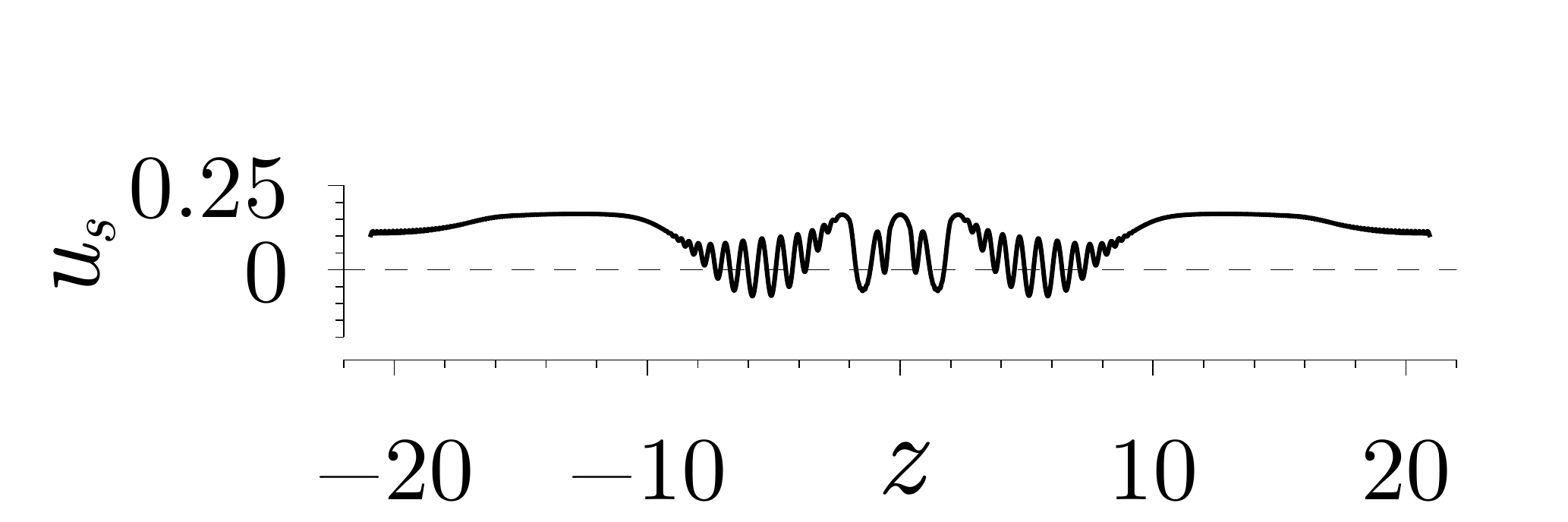} \\
($a$) $t = 38$ & ($b$) $t = 43$ & ($c$) $t = 46$  & ($d$) $t = 47.226$ \\[4pt]
  \includegraphics[width=0.24\textwidth]{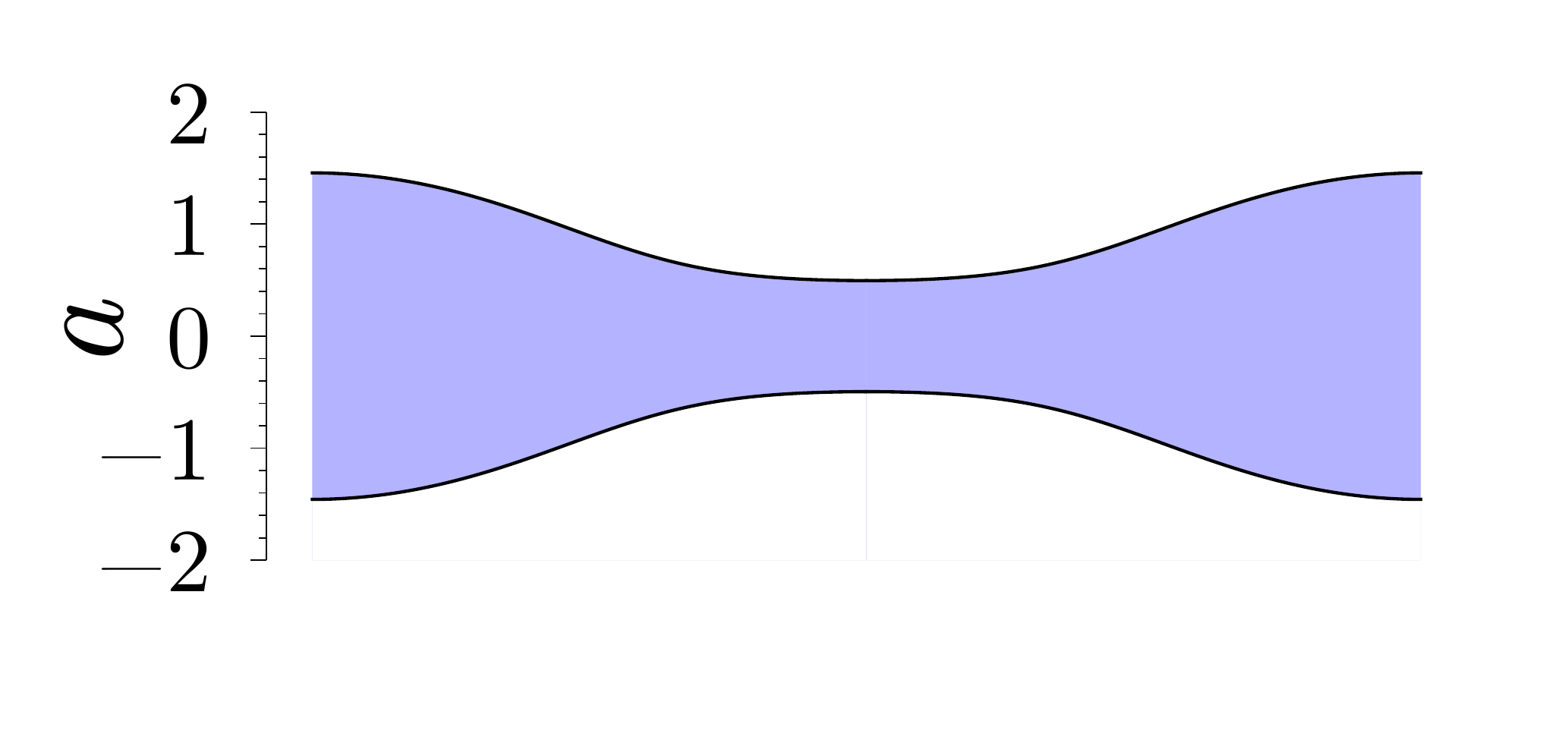} &   \includegraphics[width=0.24\textwidth]{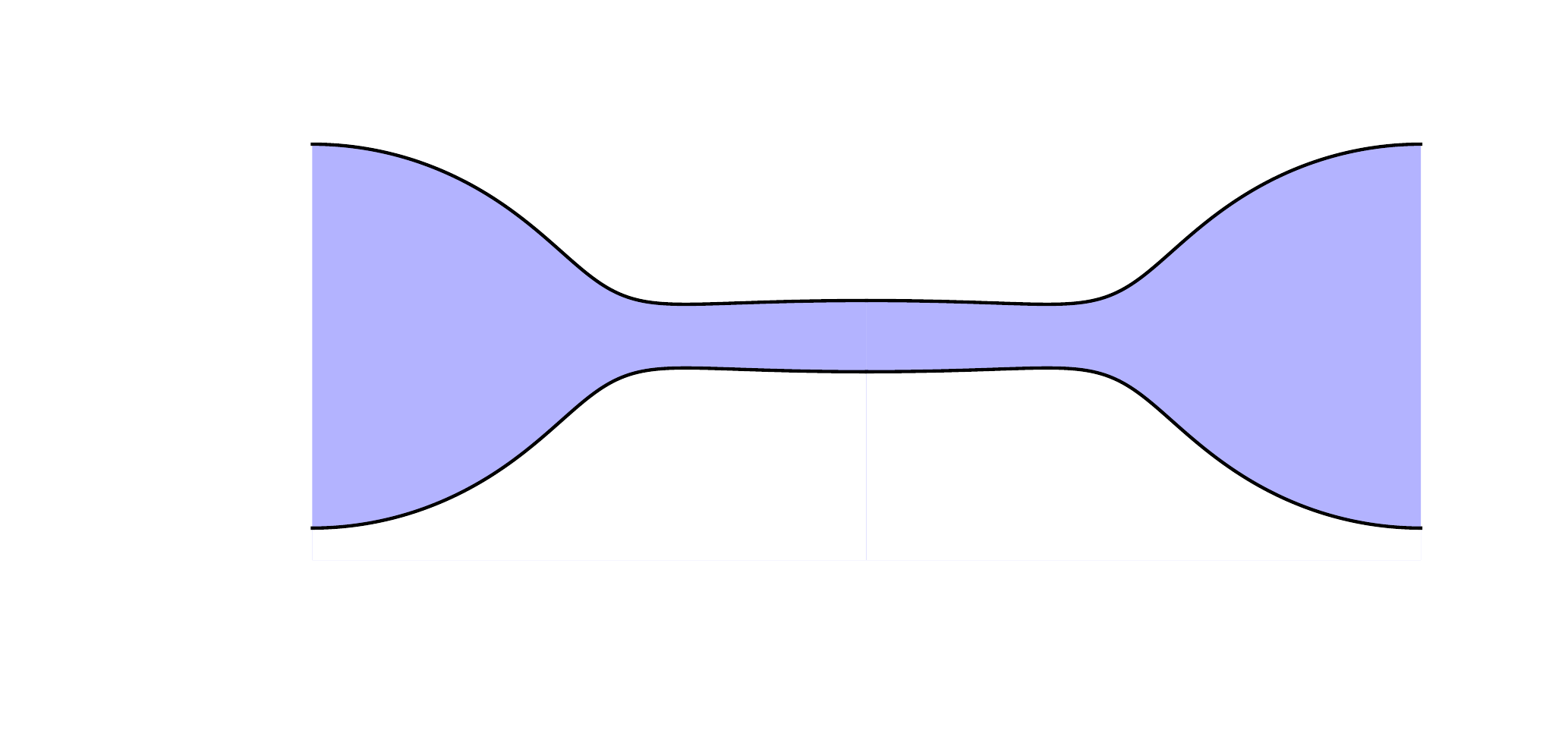} &  \includegraphics[width=0.24\textwidth]{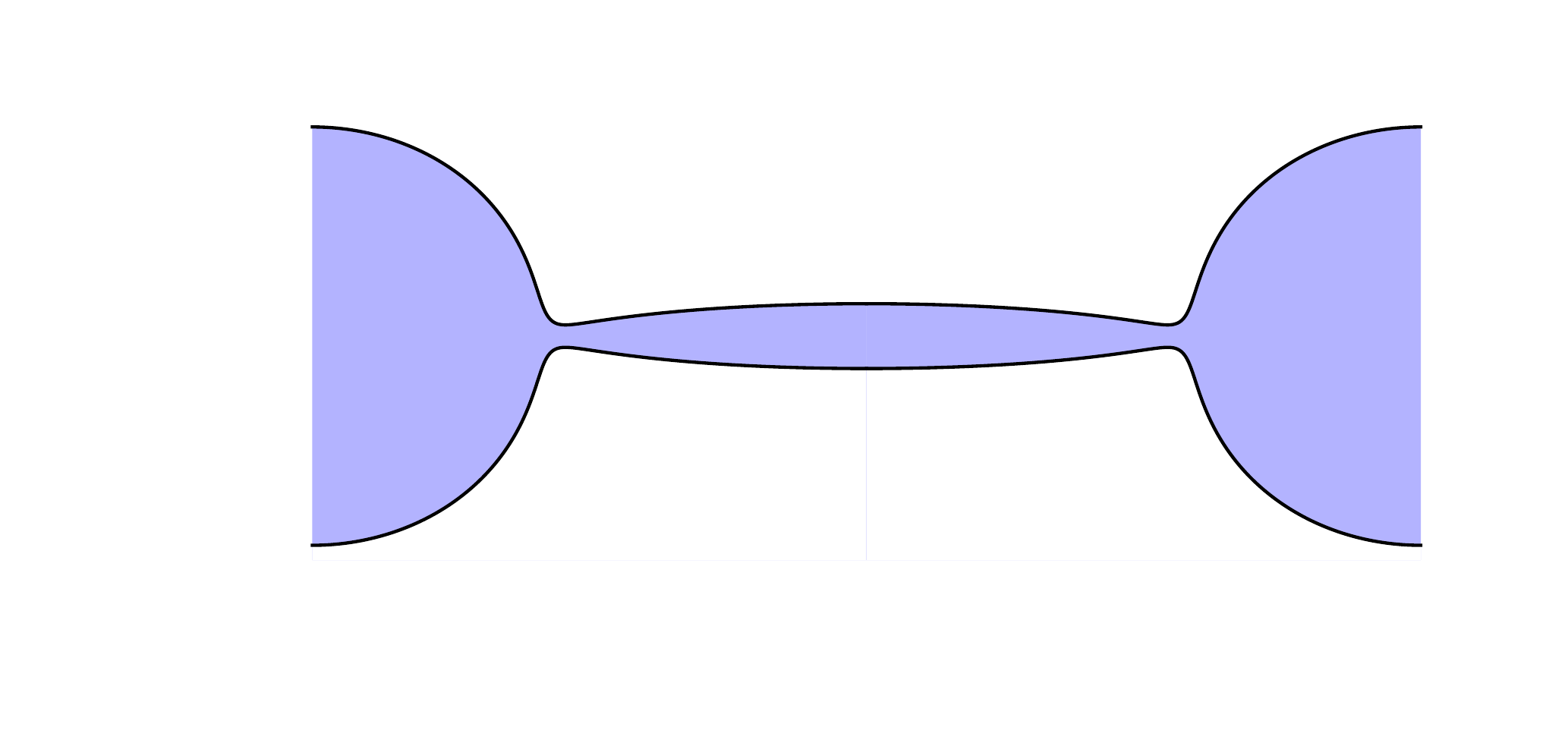}  & \includegraphics[width=0.24\textwidth]{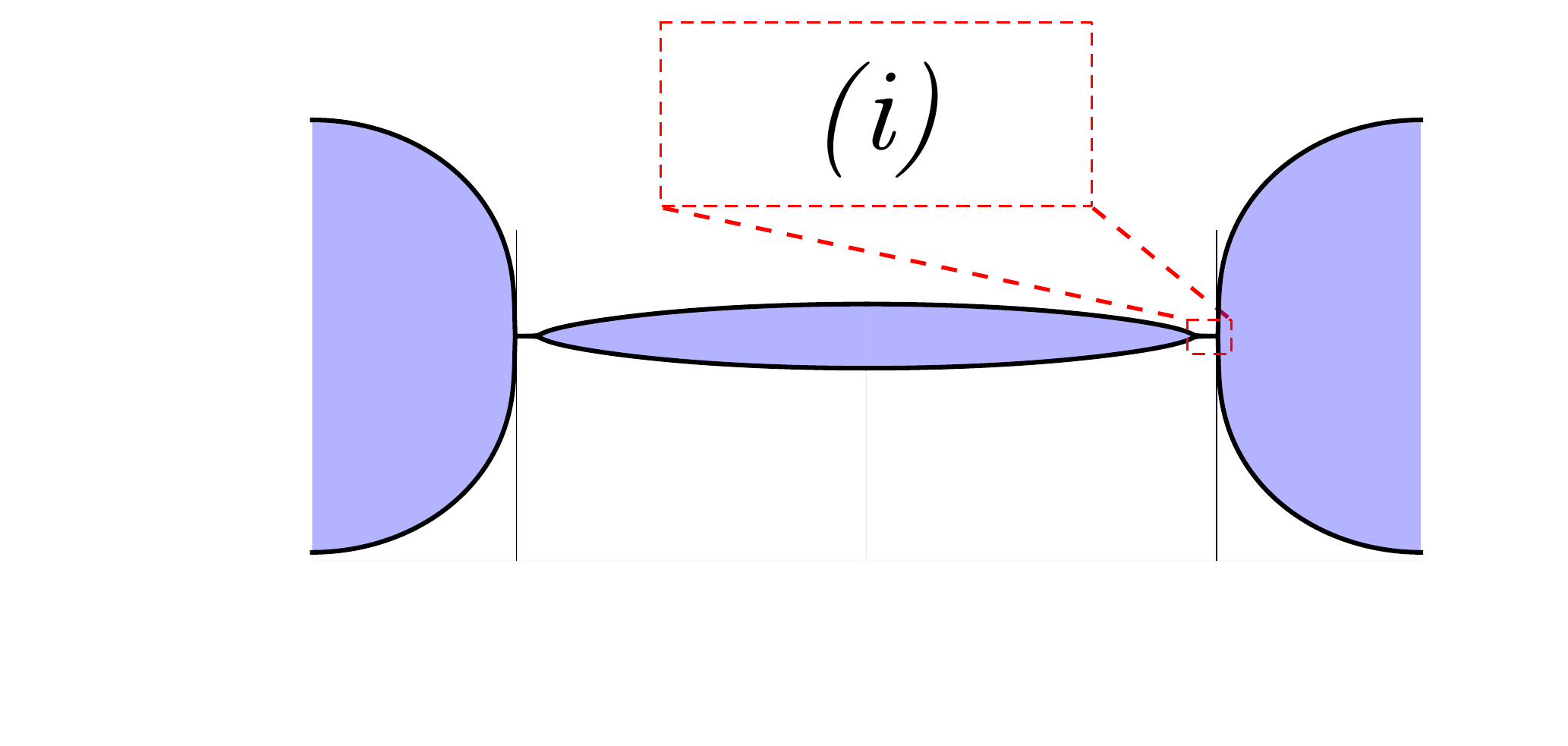}  \\
    \vspace{-0.8cm} & & & \\
      \includegraphics[width=0.24\textwidth]{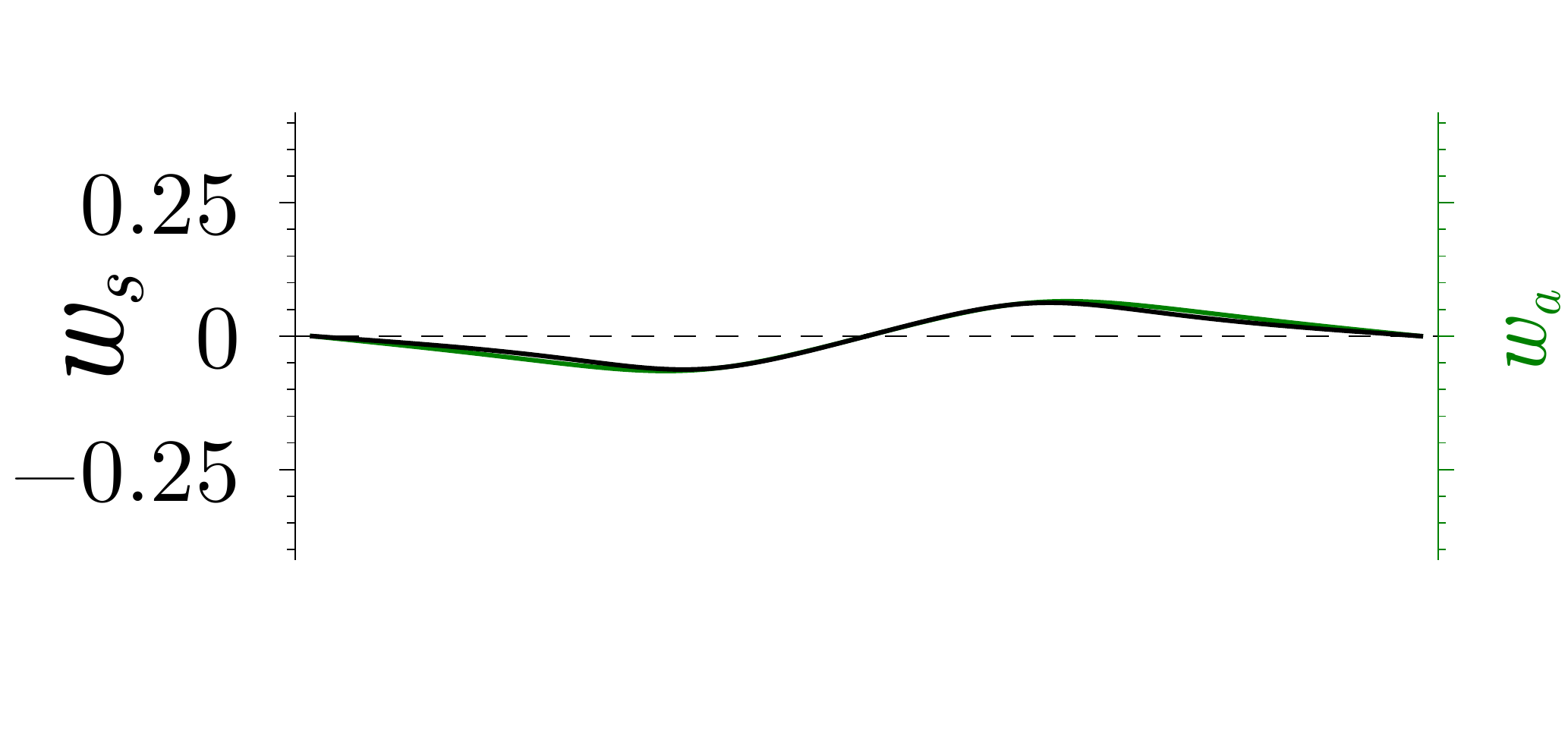} &   \includegraphics[width=0.24\textwidth]{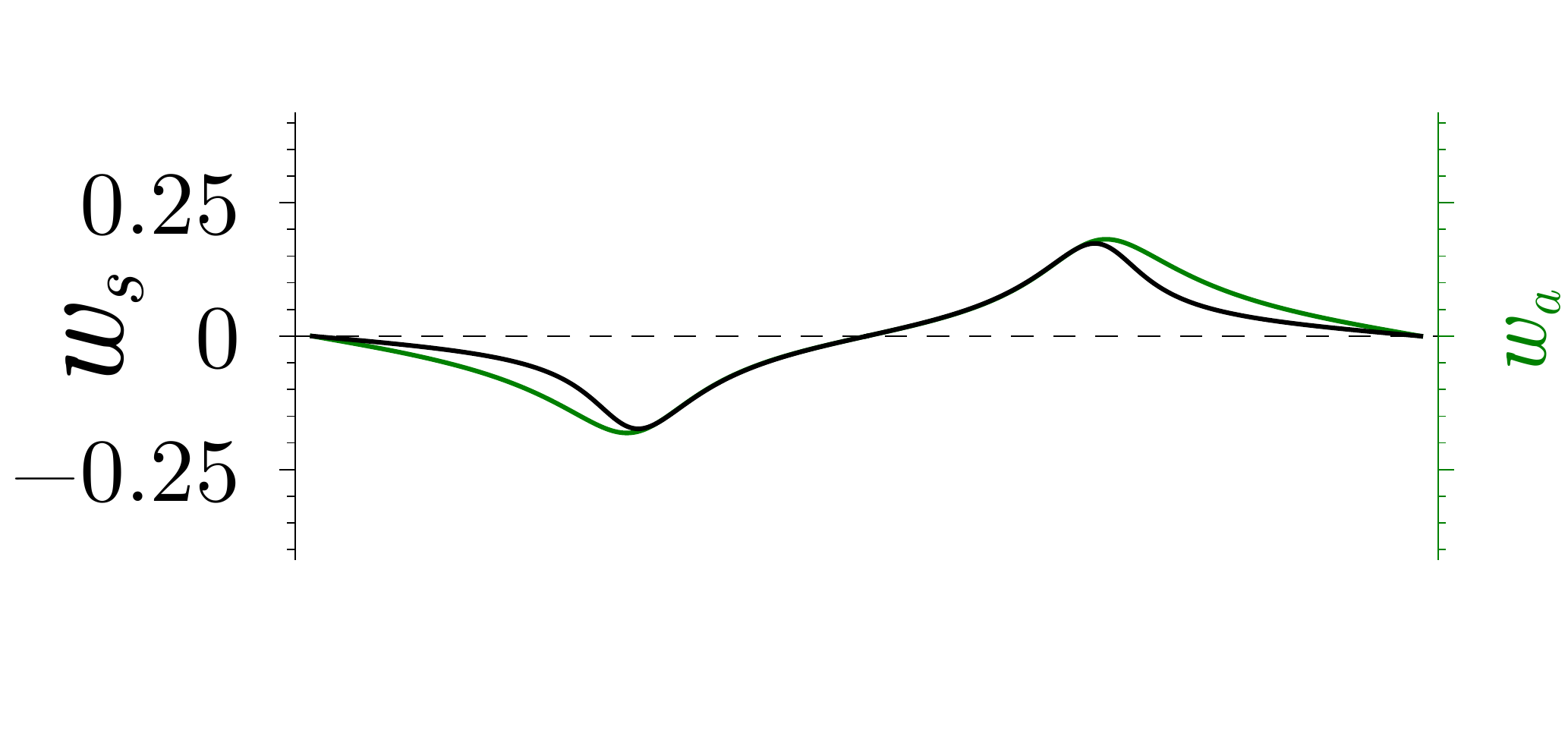} &   \includegraphics[width=0.24\textwidth]{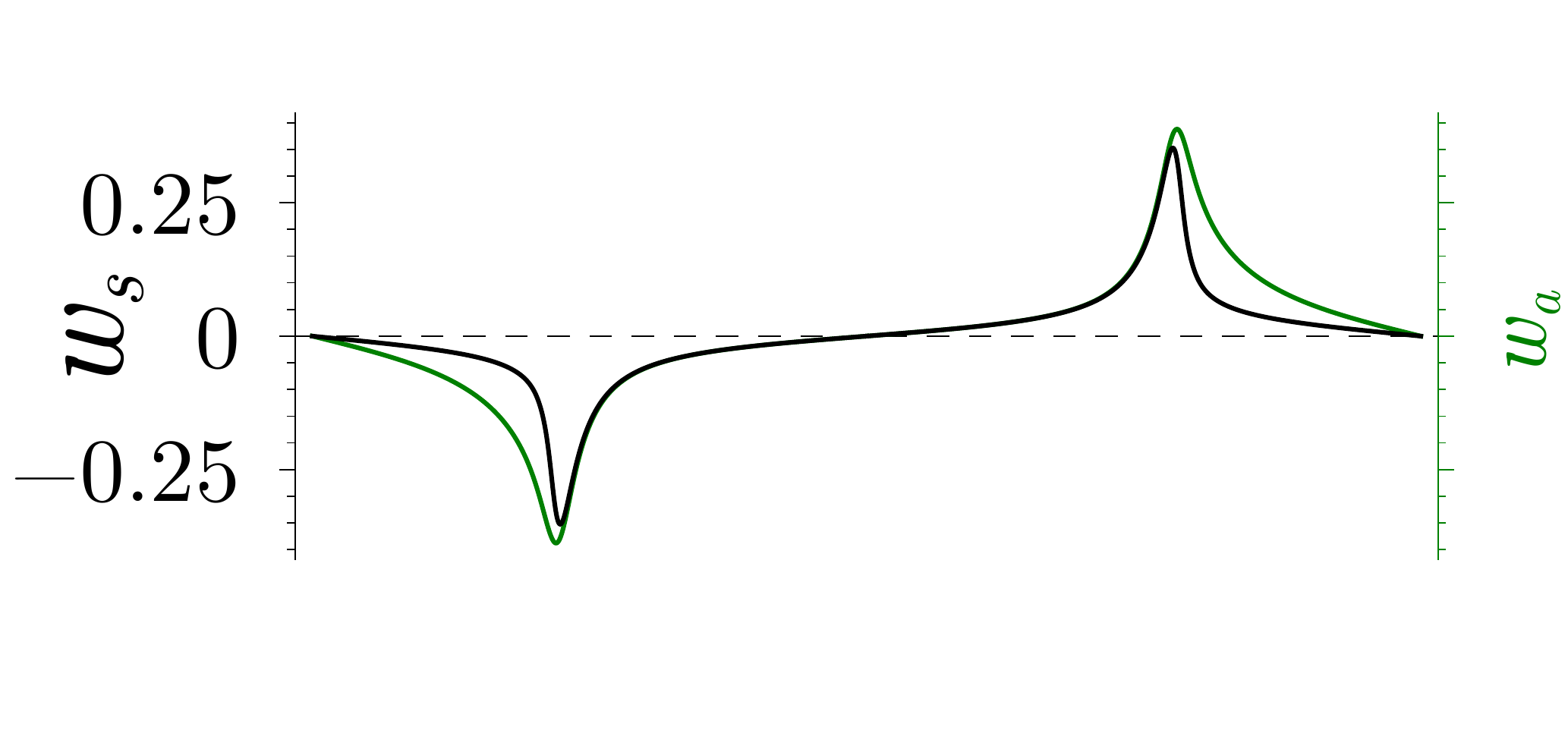} &   \includegraphics[width=0.24\textwidth]{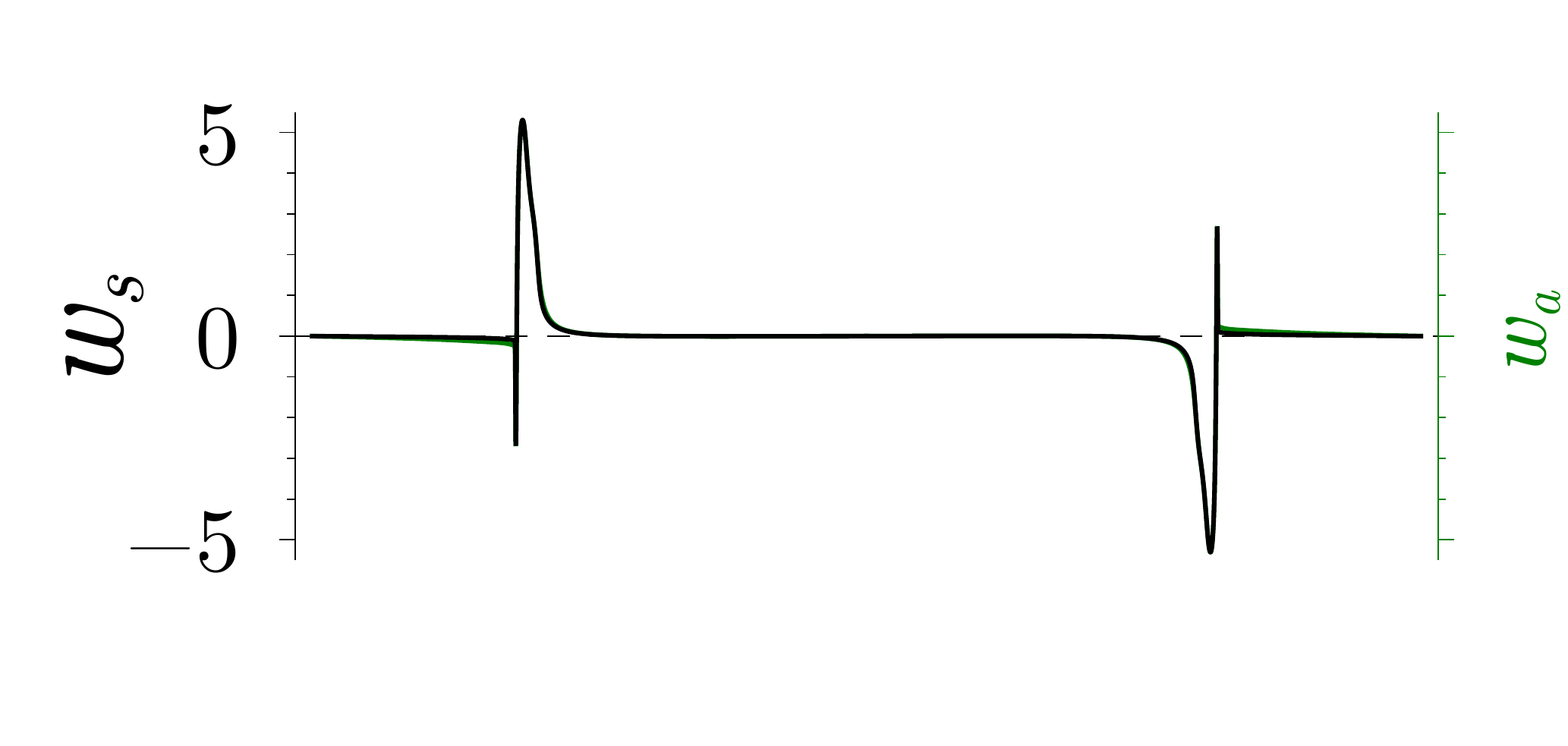} \\
        \vspace{-0.8cm} & & & \\
            \includegraphics[width=0.24\textwidth]{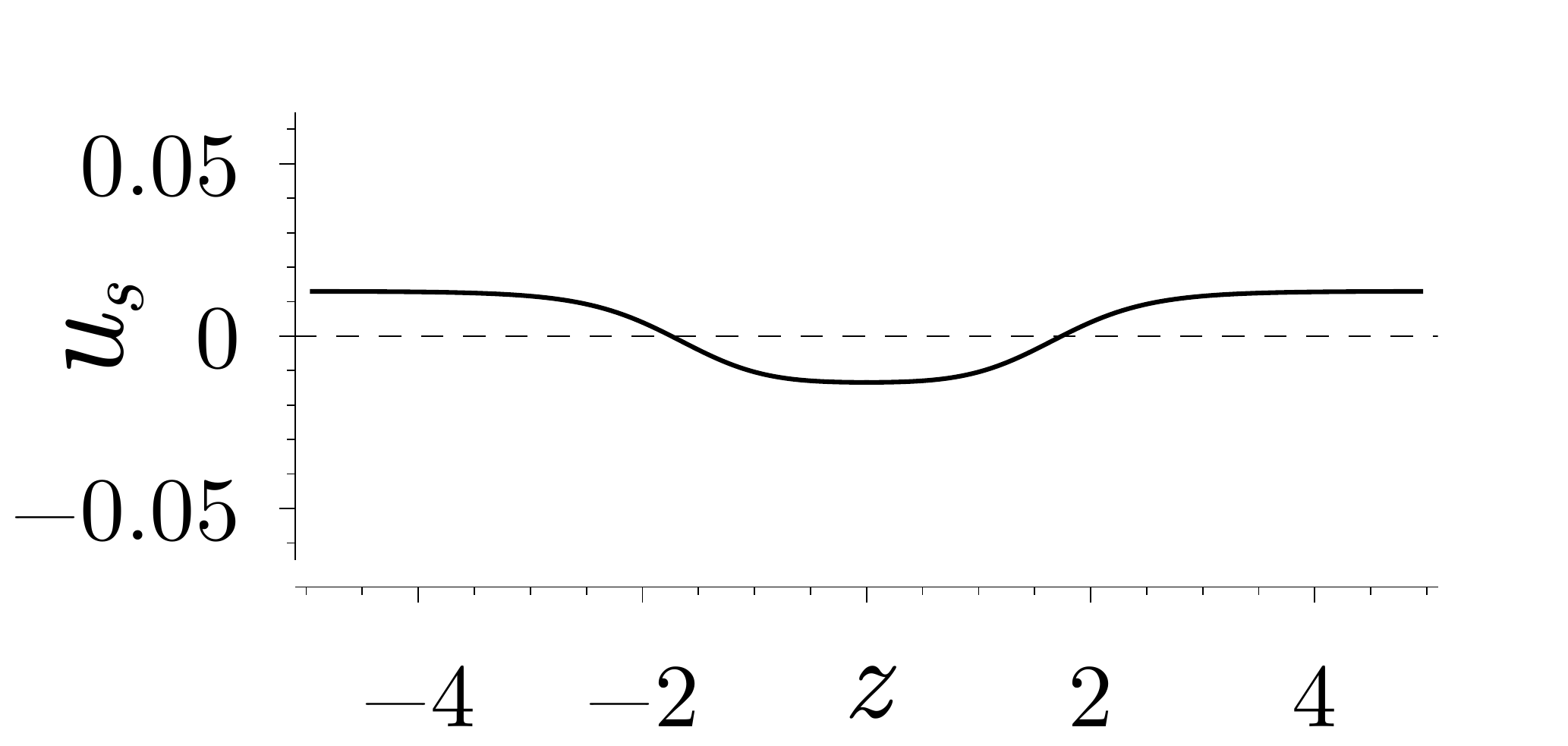} &   \includegraphics[width=0.24\textwidth]{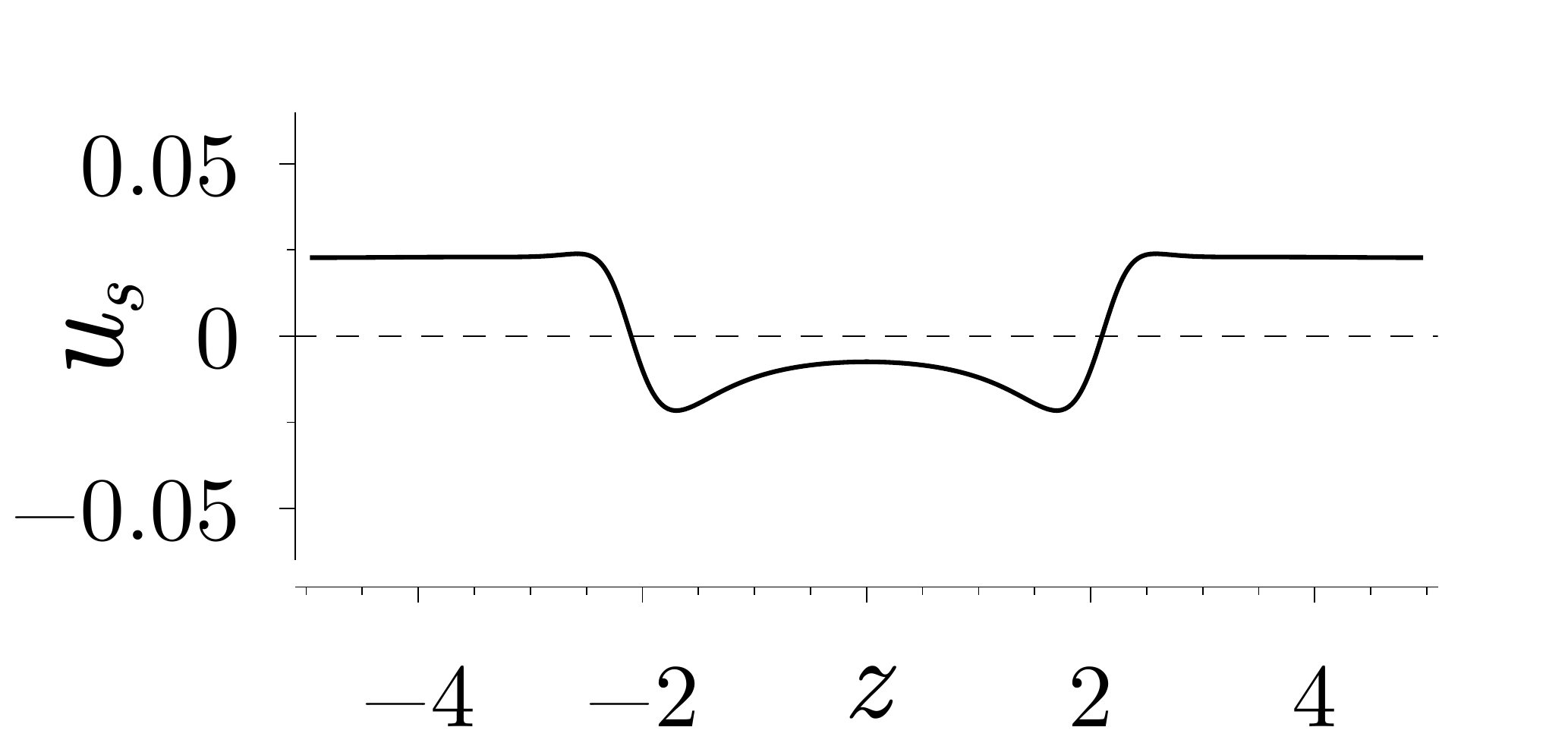} &   \includegraphics[width=0.24\textwidth]{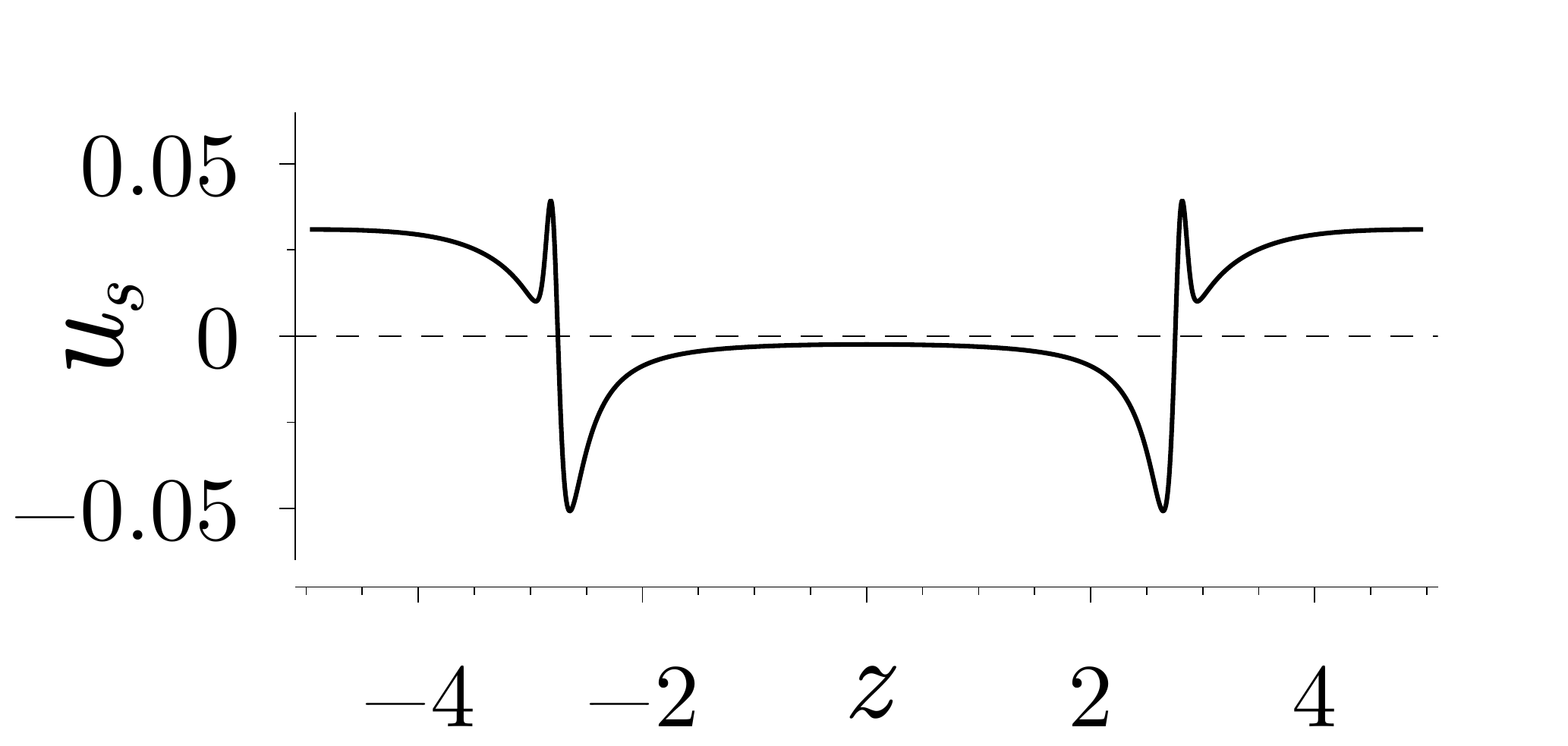} &   \includegraphics[width=0.24\textwidth]{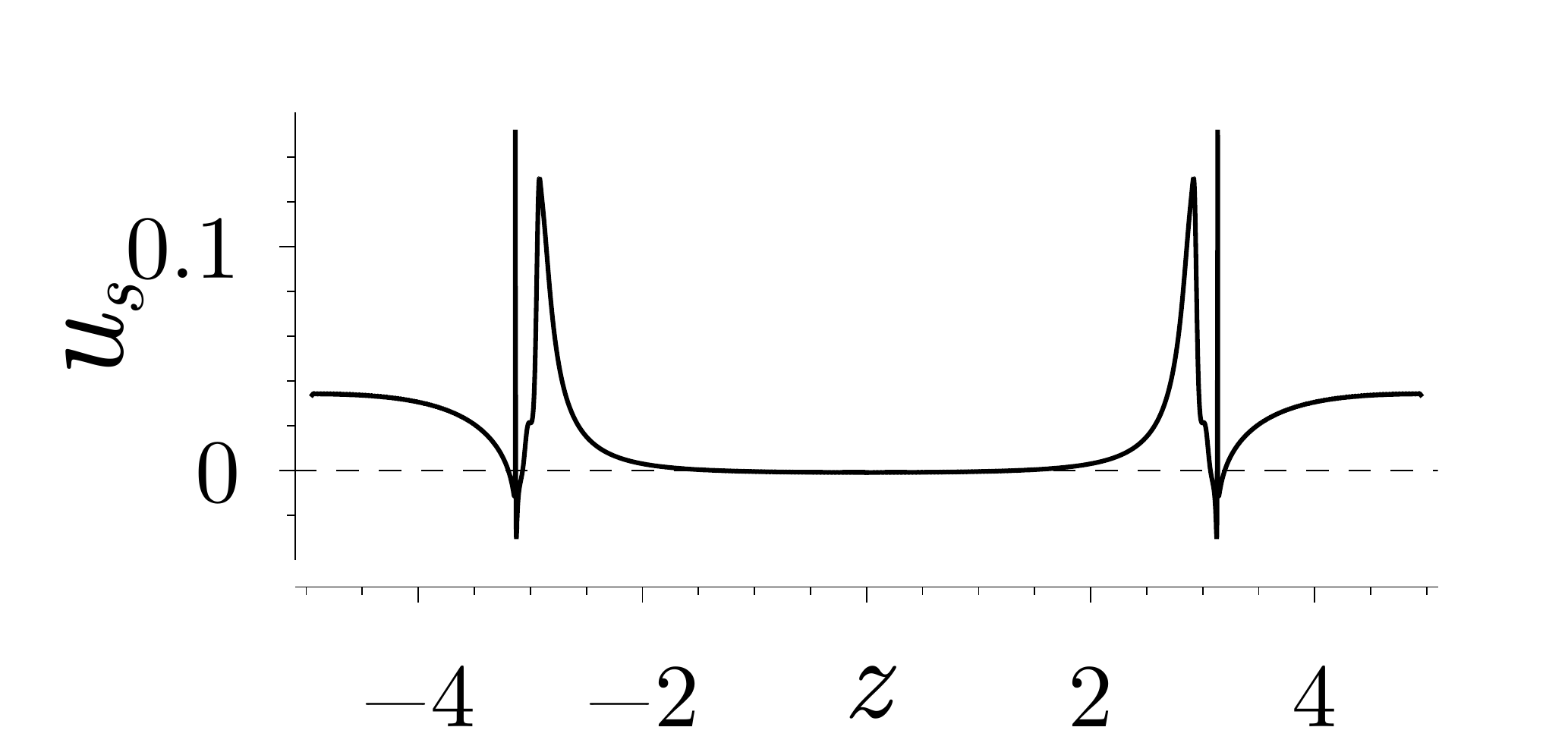} \\
($e$) $t = 230$ & ($f$) $t = 245$ & ($g$) $t = 250.8$  & ($h$) $t = 252.73$ \\[4pt]
\end{tabular}
\includegraphics[width=1\textwidth]{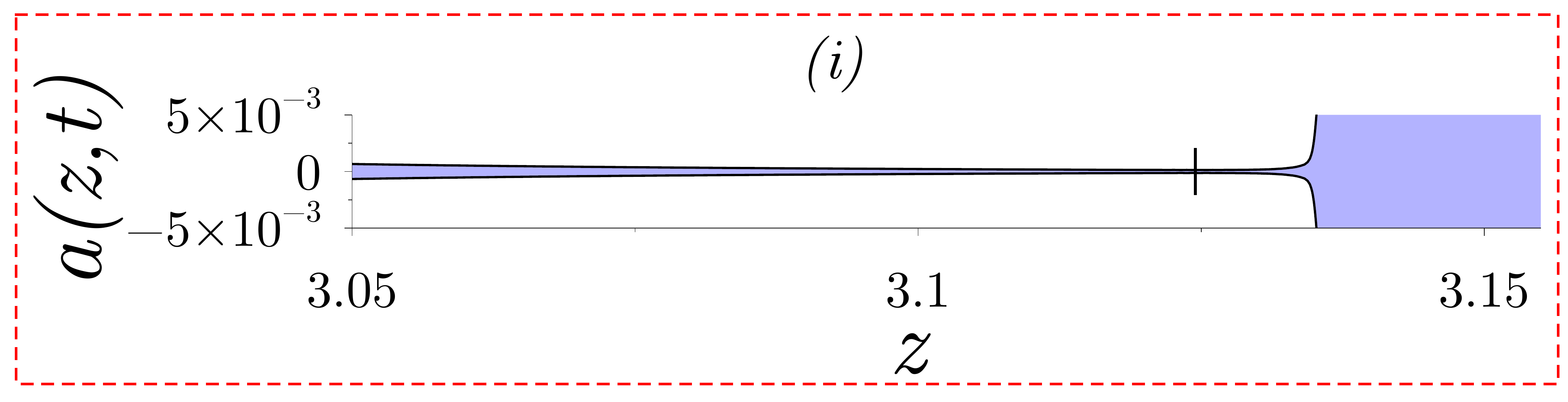}
\caption{\label{fig:figure7} (Colour online) Temporal evolution of the liquid thread radius $a$ (upper row), of the axial velocities at the free surface, $w_s$, and at the axis, $w_a$ (middle row) and the radial surface velocity $u_s$ (bottom row), for $\epsilon = 10^{-3}$, $\Ela = 0$,  ($a$--$d$) $\Lap = 0.01$, $k = k_m = 0.150$, and ($e$--$h$) $\Lap = 100$, $k = k_m = 0.635$. The vertical lines in each last snapshot indicate the axial position $z_{\min}$ of minimum radii $a_{\min}$, being $z_{\min} = 1.49$ and $a_{\min} = 3.63 \times 10^{-5}$ for $\Lap = 0.01$, and $z_{\min} = 3.12$ and $a_{\min} = 1.29 \times 10^{-4}$ for $\Lap = 100$. ($i$) Zoomed region close to the neck at the instant shown in ($h$).}
\end{figure}

\subsubsection{Analysis of the temporal evolution of clean interfaces}

To present the dynamics of satellite droplet formation, we take as reference cases the two canonical temporal evolutions of clean interfaces ($\Ela=0$) illustrated in figure~\ref{fig:figure7}, for $\Lap = 0.01$ in ($a$--$d$), close to the Stokes limit, and for $\Lap = 100$ in ($e$--$h$), an almost inviscid case close to the Euler limit (as shown in~\S\ref{subsubsec:scaling_laws}). Specifically, we plot snapshots at different times, indicated in the labels, of the jet radius $a$ (upper rows), the axial surface velocity $w_s$ (middle rows, black lines), the axial velocity at the centreline $w_a$ (middle rows, green lines) and the radial surface velocity $u_s$ (bottom rows). In both cases the initial disturbance amplitude is very small, $\epsilon=10^{-3}$, and thus the initial evolution is triggered by the Plateau--Rayleigh instability mechanism, and can be described with linearised theory. This initial stage is not shown in figure~\ref{fig:figure7} for conciseness, but it can be appreciated in figure~\ref{fig:figure2}($a$). The initial disturbance, of most amplified wavelength $k_m$, creates an axial capillary pressure gradient that induces a flow from the valley to the crest of the wave. The latter mechanism finally leads to the break-up of the liquid thread and the formation of two main drops with either a liquid thread or a satellite droplet in between.

A key feature that determines the nonlinear evolution of the destabilised thread is the fact that the axial curvature makes the capillary pressure gradient to be locally larger in the regions that connect the central part of the thread with the growing crests, as evidenced by the surface and axis velocities in the snapshot ($f$). This enhanced pressure gradient drives liquid towards the crests faster in the nearby regions than in the central part, and explains the appearance of two local minima in the jet radius for large enough values of $\Lap$, as can be clearly appreciated in snapshots ($f$,$g$) for $\Lap=100$. In addition, the axial position of the minimum radii $z_{\min}$ is advected with the flow along with the maximum pressure gradient, i.e. towards higher values of $z$ as time advances~\citep{Ashgriz1995,Castrejon2015}. These two local minima become the two neck regions where pinch-off takes place, leading to the formation of an oval-shaped satellite droplet, as can be observed in the snapshot ($h$). This scenario applies to cases where $\Lap \gg 1$ (figure~\ref{fig:figure7} $e$--$h$), for which the viscous stress is negligible, and the capillary pressure gradient is entirely transferred to liquid inertia leading to a self-accelerated process.

In contrast, when $\Lap \ll 1$ (figure~\ref{fig:figure7} $a$--$d$), the viscous dissipation inhibits the growth of higher harmonics, and larger pressure gradients are needed to overcome the viscous damping, as has already been pointed out by~\cite{Ashgriz1995}. Hence, the axial movement of the minimum radius is delayed by the viscous stress, since it weakens the capillary pressure and the concomitant liquid advection. Consequently, the central region shrinks almost uniformly until the last instants before break-up, giving rise to long and thin filaments without the formation of appreciable satellite droplets before detachment. Notice also that, for $\Lap\ll 1$, the axial velocities at the centreline, $w_a$, and at the interface, $w_s$, are almost equal (green and black lines in figures~\ref{fig:figure7} $a$--$d$, respectively), indicating that the radial profile of axial velocity inside the thread is nearly uniform at low Laplace numbers.


\begin{figure}
\centering
\includegraphics[height=0.5\textwidth]{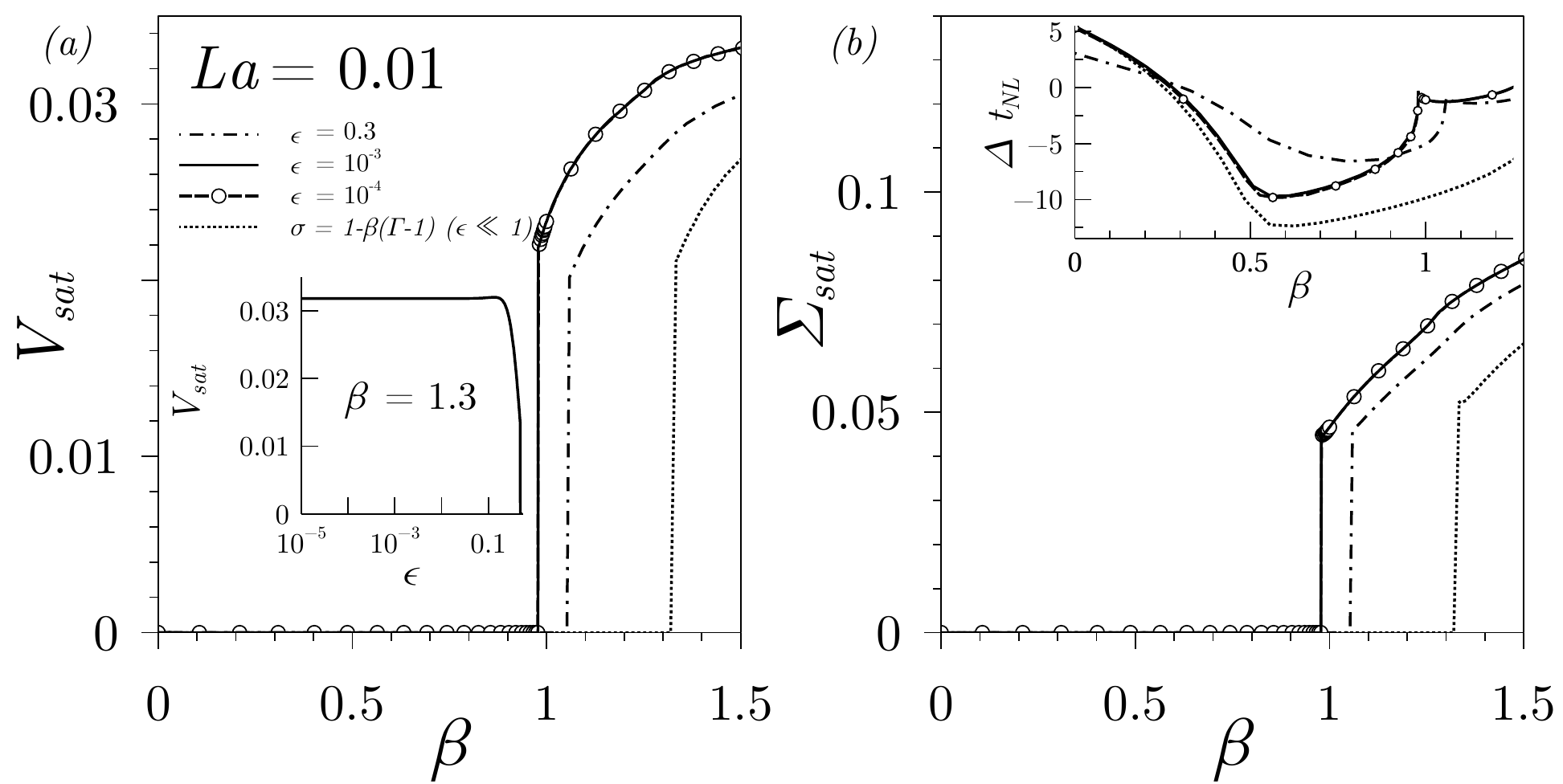}
\includegraphics[height=0.5\textwidth]{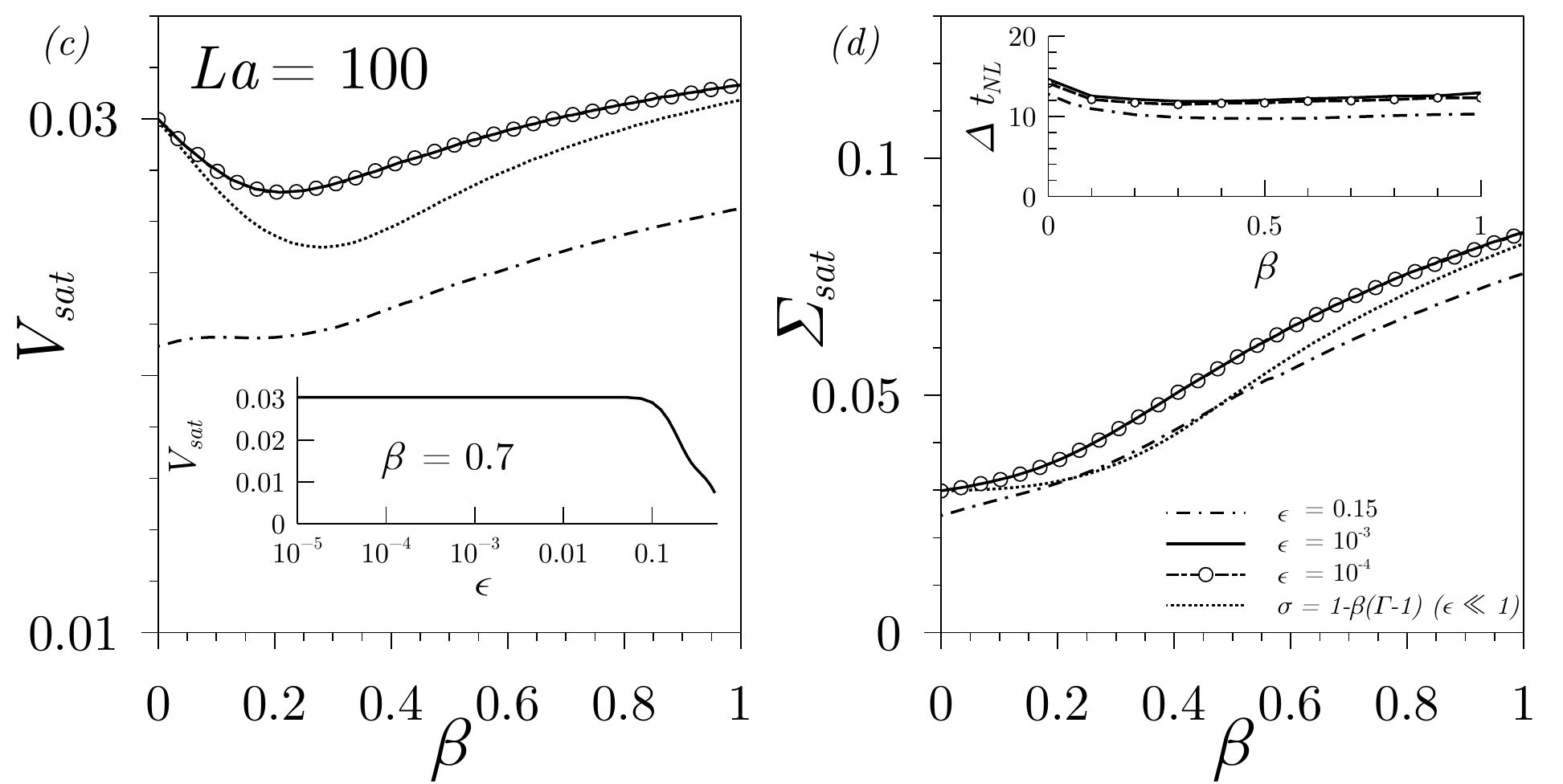}
\caption{\label{fig:figure8} (Colour online) $\Vsat$ and $\Ssat$ as a function of $\Ela$ for ($a$,$b$) $\Lap = 0.01$ and ($c$,$d$) $\Lap = 100$, and for different values of the initial perturbation amplitude $\epsilon$ indicated in the legend. The dotted line represents the results of using the linear equation of state $\sigma = 1 - \Ela (\Gamma-1)$~\citep{Dravid2006}. The insets show $\Vsat$ as a function of $\epsilon$ in logarithmic scale for ($a$) $\Ela = 1.3$ and ($c$) $\Ela = 0.7$, demonstrating that the final stage of the liquid thread just before pinch-off becomes independent of $\epsilon$ when its value is sufficiently small. The insets in ($b$, $d$) show the nonlinear correction to the break-up time $\DtNL$ as a function of $\Ela$.}
\end{figure}

\subsubsection{Analysis of the temporal evolution of surfactant-laden interfaces}

To explain the different trends and transitions observed in figure~\ref{fig:figure5}, let us first focus on the effect of $\Ela$ for the particular cases of $\Lap=0.01$ and $\Lap=100$. Figure~\ref{fig:figure8} shows $\Vsat$ ($a$,$c$), $\Ssat$ ($b$,$d$) and $\DtNL$ (insets in $b$ and $d$), as functions of $\Ela$ for $\Lap = 0.01$ in ($a$,$b$), and for $\Lap = 100$ in ($c$,$d$). In addition, we have computed the results for several values of $\epsilon$ indicated in the legend of figure~\ref{fig:figure8}($a$), with the aim of clearly establishing the limit of infinitesimal disturbances. In particular, figure~\ref{fig:figure8} shows that $\Vsat$, $\Ssat$ and $\DtNL$ become independent of $\epsilon$ provided that $\epsilon$ is small enough, as stated before. Indeed, the insets in ($a$,$c$), which show the dependence of $\Vsat$ on $\epsilon$, clearly demonstrate that the value of $\Vsat$ reaches the infinitesimal-disturbance plateau when $\epsilon \lesssim 0.1$. Figure~\ref{fig:figure8} also displays the results obtained with the linear equation of state $\sigma = 1 - \Ela(\Gamma-1)$ (dotted line), instead of the nonlinear one~\eqref{eq:sigma_gamma}. It is important to note that the use of the linear equation of state leads to substantial quantitative differences with respect to the nonlinear one~\eqref{eq:sigma_gamma}. In particular, the linear equation underestimates the values of $\Vsat$ and $\Ssat$ considerably. We note also that we compared our numerical results using the linear equation of state with those reported by~\cite{Dravid2006}, finding very good agreement. However, their results were calculated for wavenumbers $k\neq k_m$, and the satellite droplet was measured by those authors by means of the thread radius at $z = 0$ close to pinch-off, instead of using either $V_{sat}$ or $R_{sat}$.

Figures~\ref{fig:figure8}($a$,$b$) show the discontinuous transition in $\Vsat$ and $\Ssat$ that occurs when $\Ela$ is increased above the critical value $\Ela_c (\Lap = 0.01) = 0.978 \pm 0.0003$. For $\Ela > \Ela_c (\Lap = 0.01)$ a satellite drop centred at $z = 0$ is formed, trapping approximately $2.1 \%$ of the total volume of liquid and $4.5 \%$ of the total mass of surfactant.
\begin{figure}
\hspace{-1.6cm}
\setlength{\tabcolsep}{-0.5pt}
\begin{tabular}{cccc}
  \includegraphics[width=0.3\textwidth]{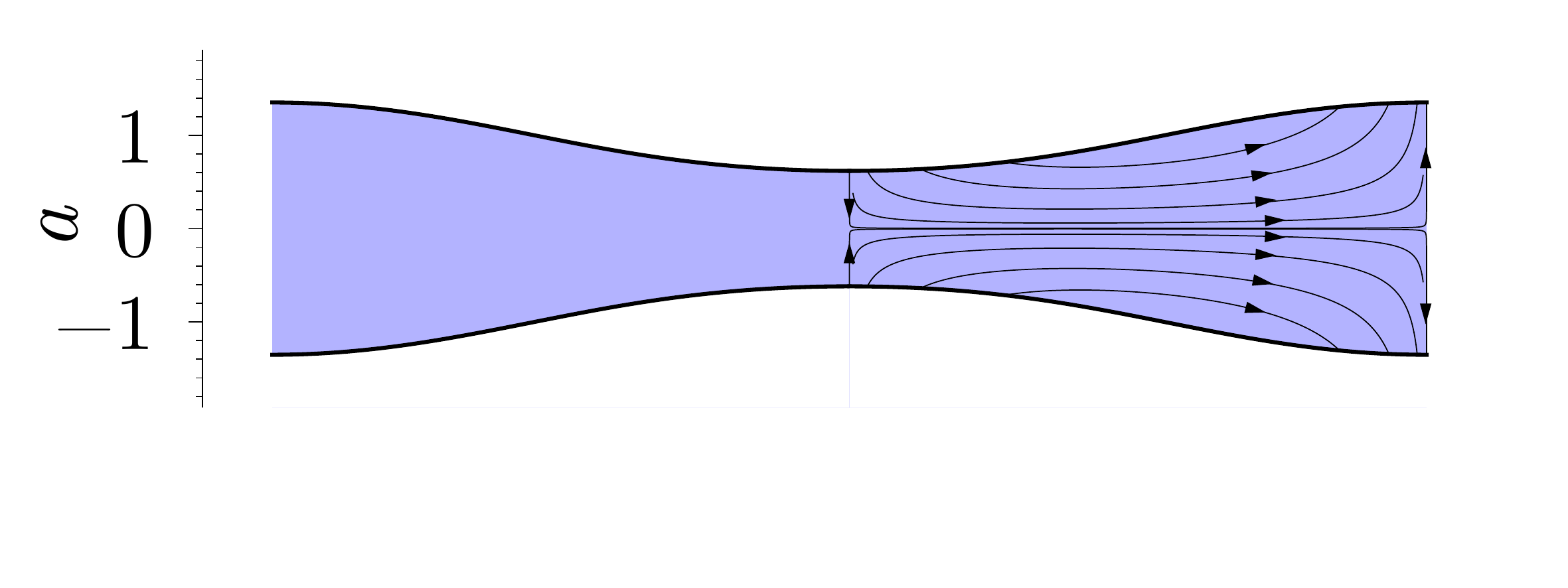} &   \includegraphics[width=0.3\textwidth]{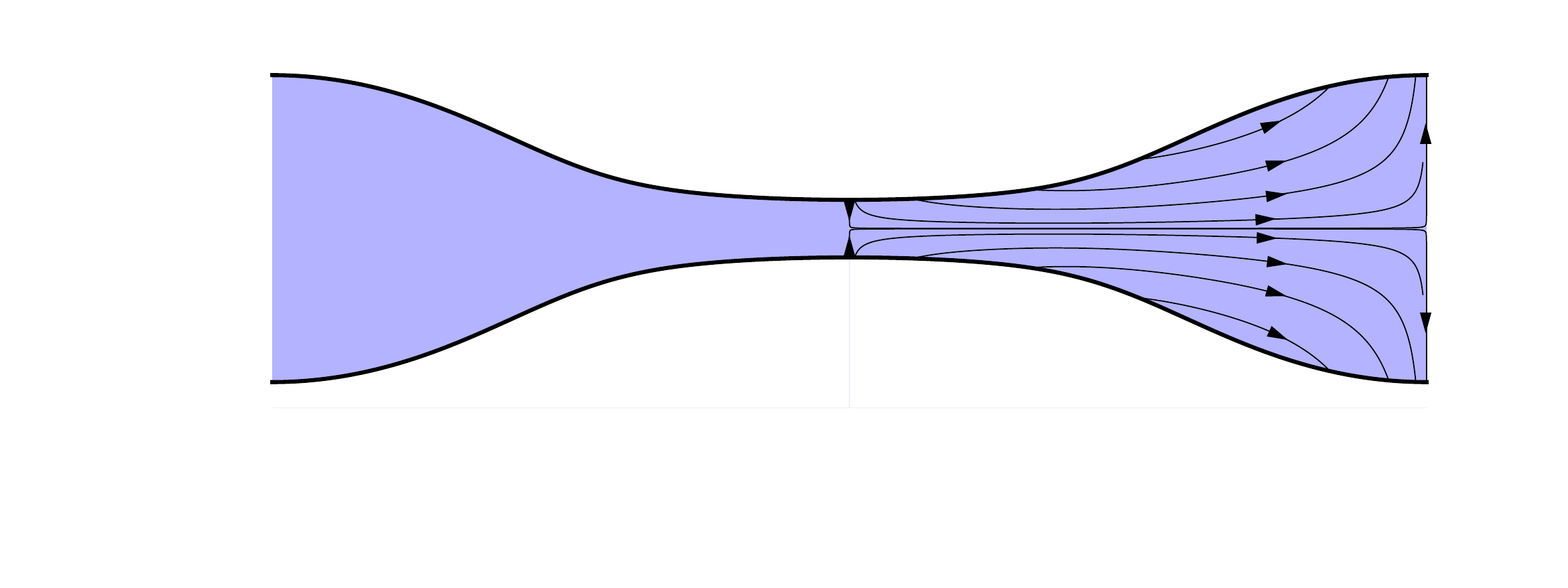} &  \includegraphics[width=0.3\textwidth]{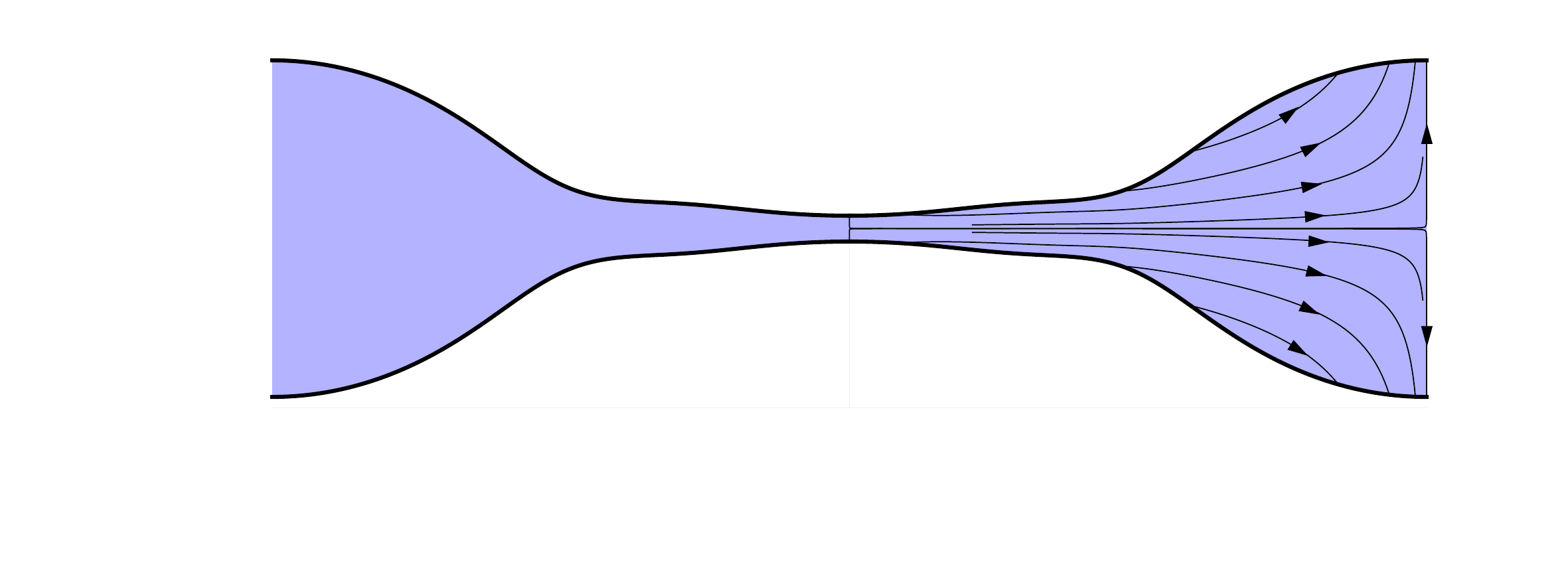}  & \includegraphics[width=0.3\textwidth]{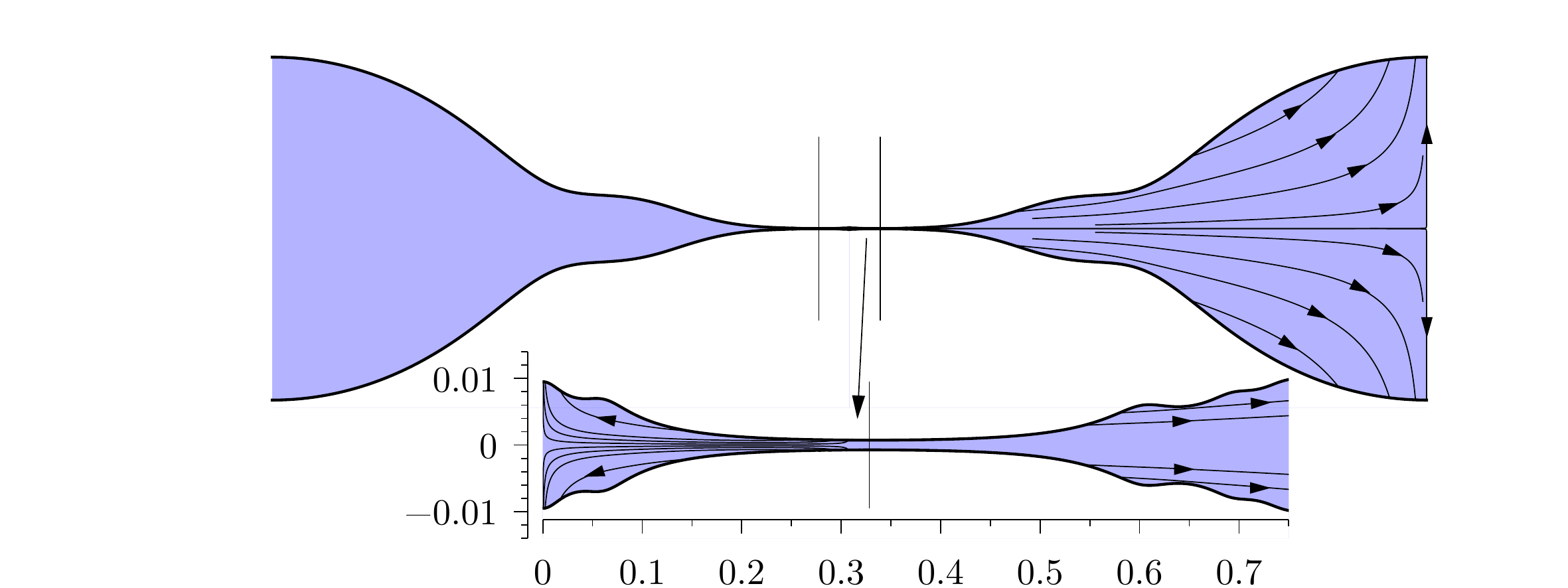}  \\
  \vspace{-0.45cm} & & & \\
   \includegraphics[width=0.3\textwidth]{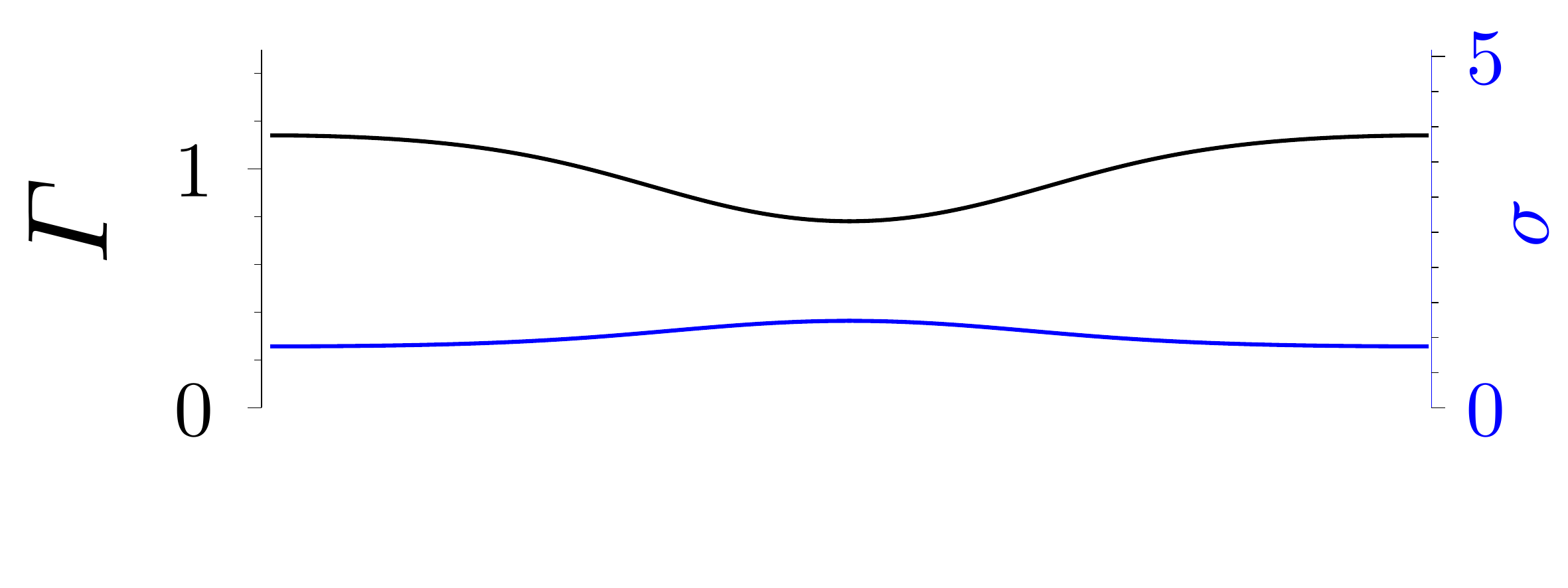} &   \includegraphics[width=0.3\textwidth]{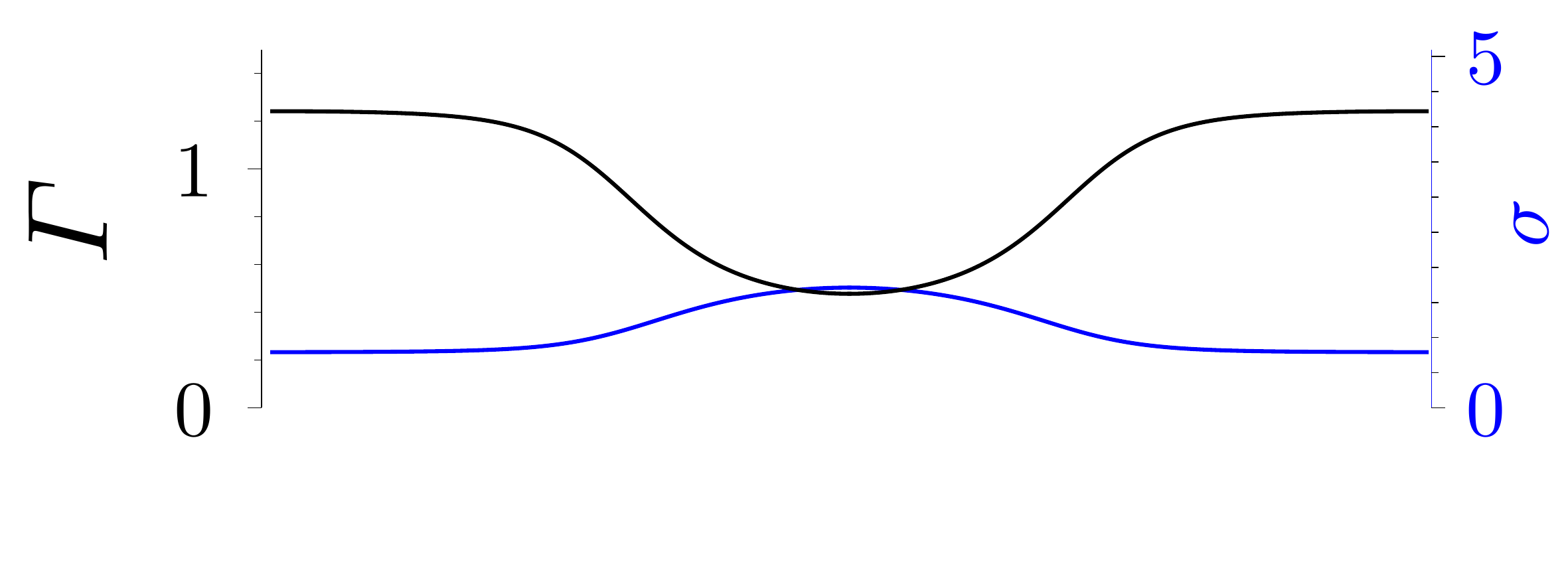} &   \includegraphics[width=0.3\textwidth]{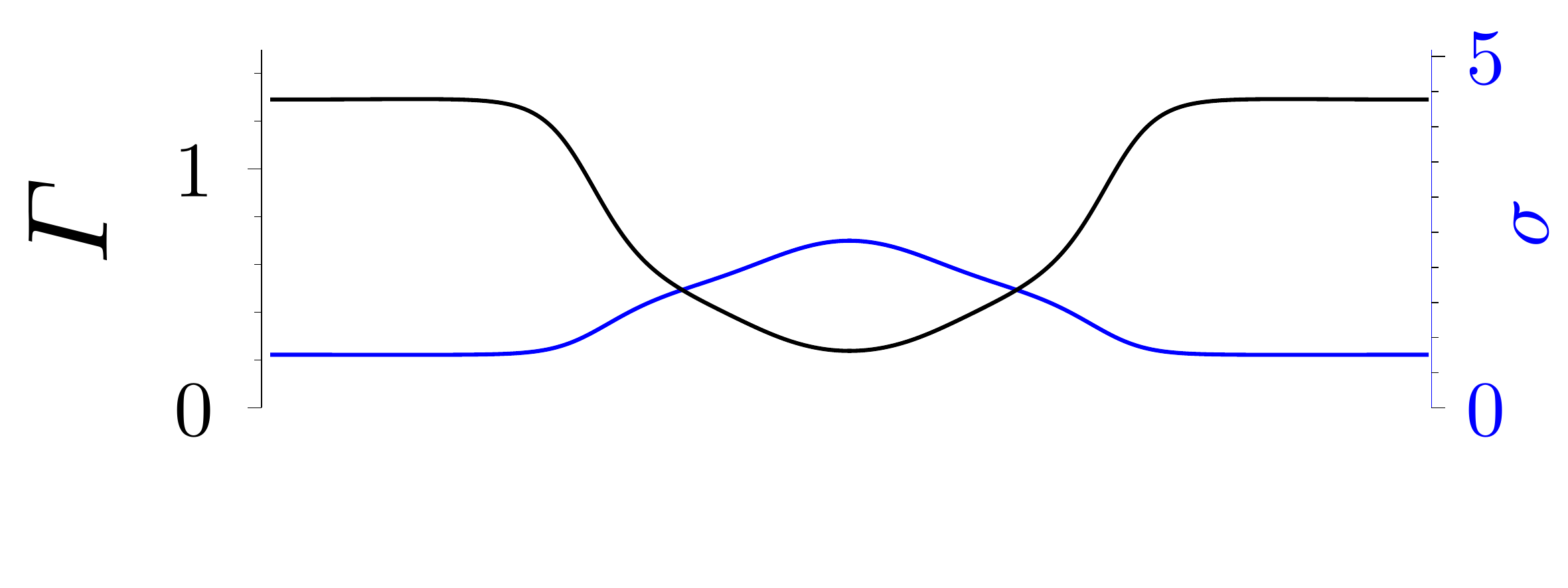} &   \includegraphics[width=0.3\textwidth]{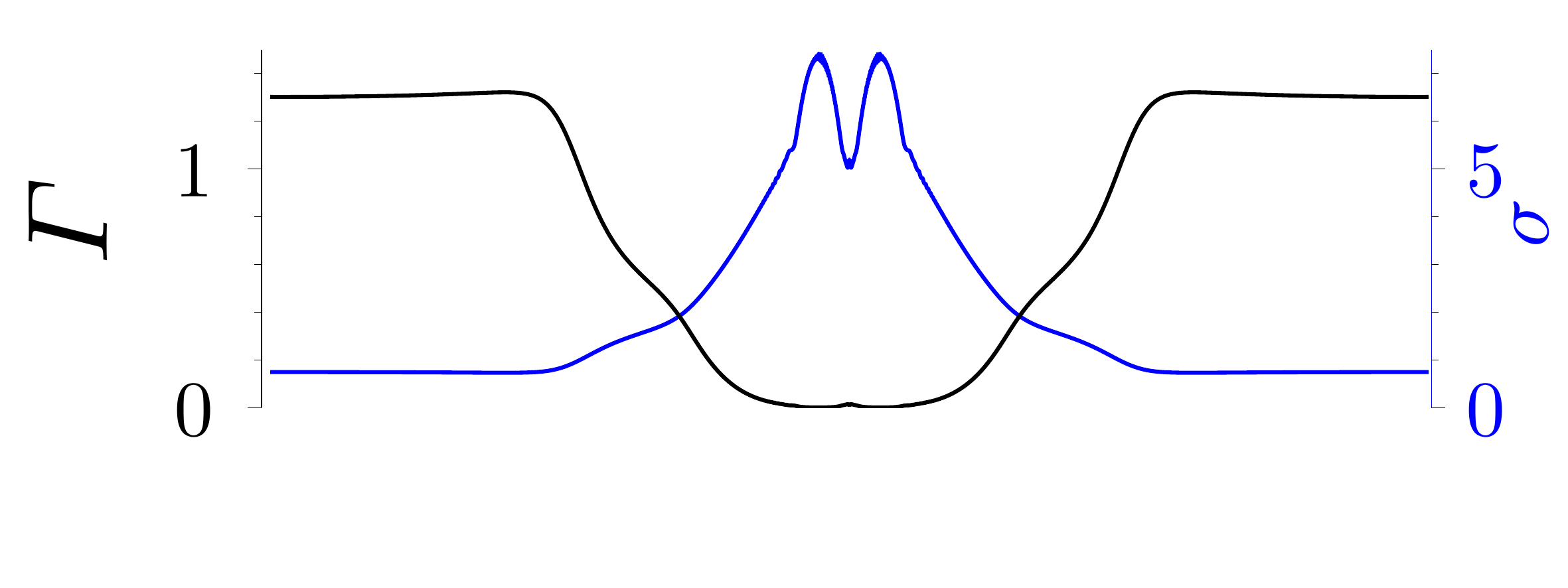} \\
          \vspace{-0.8cm} & & & \\
      \includegraphics[width=0.3\textwidth]{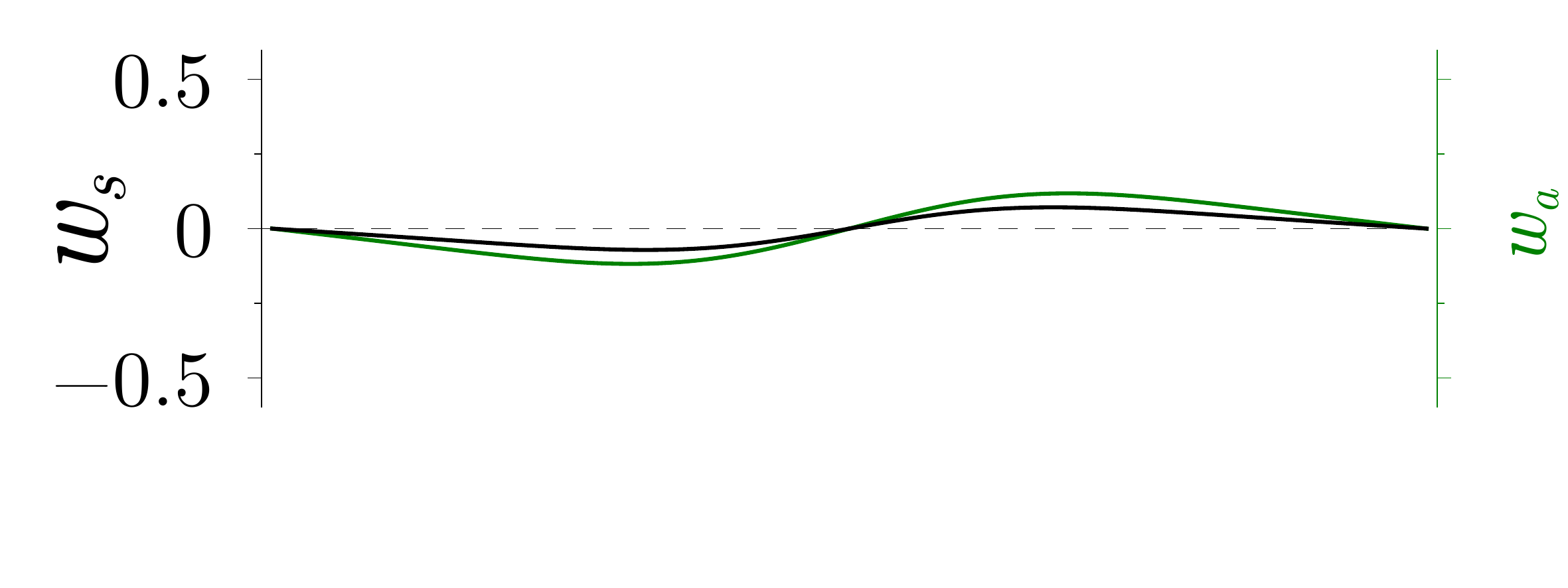} &   \includegraphics[width=0.3\textwidth]{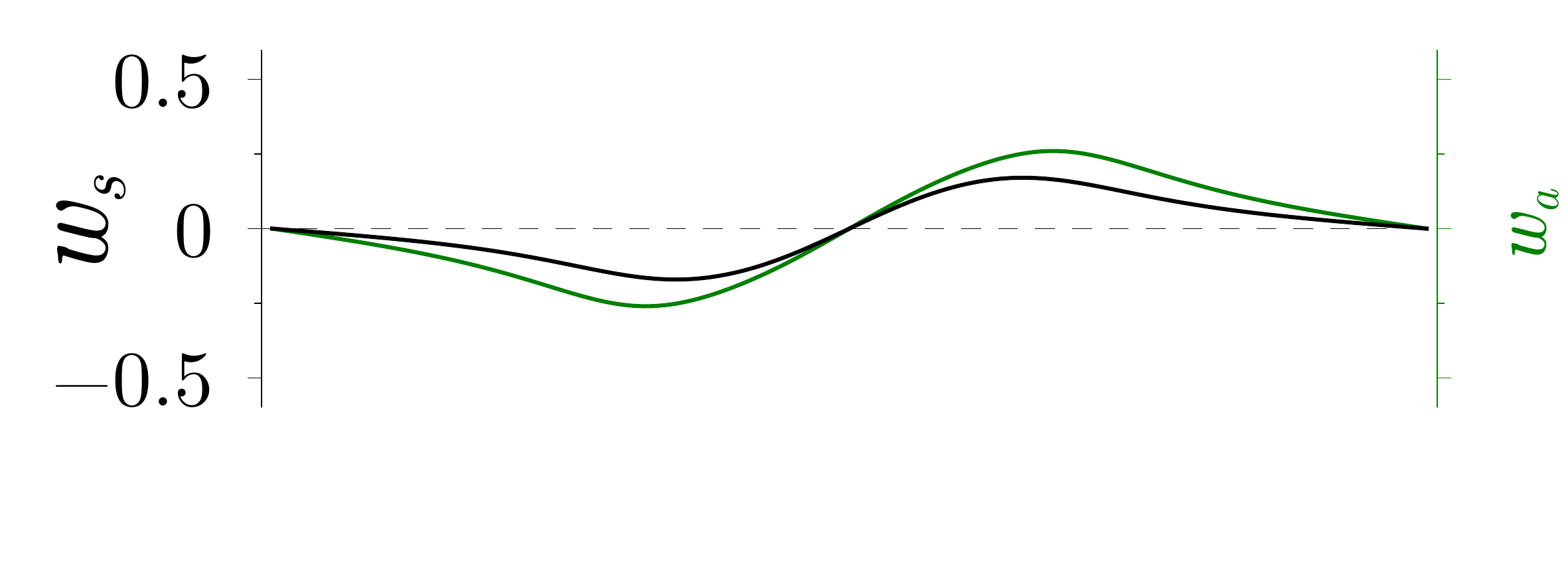} &   \includegraphics[width=0.3\textwidth]{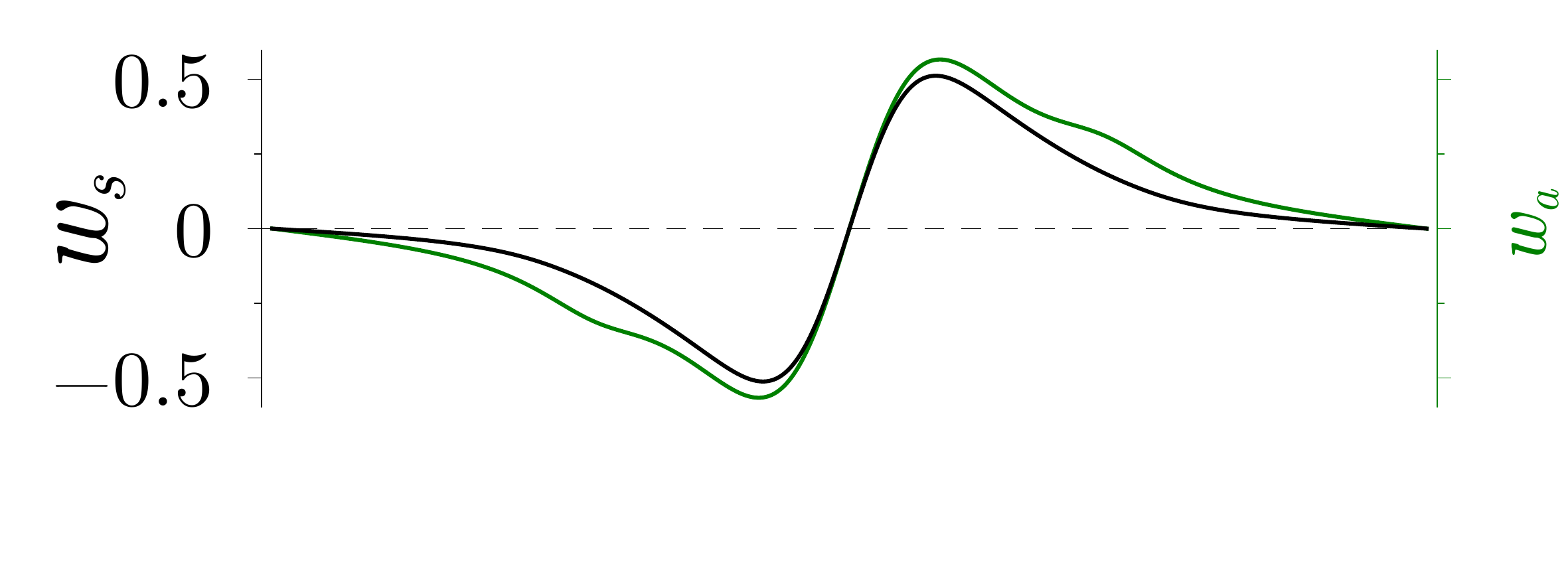} &   \includegraphics[width=0.3\textwidth]{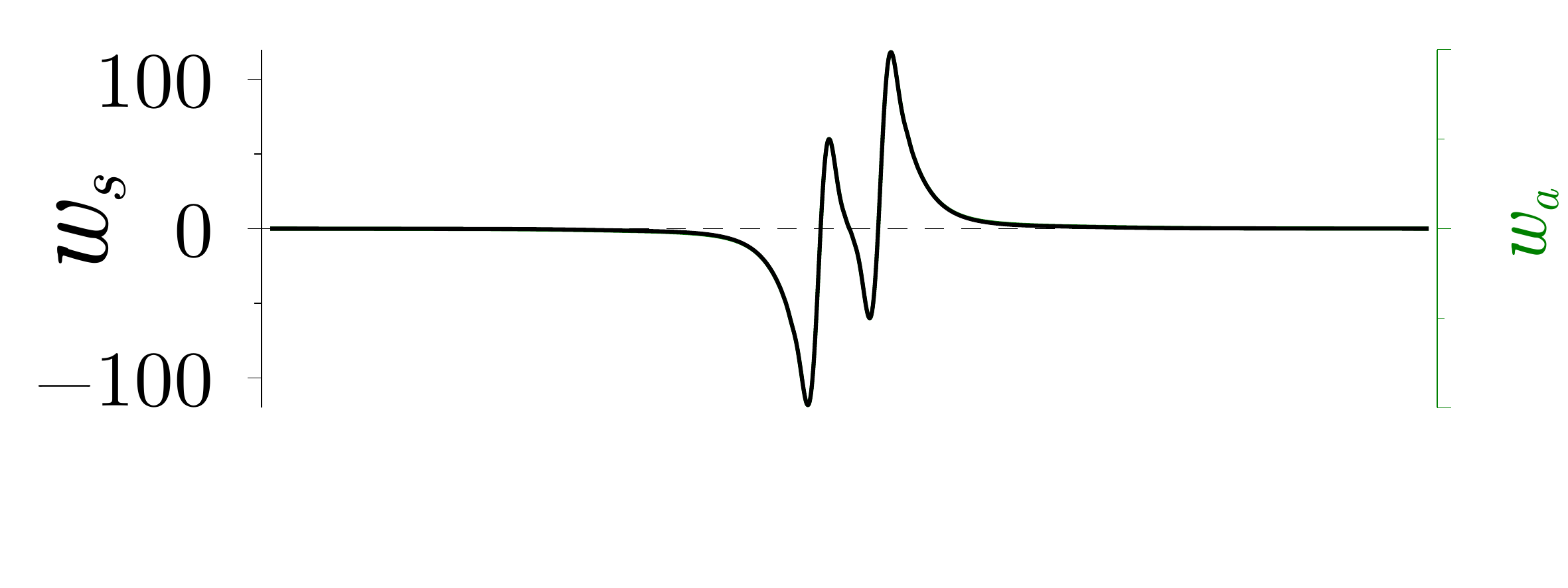} \\
             \vspace{-0.8cm} & & & \\
\includegraphics[width=0.3\textwidth]{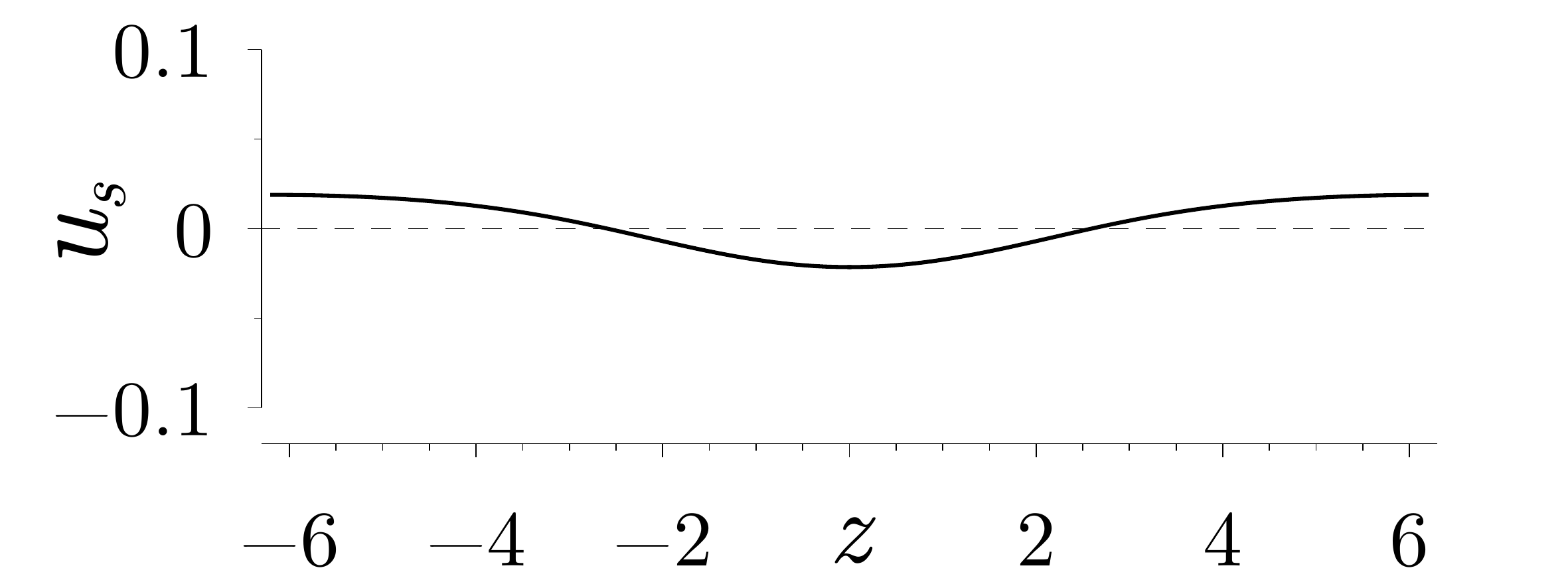} &   \includegraphics[width=0.3\textwidth]{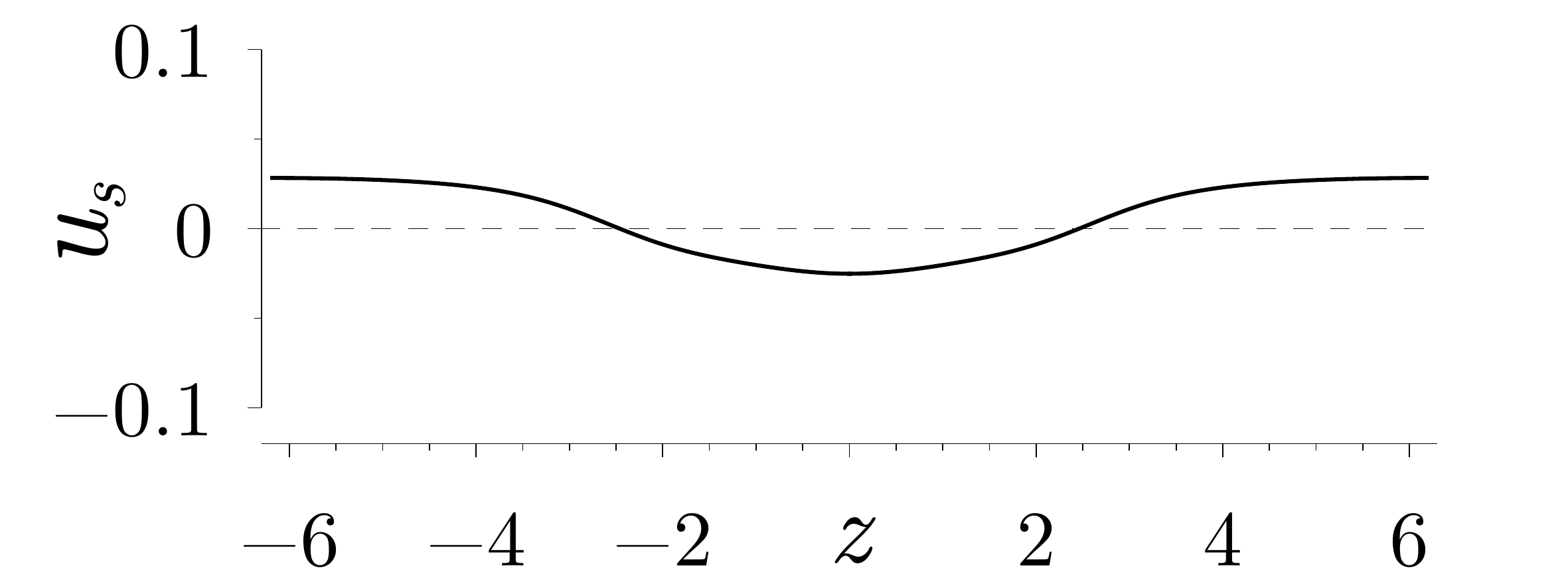} &   \includegraphics[width=0.3\textwidth]{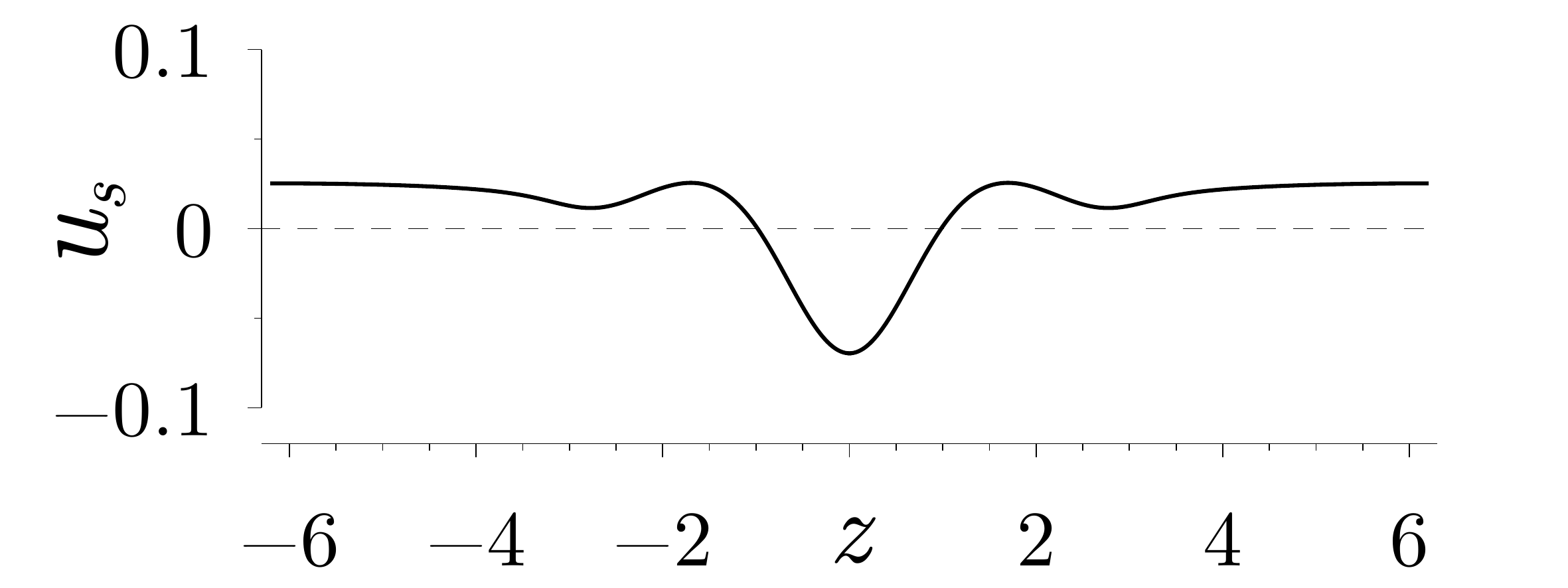} &   \includegraphics[width=0.3\textwidth]{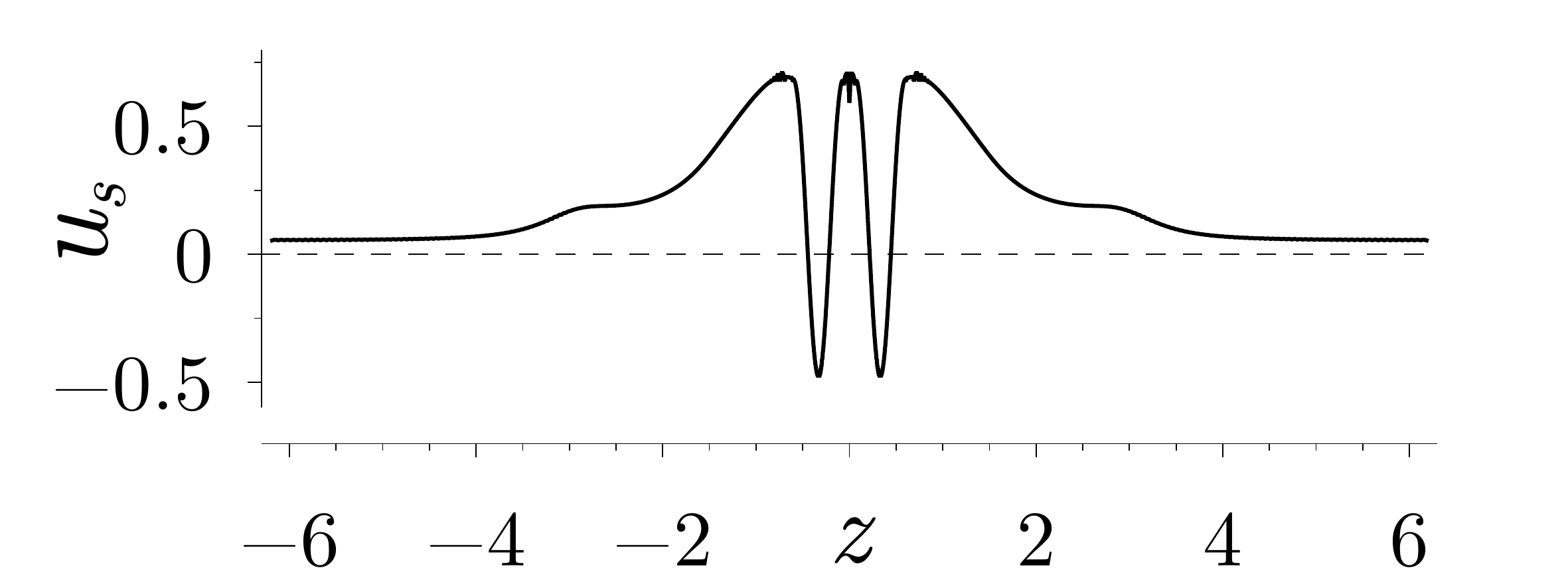} \\
($a$) $t = 100$ & ($b$) $t = 112$ & ($c$) $t = 118$  & ($d$) $t = 119.03$ \\[4pt]
\end{tabular}
\caption{\label{fig:figure9} (Colour online) Temporal evolution of the liquid thread radius $a$ (first row), surfactant concentration $\Gamma$ (second row, black lines), surface tension $\sigma$ (second row, blue lines), axial velocity at the interface $w_s$ (third row, black lines) and at the centreline $w_a$ (third row, green lines), and radial surface velocity $u_s$ (fourth row), for $\Lap = 0.01$, $\epsilon = 10^{-3}$ and $\Ela = 0.960 < \Ela_c(\Lap=0.01) $, with $k = k_m = 0.508$. The vertical line in the last snapshot of $a$ indicates the position of $z_{\min}$. Here $z_{\min} = 0.33$ and $a_{\min} = 7.29 \times 10^{-4}$.}
\end{figure}

\subsubsection{Physical explanation of the discontinuous transition}

\begin{figure}
\hspace{-1.5cm}
\setlength{\tabcolsep}{-0.5pt}
\begin{tabular}{cccc}
\begin{tikzpicture} \centering
\begin{axis}[
hide axis,
scale only axis,
height=0pt,
width=0pt,
colormap/jet,
colorbar horizontal,
point meta min=0.6526484372286867,
point meta max=10.368367806839974,
colorbar style={
    height=1mm, width= 1.5cm, xticklabel pos=upper, xtick={0.653,5.7,10.36}
}]
\addplot [draw=none] coordinates {(0,0)};
\end{axis}
\node at (1.7,-0.4) {$p$};
\end{tikzpicture} & \begin{tikzpicture} \centering
\begin{axis}[
hide axis,
scale only axis,
height=0pt,
width=0pt,
colormap/jet,
colorbar horizontal,
point meta min=0.6530578401995142,
point meta max=9.862185848266225,
colorbar style={
    height=1mm, width= 1.5cm, xticklabel pos=upper, xtick={0.653,5.4,9.86}
}]
\addplot [draw=none] coordinates {(0,0)};
\end{axis}
\node at (1.7,-0.4) {$p$};
\end{tikzpicture} & \begin{tikzpicture} \centering
\begin{axis}[
hide axis,
scale only axis,
height=0pt,
width=0pt,
colormap/jet,
colorbar horizontal,
point meta min=0.6561843482397198,
point meta max=19.34919131152207,
colorbar style={
    height=1mm, width= 1.5cm, xticklabel pos=upper, xtick={0.66,9.5,19.34}
}]
\addplot [draw=none] coordinates {(0,0)};
\end{axis}
\node at (1.7,-0.4) {$p$};
\end{tikzpicture} &  \\
  \includegraphics[width=0.3\textwidth]{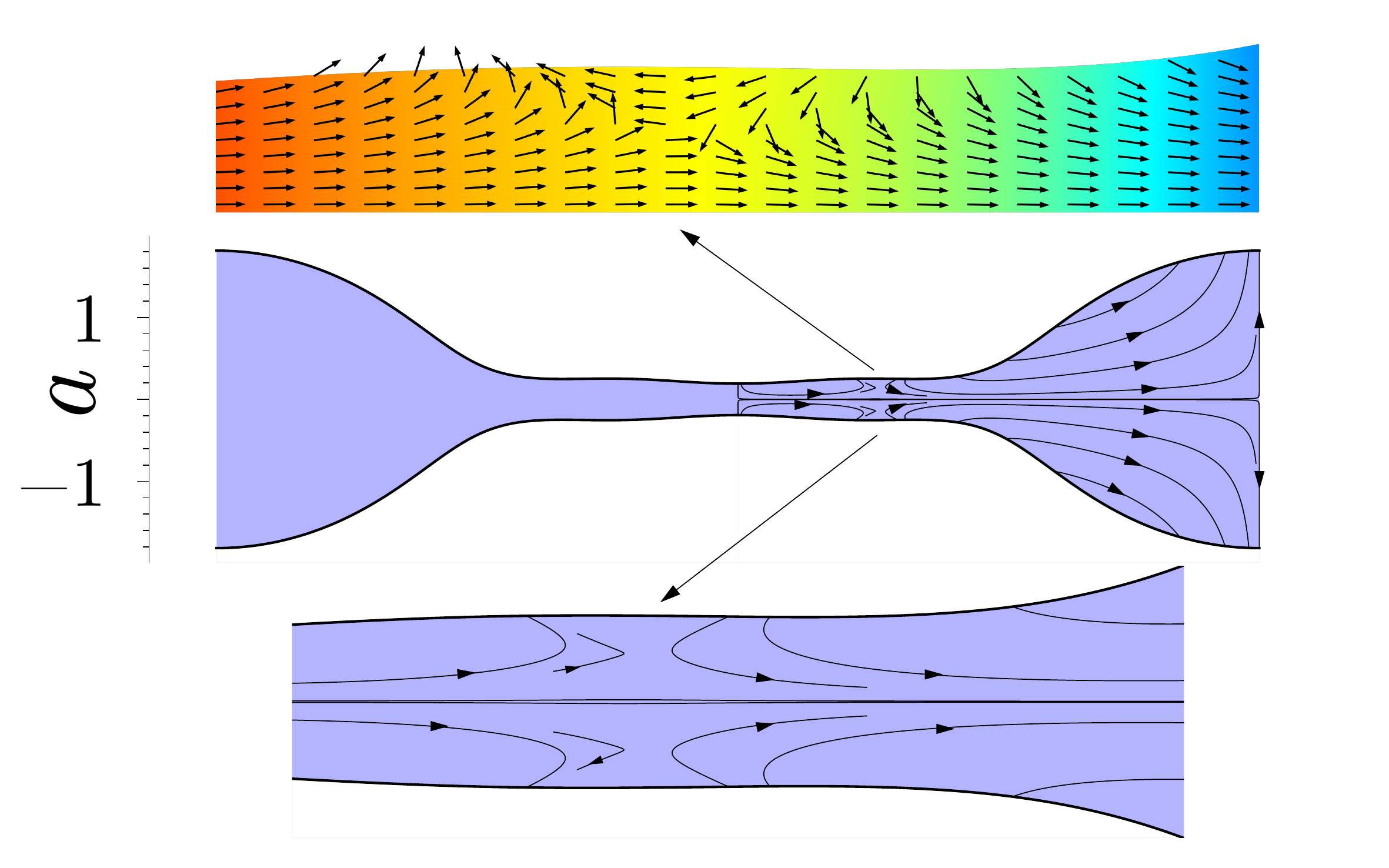} &   \includegraphics[width=0.3\textwidth]{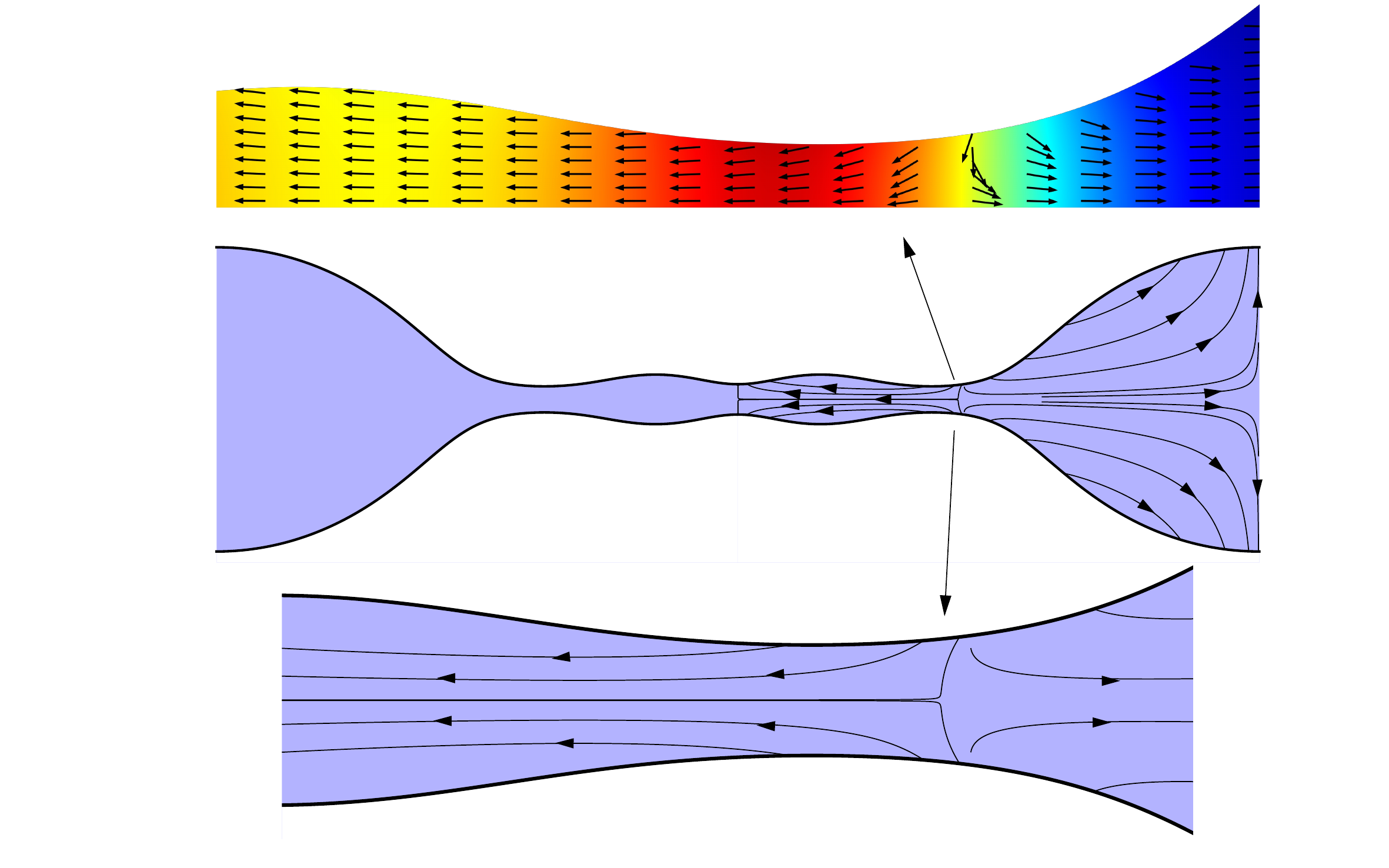} &  \includegraphics[width=0.3\textwidth]{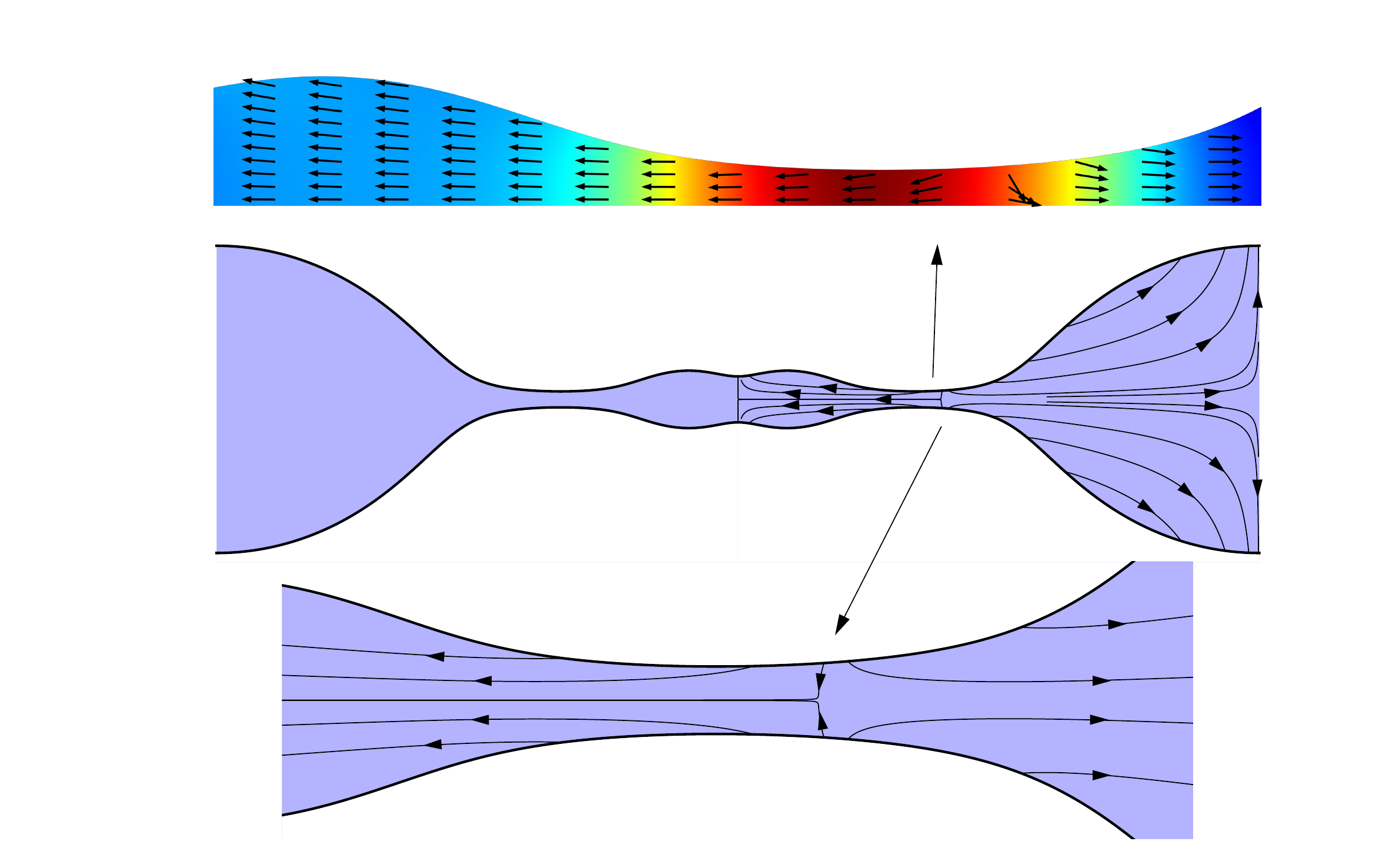}  & \includegraphics[width=0.3\textwidth]{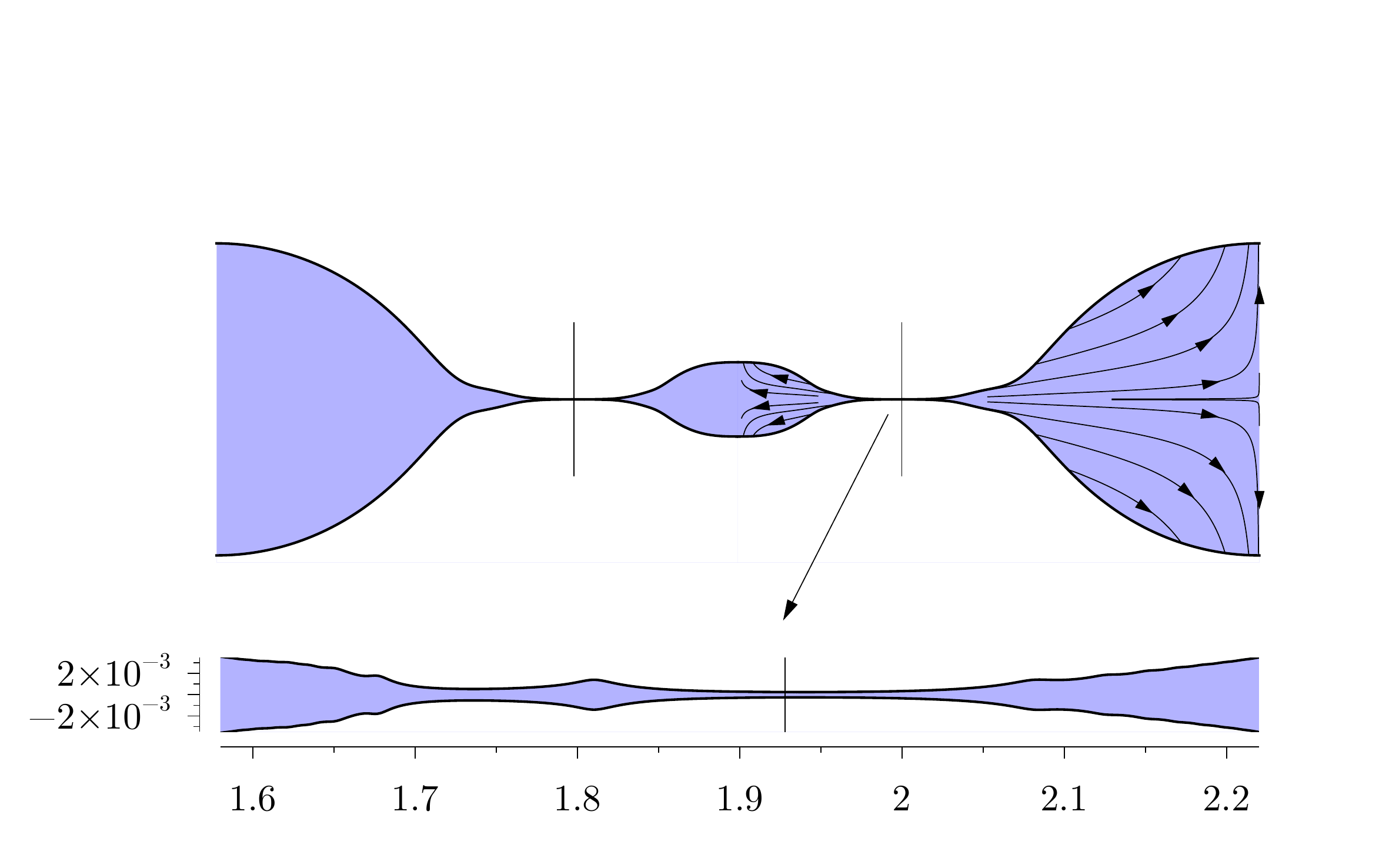} \\
       \vspace{-0.4cm} & & & \\
  \includegraphics[width=0.3\textwidth]{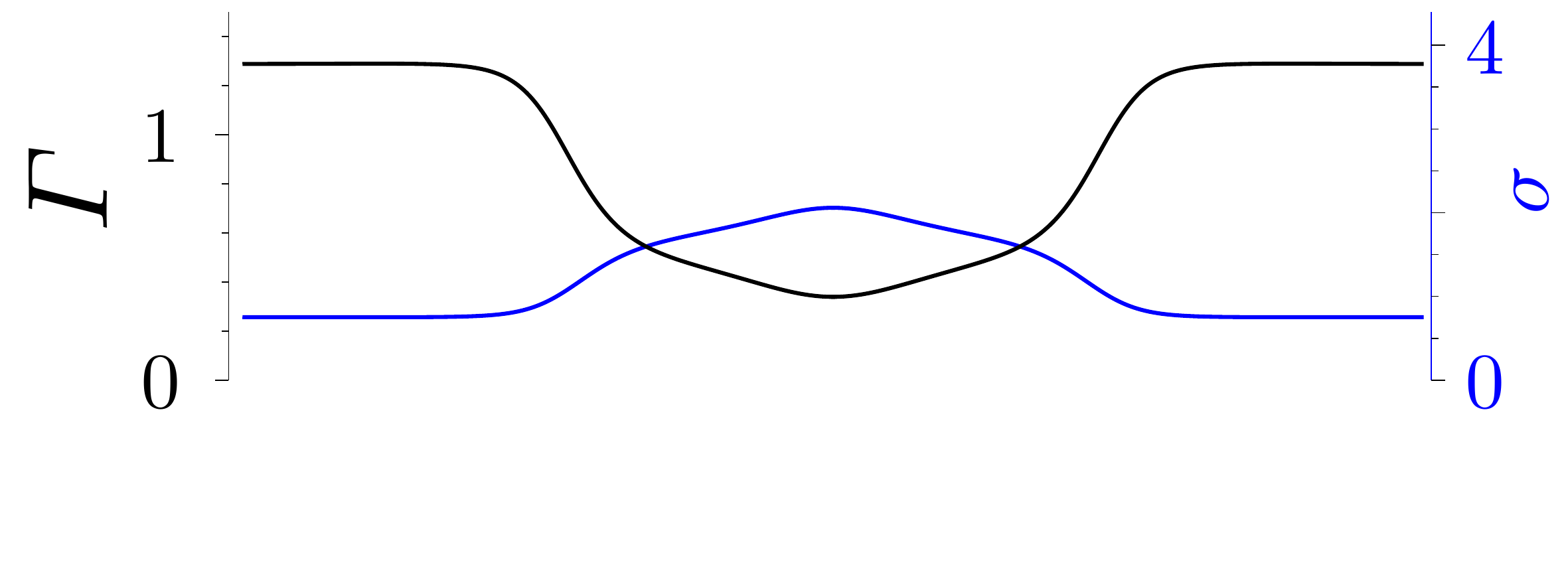} &   \includegraphics[width=0.3\textwidth]{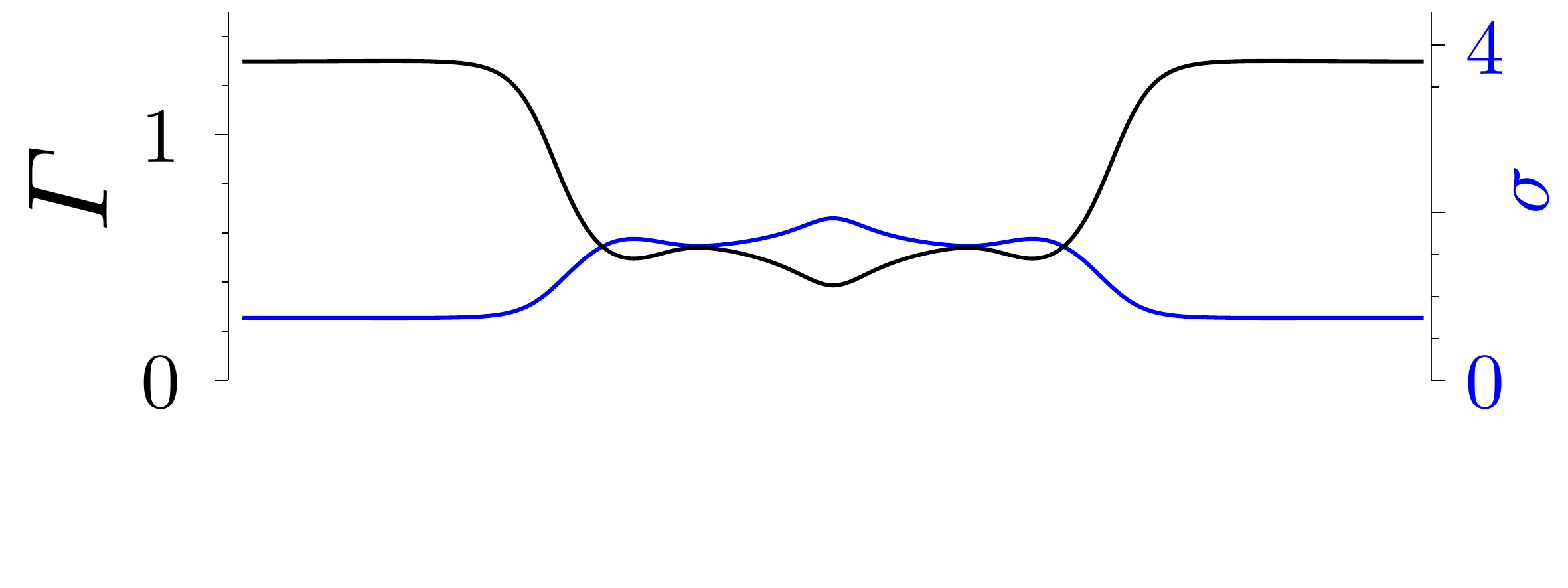} &  \includegraphics[width=0.3\textwidth]{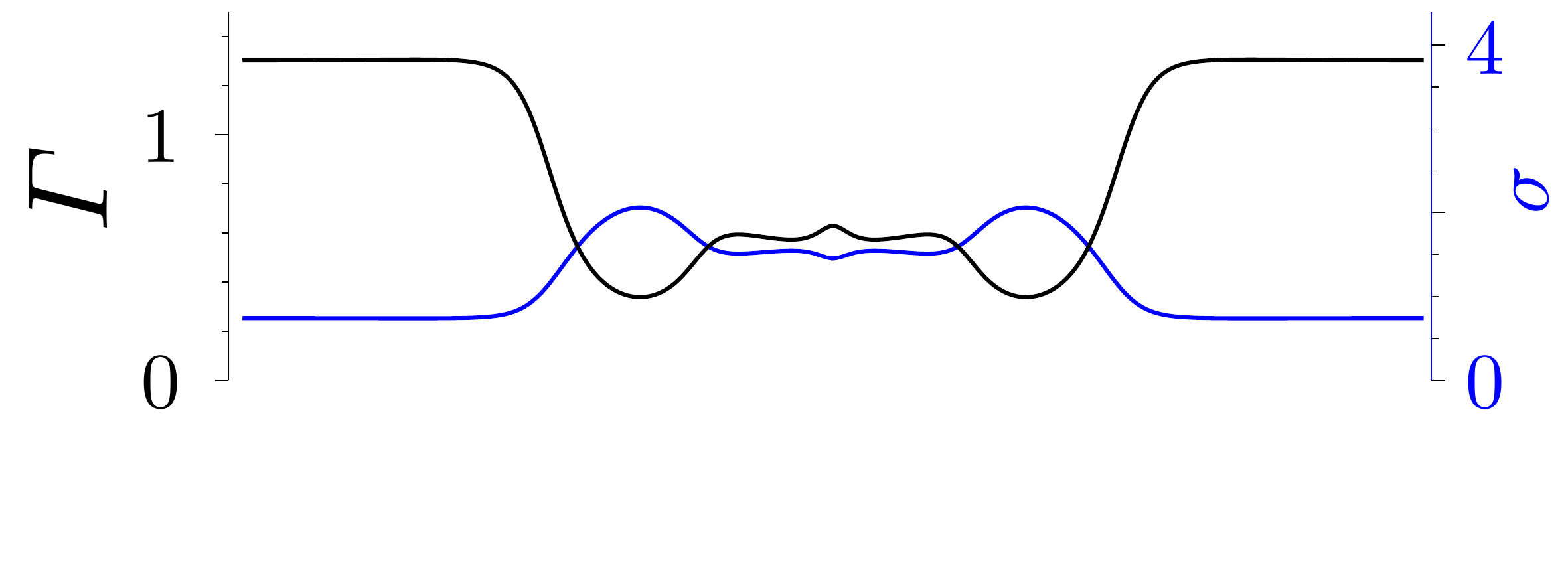}  & \includegraphics[width=0.3\textwidth]{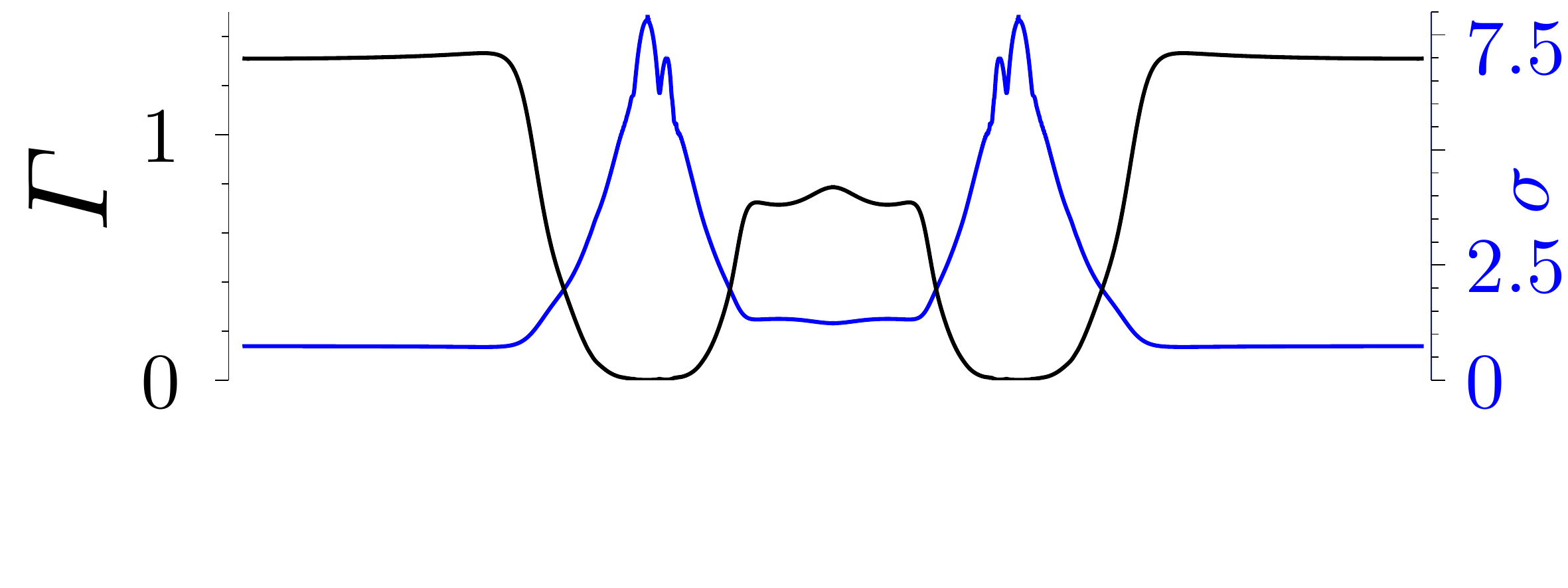}\\
       \vspace{-0.8cm} & & & \\
        \includegraphics[width=0.3\textwidth]{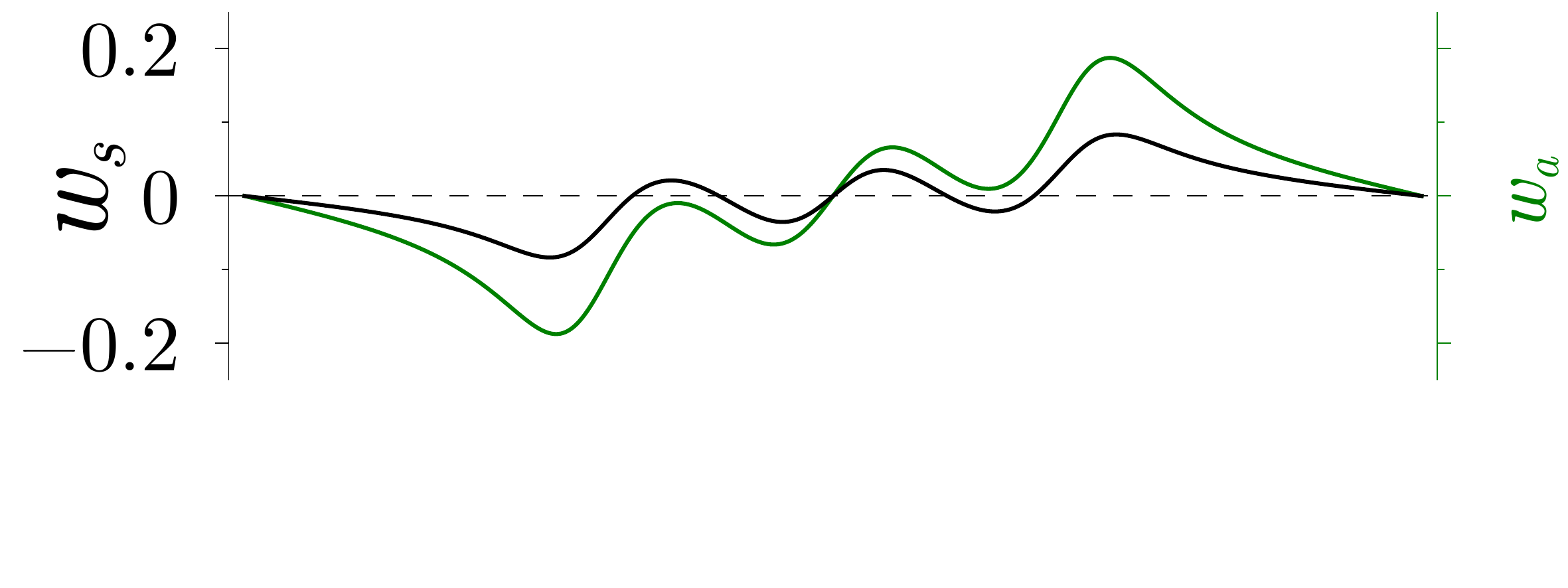} &   \includegraphics[width=0.3\textwidth]{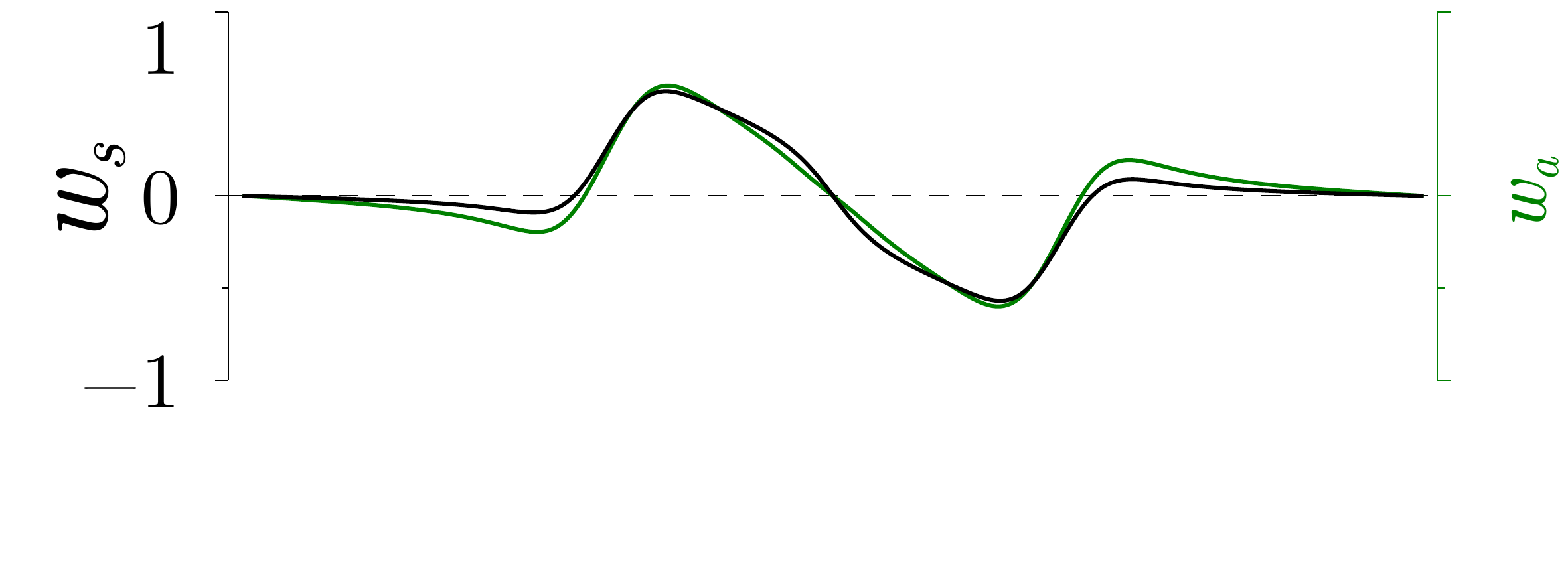} &   \includegraphics[width=0.3\textwidth]{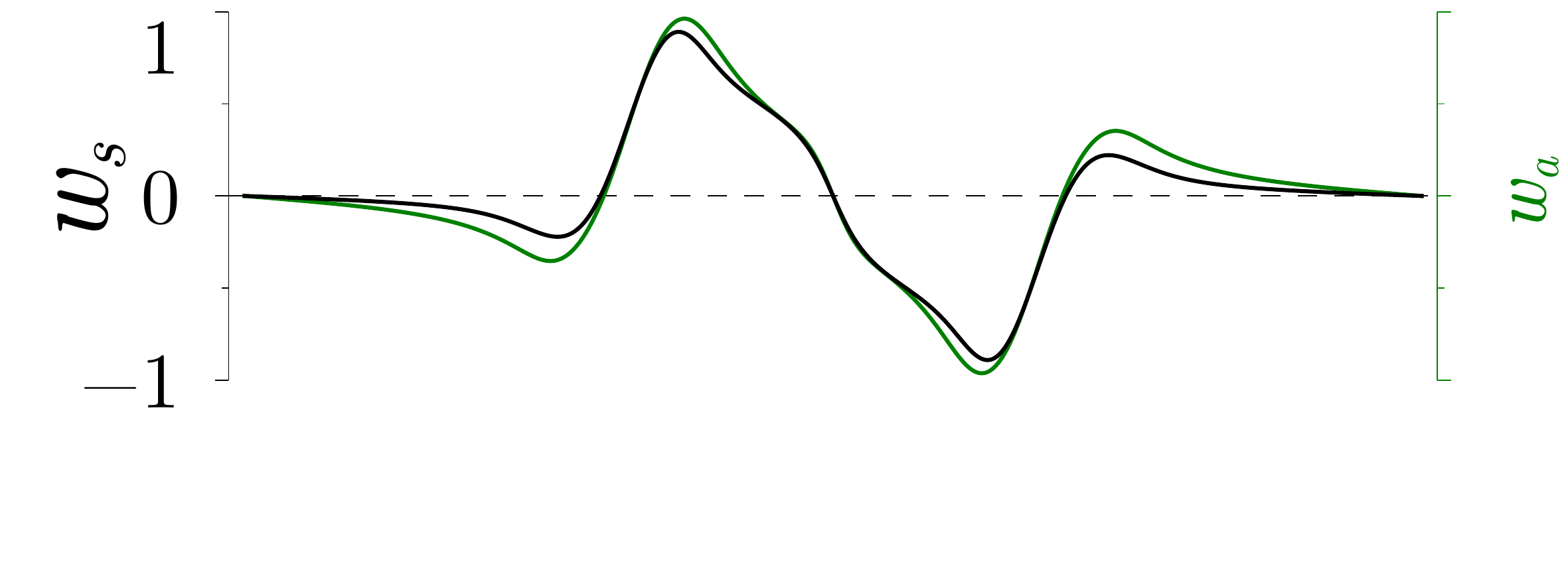} &   \includegraphics[width=0.3\textwidth]{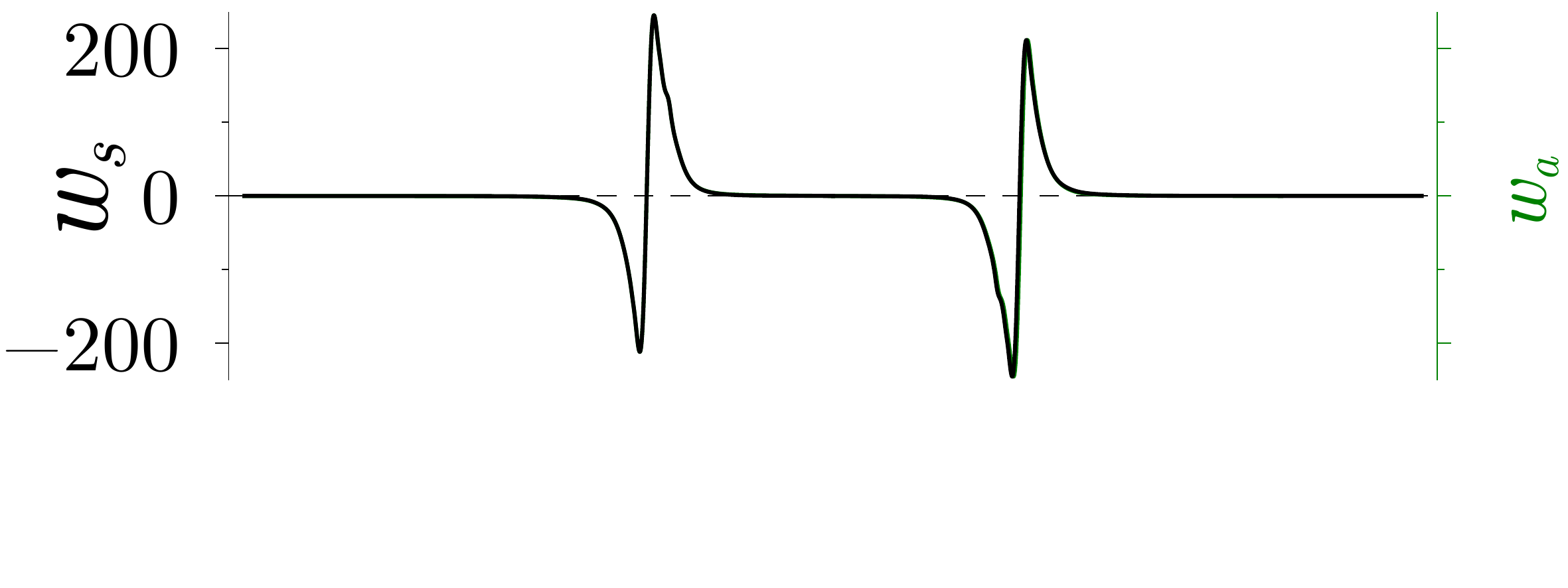} \\
             \vspace{-0.8cm} & & & \\
            \includegraphics[width=0.3\textwidth]{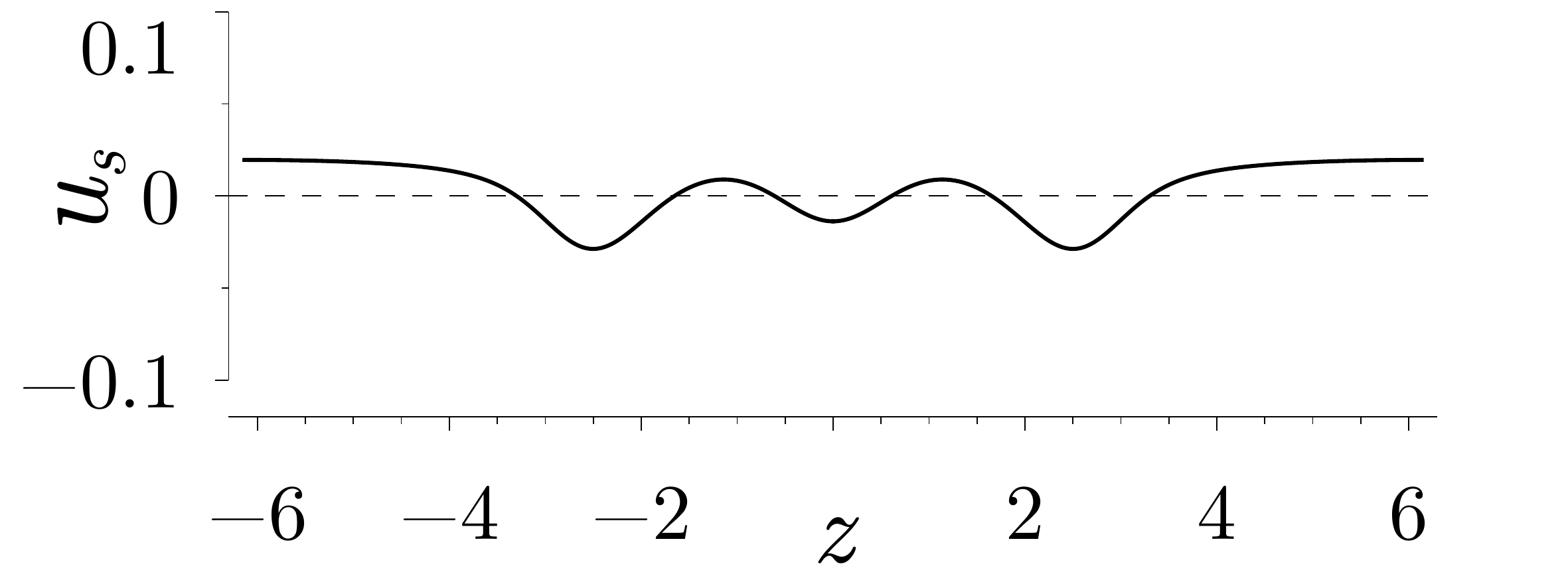} &   \includegraphics[width=0.3\textwidth]{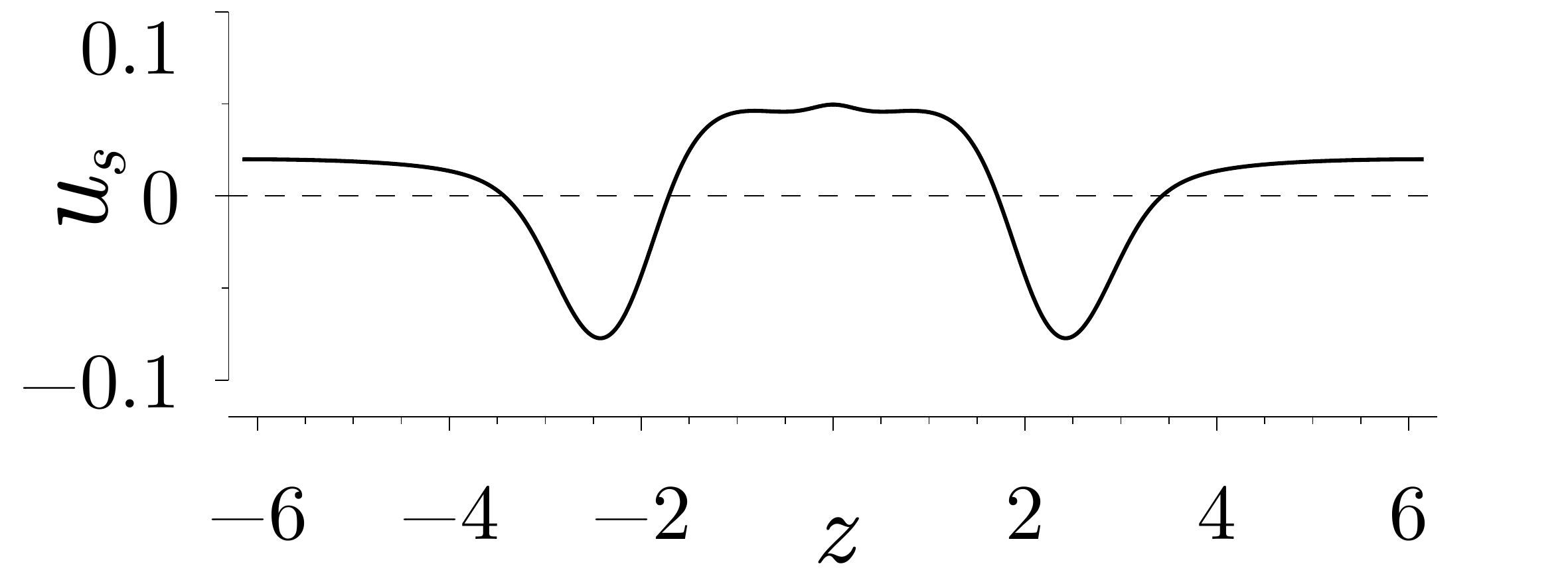} &   \includegraphics[width=0.3\textwidth]{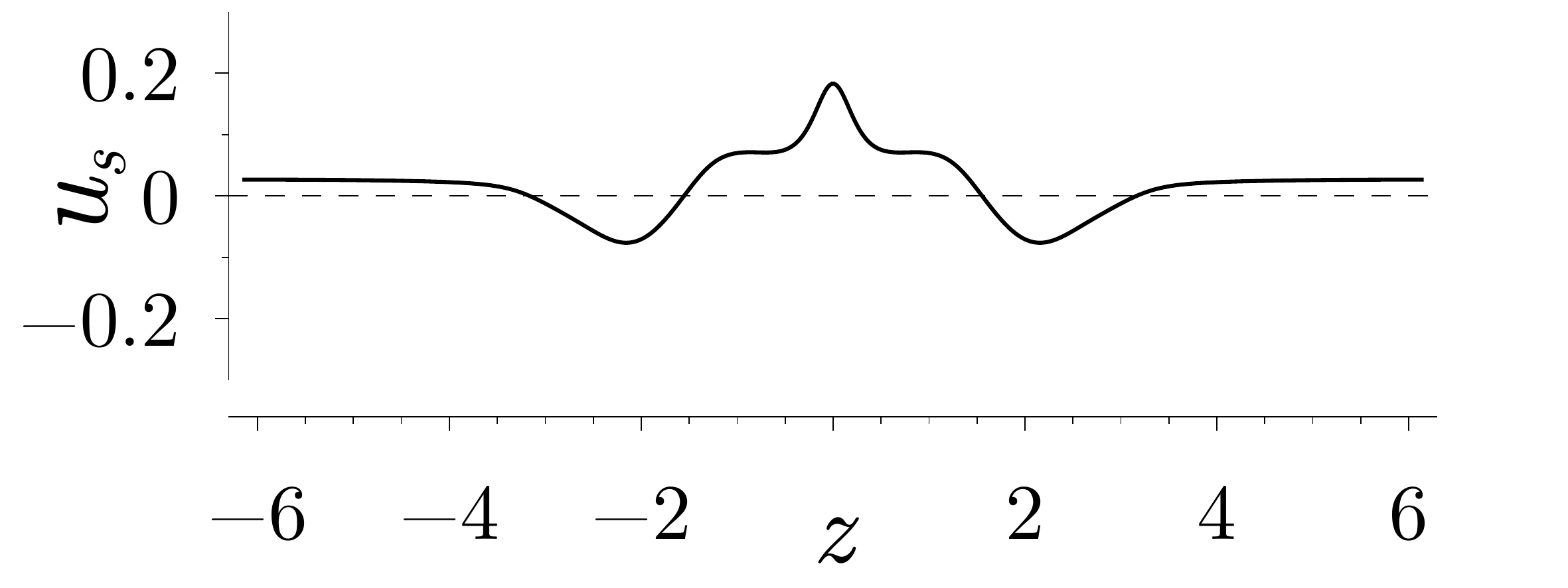} &   \includegraphics[width=0.3\textwidth]{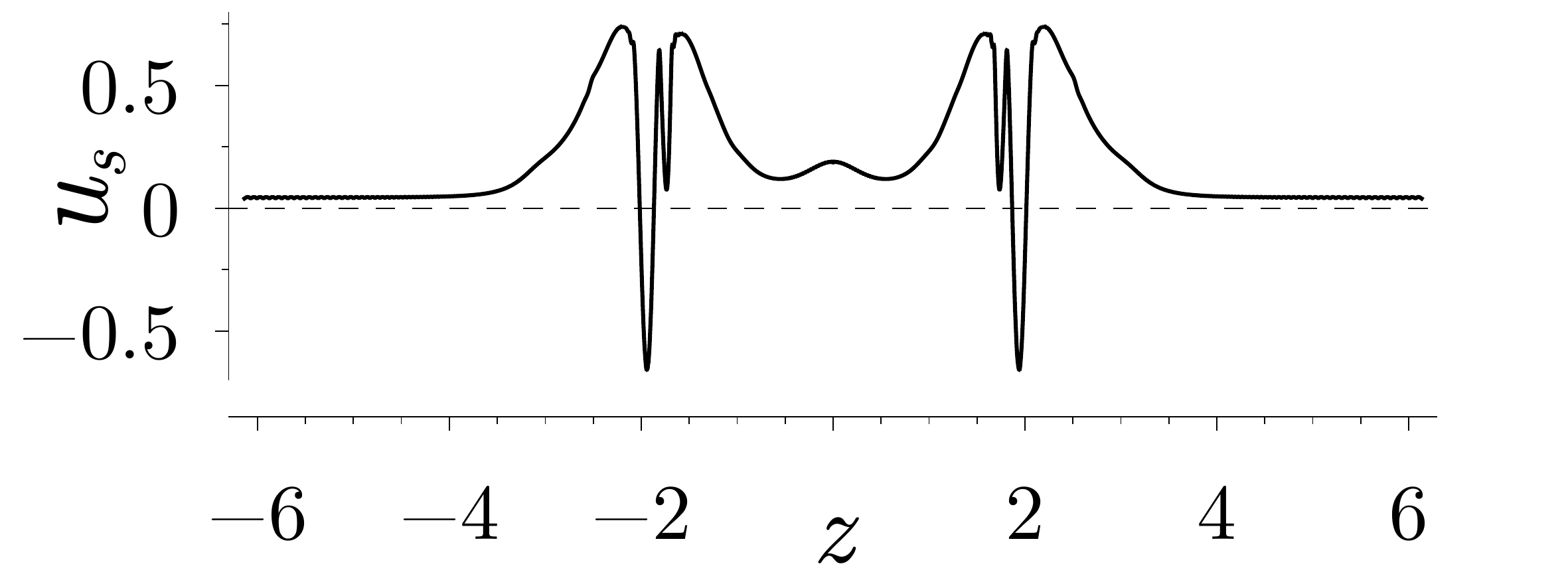} \\
($a$) $t = 119.8$ & ($b$) $t = 122$ & ($c$) $t = 122.8$  & ($d$) $t = 123.7$ \\[4pt]
\end{tabular}
\caption{\label{fig:figure10}(Colour online) Same as figure~\ref{fig:figure9} but for $\Ela = 0.979 > \Ela_c(\Lap=0.01)$, with $k = k_m = 0.512$. The insets are zooms showing the normalised velocity vector field and isocontours of the pressure field. Here $z_{\min} = 1.93$ and $a_{\min} = 2.5 \times 10^{-4}$.}
\end{figure}

To explain the abrupt transition induced by the presence of surfactants, figures~\ref{fig:figure9} and~\ref{fig:figure10} show the temporal evolution of the liquid thread for $\Lap = 0.01$ and two different values of $\Ela$, namely $\Ela = 0.960 < \Ela_c(\Lap=0.01)$, with $k = k_m = 0.508$, and $\Ela = 0.979 > \Ela_c(\Lap=0.01)$, with $k = k_m = 0.512$, respectively. In both cases, we have computed the thread radius $a$ (first row), the surfactant concentration $\Gamma$ together with the surface tension $\sigma$ (second row), the axial velocity at the interface, $w_s$, and at the centreline, $w_a$ (third row), and the radial surface velocity $u_s$ (fourth row). Time is indicated in the labels.

The presence of surfactants introduces two main effects. The advection of surfactant molecules outside the central region of the thread increases the local surface tension in this region, as can be observed in the figures~\ref{fig:figure9}($a$) and~\ref{fig:figure10}($a$). This surfactant depletion generates two opposed effects. First, the axial capillary pressure gradient is enhanced, since the value of $\sigma$ becomes larger in the central region, where $\Gamma$ is smaller, while $\sigma$ becomes smaller away from the centre, where $\Gamma$ is larger. Second, there is a stabilising effect induced by the elastic or Marangoni stress, which competes with the destabilising Plateau-Rayleigh mechanism enhanced by the first effect. Actually, the gradient of $\sigma$ generates a tangential stress at the interface directed towards increasing values of $\sigma$, which opposes the drainage flow and tends to replenish the central zone with surfactant.

In the case of $\Ela < \Ela_c(\Lap=0.01)$, figure~\ref{fig:figure9}($b$,$c$) shows that the Marangoni stress reduces the axial surface velocity, $w_s$, compared with the centreline velocity, $w_a$, the difference between both velocities being larger in the region where $\bnabla_s \sigma$ is higher. As the fluid is drained from the centre for increasing times, $\bnabla_s \sigma$ becomes larger. When $\Ela < \Ela_c(\Lap=0.01)$ the capillary pressure gradient is able to remove most of the liquid from the centre. Eventually, close to pinch-off, inertia becomes important and the flow is reverted close to $z = 0.33$, so that the rate of thinning increases in this region and $z_{\min}$ moves towards the latter axial position where the liquid thread finally detaches forming a tiny satellite droplet with $\Vsat < 10^{-5}$, as evidenced by figure~\ref{fig:figure9}($d$). Note that, during thread evolution, two bulges connecting the central and outer regions grow due to the reduction of the surface velocity, and are finally connected by a thin liquid thread close to pinch-off.

When $\Ela > \Ela_c(\Lap=0.01)$ the foregoing explanation still holds, but the elastic stress is large enough to revert the flow near the interface at early times far from break-up, as shown in figure~\ref{fig:figure10}($a$). The associated stagnation point diffuses radially inwards, and leads to a counterflow separating a region where liquid flows towards the centre and induces the formation of a satellite from another region where the incipient main drop is fed with liquid. Consequently, the thread detaches in between these two regions. If $\Ela$ increases further, the break-up time increases and the flow reversal occurs at earlier stages, so that $\Vsat$ and $\Ssat$ increase monotonically, as shown in figures~\ref{fig:figure8}($a$,$b$).

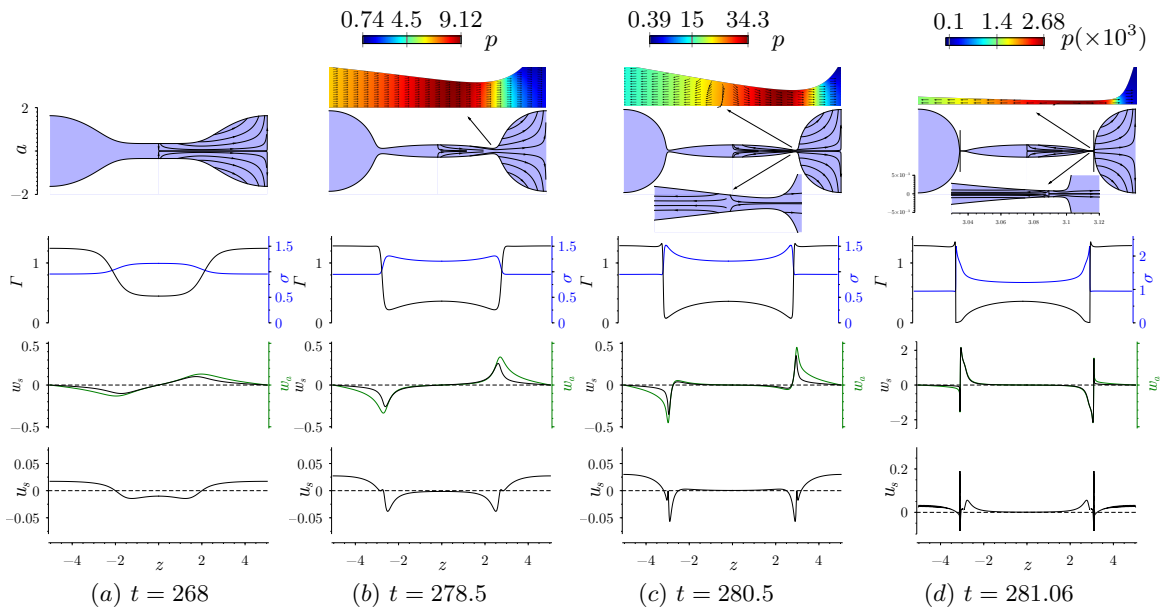
\begin{figure}
\hspace{-0.8cm}
\setlength{\tabcolsep}{-5.6pt}
\begin{tabular}{cccc}
& \begin{tikzpicture} \centering
\begin{axis}[
hide axis,
scale only axis,
height=0pt,
width=0pt,
colormap/jet,
colorbar horizontal,
point meta min=0.743389298316886,
point meta max=9.128156379671164,
colorbar style={
    height=1mm, width= 1.3cm, xticklabel pos=upper, xtick={0.74,4.5,9.12}
}]
\addplot [draw=none] coordinates {(0,0)};
\end{axis}
\node at (1.7,-0.4) {$p$};
\end{tikzpicture} & \begin{tikzpicture} \centering
\begin{axis}[
hide axis,
scale only axis,
height=0pt,
width=0pt,
colormap/jet,
colorbar horizontal,
point meta min=0.38644437301197526,
point meta max=34.306313919413526,
colorbar style={
    height=1mm, width= 1.3cm, xticklabel pos=upper, xtick={0.39,15,34.30}
}]
\addplot [draw=none] coordinates {(0,0)};
\end{axis}
\node at (1.7,-0.4) {$p$};
\end{tikzpicture} & \hspace{1cm} \begin{tikzpicture} \centering
\begin{axis}[
hide axis,
scale only axis,
height=0pt,
width=0pt,
colormap/jet,
colorbar horizontal,
point meta min=0.0005282374353695457,
point meta max=2.687734788230827,
colorbar style={
    height=1mm, width= 1.3cm, xticklabel pos=upper, xtick={0.1,1.4,2.68}
}]
\addplot [draw=none] coordinates {(0,0)};
\end{axis}
\node at (2.1,-0.3) {$p (\times 10^{3})$};
\end{tikzpicture}\\
\includegraphics[width=0.28\textwidth]{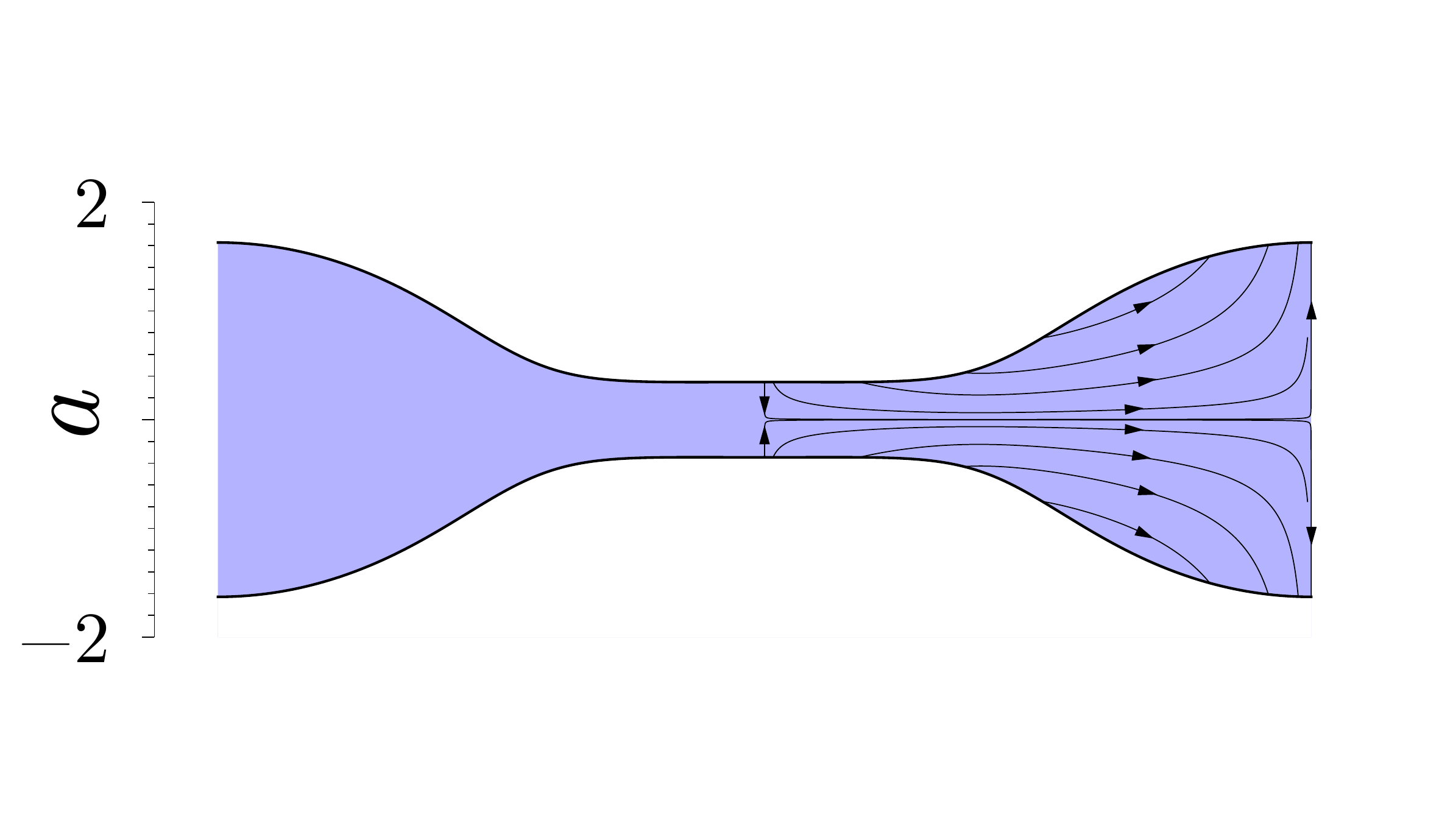} &  \hspace{0.25cm}\includegraphics[width=0.28\textwidth]{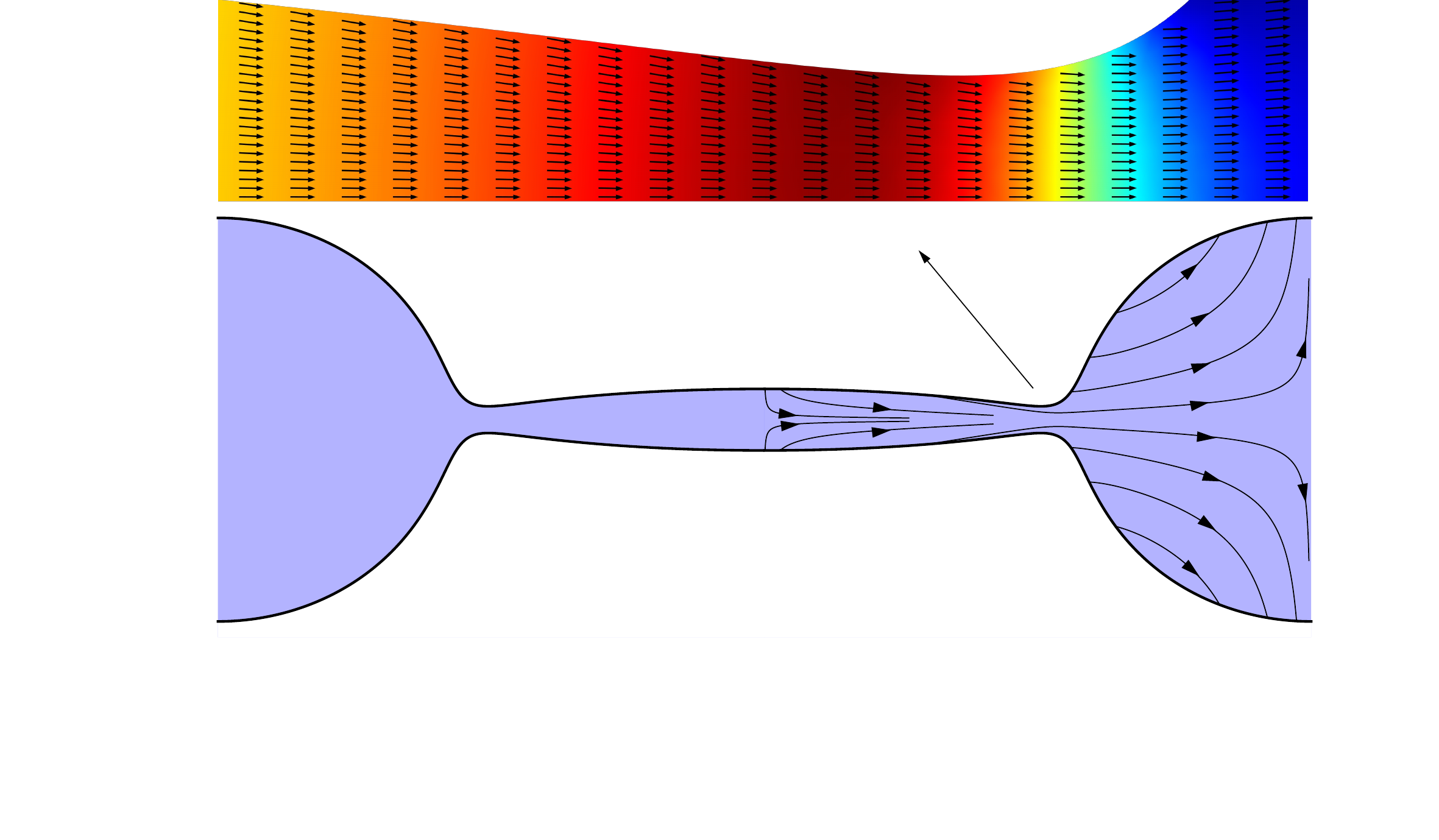}  &  \hspace{0.35cm} \includegraphics[width=0.28\textwidth]{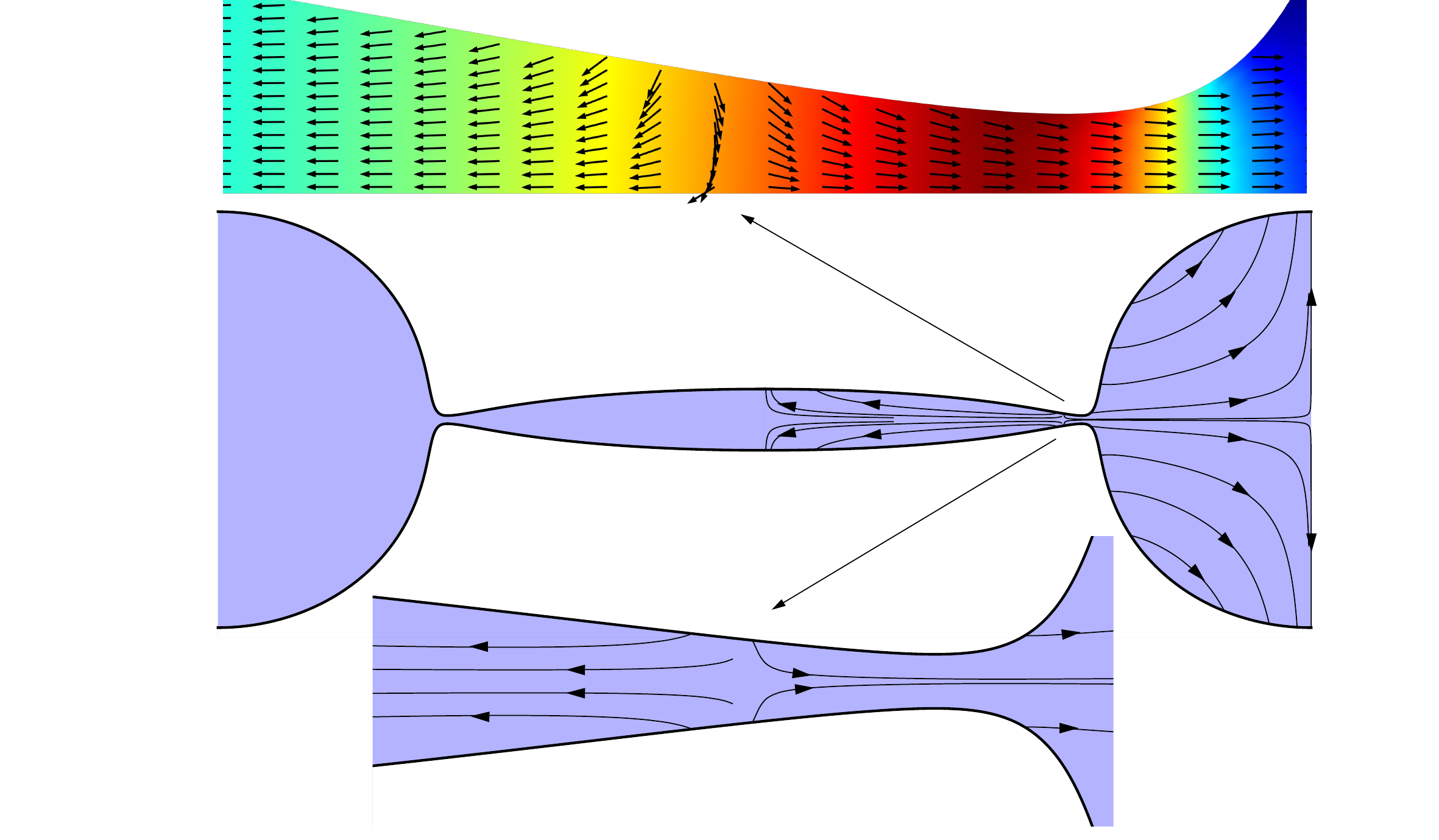}  & \hspace{0.4cm} \includegraphics[width=0.28\textwidth]{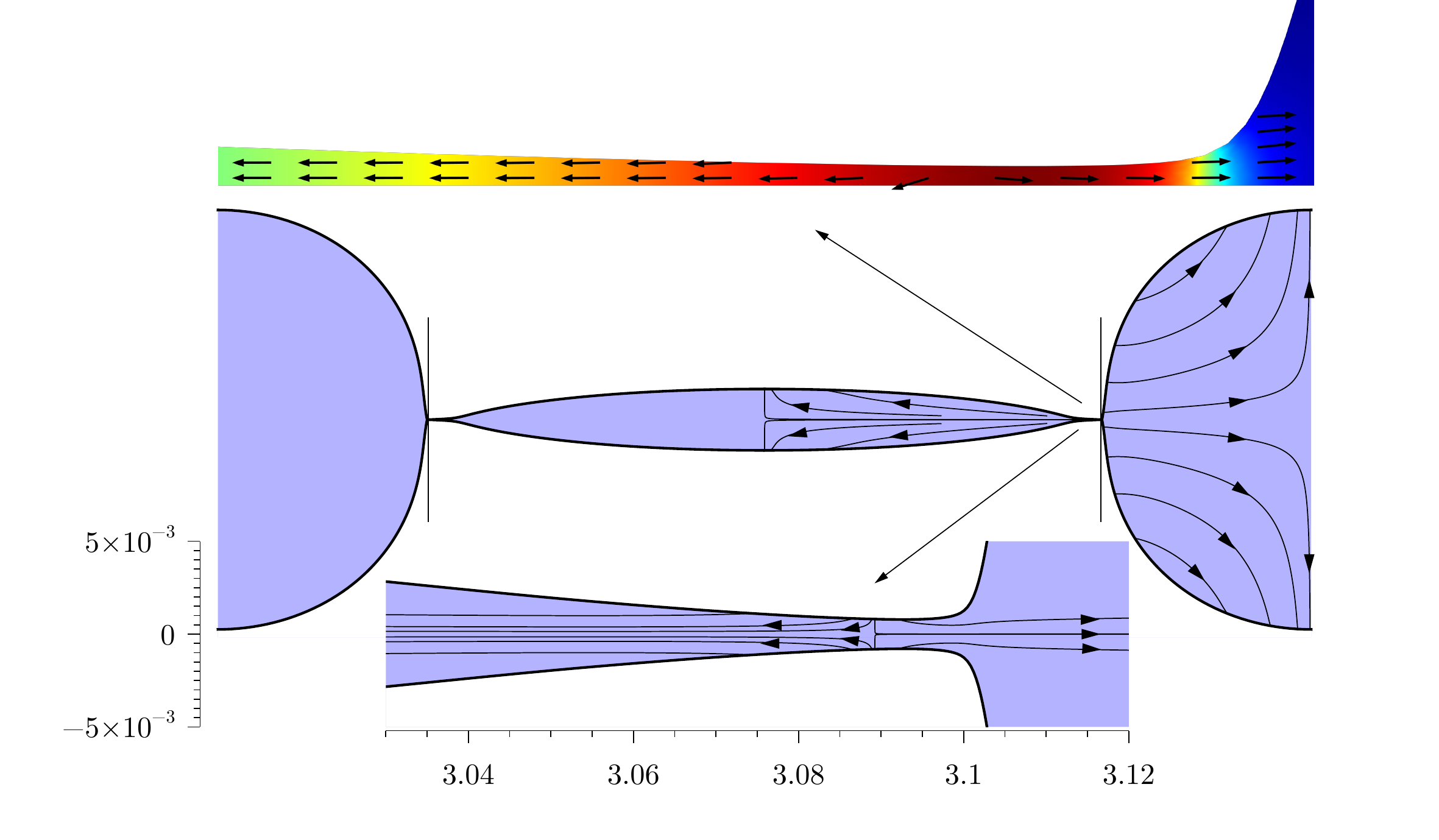}        \\
     \vspace{-0.5cm} & & & \\
  \includegraphics[width=0.28\textwidth]{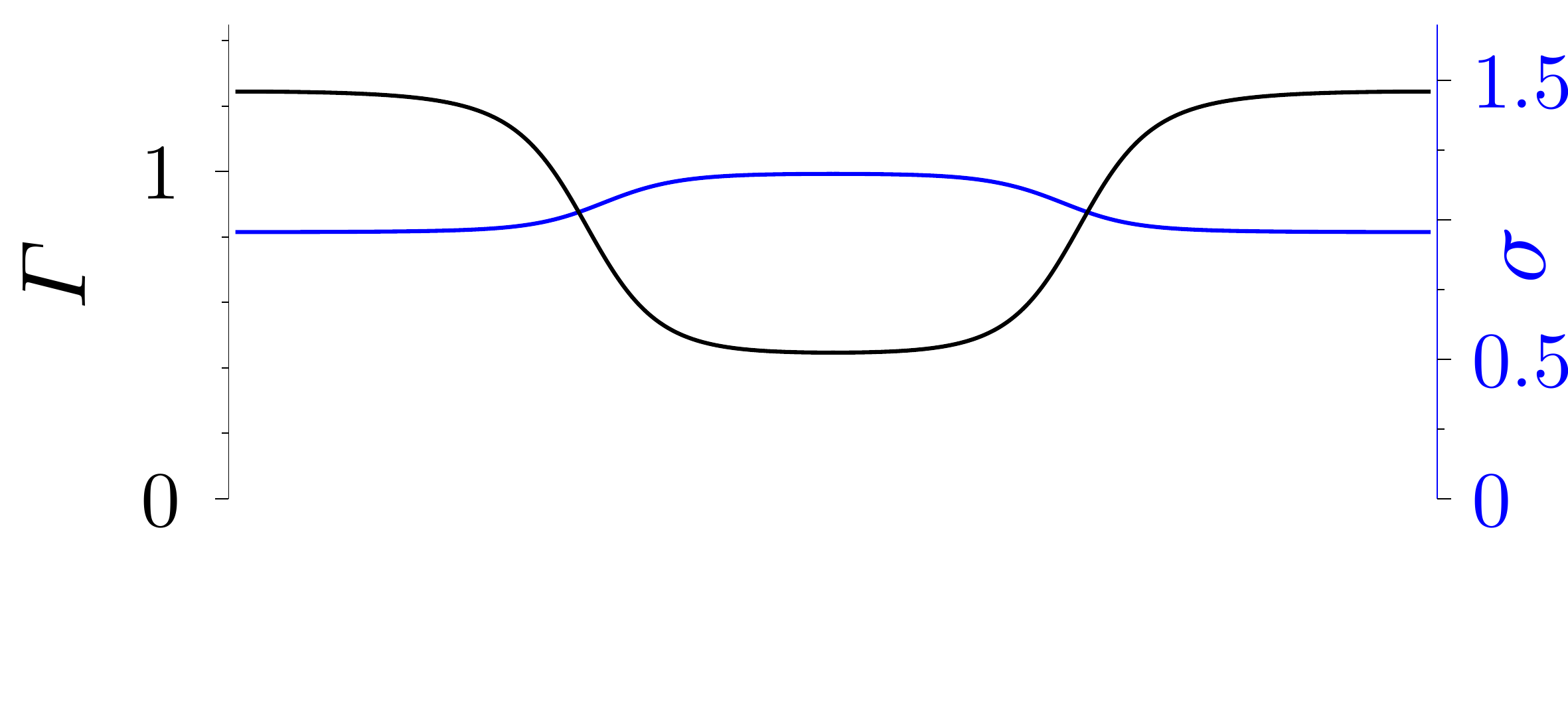}  &  \hspace{0.25cm} \includegraphics[width=0.28\textwidth]{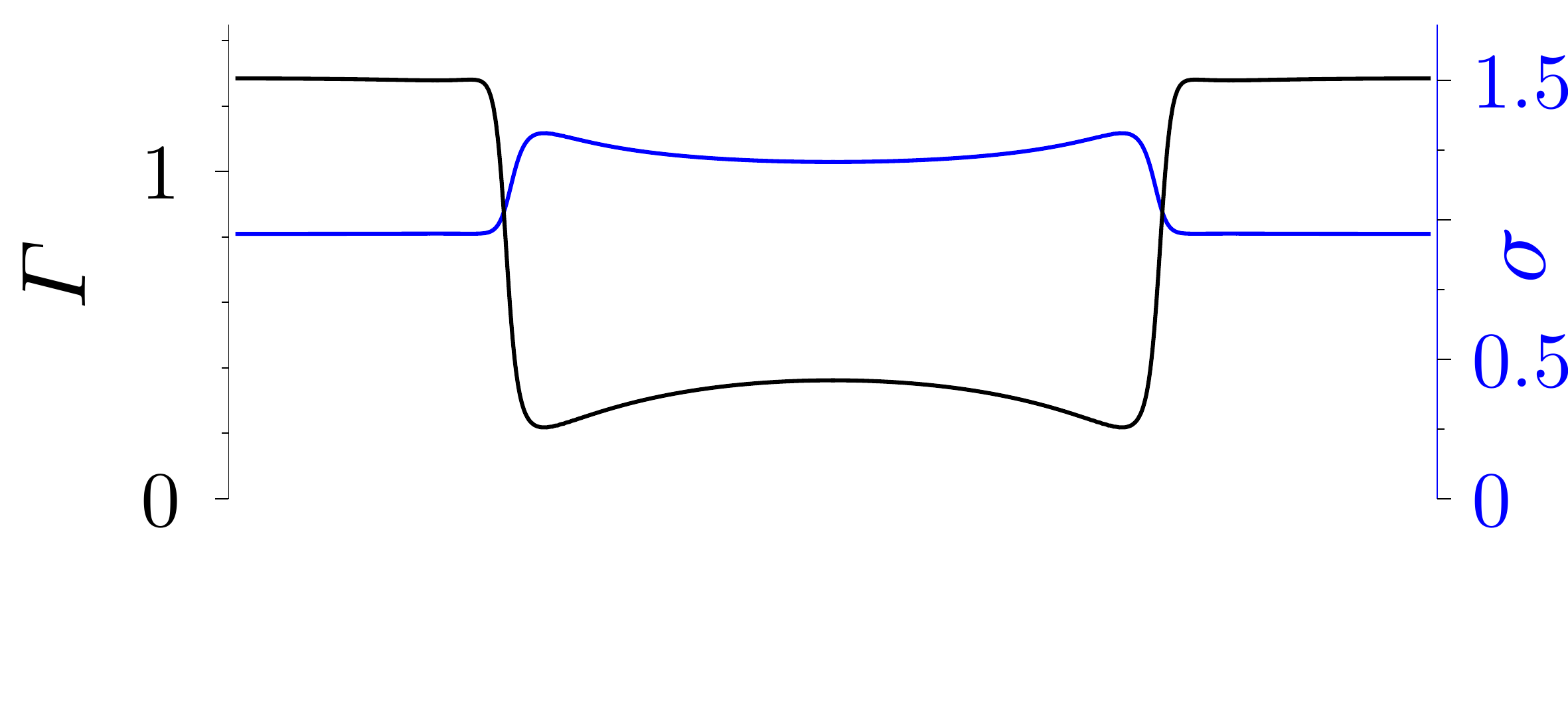}  &   \hspace{0.35cm}\includegraphics[width=0.28\textwidth]{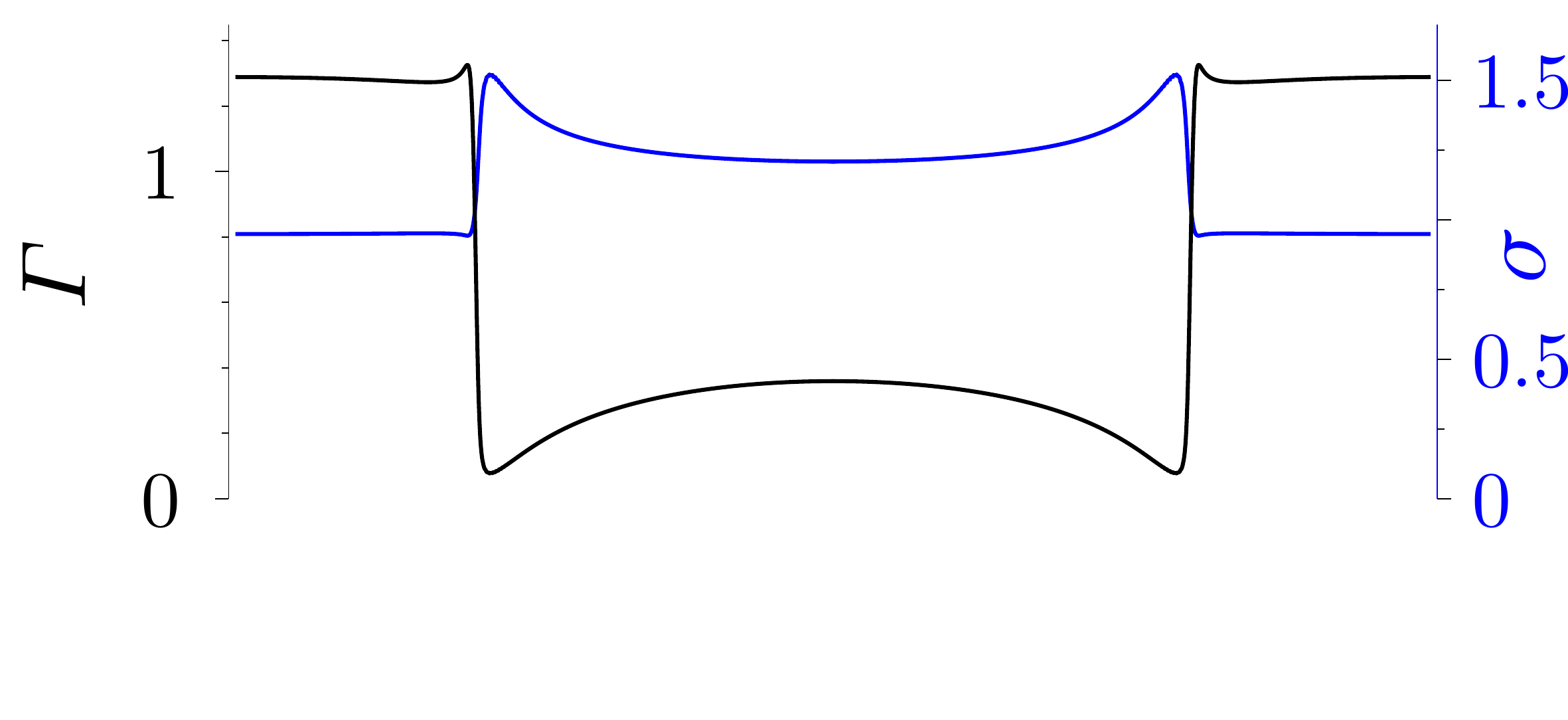} &   \hspace{0.4cm}\includegraphics[width=0.28\textwidth]{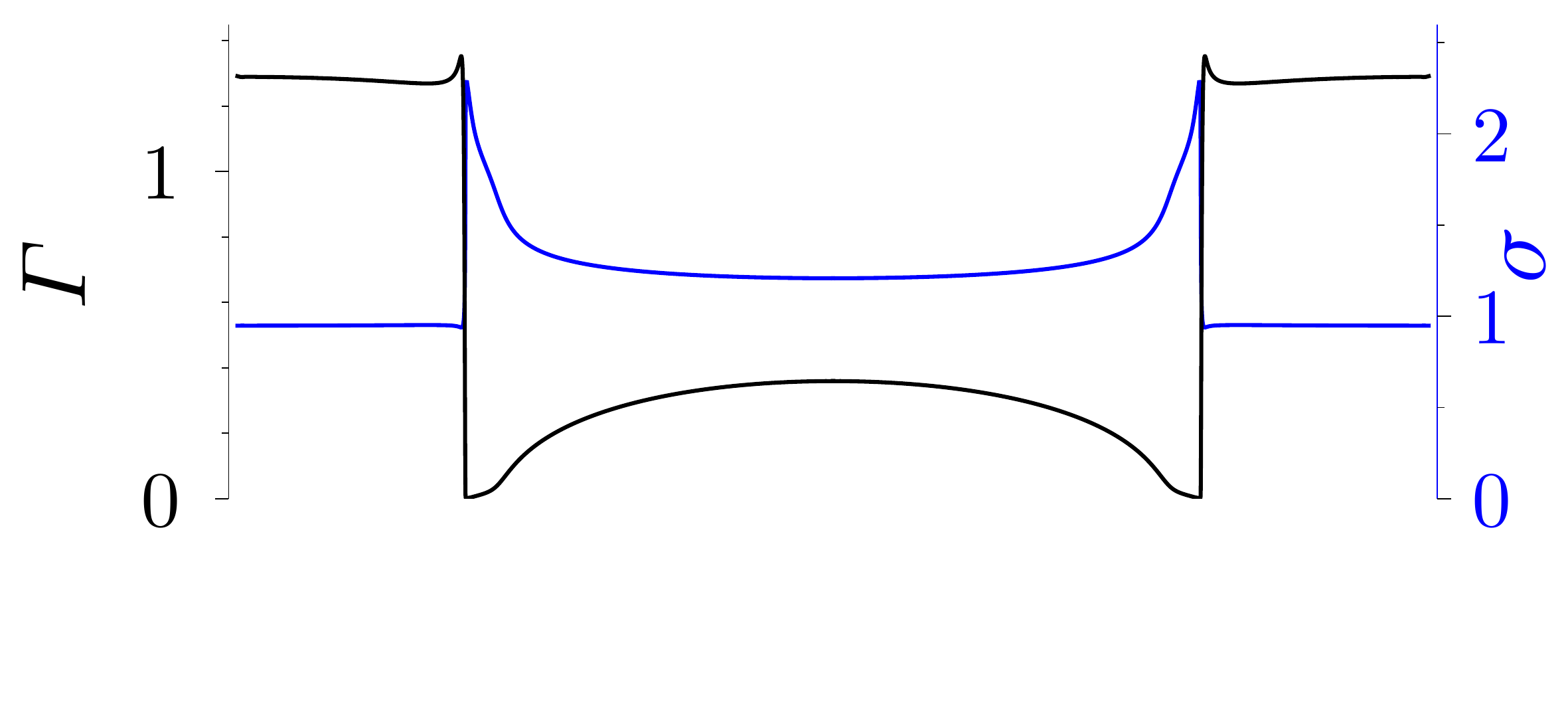} \\
       \vspace{-0.8cm} & & & \\
 \includegraphics[width=0.28\textwidth]{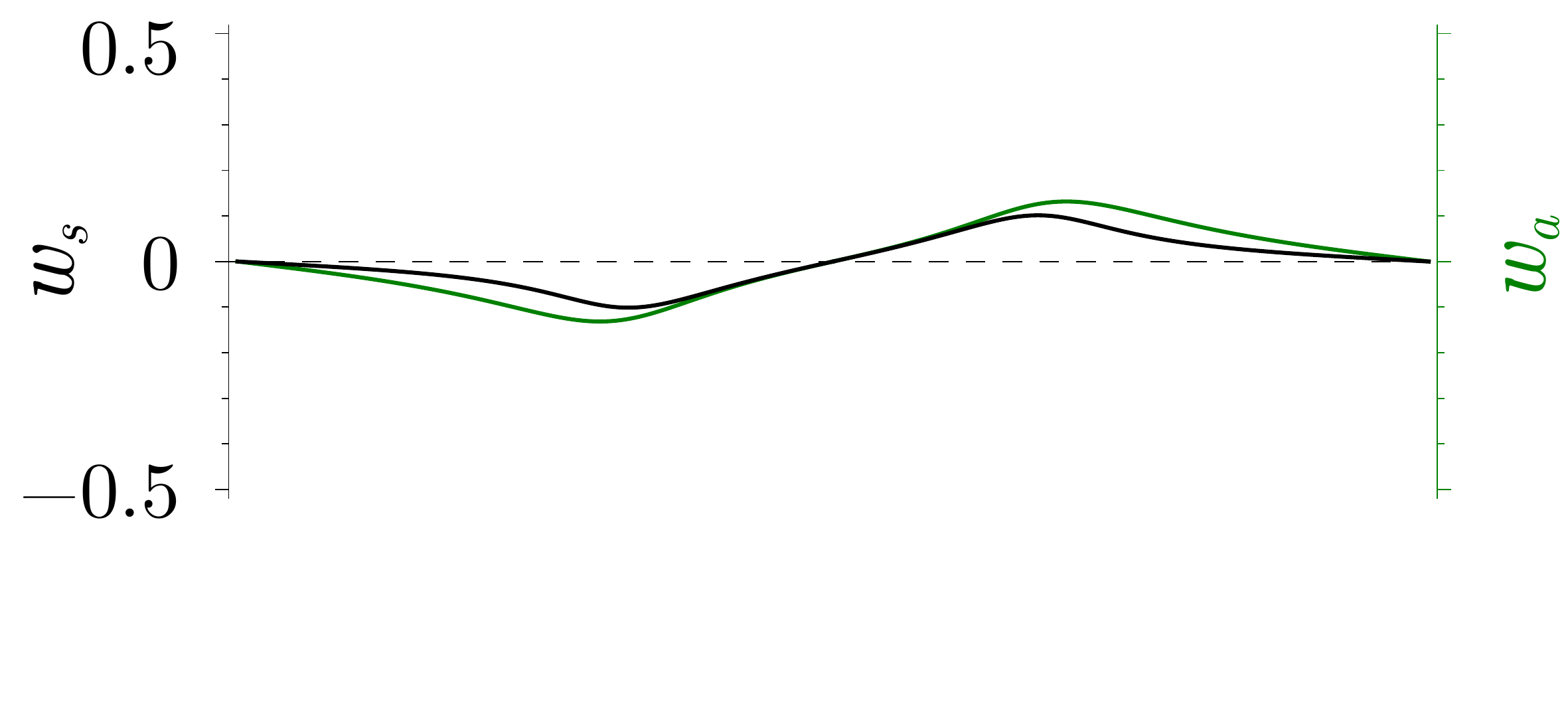} &  \hspace{0.25cm} \includegraphics[width=0.28\textwidth]{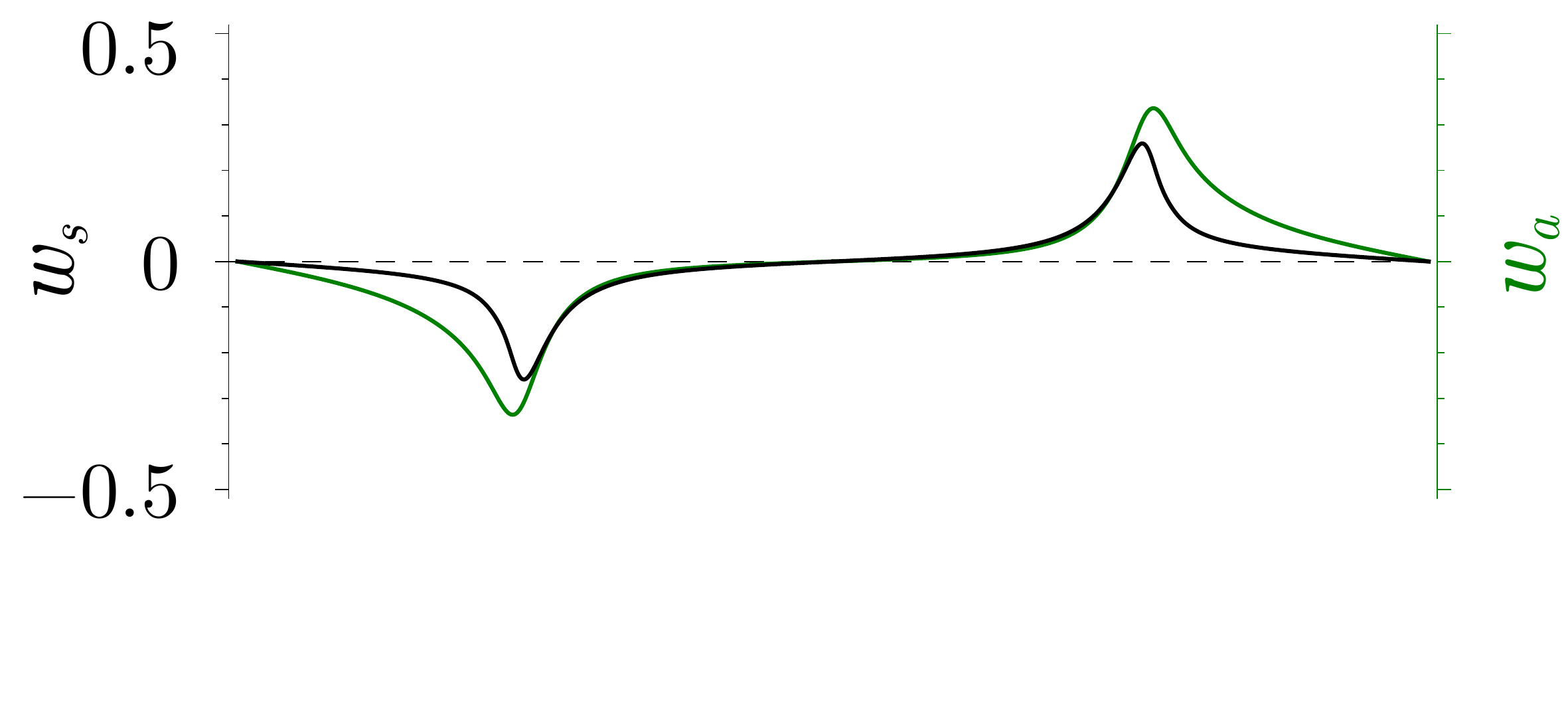} &   \hspace{0.35cm} \includegraphics[width=0.28\textwidth]{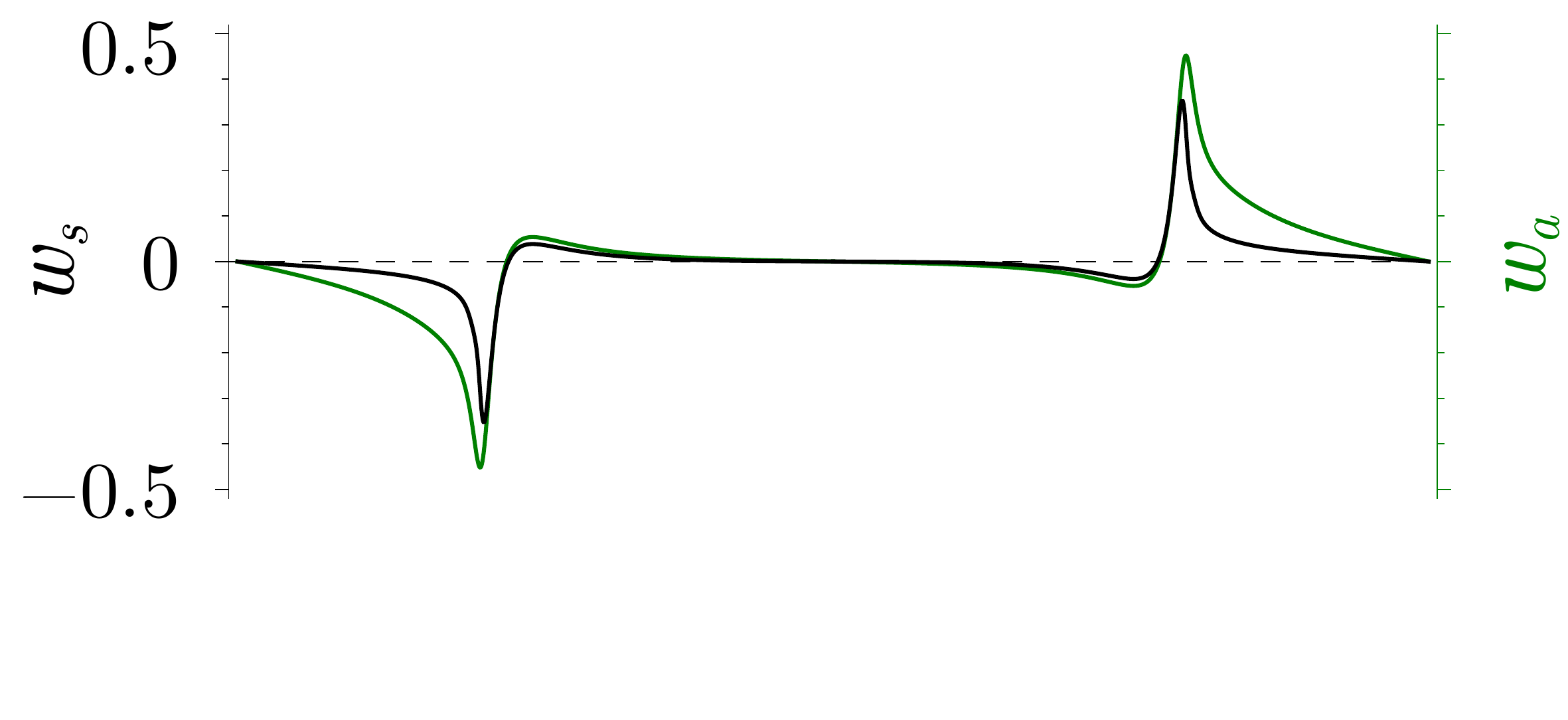} &    \hspace{0.4cm} \includegraphics[width=0.28\textwidth]{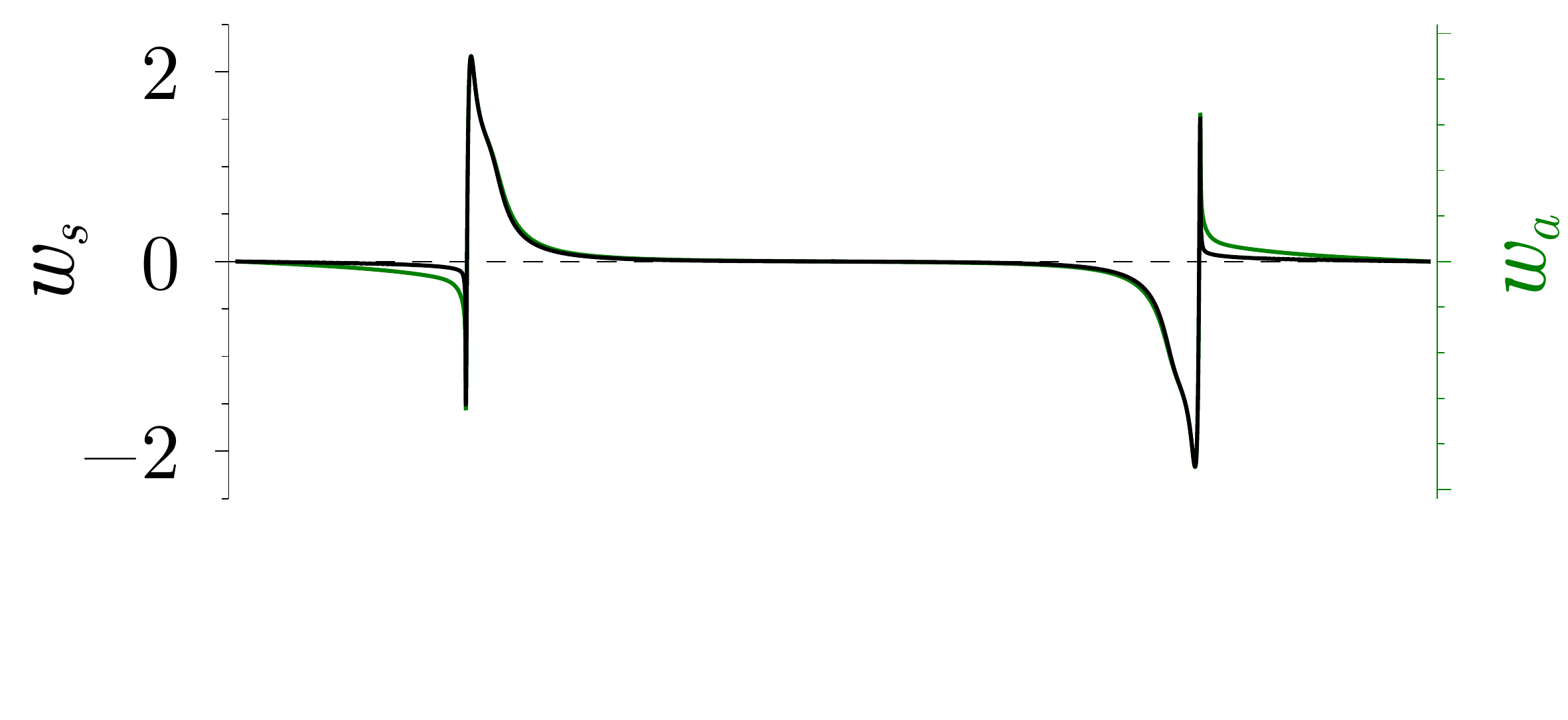}  \\
     \vspace{-0.8cm} & & & \\
\includegraphics[width=0.28\textwidth]{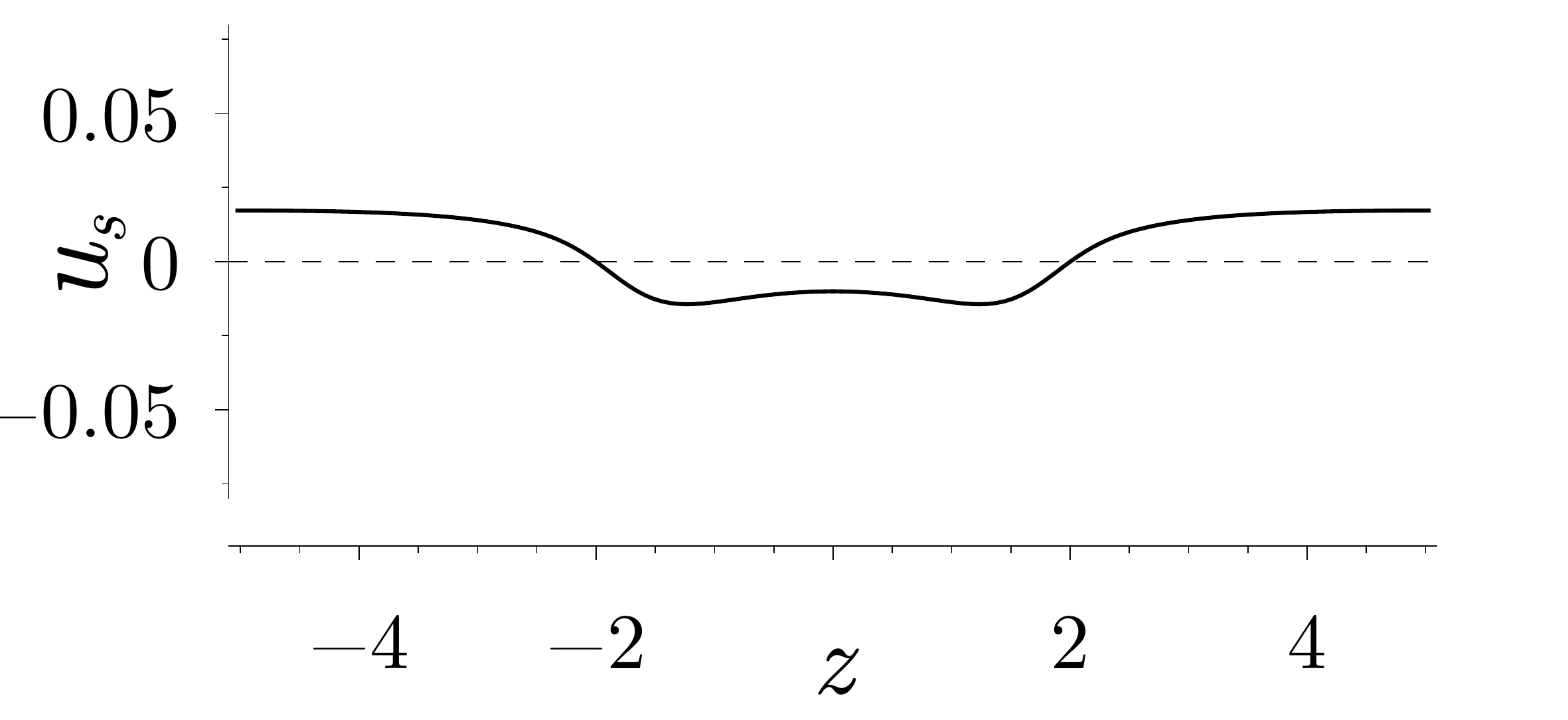} &   \hspace{0.25cm} \includegraphics[width=0.28\textwidth]{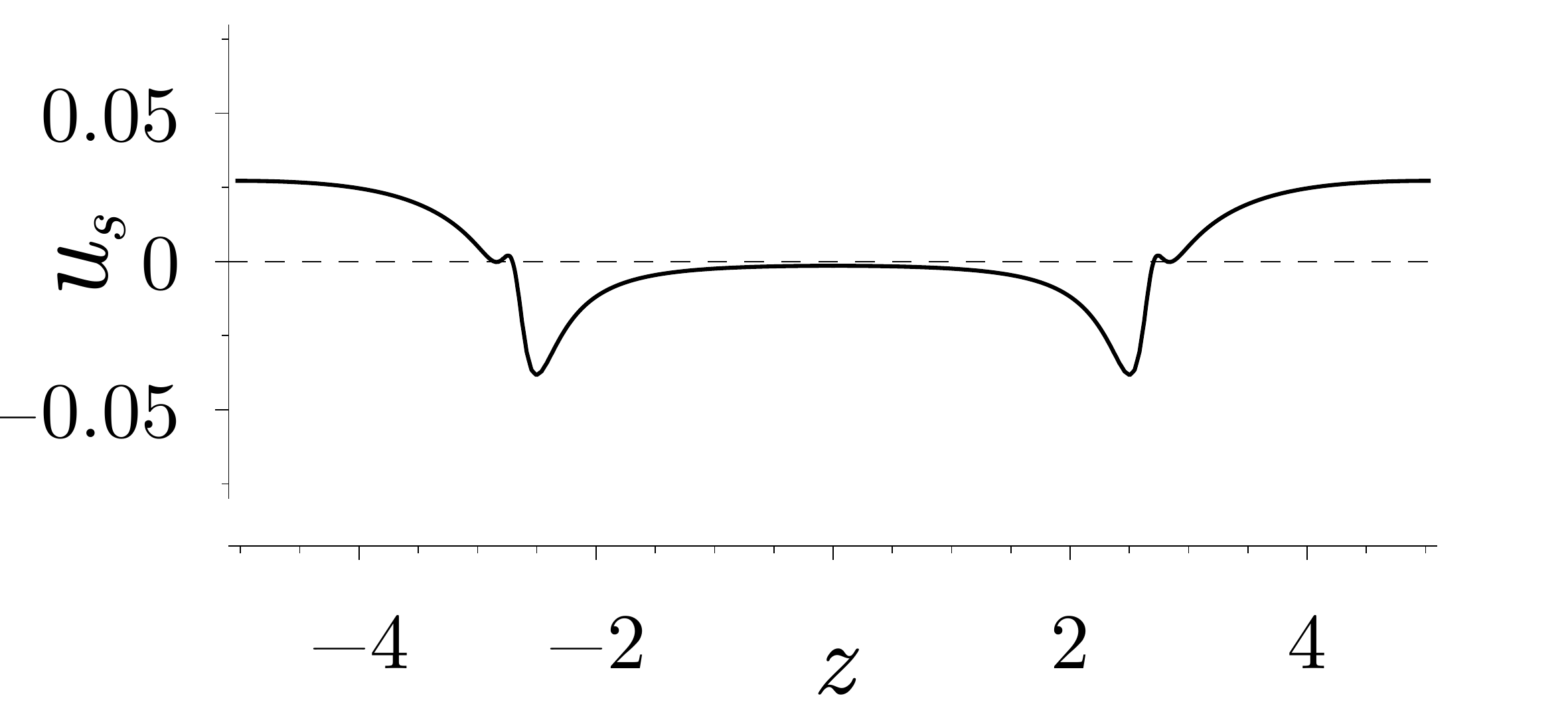} &     \hspace{0.35cm} \includegraphics[width=0.28\textwidth]{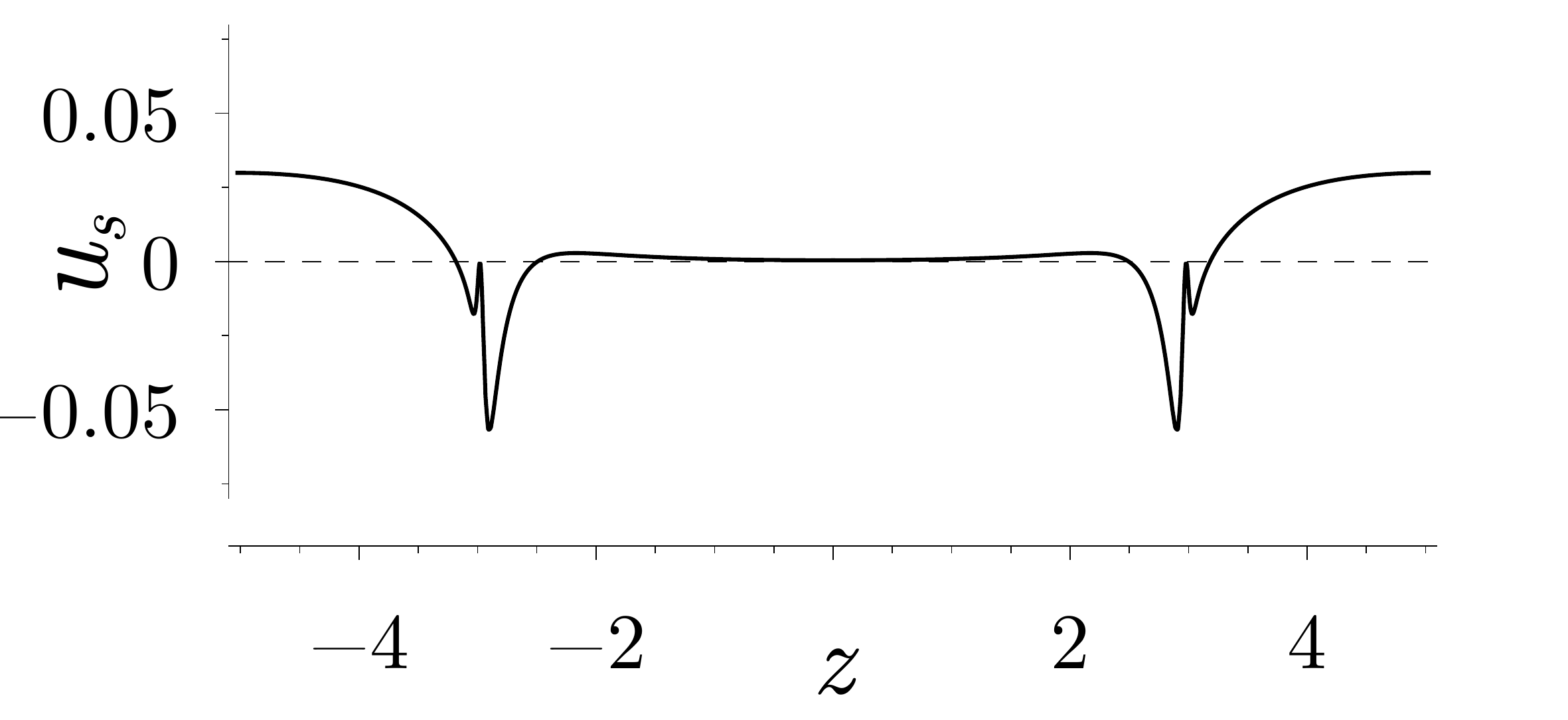} &    \hspace{0.4cm} \includegraphics[width=0.28\textwidth]{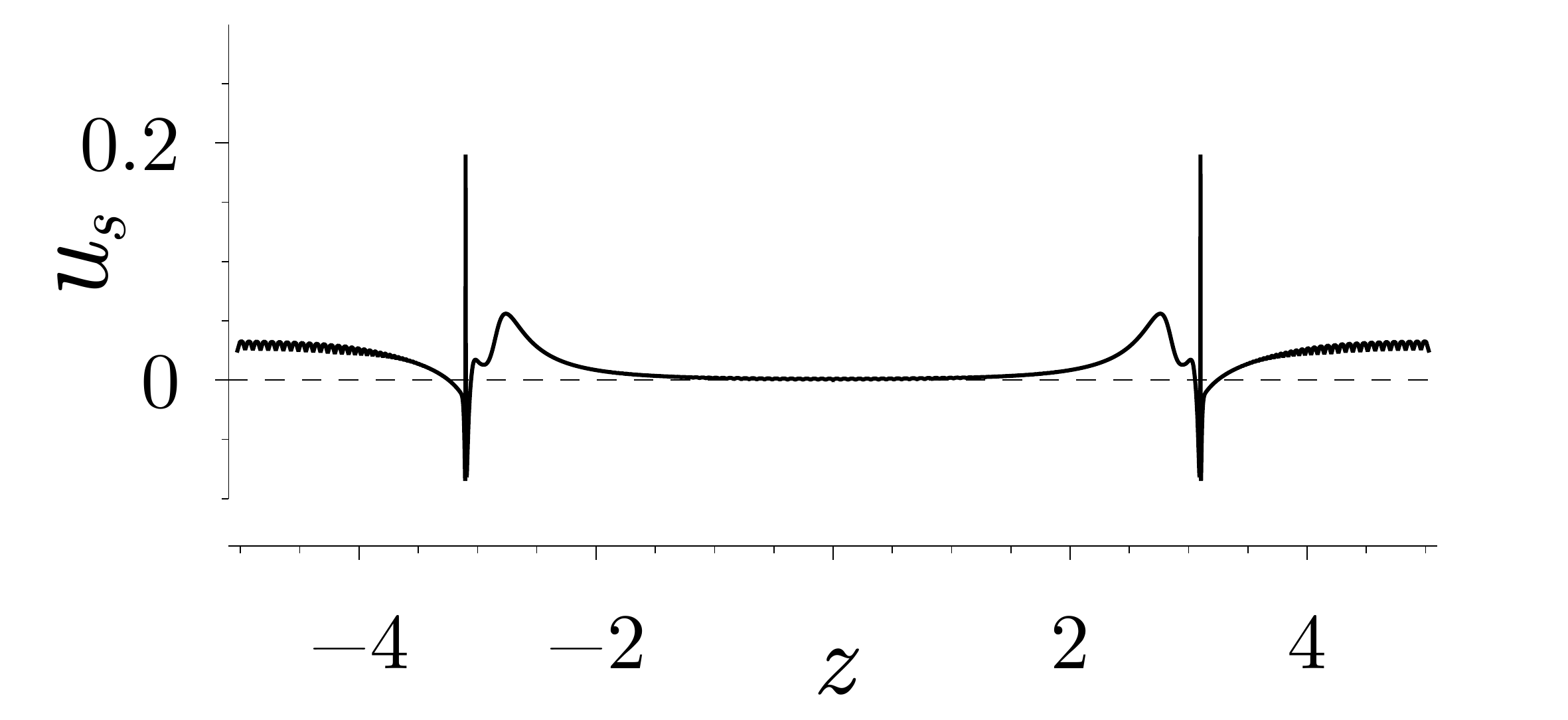} \\
($a$) $t = 268$ &   ($b$) $t = 278.5$ &  ($c$) $t = 280.5$  &     ($d$) $t = 281.06$ \\[4pt]
\end{tabular}
\caption{\label{fig:figure11} (Colour online) Same as figure~\ref{fig:figure9} but for $\Lap = 100$ and $\Ela = 0.203$, with $k = k_m = 0.625$. Here $z_{\min} = 3.09$ and $a_{\min} = 7.89 \times 10^{-4}$.}
\end{figure}

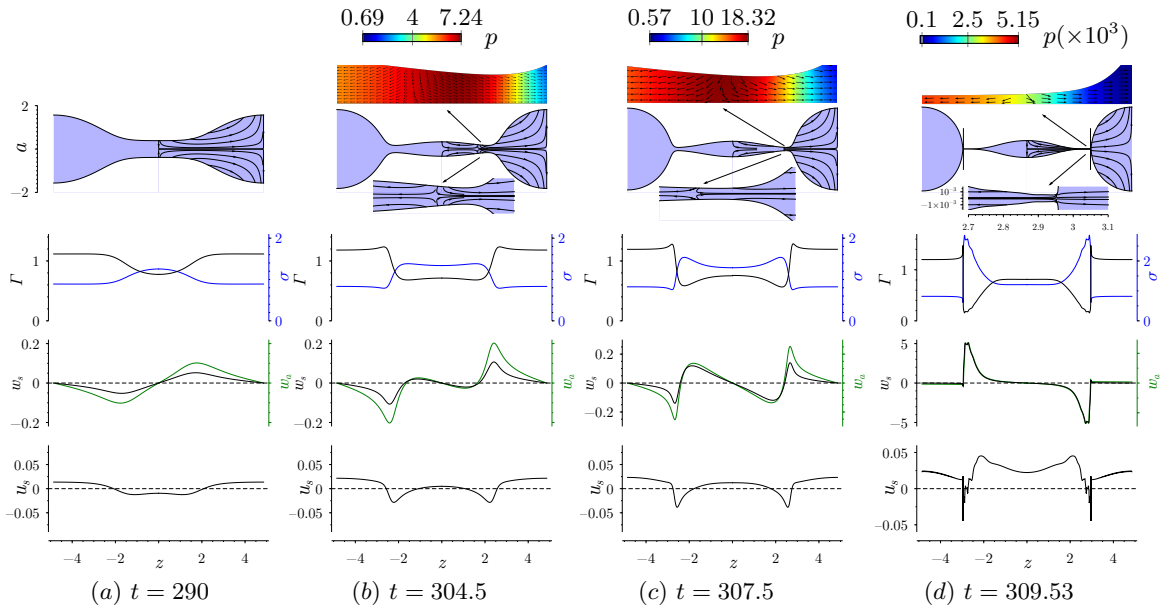
\begin{figure}
\hspace{-0.8cm}
\setlength{\tabcolsep}{-5.6pt}
\begin{tabular}{cccc}
& \begin{tikzpicture} \centering
\begin{axis}[
hide axis,
scale only axis,
height=0pt,
width=0pt,
colormap/jet,
colorbar horizontal,
point meta min=0.6907332430282015,
point meta max=7.235032501785322,
colorbar style={
    height=1mm, width= 1.3cm, xticklabel pos=upper, xtick={0.69,4,7.24}
}]
\addplot [draw=none] coordinates {(0,0)};
\end{axis}
\node at (1.7,-0.4) {$p$};
\end{tikzpicture} & \begin{tikzpicture} \centering
\begin{axis}[
hide axis,
scale only axis,
height=0pt,
width=0pt,
colormap/jet,
colorbar horizontal,
point meta min=0.5690128098404692,
point meta max=18.317504061117173,
colorbar style={
    height=1mm, width= 1.3cm, xticklabel pos=upper, xtick={0.57,10,18.32}
}]
\addplot [draw=none] coordinates {(0,0)};
\end{axis}
\node at (1.7,-0.4) {$p$};
\end{tikzpicture} & \hspace{0.4cm} \begin{tikzpicture} \centering
\begin{axis}[
hide axis,
scale only axis,
height=0pt,
width=0pt,
colormap/jet,
colorbar horizontal,
point meta min=0.0006179915465841951,
point meta max=5.150913782641715,
colorbar style={
    height=1mm, width= 1.3cm, xticklabel pos=upper, xtick={0.1, 2.5,5.15}
}]
\addplot [draw=none] coordinates {(0,0)};
\end{axis}
\node at (2.2,-0.3) {$p (\times 10^{3})$};
\end{tikzpicture}\\
  \includegraphics[width=0.28\textwidth]{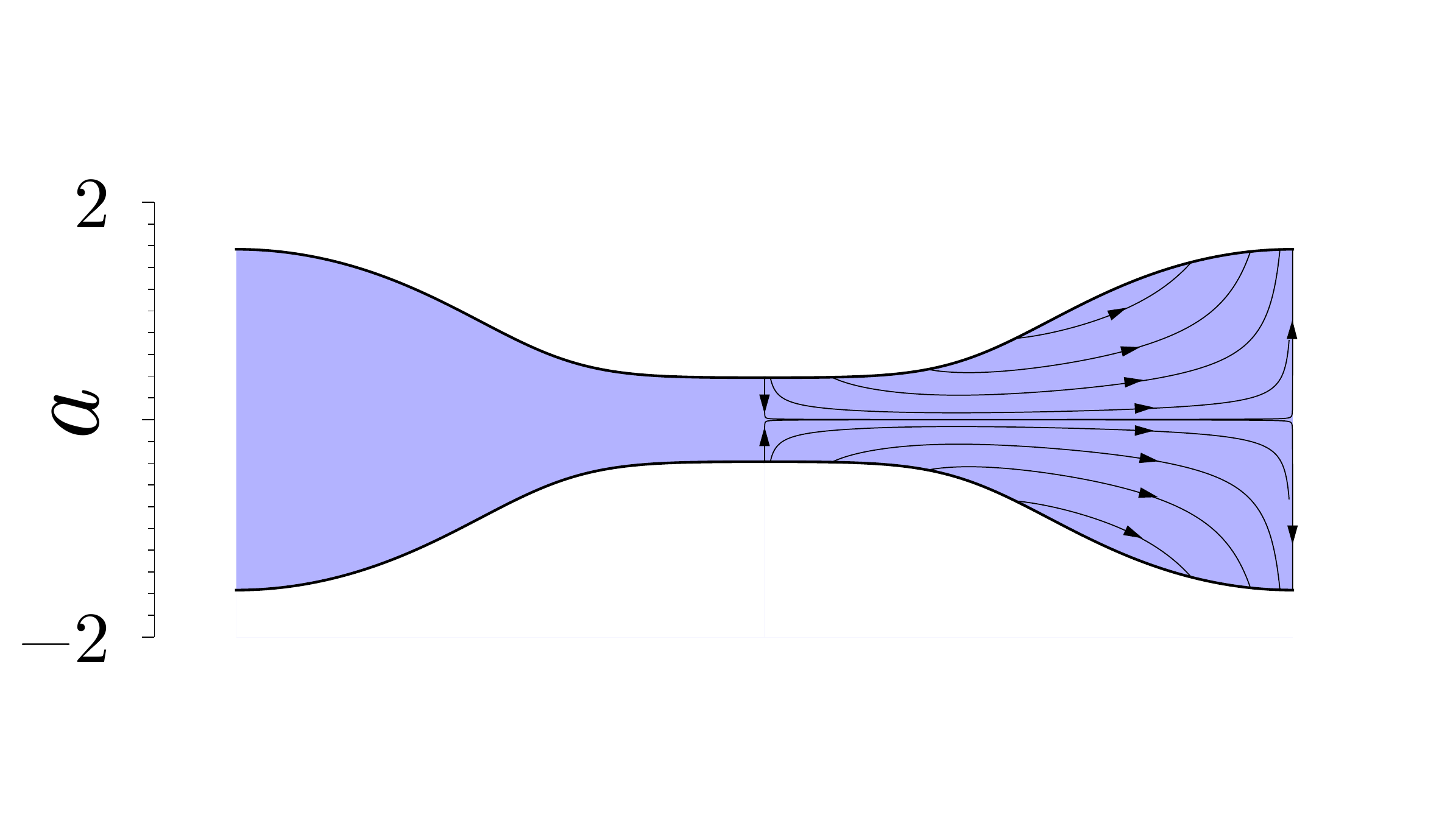} &  \hspace{0.25cm} \includegraphics[width=0.28\textwidth]{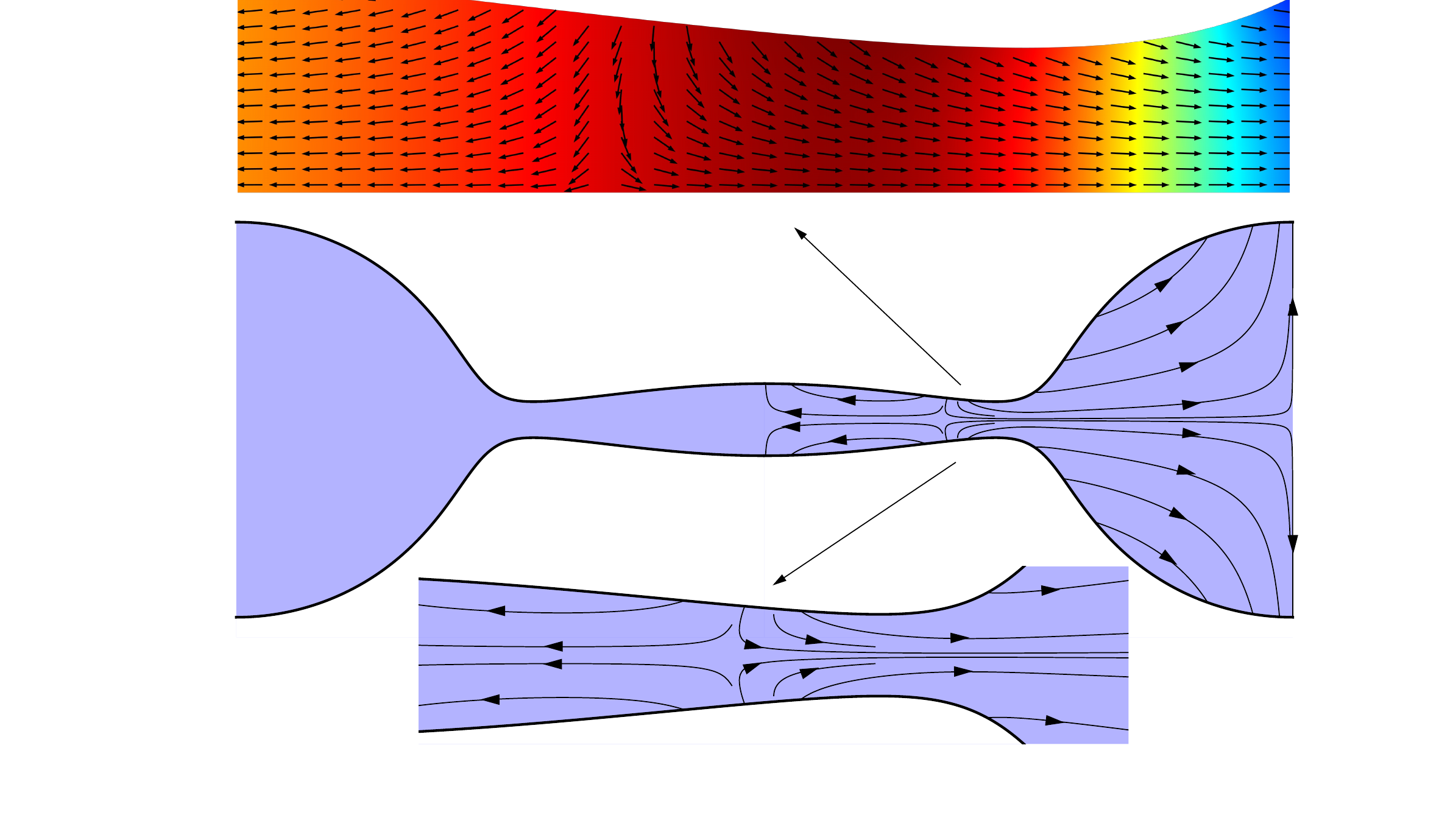} & \hspace{0.35cm} \includegraphics[width=0.28\textwidth]{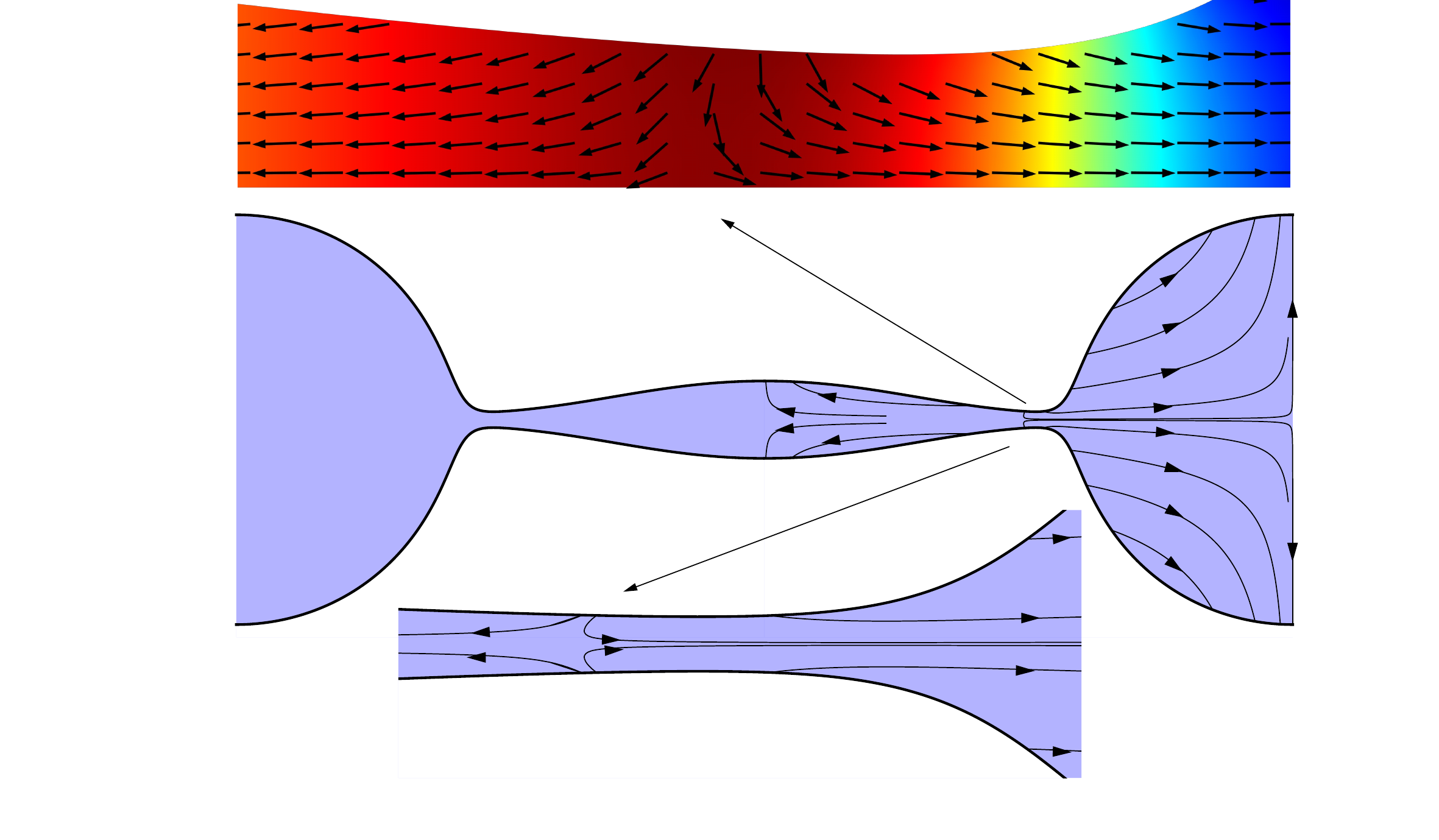}  &\hspace{0.4cm} \includegraphics[width=0.28\textwidth]{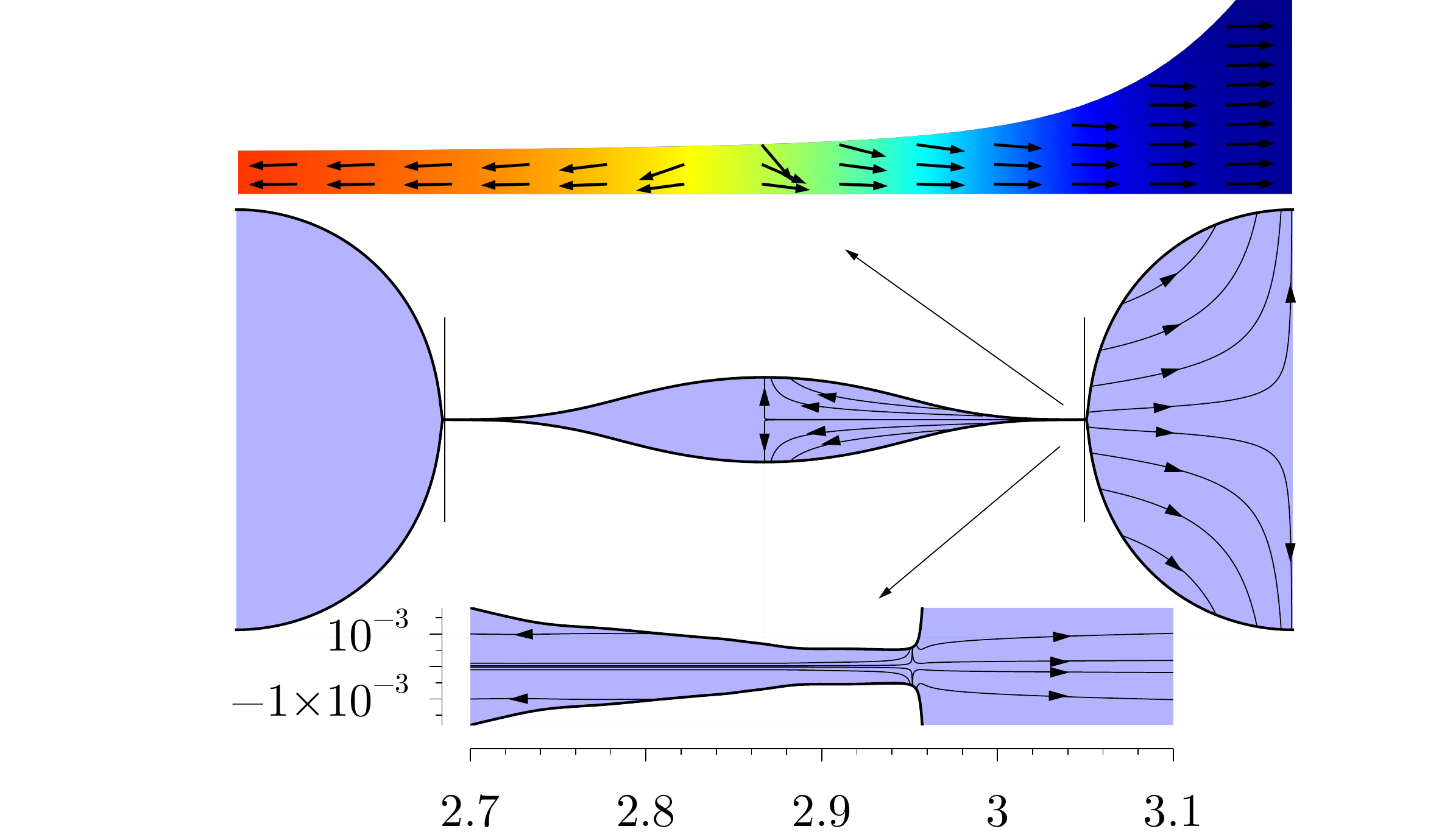} \\
       \vspace{-0.5cm} & & & \\
   \includegraphics[width=0.28\textwidth]{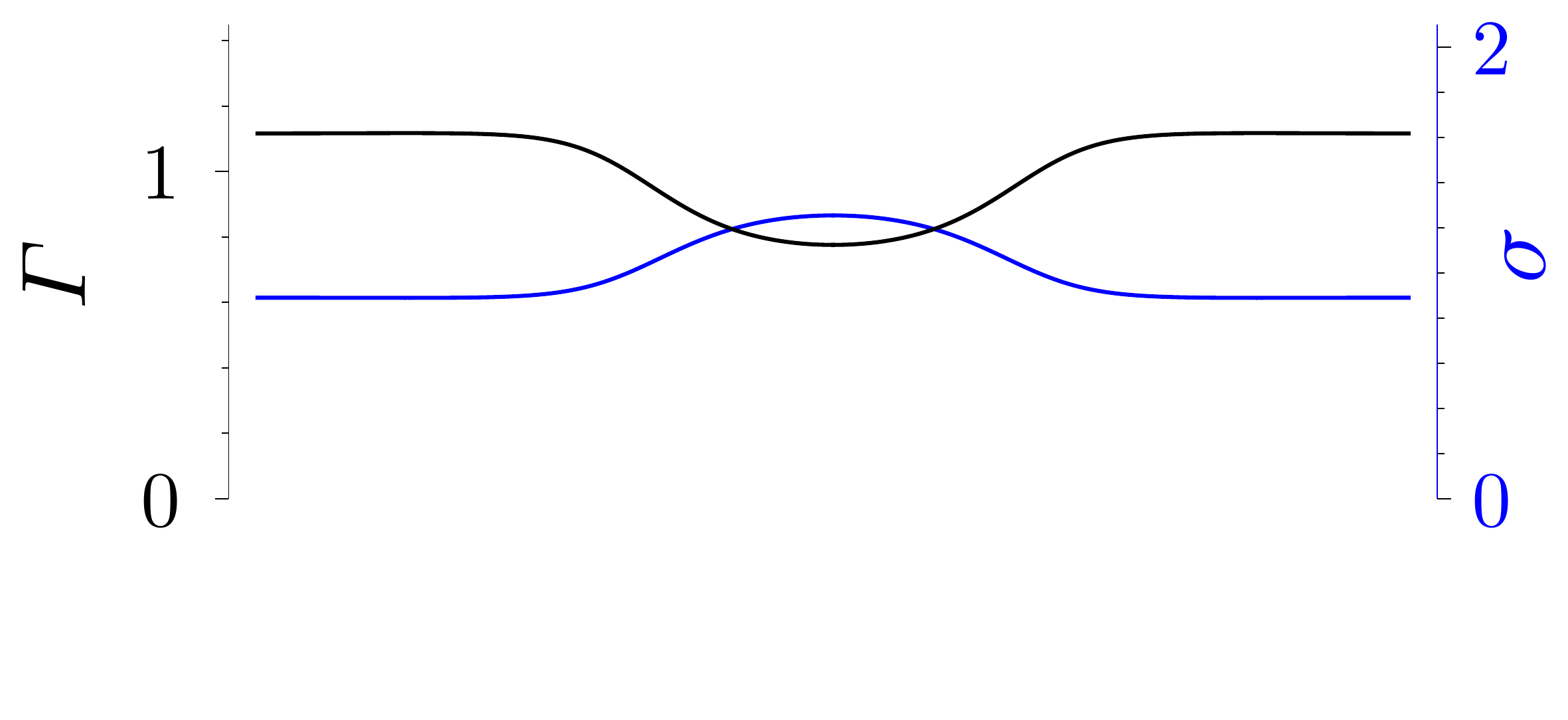} &  \hspace{0.25cm} \includegraphics[width=0.28\textwidth]{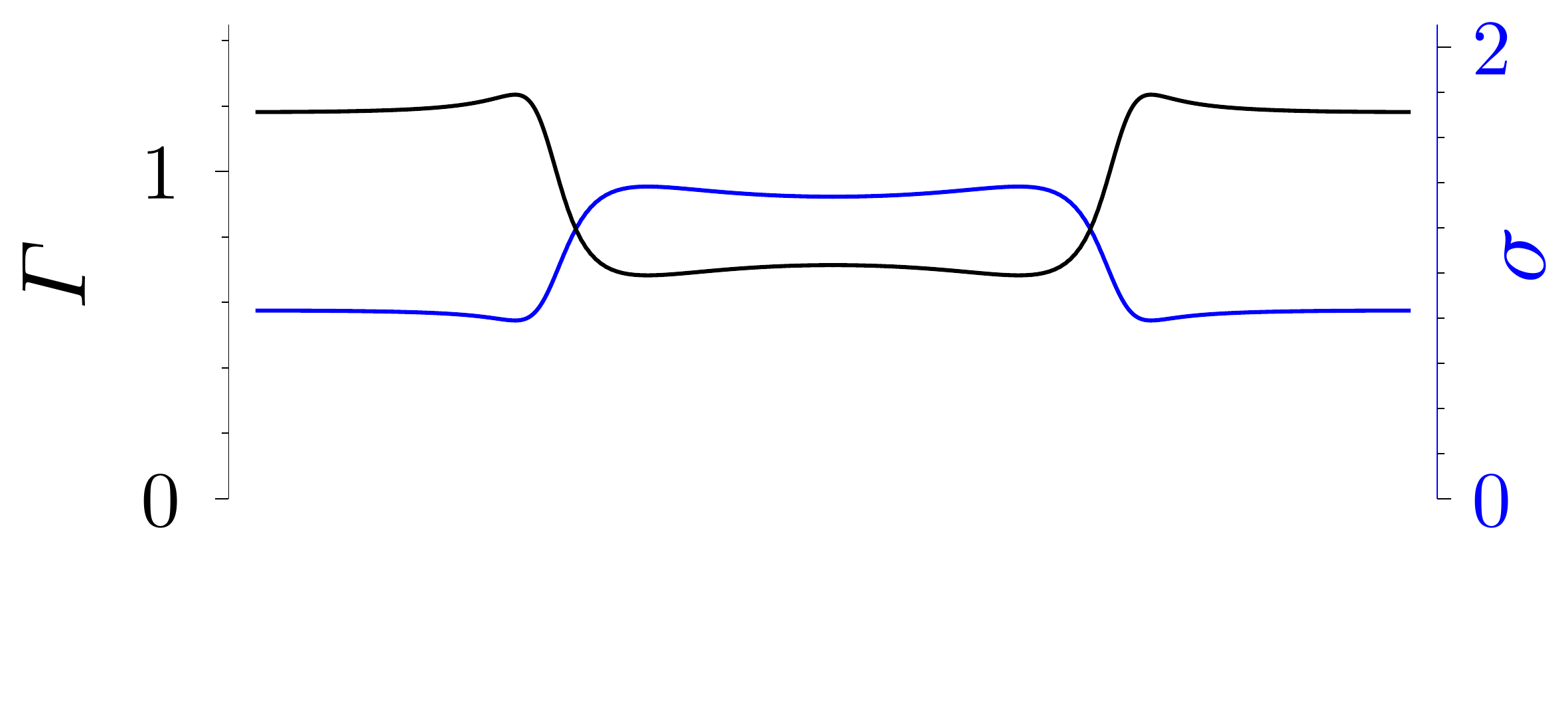} &  \hspace{0.35cm} \includegraphics[width=0.28\textwidth]{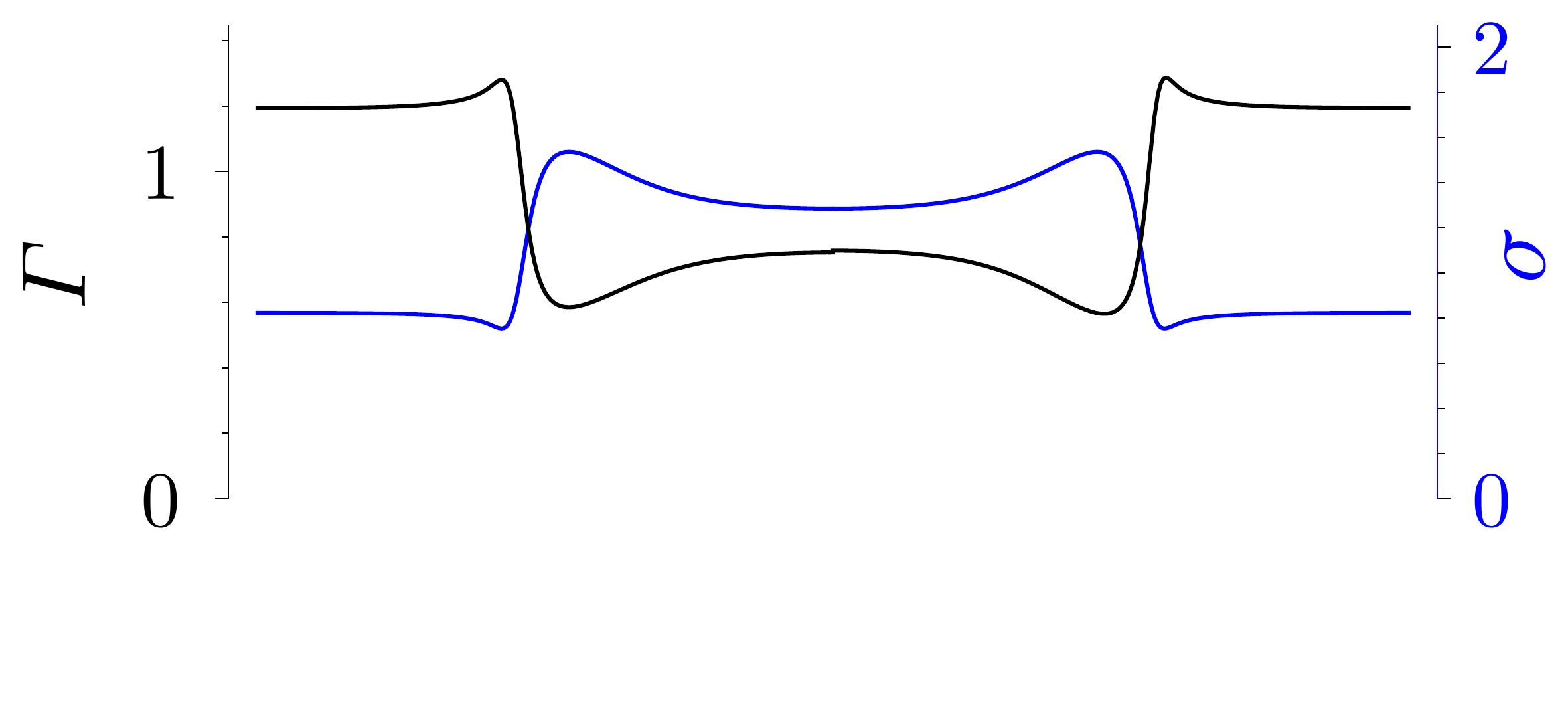} &  \hspace{0.4cm} \includegraphics[width=0.28\textwidth]{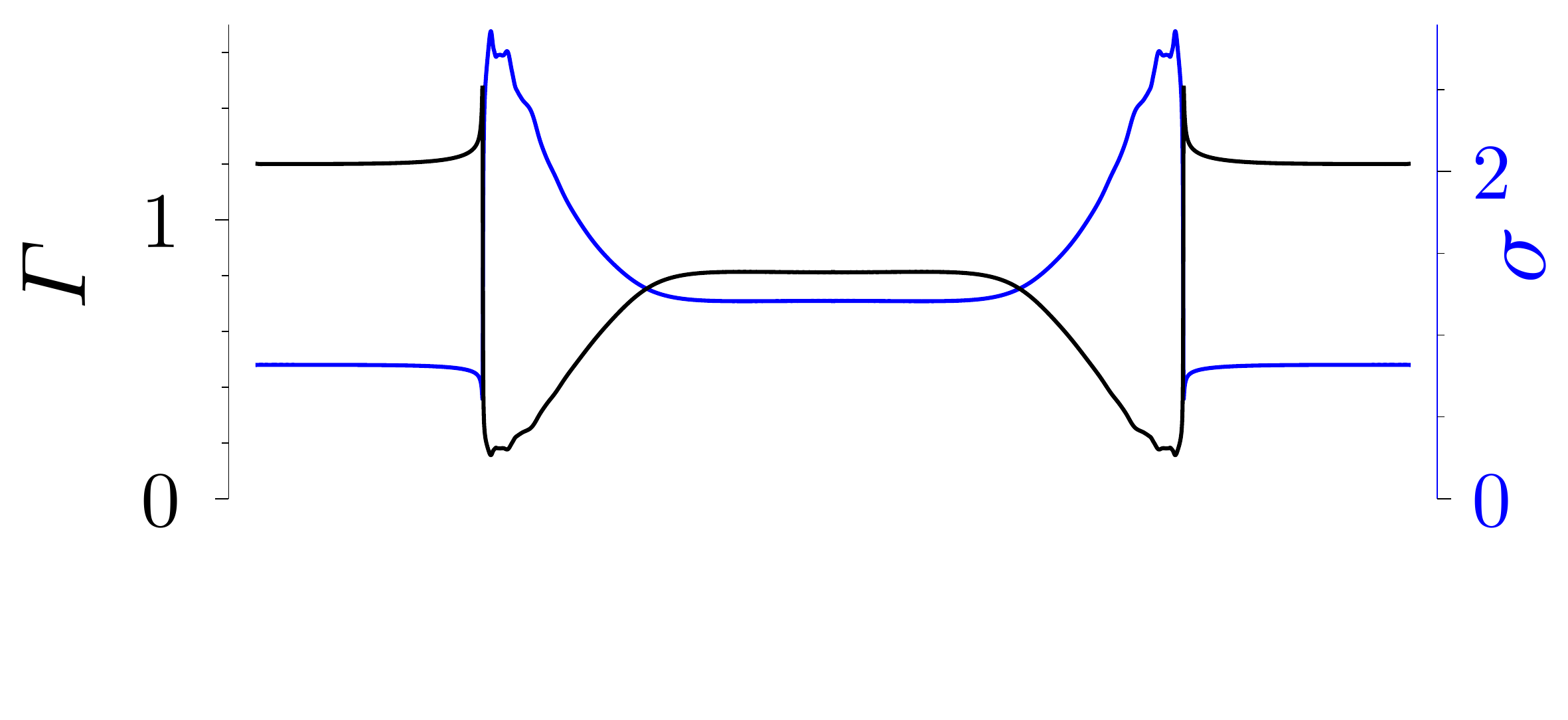} \\
        \vspace{-0.8cm} & & & \\
      \includegraphics[width=0.28\textwidth]{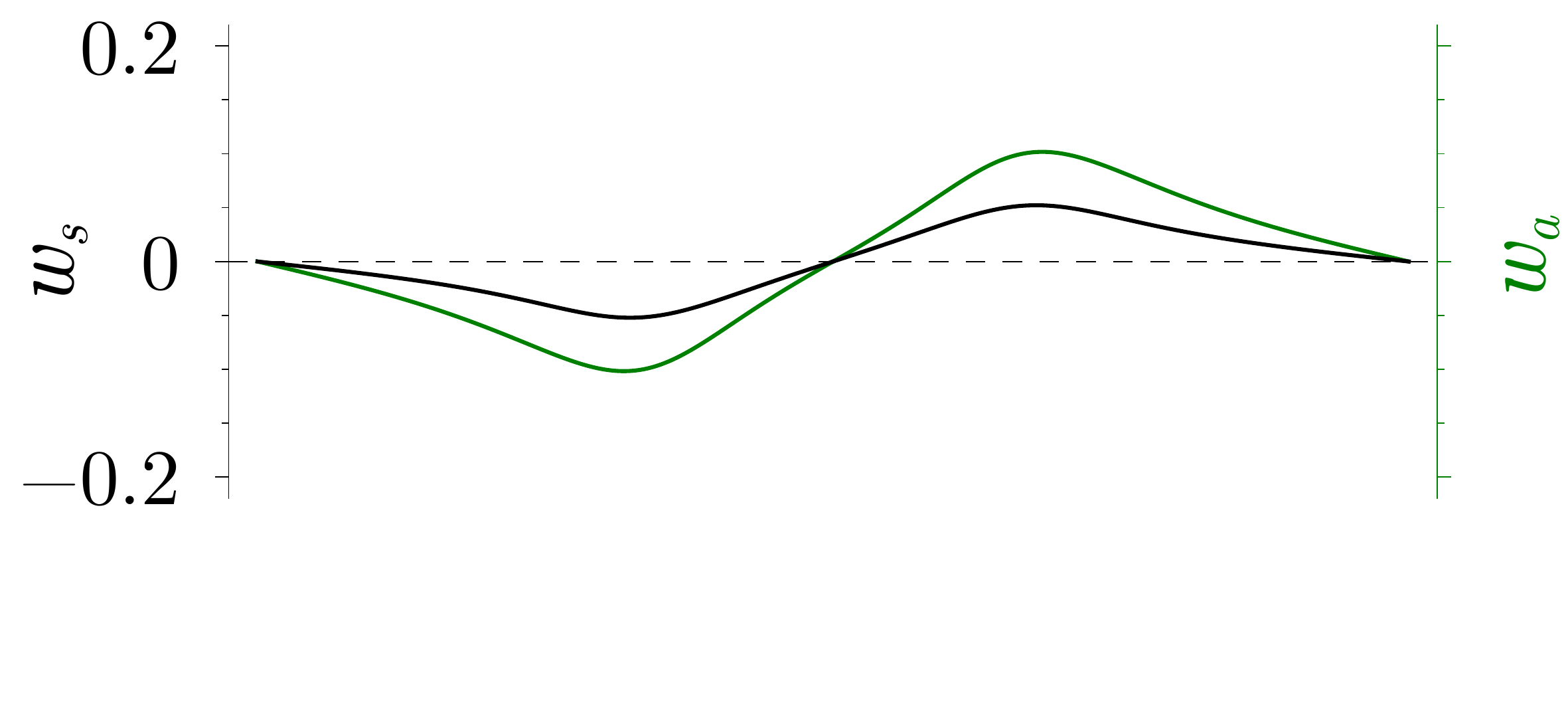} &  \hspace{0.25cm} \includegraphics[width=0.28\textwidth]{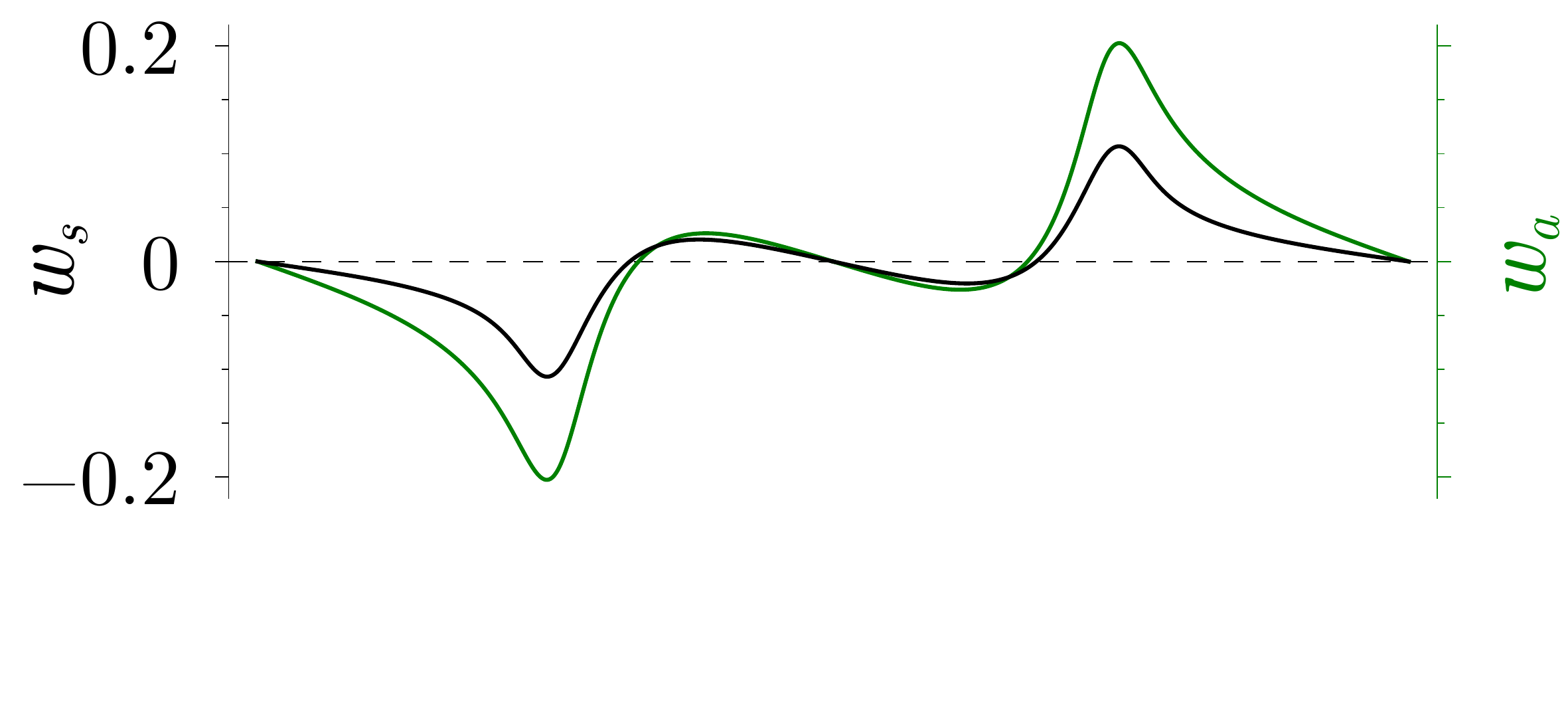} &  \hspace{0.35cm} \includegraphics[width=0.28\textwidth]{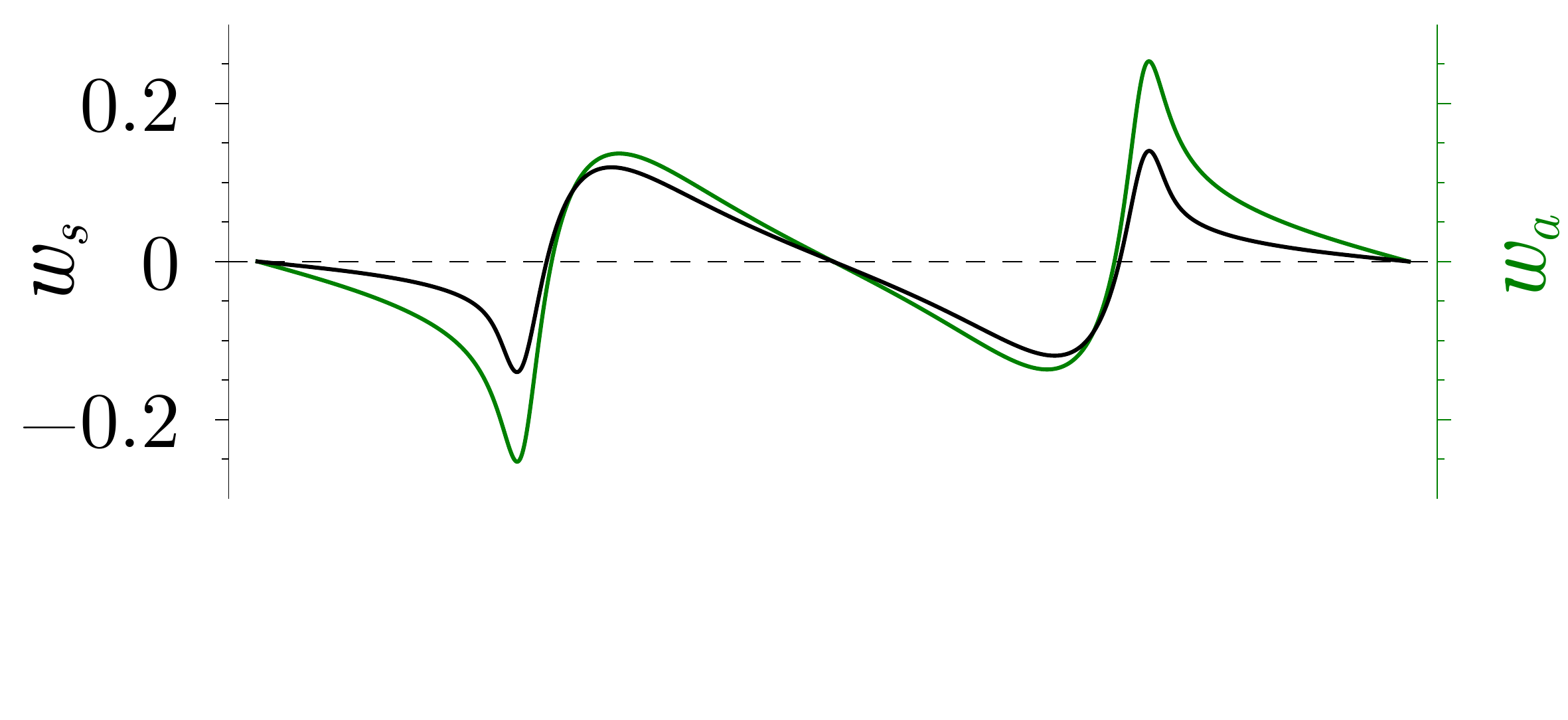} &  \hspace{0.4cm} \includegraphics[width=0.28\textwidth]{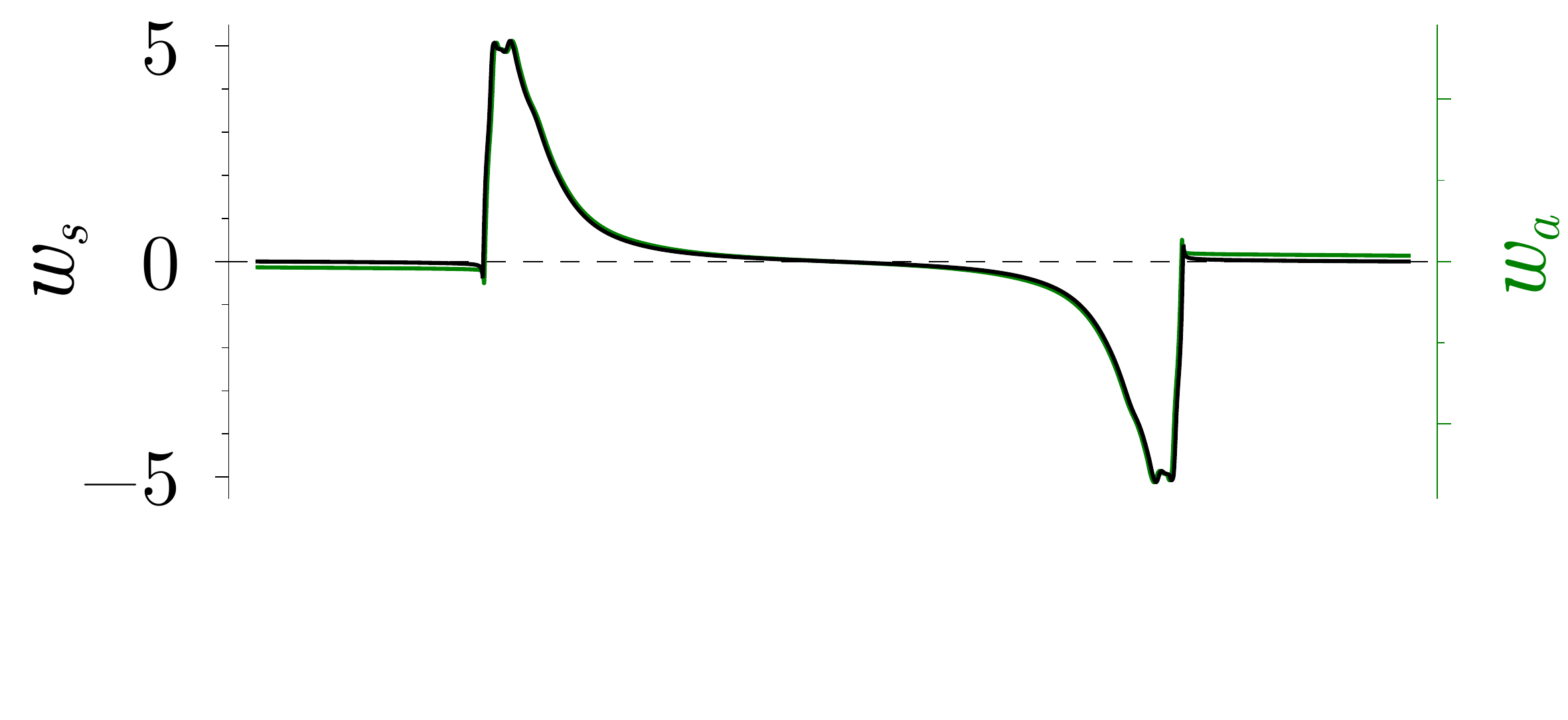} \\
           \vspace{-0.8cm} & & & \\
            \includegraphics[width=0.28\textwidth]{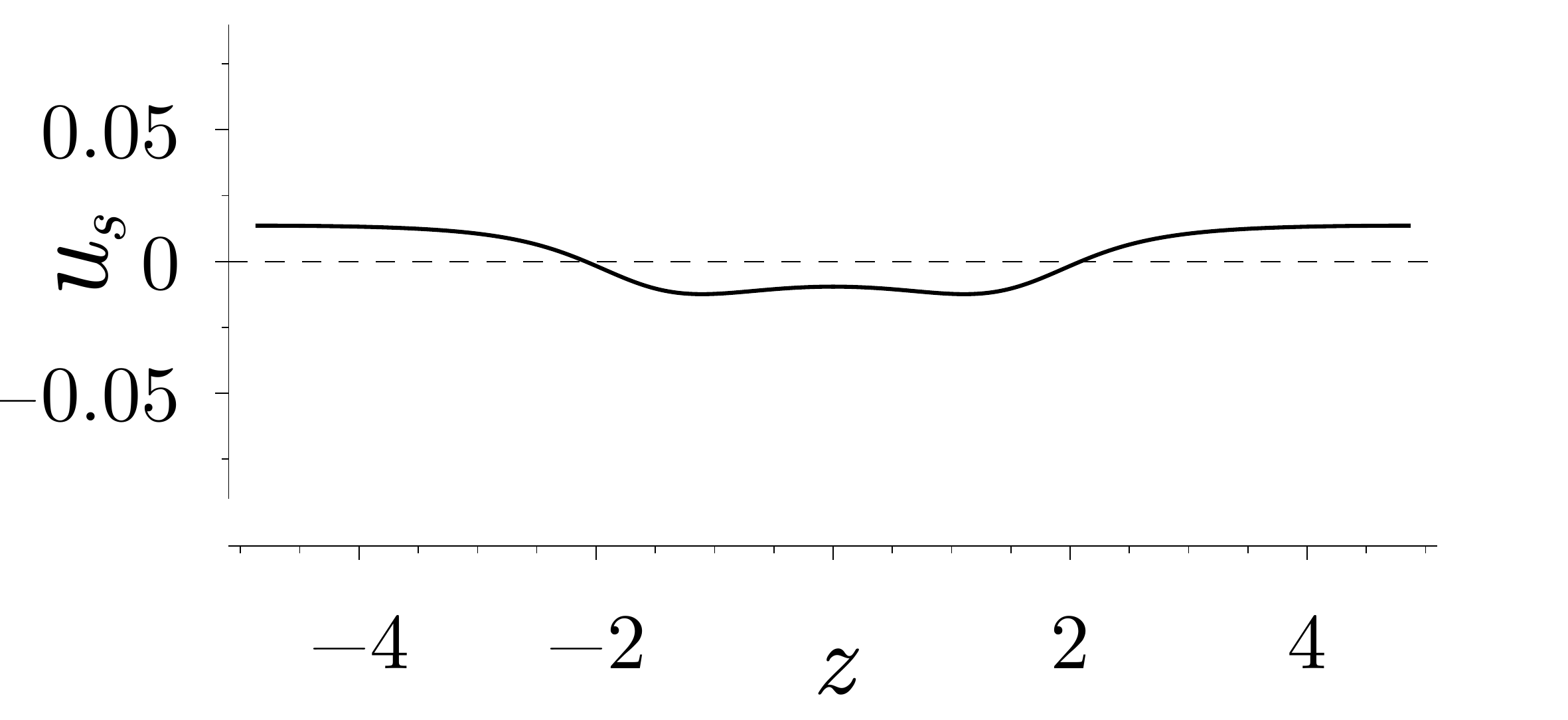} &  \hspace{0.25cm} \includegraphics[width=0.28\textwidth]{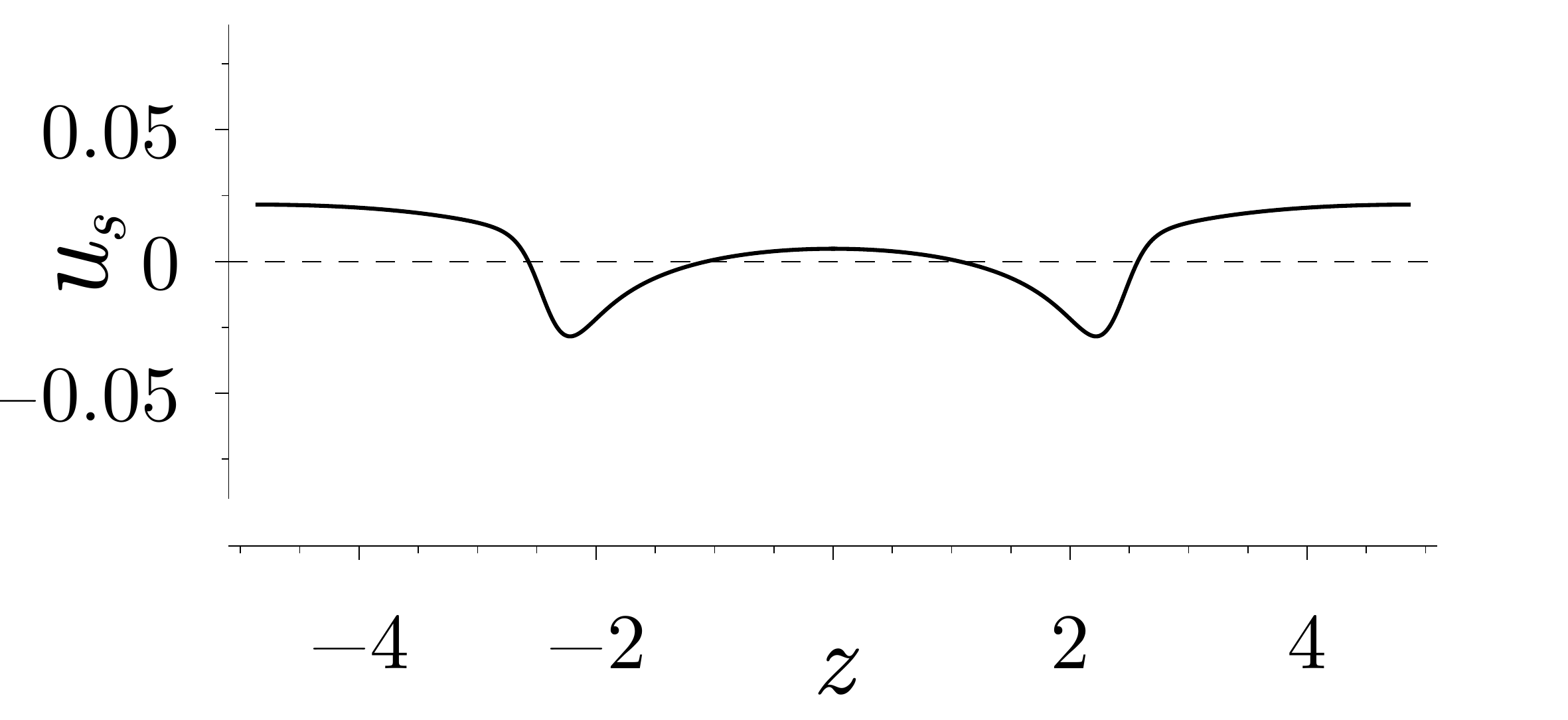} &  \hspace{0.35cm} \includegraphics[width=0.28\textwidth]{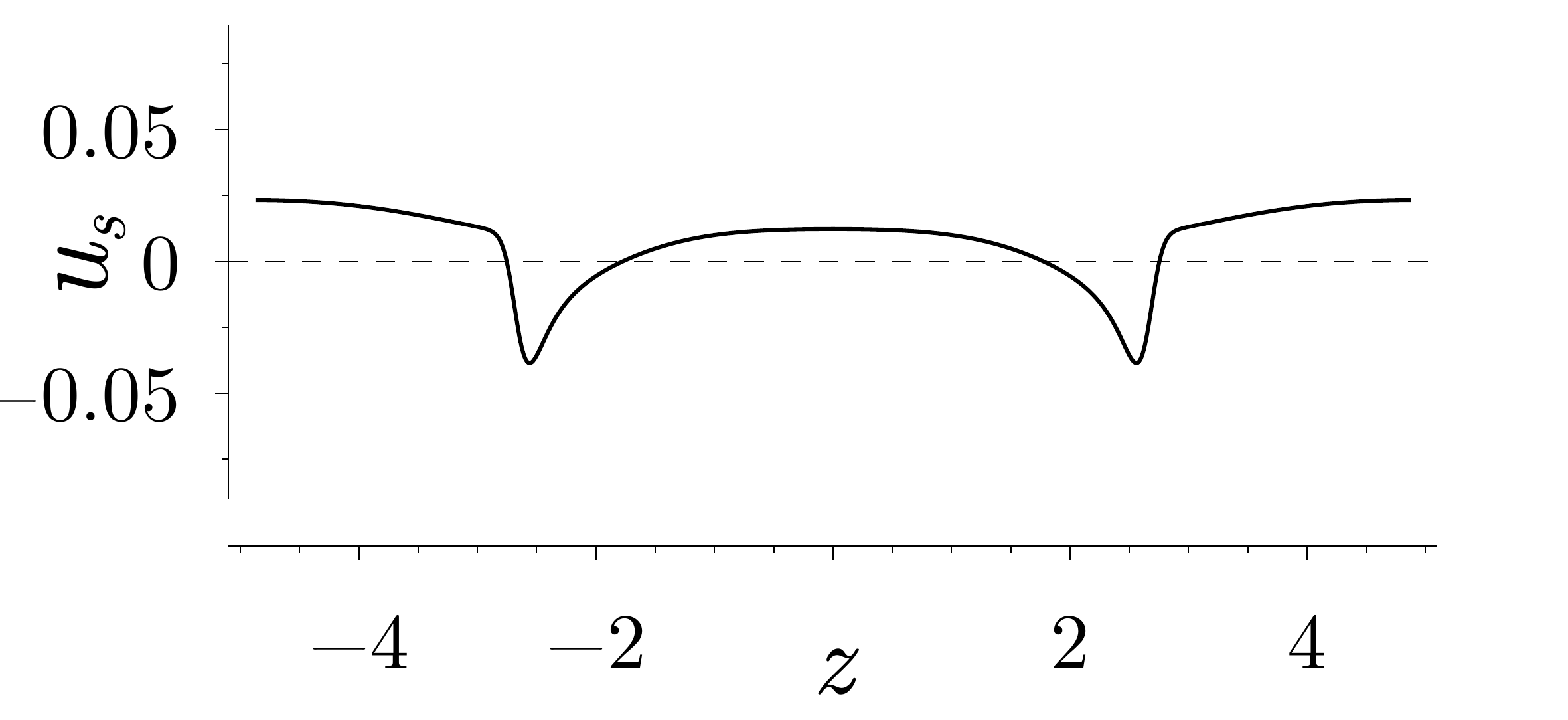} &  \hspace{0.4cm} \includegraphics[width=0.28\textwidth]{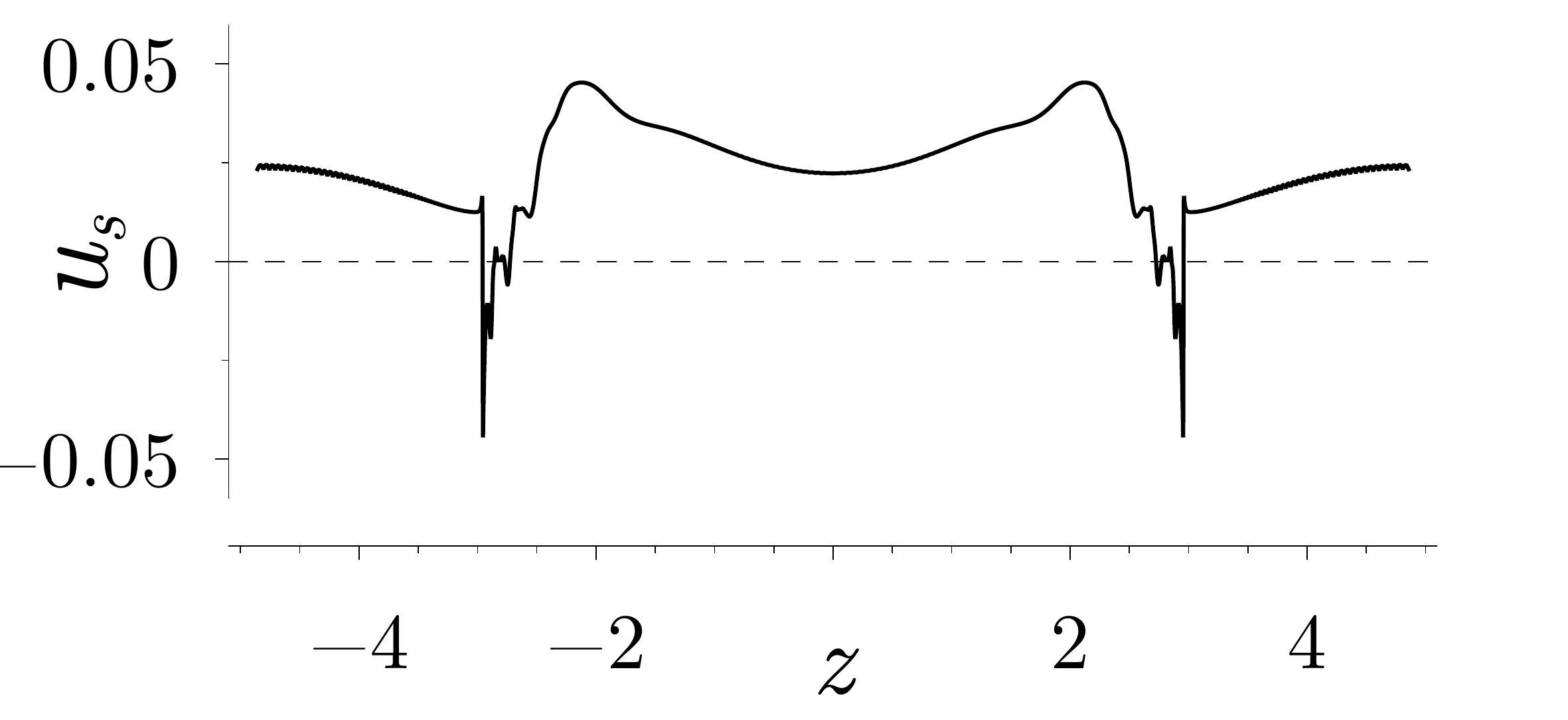} \\
($a$) $t = 290$ & ($b$) $t = 304.5$ & ($c$) $t = 307.5$  & ($d$) $t = 309.53$ \\[4pt]
\end{tabular}
\caption{\label{fig:figure12} (Colour online) Same as figure~\ref{fig:figure11} but for $\Ela = 1$ with $k = k_m = 0.647$. Here $z_{\min} = 2.94$ and $a_{\min} = 5.14 \times 10^{-4}$.}
\end{figure}

When $\Lap = 100$, figures~\ref{fig:figure8}($c$,$d$) show that the effect of surface elasticity is much weaker in the case of dominant inertia, as was anticipated both in figure~\ref{fig:figure5} and also by the shapes shown in the upper row of figure~\ref{fig:figure4}. The small influence of insoluble surfactants in the inviscid limit, $\Lap \gg 1$, had been already noted in the linear stability analyses of~\cite{Whitaker76},~\cite{Hansen99} and~\cite{Timmermans02}. Indeed, the effect of Marangoni stresses is confined to a thin boundary layer at the free surface, where the viscous stress rapidly restores any imbalance of $\sigma$, and which does not have any influence in the bulk liquid motion. Consequently, for $\Lap = 100$, the satellite volume, $\Vsat$, varies only slightly with respect to the value of a clean liquid thread, $\Vsat(\Ela = 0,\Lap=100)\simeq 0.03$, with a minimum at $\Ela \simeq 0.203$, whereas $\Ssat$ increases monotonically as $\Ela$ increases. To explain this result, figures~\ref{fig:figure11} and~\ref{fig:figure12} show two sets of snapshots of $a$, $\Gamma$, $\sigma$, $w_s$, $w_a$ and $u_s$ for $\Ela = 0.203$, at which $\Vsat$ is minimum, and for $\Ela = 1$, respectively.

In the weak-elastic limit, $\Ela \to 0$, a satellite droplet with volume $\Vsat\simeq 3\%$ is formed at pinch-off, as already shown in figures~\ref{fig:figure5} and~\ref{fig:figure8}($c$). The satellite volume decreases as $\Ela$ increases in the range $0 < \Ela \lesssim 0.203$. Indeed, when $\Ela$ increases, the Marangoni rigidification of the interface slows down the pinch-off process by decreasing the interfacial velocities, as evidenced by the time evolution of $w_s$, $w_a$ and $u_s$ in figure~\ref{fig:figure11} with respect to figure~\ref{fig:figure7}($e$--$h$). The latter behaviour, together with the fact that the pressure gradient is locally enhanced due to the variations of $\sigma$, explain why a larger volume is drained out of the satellite droplet compared to the case of a clean interface. However, the Marangoni stress that opposes the drainage flow away from the centre reduces the advection of surfactant towards the main drops, and thus the value of $\Ssat$ increases, as shown in figure~\ref{fig:figure8}($d$). The snapshot in figure~\ref{fig:figure11}($c$) shows that the flow is reversed near the neck region, as happens for a clean interface (see e.g. figures~\ref{fig:figure7}$h$ and~\ref{fig:figure7}$i$). However, in the elastic regime the flow reversal takes place earlier than in the clean interface limit. This behaviour at high values of $\Lap$ and low values of $\Ela$ was previously noticed by~\cite{Kamat2018}, who showed that the stagnation point occurs at earlier stages in surfactant-laden interfaces compared with clean interfaces, due to the strong Marangoni stress in the neck region.

A representative case of $\Lap = 100$ and $\Ela > 0.203$ is shown in the snapshots of figure~\ref{fig:figure12} for $\Ela = 1$. The main change with respect to the preceding case is the fact that for $\beta=1$ the Marangoni stress is strong enough to revert the surface flow at earlier stages, as shown in panels ($b$) and ($d$). Therefore, the stagnation point appears earlier than in the case of figure~\ref{fig:figure11}, and diffuses almost instantaneously in the radial direction, leading to a satellite droplet with larger values of the normalised volume and of the surfactant mass. It can thus be deduced that the minimum value of $\Vsat$ displayed in figure~\ref{fig:figure8}($c$) appears due to a competition between the two aforementioned opposite effects induced by the presence of surfactants.

For $\Lap<7.5$, the two effects described previously coexist when $\Ela$ is increased, as shown by the isocontours of $\Vsat$ in figure~\ref{fig:figure5}. For instance, when $\Lap = 1$, $\Vsat$ first decreases as $\Ela$ increases, and when the elastic stress is strong enough, the flow is reversed and the discontinuous transition occurs. Note that, in the latter case, inertia is important since $\Lap$ is of order unity, and a small but finite satellite droplet exists in the clean limit, $\Ela \to 0$ (see e.g. the second row of figure~\ref{fig:figure4}), where $\Vsat = 0.394$ \% (a value significantly larger than in the limit $\Lap \ll 1$, as shown in the isocontours of figure~\ref{fig:figure5}). Hence, the main difference with respect to the limit $\Lap \gg 1$ is that in this case, since $\Vsat(\Ela \to 0)$ is small, the increase of $\Ela$ reduces the satellite volume and may even make it negligible. For $\Lap < 7.5$, $\Ssat$ also decreases monotonically together with $\Vsat$ when $\Ela < \Ela_c$, which can be explained by the fact that $\Vsat$ is already small when $\Ela = 0$, so that $\Ssat$ necessarily decreases when $\Ela$ is increased.

Let us recall at this point that the critical elasticity, $\Ela_c(\Lap)$, decreases as $\Lap$ increases within the range  $0 < \Lap < 7.5$, as shown in figures~\ref{fig:figure3} and~\ref{fig:figure5}. The reason for the latter trend is the fact that the advection of surfactant away from the central region is enhanced by the liquid inertia, so that $\bnabla_s \sigma$ also increases, and thus the value of $\Ela$ for which the elastic stress reverts the flow is smaller. Furthermore, the value of $\Vsat(\Lap,\Ela \to 0)$ increases as $\Lap$ becomes larger, and therefore the jumps experienced by $\Vsat$ and $\Ssat$ at the discontinuous transition, $\Ela = \Ela_c$, decrease, as deduced from the inset of figure~\ref{fig:figure3}. Finally, for $\Lap > 7.5$, the discontinuous transition disappears.

\begin{figure}
\centering
\includegraphics[height=0.75\textwidth]{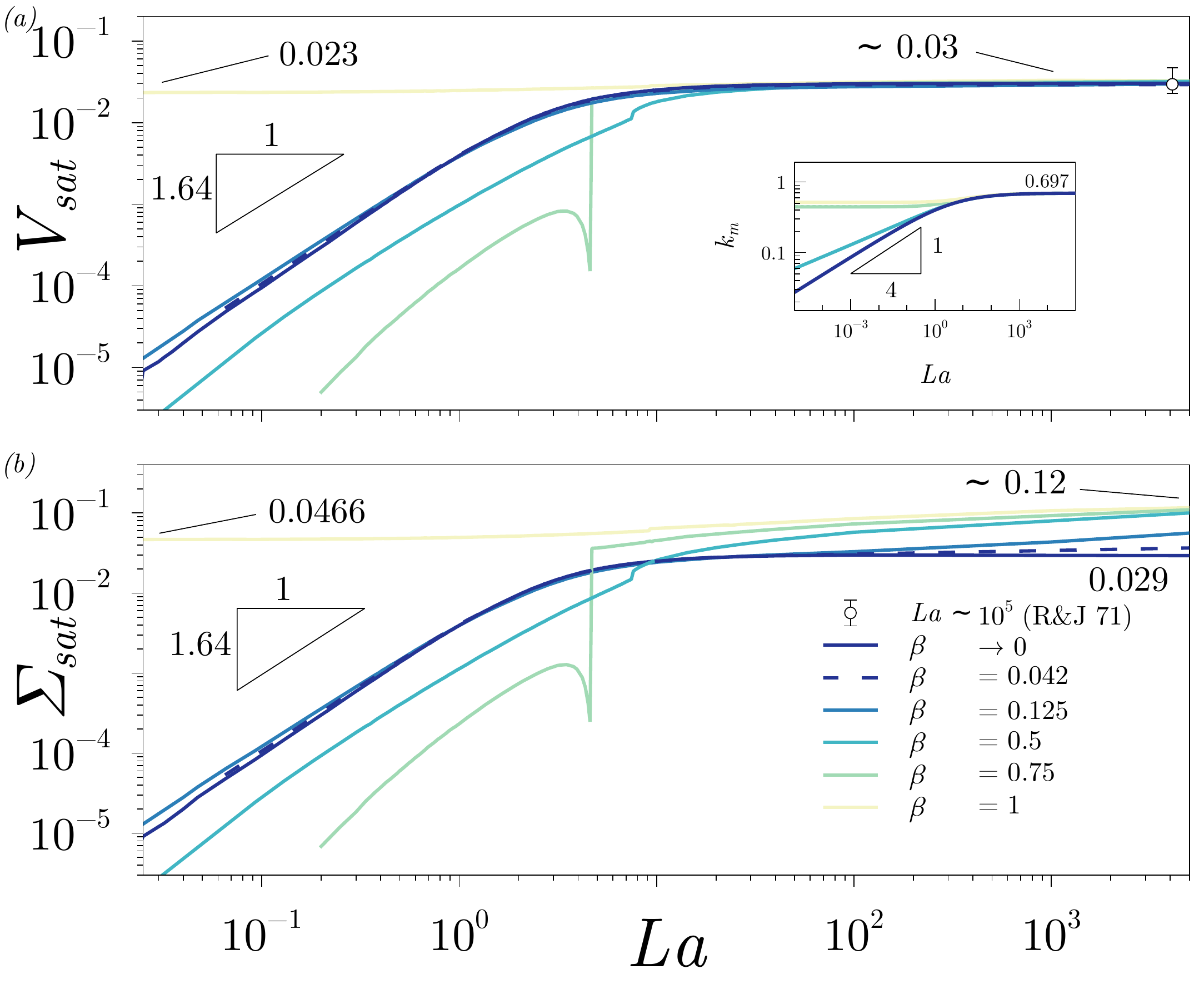}
\caption{\label{fig:figure6} (Colour online) Normalised satellite's volume $V_{sat}$, and normalised mass of surfactant trapped at its interface $\Ssat$, as a function of the Laplace number $\Lap$ in log-log for different values of $\Ela$ indicated in the legend. The inset shows the dependence of the maximum amplification wavenumber $k_m$ with respect to $\Lap$ in log-log. The circle with error bars corresponds to the experiments of the natural break-up of a liquid jet of water performed by~\cite{Rutland1970}.}
\end{figure}

\subsubsection{Scaling laws for $\Vsat$ and $\Ssat$ as functions of $\Lap$}\label{subsubsec:scaling_laws}

Figure~\ref{fig:figure6} shows $\Vsat$ and $\Ssat$ as functions of $\Lap$ for different values of $\Ela$ indicated in the legend. The circle with error bars corresponds to the experiment of~\cite{RutlandJameson} of the natural break-up of a liquid jet of clean water, which is in close agreement with our numerical result for $\Ela = 0$. The inset displays the most unstable wavenumber, $k_m$, as a function of $\Lap$, showing the inviscid plateau $k_m \simeq 0.697$ for $\Lap \gg 1$~\citep{Rayleigh1}, as well as the power-law dependence for small values of $\Lap$. The latter power law can be deduced from the long-wave approximation of the dispersion relation~\eqref{eq:DR} or, equivalently, from the leading-order one-dimensional model deduced by~\cite{EggersDupont} and~\cite{GyC}. In the clean case, $\Ela = 0$, the leading-order one-dimensional results are $k_m \sim (2+3 \sqrt{2} \Lap^{-1/2})^{-1/2}$ and $\omega_m \sim (2 \sqrt{2} + 6 \Lap^{-1/2})^{-1}$~\citep{EGG08}. The latter long-wave result provides very accurate results in the whole range of $\Lap$, since $k \in (0,1)$ accomplishes the slenderness assumption. In the inviscid limit, $\Lap \to \infty$, both $\omega_m$ and $k_m$ are slightly overestimated by the one-dimensional model, namely $\omega_m \to 2^{-3/2}$ and $k_m \to 2^{-1/2}$. However, in the Stokes limit, $\Lap \ll 1$, the values of $\omega_m \to 1/6$ and $k_m = 3^{-1/2} 2^{-1/4} \Lap^{1/4}$ are in excellent agreement with the exact linear theory. When $\Ela > 1/2$, the elastic stress regularises $k_m$ in the limit of $\Lap \to 0$, as analysed in detail by~\cite{Timmermans02} (see also the isocontours of $k_m$ in figure~\ref{fig:figure2}$c$).


In the limit of a clean interface, $\Ela = 0$, $\Vsat$ increases monotonically with $\Lap$, as previously shown in figures~\ref{fig:figure4} and~\ref{fig:figure5}, and explained in figure~\ref{fig:figure7}. In particular, our numerical results reveals that the satellite volume scales as $\Vsat = 0.00421 \Lap^{1.64}$ when $\Lap \lesssim 2$, and thus $\Vsat \to 0$ and $\Ssat \to 0$ as $\Lap \to 0$. When $\Lap$ is finite, a satellite drop is always formed, since the liquid thread always experiences a transition to the inertial--viscous regime~\citep{Eggers1993,Castrejon2015} and thus $z_{\min}$ moves from $z=0$ towards higher values when $t$ is close enough to $t_b$. The elongated satellite droplet formed when $\Lap \ll 1$ can break up into more droplets after pinch-off as it relaxes, depending on the value of $\Lap$~\citep{Notz2004, Castrejon2012, Wang2019, Anthony2019}, unless $\Lap \to 0$~\citep{EggersFontelos2005}. Alternatively, using the expression for the equivalent radius $R_{\text{sat}}$ developed in~\S\ref{subsec:parametric}, which depends on $k_m$, and since $k_m = 3^{-1/2} 2^{-1/4} \Lap^{1/4}$ within the range of $\Lap$ for which $\Vsat$ exhibits power-law scaling, it is deduced that $R_{\text{sat}} = 0.34 \Lap^{0.463}$.

When $\Lap \gtrsim 10$, the value of $\Vsat$ reaches a plateau of about 3\%, as already discussed in the context of figures~\ref{fig:figure5} and~\ref{fig:figure8}($c$). Equivalently, since $k_m \simeq 0.697$ when $\Lap \gg 1$, $R_{\text{sat}} \simeq 0.588$ in the inviscid limit, in excellent agreement with the experiments of~\cite{RutlandJameson} (circle with error bars in figure~\ref{fig:figure6}$a$), and also with the numerical simulations of~\cite{Ashgriz1995}. In the weak-elasticity limit, $\Ela < 0.05$, the behaviour of $\Ssat$ is identical to that of $\Vsat$, displaying the same scaling law within the same range in $\Lap$, and also reaching an inviscid plateau of about 2.9\% when $\Lap \gtrsim 10$. This scaling law for $\Vsat$ and $\Ssat$ prevails when $\Ela < \Ela_c$ and $\Lap < \Lap_c$, although the prefactor changes with $\Ela$ as shown in figure~\ref{fig:figure6}. In particular, when $\Ela$ increases the prefactor is smaller and thus $\Vsat$ and $\Ssat$ reach smaller values as $\Lap \to 0$. This can be explained by the translation of $z_{\min}$, which is inhibited as $\Ela$ becomes higher and thus the surface stress exerted at the interface increases.

Figure~\ref{fig:figure6} also shows that, when $\Lap \gg 1$, the 3\% plateau reached by $\Vsat$ is barely affected by $\Ela$ since, as explained previously, inertia dominates and the elastic stress cannot induce any substantial change in the bulk motion~\citep{Whitaker76,Hansen99,Timmermans02}. As inertia increases, the influence of $\Ela$ on $\Vsat$ becomes even weaker than in the case of $\Lap = 100$ displayed in figure~\ref{fig:figure8}($c$). Although the satellite shape at pinch-off is the same because viscosity cannot balance the elastic stress and transmit it to the bulk, $\Ssat$ reaches different inviscid limits as $\Ela$ increases. In particular, the 2.9\% inviscid plateau reached by $\Ssat$ in the weak-elasticity limit increases with $\Ela$, the reason being the same as in the case of figure~\ref{fig:figure8}($d$).

When $\Ela = 1 > \Ela_c(\Lap)$ for arbitrary values of $La$, a satellite is always formed with $\Vsat \gtrsim 2 \%$ and $\Ssat \gtrsim 4 \%$, and both increase smoothly with $\Lap$ as inertia becomes more important. In fact, $\Vsat$ and $\Ssat$ reach respective plateaus in both the Stokes and Euler limits. For $\Lap \to 0$, the values of $\Vsat$ and $\Ssat$ are about 2.30\% and 4.67\%, respectively, whereas in the limit $\Lap \to \infty$, their values are 3\% and 12\%, approximately. This trend prevails provided that $\Ela > \Ela_c(\Lap \to 0) = 0.978$, as shown in the isocontours of $\Vsat$ and $\Ssat$ in figure~\ref{fig:figure5}. Hence, in the elasticity-dominated regime where $\Ela > \Ela_c$, the effect of inertia on $\Vsat$ and $\Ssat$ is weaker, although the column for $\Ela=1$ in figure~\ref{fig:figure4} reveals that the shape of the thread at pinch-off changes substantially. As already mentioned, inertia tends to form oval-shaped satellites, which are more likely to break up in the relaxation process after pinch-off, whereas the surface elasticity tends to form spherical satellites which will not experience secondary break-up events.



\section{Conclusions}\label{sec:conclusions}

In this paper we have reported an exhaustive numerical study of the unforced break-up of free axisymmetric threads of Newtonian liquid whose interface is coated with insoluble surfactants. Our main objective was to describe and explain how the presence of these molecules affects the nonlinear dynamics of the liquid thread and the satellite drop formation regimes when the dynamics is triggered by the most dangerous initial disturbance. Under these conditions we have shown that, when the initial perturbation amplitude is sufficiently small, the flow depends on two dimensionless parameters, namely the Laplace number $\Lap$ and the elasticity parameter $\Ela$. Our numerical simulations have allowed us to characterise the influence of these two parameters on the satellite volume $\Vsat$, the mass of surfactant trapped at its interface $\Ssat$, the nonlinear correction to the linear break-up time $\DtNL$, and the satellite sphericity $\Spher$, all of them computed at times very close to break-up. It is important to emphasise that our numerical simulations do not contemplate the post break-up behaviour of the threads and satellites, including their relaxation or eventual secondary break-up events. Indeed, an accurate analysis of the dynamics beyond break-up is an important though technically challenging task, which is out of the scope of the present study. Clearly, a future task to be pursued would be to extend the present results by performing numerical simulations that are able to compute the post pinch-off dynamics to reveal the ultimate state of the unstable liquid thread.

We have found a discontinuous transition at a critical elasticity number $\Ela = \Ela_c(\Lap)$ within the range $0 < \Lap < 7.5$, at which $\Vsat$ and $\Ssat$ change abruptly. We have explained this behaviour in terms of a competition between the Plateau-Rayleigh instability mechanism and the elastic or Marangoni stresses that arise due to interfacial surface tension gradients. When $\Ela$ is high enough, the elastic stress that opposes the flow induced by the capillary pressure gradient is able to revert it at the interface. Afterwards, the surface stagnation point diffuses radially inwards, and finally a net flux of liquid swells the central region forming a satellite droplet prior to pinch-off.


When $\Lap < 7.5$, $\Vsat$ and $\Ssat$ increase from a non-zero satellite droplet for $\Ela < \Ela_c$, to a larger value when $\Ela > \Ela_c$. When $\Lap \lesssim 0.2$, the critical elasticity number reaches a plateau, $\Ela_c = 0.978$. Finally when $\Lap > 7.5$ the abrupt transition disappears. In between, $\Ela_c$ decreases monotonically with $\Lap$, since inertia enhances the gradients of surface tension.

For a clean liquid thread, $\Ela \to 0$, we have provided a new scaling law for the normalised satellite volume, namely $\Vsat = 0.00421 \Lap^{1.64}$, which is valid for $\Lap \lesssim 1$. We have shown the existence of a regular weak-elasticity limit, $\Ela < 0.05$, for which the latter scaling law holds, and for which the normalised mass of surfactant carried by the satellite, $\Ssat$, exhibits the same scaling law as $\Vsat$. In this limit, when inertia is sufficiently dominant, namely $\Lap \gtrsim 10$, both $\Vsat$ and $\Ssat$ reach respective limits of about 3\% and 2.9\%, the value of 3\% being in close agreement with previous experiments~\citep{RutlandJameson} and numerical simulations~\citep{Ashgriz1995}.

When $\Lap = 100$ the 3\% inviscid plateau in $\Vsat$ varies slightly with $\Ela$, and displays a minimum within the range $0.2 \lesssim \Ela \lesssim 0.4$, whereas the 2.9\% inviscid plateau of $\Ssat$ increases monotonically. The existence of this minimum has been explained by the competition between two opposed effects induced by the presence of surfactants: (I) the reduction of the surface tension $\sigma$ when $\Gamma$ increases, which enhances the capillary pressure gradient, and (II) the Marangoni stress exerted at the interface due to the gradients of $\sigma$. The initial decrease of $\Vsat$ when $\Ela$ grows is due to (I), whereas the increase above the minimum value is due to (II), which is able to revert the flow at earlier stages of the thread evolution when $\Ela$ is sufficiently high. The decrease of $\Vsat$ with $\Ela$ also coexists with the discontinuous transition for $\Lap < 7.5$. Additionally, the increase of $\Ssat$ when $\Lap > 7.5$ is explained by the reduction of the interfacial velocity due to (II), which tends to accumulate surfactant molecules at the satellite.

When $\Lap \gg 1$, the effect of surface elasticity is very weak and the 3 \% plateau of $\Vsat$ does not vary with $\Ela$, since its effect is confined to a thin Marangoni boundary layer at the interface, where viscous dissipation tends to restore a modified but constant value of $\sigma$. The most important effect of $\Ela$ in the inviscid limit is the fact that $\Ssat$ increases with $\Ela$ for the reason explained in the previous paragraph.


Here, we have considered a nonlinear equation of state for $\sigma(\Gamma)$ that is deduced from the equilibrium thermodynamics of the interface together with the conservation of molecules in the insoluble limit. We have shown that using this nonlinear equation leads to substantial quantitative differences with respect to the use of its linearised version~\citep{Dravid2006}. These differences call for a careful experimental study of the present jet flow configuration or a similar one, e.g. a cylindrical liquid bridge between two static discs whose length is above the critical one for spontaneous break-up. The latter configuration has been recently studied experimentally by~\cite{Kovalchuk2018} for concentrations above the critical micelle concentration. An experimental campaign would also be needed to probe the validity of the insoluble approximation. In fact, it would be interesting to extend the present numerical study to the soluble case, contemplating both bulk diffusion and sorption kinetics. To that end, the bulk diffusion equation together with appropriate adsorption and desorption kinetic equations should be coupled to the equations integrated in the present work~\citep{karapetsas2013primary}. We believe that the numerical techniques employed herein should be able to properly tackle the soluble problem with minor modifications.

A natural and important extension of the present work is tackling the forced jet problem, in which the wavenumber $k$ is not restricted to the most unstable one, and the amplitude $\epsilon$ is not necessarily small. Another feature that deserves future work is the effect of surface diffusion on the satellite drop formation regimes described herein, especially in cases where $\Lap\lesssim O(1)$, for which the surface diffusion time could be of the order of the thread break-up time. Similarly, for small-scale threads, the surface shear and dilational viscosities could also play an important role~\citep{Boussinesq1913,Scriven60,MartinezSevilla2018}. The present numerical analysis should also be extended together with experiments to reveal the conditions under which diffusive and surface viscous effects become relevant, and how they affect the transitions described in the present work.

\begin{acknowledgements}
\section*{Acknowledgements} \label{app:acknow}
A.M.-C. and A.S. thank the Spanish MINECO, Subdirecci\'on General de Gesti\'on de Ayudas a la Investigaci\'on, for its support through projects DPI2014-59292-C3-1-P and DPI2015-71901-REDT, and the Spanish MCIU-Agencia Estatal de Investigaci\'on through project DPI2017-88201-C3-3-R. These research projects have been partly financed through FEDER European funds. A.M.-C. also acknowledges support from the Spanish MECD through grant FPU16/02562 and its associated programme Ayudas a la Movilidad 2017 during his stay at TIPs--ULB, Brussels. J.R.-R. and B.S. thank the FRS-FNRS for financial support, in particular under the umbrella of the Wolflow project.

\end{acknowledgements}


\begin{appendix}
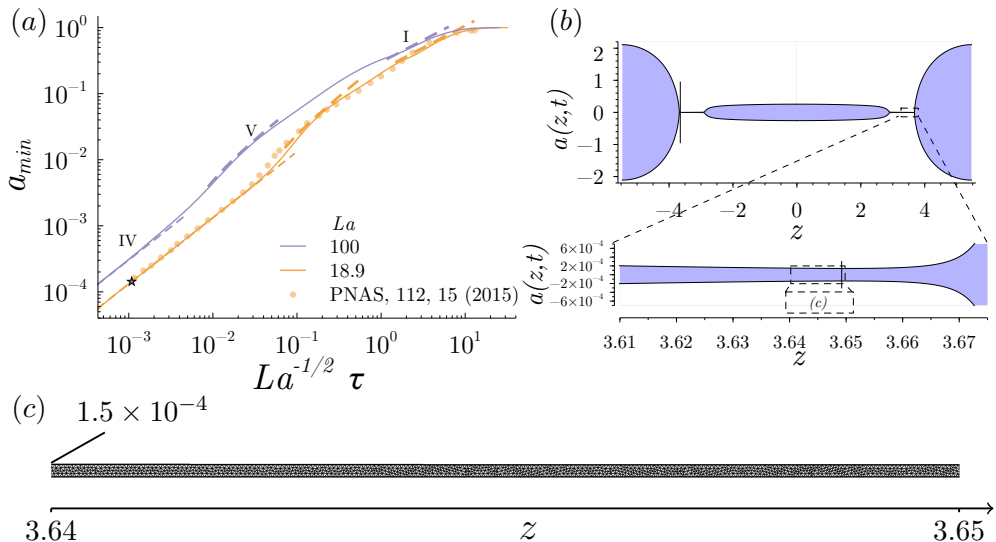
\begin{figure}
\begin{tikzpicture}
 \node at (-2,0) {
        \includegraphics[width=70mm]{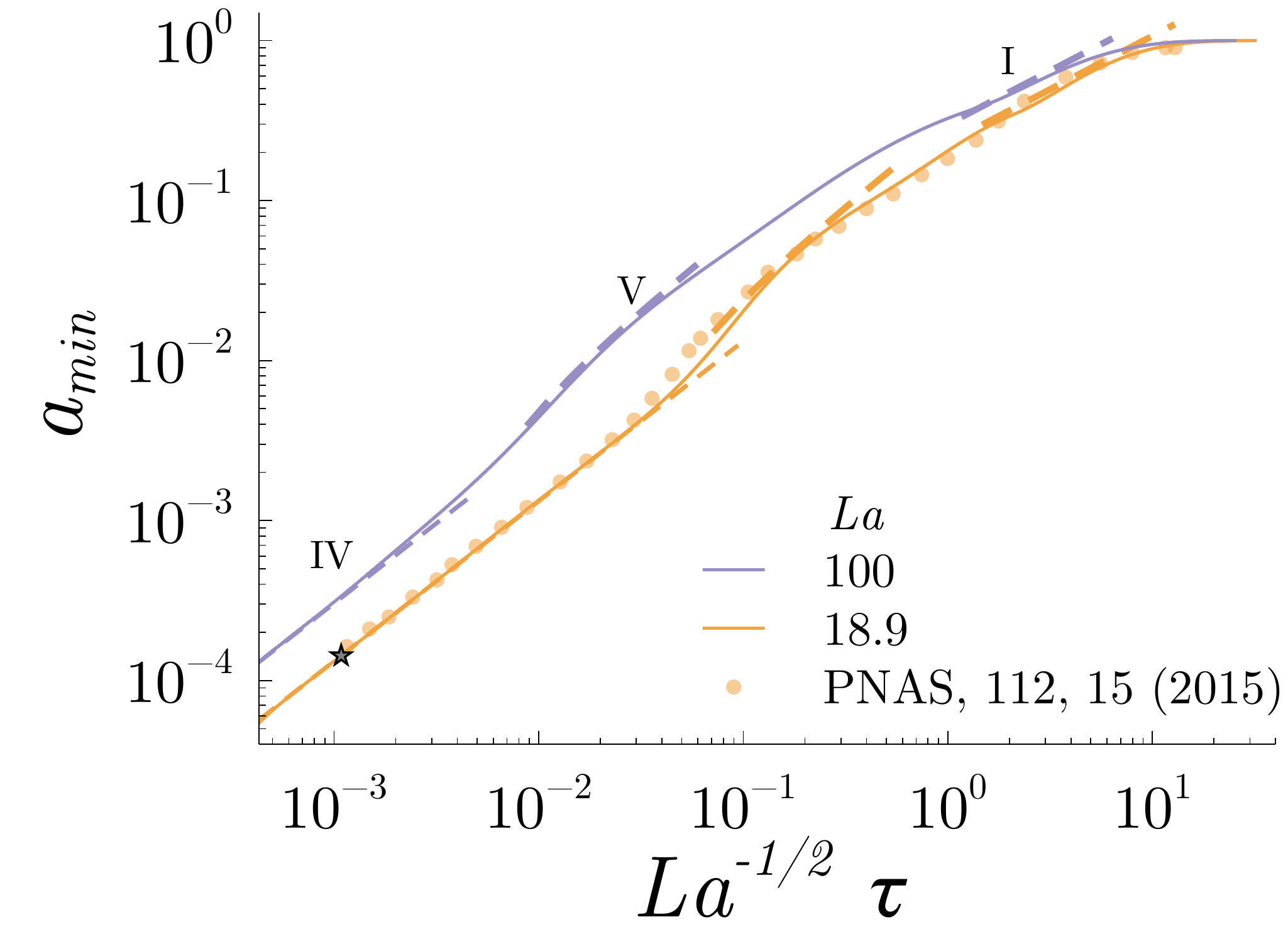}
    };
 \node at (5,0) {
        \includegraphics[width=70mm]{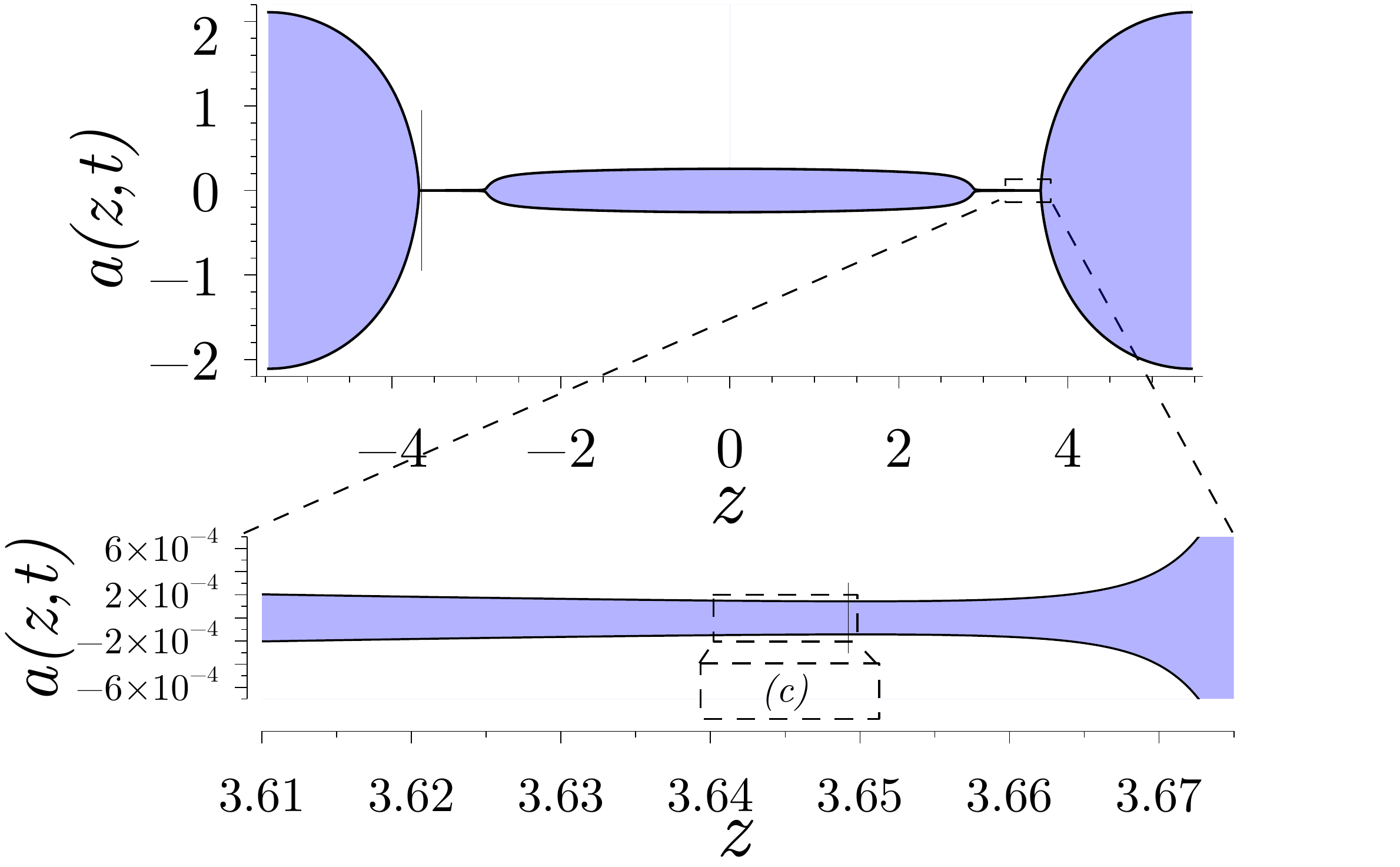}
    };
     \node at (1.3,-3.5) {
        \includegraphics[width=120mm]{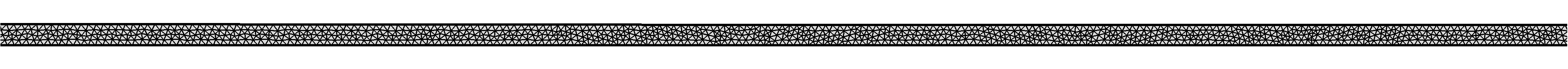}
    };
\draw[->,line width=0.25mm] (-4.71,-4) -- coordinate (x axis mid) (7.75,-4);
\draw[line width=0.25mm] (-4.71,-4) -- (-4.71,-4.05)
    	node[anchor=north] {\large 3.64};
    	\draw[line width=0.25mm] (7.295,-4) -- (7.295,-4.05)
    	node[anchor=north] {\large 3.65};
    	\node  at (1.6,-4.3) {\Large $z$}; 
    	\node  at (-5,2.4) {\large ($a$)}; 
    	\node  at (2.1,2.4) {\large ($b$)}; 
    	\node  at (-5,-2.7) {\large ($c$)}; 
    	\draw[line width=0.25mm] (-4.71,-3.4) -- (-4,-3);
        \node  at (-3.5,-2.7) {\large $1.5 \times 10^{-4}$}; 

\end{tikzpicture}
\caption{\label{fig:figapp} (Colour online) ($a$) Minimum thread radius $a_{\min}$ as a function of the time to break-up $\tau$ for two different values of the Laplace number, namely $\Lap = 18.9$ and $100$, and $\Ela = 0$. The dashed lines indicate the scaling laws in the different regimes, and the symbols correspond to the results extracted from the numerical simulations of~\cite{Castrejon2015}. ($b$) Shape of the thread for the case $\Lap = 18.9$ at $t = 138.017$, where $a_{\min} = 1.43 \times 10^{-4}$ and $z_{\min} = 3.65$, and which corresponds to the star symbol in ($a$) for $\tau = 1.08 \times 10^{-3}$. The zoomed region shows the micro-filament formed just prior to pinch-off. ($c$) Local mesh in the micro-filament region.}
\end{figure}

\section{Validation of the numerical method} \label{app:appendix}

To demonstrate the performance of our numerical technique, in this appendix we report numerical simulations aimed at comparing our results with the well-known scaling laws of $a_{\min}$ as a function of the time to break-up, $\tau = t_{b} - t$, for two different values of the Laplace number, namely $\Lap = 18.9$ and $100$, in the case of a clean interface, $\Ela = 0$. Figure~\ref{fig:figapp} shows $a_{\min}$ as a function of $\tau$, where the dashed lines represent the different scaling laws, and symbols have been extracted from the results of~\cite{Castrejon2015} for the particular case of $\Lap = 18.9$. Since $\Lap$ is moderately high in both cases, intertial and capillary forces balance initially, providing $a_{\min} \sim \tau^{2/3}$~\citep{Keller1983,Day1998,EggersFontelos2015}, a regime usually referred to as the inertial (I) regime. However, as the thread thins, viscous forces come into play, as shown numerically and experimentally by~\cite{Castrejon2015}, leading to the linear behaviour $a_{\min} = 0.0709 \tau$~\citep{Papageorgiou1995}, which is known as the viscous (V) regime. Finally, when $a_{\min}$ is sufficiently small, inertial, capillary and viscous forces balance, leading to what is usually known as the inertial-viscous (IV) regime, in which $a_{\min} = 0.0304 \tau$~\citep{Eggers1993}. As revealed by figure~\ref{fig:figapp}($a$), our numerical method is in excellent agreement with these scaling laws close to pinch-off, and with the numerical computations of~\cite{Castrejon2015}, thereby validating our numerical framework. Finally, although not shown here for conciseness, we have checked that the unphysical singularity of the equation of state~\eqref{eq:sigma_gamma} as $\Gamma\to 0$ leads to a spurious deviation from the asymptotic IV regime, which precludes its use in correctly predicting the smallest scales prior to pinch-off for $\beta\neq 0$. To that end, a different equation of state that provides the clean-interface constant value of $\sigma$ as $\Gamma\to 0$ must be used~\citep{Mcgough2006,Kamat2018}.



\end{appendix}

\bibliographystyle{jfm}
\bibliography{free_jet}

\end{document}